\documentclass[manuscript]{aastex}

\newcommand{\HI}{\ion{H}{1}\,\,}
\newcommand{\NHI}{$N_{\mathrm{H\,I\,}}$}
\newcommand{\cmm}{cm$^{-2}\:$}
\newcommand{\cmmm}{cm$^{-3}\:$}
\newcommand{\kms}{km~s$^{-1}\:$}
\newcommand{\Msun}{$M_{\odot}$}
\newcommand{\be}{\begin{equation}}
\newcommand{\ee}{\end{equation}}
\newcommand{\lbr}{\left<}
\newcommand{\rbr}{\right>}

\newcommand{\VLSR}{V_{\mathrm{LSR}}}
\newcommand{\fcal}{$f_{\mathrm{cal}}\:$}

\usepackage{graphicx}
\usepackage[caption=false]{subfig}
\usepackage{fullpage}
\usepackage{amssymb}
\usepackage{amsmath}
\usepackage{hyperref}

\begin{document}

\title{High-Resolution Images of Diffuse Neutral Clouds in the Milky Way. I. Observations, Imaging, and Basic Cloud Properties.}


\author{Y.~Pidopryhora\altaffilmark{1}}
\affil{School of Physical Sciences, University of Tasmania, Private Bag 37, Hobart, 7001, Tasmania, Australia}
\email{Yurii.Pidopryhora@utas.edu.au}

\author{Felix J. Lockman\altaffilmark{2}}
\affil{National Radio Astronomy Observatory, Green Bank, WV 24944}
\email{jlockman@nrao.edu}

\author{J.M.~Dickey}
\affil{School of Physical Sciences, University of Tasmania, Private Bag 37, Hobart, 7001, Tasmania, Australia}
\email{John.Dickey@utas.edu.au}

\author{M.P.~Rupen\altaffilmark{2,3}}
\affil{National Radio Astronomy Observatory, Socorro, NM 87801}
\email{Michael.Rupen@nrc-cnrc.gc.ca}

\altaffiltext{1}{Current Address: Argelander-Institut f{\"u}r Astronomie, Auf dem H{\"u}gel 71, D-53121, Bonn, Germany}
\altaffiltext{2}{The National Radio Astronomy Observatory is a facility of the National Science Foundation operated under a cooperative agreement by Associated Universities, Inc.}
\altaffiltext{3}{Current Address: Dominion Radio Astronomy Observatory, National Research Council,
PO Box 248, Penticton, BC, V2A 6J9 Canada}

\begin{abstract}
A set of diffuse interstellar
clouds in the inner Galaxy  within a few hundred pc of the Galactic plane
has been observed at an angular resolution of $\approx 1\farcm0$ combining data from the
NRAO Green Bank Telescope and the
 Very Large Array.  At the distance of the clouds the linear resolution ranges
from $\sim 1.9$~pc  to $\sim 2.8$~pc.  These clouds have been selected to be
 somewhat out of the Galactic plane and are thus not confused with unrelated emission, but in
other respects they are a Galactic population. They are  located near the tangent points in
the inner Galaxy, and thus at a quantifiable distance:  $2.3 \leq R \leq 6.0$~kpc
from the Galactic Center, and  $-1000 \leq z \leq +610$~pc from the Galactic plane.
These are the first images of the  diffuse neutral \HI clouds
that may constitute a considerable fraction of the ISM.
Peak \HI column densities range from \NHI $=  0.8-2.9 \times 10^{20}$~\cmm.
Cloud diameters vary between about 10 and 100 pc, and their \HI mass
spans the range from less than a hundred to a few thousands~\Msun.
The clouds show no morphological consistency of any kind except that their shapes
are highly irregular. One cloud may lie within the hot wind from the nucleus of the Galaxy, and
 some clouds show evidence of two distinct thermal phases as would be  expected from equilibrium models of the
interstellar medium.
\end{abstract}



\section{Introduction}

The concept of a diffuse interstellar cloud is more than 50 years old, yet there are few observations that
support  the most basic aspects of the standard picture.
The strongest evidence for discrete clouds is kinematic:  there are usually distinct
absorption lines at different velocities in spectra toward
stars (e.g., \citet{Munch1952, Hobbs1978,Redfield2008}).  But spectral features can be produced not
only by spatial structures, i.e., clouds, but in a continuous turbulent medium as well \citep{LazarianPogosyan2000}.
In contrast to the often well-defined clouds of molecular gas like the Infrared-dark Clouds
\citep[e.g.,][]{Rathborne2010},
there is little support for the existence of discrete clouds in 21cm emission
observations, which suggest instead that the atomic interstellar medium (ISM) consists of fragments of filaments
and ``blobby sheets'', many of which may be a consequence of turbulence
\citep{KulkarniHeiles1987,DL1990,Miville-Deschenes2003, HeilesTroland2003,KalberlaRev2009}.
  In most direction 21cm \HI
emission maps are  highly confused, leading to considerable ambiguity in determining the
morphology and boundries of interstellar clouds, even assuming that they do exist.
A mammoth study of \ion{Na}{1} and \ion{Ca}{2}
absorption lines toward nearly 2000 stars within 800 pc of the Sun
by \citet{Lallement2003} and \citet{Welsh2010}
has revealed the three-dimensional structure of the local interstellar medium,
``cell-like cavity structures'' a fragmented ``wall'' of neutral gas,
and what appears to be  clouds ``physically linked to the
wall of denser gas'', but it is difficult to know how to generalize this result to the broader ISM.

The situation is quite different, however, in the lower halo of the inner Galaxy,
where there is a population of discreet \HI clouds whose velocities are consistent with circular rotation, but whose location several hundred pc from the place separates them from unrelated emission \citep{Lockman2002, Lockman2004}\footnote{First detections of a few prominent representatives of this population date back to \citet{Prata64}, \citet{Simonson71}, and \citet{Lockman1984}. A very detailed history and bibliography of both observational and theoretical early studies of interstellar clouds and \HI halo can be found in Chapter 1 of \citet{PhD}.}. Similar clouds can be seen at low Galactic latitude when their random velocity is large enough to remove confusion \citep{Stil2006};  others are detected in the outer Galaxy \citep{Strasser2007, Stanimirovic2006, DedesKalberla2010}. The clouds in the inner Galaxy are likely the product of \HI supershells as their abundance and scale height are linked to the large-scale pattern of
star formation in the disk \citep{Ford2008, Ford2010}.

While many aspects of these ``disk-halo'' clouds are poorly understood, they can be used as test particles sensitive
 to the physical conditions in their surroundings, and thus give information about
interstellar processes not easily gotten from the highly blended spectra typical of most observations. The disk-halo
 clouds in the inner Galaxy are so abundant that a number of them lie near the terminal velocity in their direction,
and thus near the tangent point, whose distance is determined from simple geometry. Their location, mass and
size can be estimated with quantifiable errors.

There have been numerous theoretical studies of the expected properties of the diffuse ISM as a
function of location in the Galaxy, distance from the Galactic plane, sources of heating, etc.
Strong theoretical predictions have been made, especially about the existence of two thermal
phases in pressure equilibrium  \citep{FGH1969, Wolfire1995a,Wolfire1995b, KoyamaOstriker2009}.
The disk-halo clouds offer the perfect laboratories for testing these predictions.

We selected a set of disk-halo clouds using observations with the Robert C.~Byrd Green Bank Telescope
(GBT) and measured them with the Very Large Array (VLA) in three different
array configurations. The clouds were selected to cover a range of longitude and latitude and to be located near the tangent points of the inner Galaxy, and thus at a known distance.

This paper is the first in a series about these clouds.  Here we discuss the observations and data reduction for the
GBT and VLA D-array data only, taken at an angular resolution of 0\farcm9-1\farcm5  providing a linear resolution of 1.9 to 2.8 pc. We concentrate on understanding all sources of uncertainty.

\section{Selection of Targets}
\label{sec:targets}

\begin{figure}
\hspace{-1.0in}
\includegraphics[width=9.0in,height=9.0in]{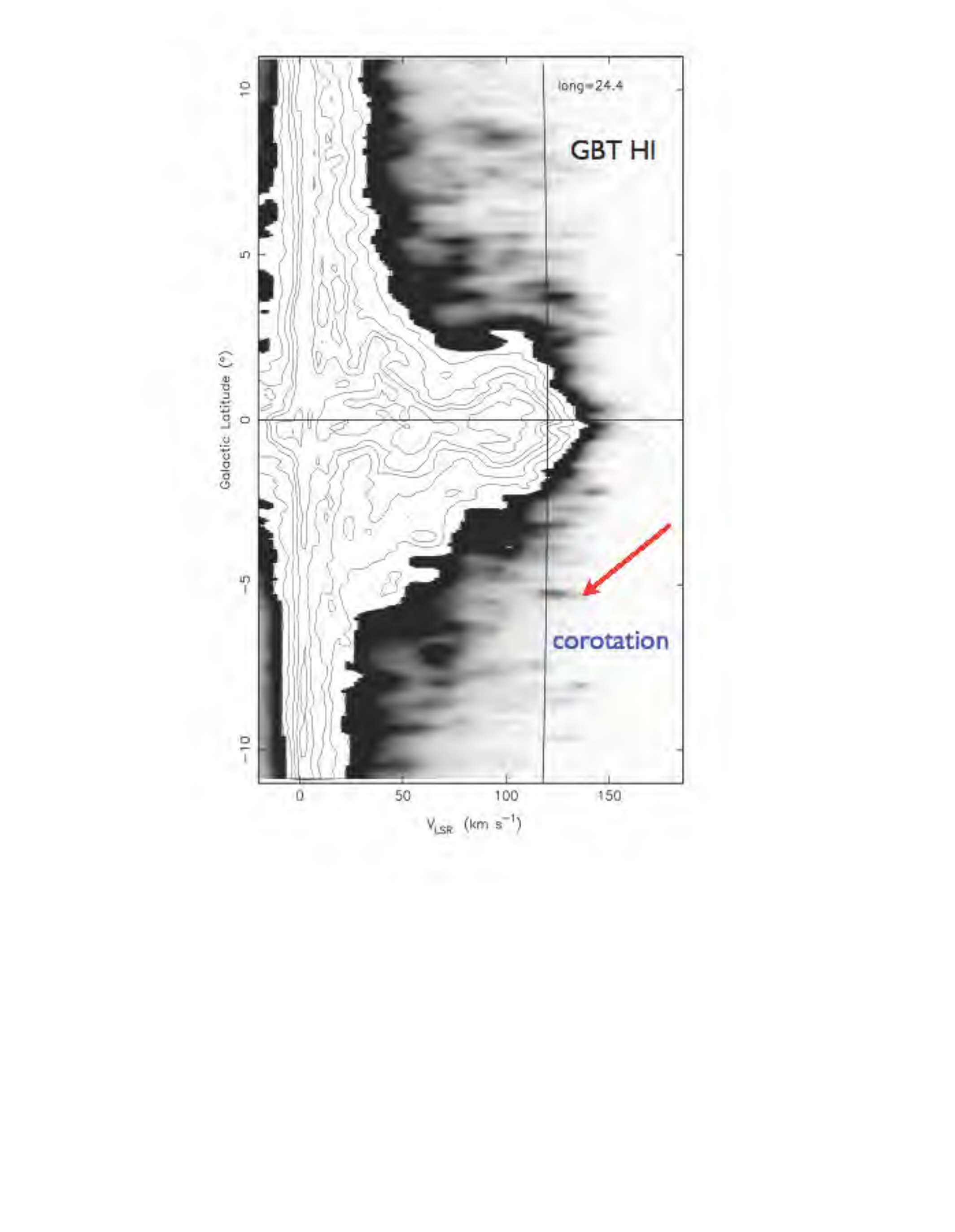}
\vspace{-3.0in}
\caption{Velocity-latitude image of \HI emission from GBT observations after \citet{LP}.  The cloud at $24.3-5.3$ is marked with
an arrow.  The expected maximum $\VLSR$ from Galactic rotation \citep{Clemens1985}  is marked by the slightly curved vertical line.
The location of G$24.3-5.3$ slightly beyond the maximum velocity indicates that it must lie near the
tangent point in its direction.  Note that it is separated in position and velocity from other \HI emission and is thus
relatively unconfused. }
\label{fig:GBTvb}
\end{figure}

Targeted 21cm \HI surveys of regions in the inner Galaxy made with the GBT
provided a list of diffuse clouds that might be suitable
 for high-resolution imaging \citep{Lockman2002,Lockman2004}.  From
these we selected a set using the
following criteria: 1) The clouds have LSR velocities at or beyond the terminal velocity in their direction
ensuring that their distance could be determined (see \S\ref{subsec:distance} ); 2) the clouds cover a
 range of Galactic longitude and latitude ensuring that different environments were probed, though all are in the
first quadrant of Galactic longitude;
3) the clouds are relatively isolated in position and velocity to minimize potential confusion;
4) their 21cm emission as observed with the GBT is bright enough to be detectable with the VLA in a few hours.
An example of the GBT observations used to select the clouds is given in Figure~\ref{fig:GBTvb}.
Table~\ref{tab:table0} gives the cloud designation, field centers, and the 21cm line peak brightness
 temperature, FWHM,  and velocity as determined from the GBT observations.

\subsection{Distance}
\label{subsec:distance}
The kinematics of the disk-halo cloud population studied here is dominated by the circular
rotation of the Milky Way with a cloud-cloud velocity dispersion $\sigma_{cc} \approx 16$~\kms
\citep{Lockman2002, Ford2008, Ford2010}.
Toward the inner Galaxy, the maximum velocity permitted by
Galactic rotation at $b = 0\arcdeg$  is called the terminal velocity, $V_t$, and
arises from the tangent point where the distance from the Galactic
center is $R_t = R_0\ \sin{\ell}$. Here we use $R_0 = 8.5$~kpc, the IAU recommended value
\citep{Kerr_R0}. The terminal velocity
can be measured from observations of species such as \HI or CO
or can be approximated with a rotation curve (e.g., \citet{BurtonLiszt1993, Clemens1985,
McClure-GriffithsDickey2007,Dickey2013}) .
In the first quadrant of Galactic longitude $V_t > 0$ and
 an object with V$_{LSR} \gtrsim V_t$ must thus lie near the tangent point where its
distance is known from  geometry.  In the current sample all clouds have velocities  $\VLSR \gtrsim V_t$
so we calculate the tangent-point distance projected on the Galactic plane $d_p  = R_0 \ \cos{\ell}$.
A cloud's distance from the Galactic plane is then $z = d_p \tan{b}$ and the distance to the cloud center
 $d = (d_p^2 + z^2)^{\onehalf}$.

\subsubsection{Distance Uncertainties}

An estimate of the  uncertainties in a cloud's distance can be derived from the
change in distance that would correspond to a change in the cloud's $\VLSR$  of
$\sigma_{cc}$  in that particular direction for
an assumed rotation curve.   As almost all of the clouds have $\VLSR > V_t$, in some cases
by as much as $2\sigma_{cc}$, we conservatively estimate errors by the change in distance were
the cloud to have a velocity ($V_t - \sigma_{cc}$) or ($\VLSR - \sigma_{cc}$), whichever produces
the larger change in distance.
Errors calculated this way using the rotation curves of \citet{BurtonLiszt1993}  and \citet{Dickey2013}
 agree to within a few percent, and are given in Table~\ref{tab:table0}.  We
note that the greater a cloud's velocity beyond $V_t$, the more likely that it lies near the
tangent point \citep{Ford2010}.  Thus for many of our clouds, especially G$44.8-7.0$
which lies nearly 30 \kms\ past $V_t$,  our error estimates are probably overstated.

The final two columns of Table~\ref{tab:table0} give
 derived distances and errors as well as the distance of each
cloud from the Galactic plane.

\section{Observations and Basic Data Reduction}

\label{sec:obs}

\subsection{Green Bank Telescope Observations}
\label{sec:mapobs}

The GBT was used to map \HI emission around the disk-halo clouds to
measure their overall properties, to determine whether they might be suitable for high-resolution imaging,
 and to provide the short spacing data for image reconstruction
(see e.~g., \cite{Stanimirovic}).
Maps were made over an area around each cloud of
 $1\fdg5 \times 1\fdg5$ or $2\degr \times 2\degr$ depending on the extent of the cloud.
Spectra were taken every $3\farcm5$ in Galactic longitude and latitude, somewhat finer than the Nyquist sampling interval for the GBT's $9\farcm1$ beam (FWHM). Observations were repeated several times over a period of a few months.
In-band frequency switching gave a useable velocity coverage of 400 km s$^{-1}$ around
zero velocity (LSR) at a channel spacing of 0.16 \kms.

\subsection{Very Large Array Observations}

\begin{figure}
\includegraphics[width=\textwidth]{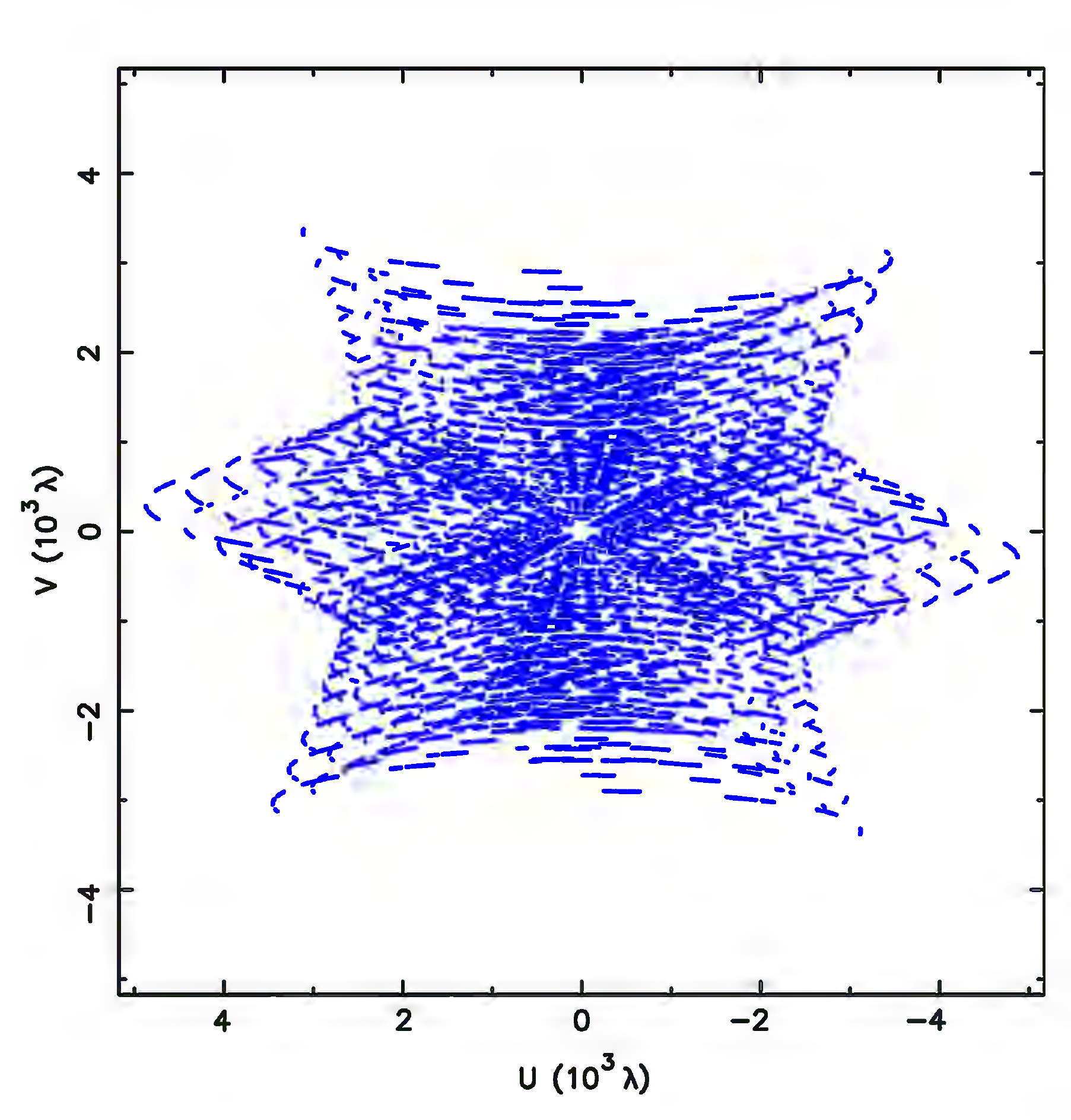}
\caption{$uv$-coverage of the VLA observations of one of the clouds, G$26.9-6.3$. Two epochs of June 29 and July 24 2004 combined, total integration time 215 min., source declination $\delta \approx -8\arcdeg$.}
\label{fig:UVdiagram}
\end{figure}

A sample of 20 \HI disk-halo clouds was observed in 21~cm line emission spectroscopy
with the Very Large Array (VLA) in D configuration during 2003 and 2004.
The spectra had 256 frequency channels separated by $ 0.64$~\kms in equivalent
Doppler velocity centered on the peak velocity of each cloud
\citep{AAS04, LP, PhD}.
Fifteen of the clouds were observed in single pointings, and 5 as mosaics.
Two of these
were also observed in C configuration with the identical spectroscopic setup (\cite{TwoClouds2012}) and
two more (G$21.2+2.2$ and G$35.6+3.9$) were observed in B configuration at both \HI and OH frequencies
in an attempt to detect absorption against bright background continuum sources.  Here we present
results on just the ten clouds observed only in D configuration with single pointings, deferring discussion of the
others to a later paper. Parameters of the VLA observations are given in Table~\ref{tab:VLAobserve}. An example of the $uv$-coverage for one of the clouds is given in Fig.~\ref{fig:UVdiagram}.

\subsection{Green Bank Telescope Data Reduction}

The spectra were calibrated and  corrected for stray radiation as described in \citet{Boothroyd2011},
 and a second-order polynomial was fit to emission-free regions of each spectrum to correct for
residual instrumental effects. For each cloud the data were assembled into a cube on a 105\arcsec\
 grid. There was occasional narrow-band interference that was stable in frequency,
so spectra were interpolated over the affected channels. The final GBT data cubes had a brightness temperature noise $\approx 0.1$ K in a 0.16 \kms\ channel.

Each GBT image cube was converted to the same coordinate system as the VLA,
cropped to fit the exact VLA field size and interpolated to the matching grid and
sequence of spectral channels with the Miriad~\citep{MiriadReference} task REGRID. The GBT data were taken while the telescope was
moving and thus have an effective resolution of
$\approx 9\farcm6 \times 9\farcm1$, with the major axis along Galactic longitude, the scanning direction.
In the VLA's equatorial coordinate system this resulted in slightly different beam
position angles for each field.
In order to smooth out  gridding artifacts each GBT image was also convolved with a circular beam function
of 200\arcsec,  approximately 1/3 of the original beam size, so the
final GBT angular resolution is  approximately 10\arcmin\  FWHM.

\subsection{Very Large Array Data Reduction}

After calibration and study of preliminary dirty images of each VLA field,
the continuum was subtracted in the $uv$ domain using AIPS UVLIN task based on
selected line-free channels. Naturally-weighted dirty images of continuum-free data were then cleaned channel by channel with AIPS SDCLN with no cleaning mask applied. The residual flux threshold was set to 0.7~mJy/beam for all clouds, corresponding to 0.2--0.3~$\sigma$. In the final step the correction for the VLA primary beam  was applied to the clean image cubes with the AIPS task PBCOR. The imaging synthesized beam size was different for each field as described below; their values are given in Table~\ref{tab:beams}.

\section{Further Data Processing, Noise Levels and Errors}

\label{sec:data_proc}

\subsection{Combining the Interferometric and Single-Dish Data}
\label{sec:combining}
For several reasons we have chosen to use Miriad's IMMERGE  to combine the interferometric and single-dish data:
1) it uses a well understood algorithm that is easy to control; 2) it runs quickly
and is easily applied to large image cubes; 3) we have developed a calibration technique described below that ensures
accuracy of the results.

Another approach was also tried for two clouds not of the set described in this paper \citep{TwoClouds2012}: using a maximum entropy (MEM) algorithm (AIPS VTESS task) with the GBT image used as the default, but it was found to be less efficient. For a detailed review of all possible methods see \citet{StanimirovicPhD}.

Miriad's task IMMERGE uses a linear method sometimes known as `feathering'  to combine
interferometric and single-dish data \citep{MiriadManual,StanimirovicPhD}. Essentially this is just
merging clean interferometric and single-dish images after Fourier-transforming
them into the $uv$ domain, the result covering the whole combined spatial frequency range.
For this procedure to be
meaningful the Fourier images of interferometric and single-dish data should match  within the
overlapping ranges of their common spatial frequencies. Due to the completely different natures of
the two original datasets and thus unavoidable discrepancies,
 this has to be ensured by varying their common calibration scaling factor \fcal, which is the main
control parameter of the IMMERGE task. In the case that
 the calibration of both data sets was done properly this factor
should be close to unity, but its exact value has to be determined empirically in each particular case. If a
compact source of 21cm emission were present in the field of view, both unresolved by the VLA
and unconfused with other emission by the GBT, determining \fcal
would be as simple as dividing the interferometric flux density of this source by its single-dish flux
density. Unfortunately, such sources are rare and not present in our data, so a more complex
strategy of determining \fcal was used.

There is another important control parameter called `tapering'. If IMMERGE tapering is applied,
the Fourier image of the interferometric data is smoothly continued into the low spatial-frequency region
 \citep{MiriadManual}.
This fixes possible edge effects in the Fourier transformation,
but introduces an additional non-linear distortion of the data.
As well as determining the best value of \fcal, it is necessary to decide if tapering should be applied.

\subsection{Derivation of the Optimal \fcal}

\begin{figure}
\includegraphics[width=\textwidth]{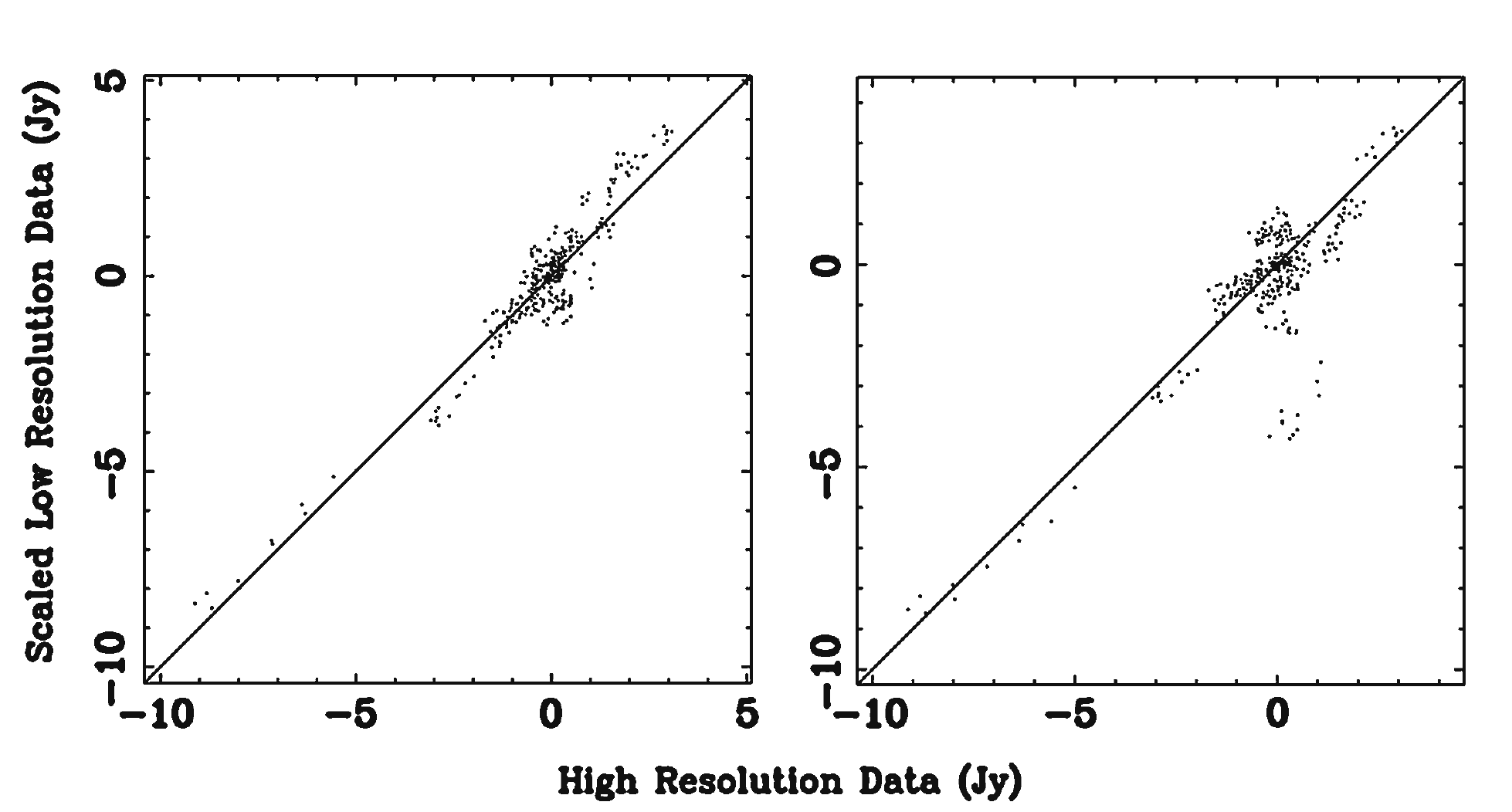}
\caption{Linear fit to pixel-by-pixel comparison of single-dish and interferometer data of cloud G$17.5+2.2$
generated by Miriad's task IMMERGE in the process of solving for the optimal \fcal value.
The left panel is without tapering, the right panel uses tapering.
Data are used from the position of the brightest line
averaged over a region $ 15\farcm3 \times 15\farcm3$.
The overlapping frequency range in the Fourier domain was set to be 20 -- 50~meters
$\approx 95$ -- $240\lambda$. The values
for  the high resolution data were calculated from the interferometric data convolved with the
single-dish beam to compensate for the difference in resolution. The values for the low resolution,
single-dish  data were scaled by \fcal.  The value of
\fcal determined from the left panel is 0.68, from the right is 0.62.
In the tapered case it is clear that the assumption of linearity fails,
while in the non-tapered case the bulk of the points close to zero favor a somewhat different slope.}
\label{fig:IMMERGEfcal}
\end{figure}

IMMERGE has a built-in method of matching the two datasets and deriving
 \fcal, provided that the overlapping range of spatial frequencies is defined.
Figure~\ref{fig:IMMERGEfcal} shows examples of application of this method.
The interferometric image is convolved with the single-dish beam, then both images
are Fourier-transformed and compared to each other within the overlapping spatial
frequency range. The value of \fcal is selected by scaling the single-dish data until the
slope of the line fit to the data is exactly 1. Based on several trials varying IMMERGE
parameters with different
samples taken from our data, we have determined that:
1) tapering distorts the data and makes a reasonable linear fit impossible; 2) due to a
large scatter of the values the best linear fit usually does not characterize the data well and \fcal derived
from it should be treated only as an estimate. Thus for our purposes we  use
IMMERGE without tapering and we determine
the optimal \fcal by other methods.

It should be noted that describing the discrepancy between single-dish and interferometric
data with a single linear parameter is only an approximation of very complex behavior.
One immediately notices that the optimal \fcal seems to be different for different
channels \citep{StanimirovicPhD}. In particular, it is sensitive to the signal-to-noise ratio and spatial
signal distribution of brightness in each channel. This is understandable as
ideally \fcal should be determined from a point source unresolved by both telescopes. For a cloud comprised of diffuse gas that smoothly fills the field of view and is detectable by the single-dish but is completely invisible to the interferometer, the estimated \fcal approaches zero. In the cloud data we usually have some combination
of these two extreme cases, and what seems to be the optimal \fcal may fluctuate significantly. But
allowing it to vary with frequency would
introduce an unknown non-linear distortion.
Based on the general assumptions of this model, and as all clouds observed only with the
VLA D-array present the same variety of interferometric data (similar beam sizes, noise levels,
$uv$-coverage etc.), we sought a single value of \fcal that would work well not only for
every channel of a particular cloud, but for all ten clouds.

One criterion for testing the goodness of a particular value of \fcal is that
the spectrum of the merged cube averaged  over an area somewhat larger than the GBT beam, but small
enough not to be effected by the VLA primary beam pattern,
should match the average taken over the same region in the single-dish data alone.
We have found that averaging over
 an area $15\farcm3 \times 15\farcm3$ at the center of the field works well for all clouds
observed with the VLA  D-array.   Requiring that the average spectrum
 over this area given by IMMERGE
matches the single-dish mean profile is sufficient to derive an
optimal \fcal\ = 0.87. In fact, this method may be preferable
to any other since it directly preserves the single-dish flux.

Figure~\ref{fig:IMM} shows the comparison
of mean line profiles of the VLA, the GBT and the best IMMERGE combined
images for all 10 targets using the single value of \fcal\ = 0.87,
which was adopted for all subsequent work.

\subsection{Testing the Accuracy of the Image Produced by IMMERGE}

\begin{figure}
\includegraphics[width=\textwidth]{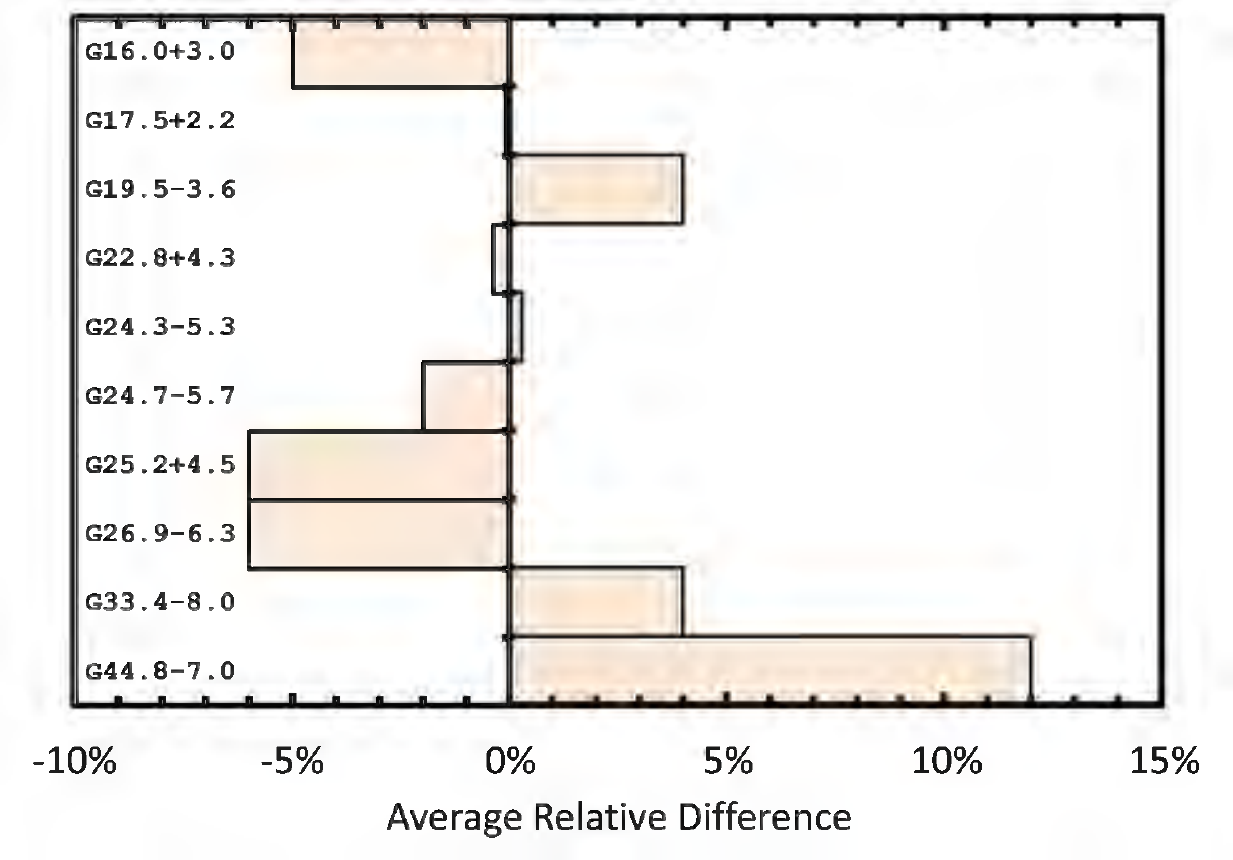}
\caption{Quantitative measure of the difference between GBT and IMMERGE spectra defined by eq.~\ref{fla:average_profiles}.
The small differences indicate that the data reduction procedure has restored the
missing flux in the line.
}
\label{fig:values_of_D}
\end{figure}

The average difference between the observed GBT and final IMMERGE line profiles can be calculated
\be
\lbr D \rbr =  \frac{\sum\limits_i \left( P_{\mathrm{GBT}}(i) - P_{\mathrm{IMMERGE}}(i) \right) }{\sum\limits_i P_{\mathrm{GBT}}(i)}
\label{fla:average_profiles}
\ee
where $P_{\mathrm{GBT}}(i)$ and $P_{\mathrm{IMMERGE}}(i)$ are the spectral values for individual
channels and the summing is done over velocities of interest for each cloud.
Figure~\ref{fig:values_of_D} shows values of $\lbr D \rbr$ for all 10 clouds.
Taking into account the peculiarities of profiles that make their exact match impossible, e.~g., the
presence of large amounts of Galactic diffuse gas invisible to the interferometer as in
clouds G$33.4-8.0$ and G$44.8-7.0$, these
values set an upper limit to the  error possibly introduced by IMMERGE in the process of
merging the VLA and GBT data. For most of the clouds  this error is  only a few per cent,
indicating that the resulting cubes are scaled accurately.

\subsection{Converting to Galactic Coordinates}

The final image cubes were regrided to Galactic coordinates as the last step of the data reduction.
To ensure a smooth transformation the pixel size was decreased. The final cubes are $512 \times
512$  with a pixel size of 5\farcs7 over the 49\arcmin\ diameter field.
The synthesized beam sizes and resulting gains
for the data cubes are given in Table~\ref{tab:beams}.

\subsection{The Noise Pattern}
\label{subsec:noise_pattern}

\begin{figure}
\includegraphics[width=\textwidth]{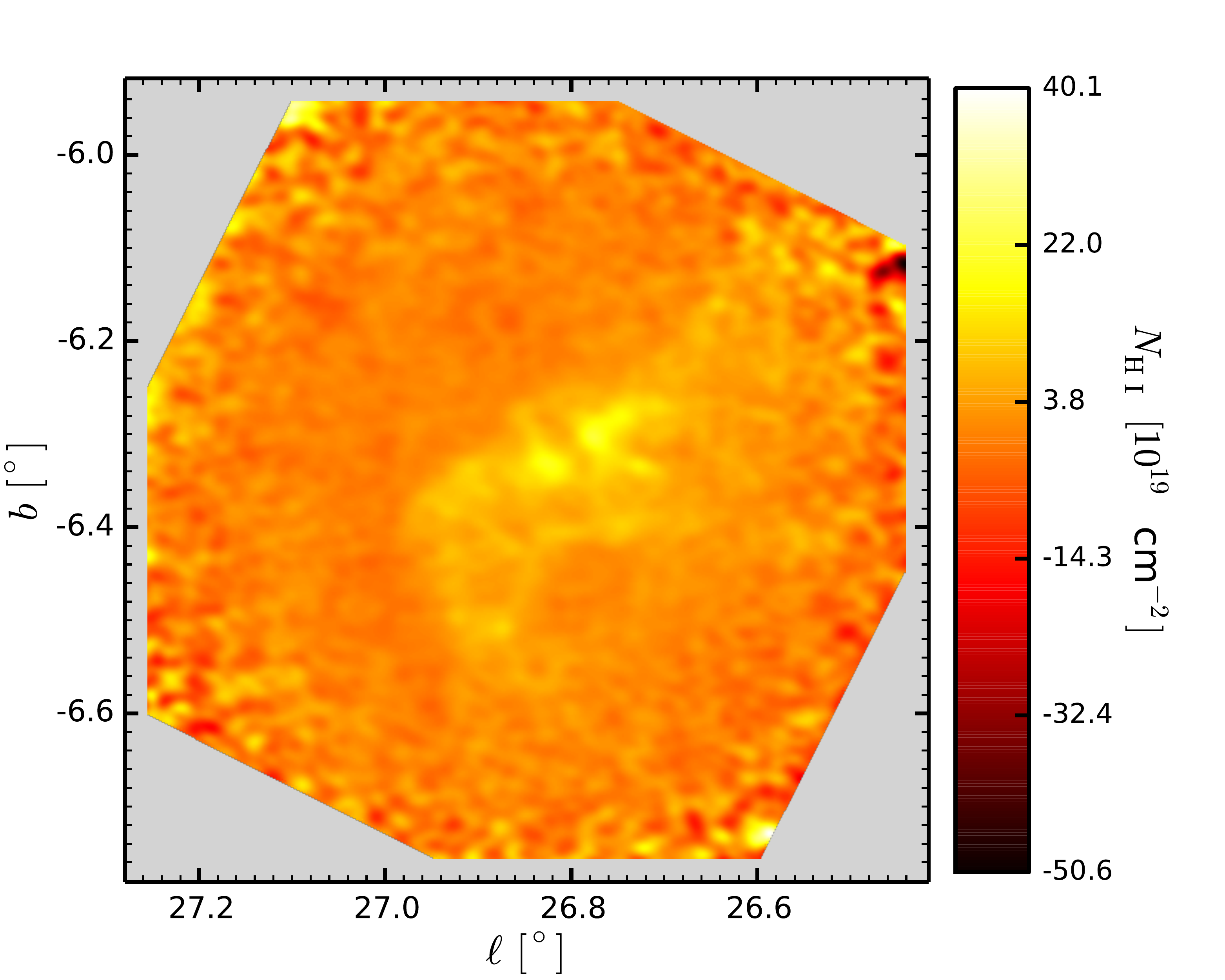}
\caption{An example of an \HI column density map  with no
noise threshold applied, showing the distinct
radial increase in noise pattern arising from  the VLA primary beam correction.
This map of cloud G$26.9-6.3$ was made from 51 spectral channels covering the
velocity range $108.47 \leq \VLSR \leq 140.68$~\kms.}
\label{fig:RAW_NHI_map}
\end{figure}

The detection threshold for the 21-cm line emission depends on the noise
and its distribution across the map areas.
Examining the noise distribution is also a useful test of the effects of the merger of the GBT and
VLA data.  We have measured the noise in the cubes at various stages of the analysis.
An essential step of the VLA data reduction is the primary beam attenuation correction,
PBCOR, which involves
scaling the data to correct for the attenuation from  the primary
beam response of the 25m dishes of the VLA.  This correction must be done before using
IMMERGE to determine \fcal as described in section \ref{sec:combining}.
The correction for the primary beam response produces images whose noise
is a strong function of radius from the pointing center.
 This is illustrated in Fig.~\ref{fig:RAW_NHI_map}, which shows the  VLA+GBT column density map for
G$26.9-6.3$.  The high noise near the edges of the map shows the effect of the
primary beam correction on the interferometer data, which persists
after merger with the GBT data. (The GBT maps have nearly uniform sensitivity everywhere.)
The noise distribution has a characteristic shape $\sigma(r)$ as a function of radius, $r$, from the field center.
The noise at  the map edge is so high that it dominates the brightness scale of the image.

\begin{figure}
\includegraphics[width=\textwidth]{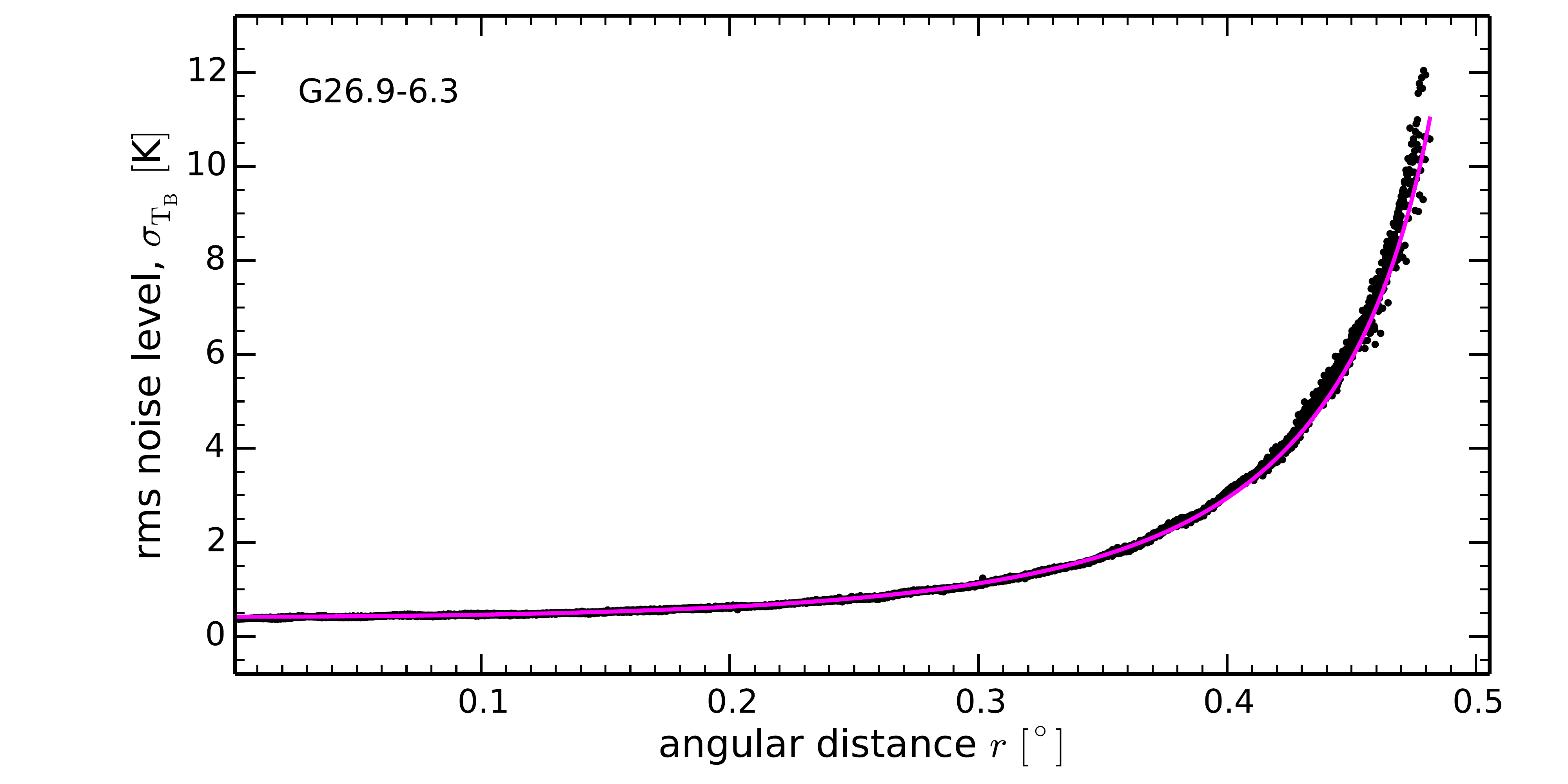}
\caption{The rms noise value $\sigma_{\mathrm{T_B}}(r)$ for the final G$26.9-6.3$
 VLA+GBT cube as a function of angular distance $r$ from the center of the field of view.
The values of scattered points in this plot were calculated from the same data cube as the map in
Fig.~\ref{fig:RAW_NHI_map}, only in a range of 41 line-free channels $150.34 \leq \VLSR \leq
176.11$~\kms.
The magenta curve fit was obtained by multiplying each point by the corresponding beam gain factor
from eq.~\ref{fla:PerleyBeam}, averaging, then multiplying the resulting mean value
$\sigma_0 = 0.42$~K by the same gain factor (eq.~\ref{fla:noise}).}
\label{fig:rawIMMERGEnoise}
\end{figure}

To set a robust noise threshold for detection of the line, and to determine the errors
in measured parameters based on the
data, we need to understand the function $\sigma(r)$.
The VLA primary beam gain factor is approximated in PBCOR by \citep{PerleyBeam}:
\begin{align}
\label{fla:PerleyBeam}
G(X) & = 1 + G_1 X^2 + G_2 X^4 + G_3 X^6, \\
X & = fr, \nonumber
\end{align}
where $f$ is the frequency of observation in GHz and $r$ is the angular radius in arcmin. The coefficients $G_i$ are
roughly constant for each VLA band; for our observations their values are:
\begin{align}
G_1 & = -1.343 \times 10^{-3}, \nonumber \\
G_2 & = +6.579 \times 10^{-7},\\
G_3 & = -1.186 \times 10^{-10}. \nonumber
\end{align}
PBCOR divides the spectrum at each point by $G(fr)$ from eq.~(\ref{fla:PerleyBeam}), thus amplifying both signal and noise,
since $0 < G(fr) < 1$.  The various clouds have slightly different values of the center frequency, $f$, depending on
their radial velocities, and different values of $\sigma_0$, which are almost entirely due to the noise in the VLA
data, as explained in the next section.  For setting detection thresholds and computing errors in the column density
and mass at different points in each cloud, we use:
\be
\label{fla:noise}
\sigma(r) = \sigma_0 \ \cdot \ [1 + G_1 (fr)^2 + G_2 (fr)^4 + G_3 (fr)^6]
\ee
Figure~\ref{fig:rawIMMERGEnoise} shows the measured values of $\sigma(r)$ using channels with no line
emission as a function of
distance from the center of the map shown in Fig.~\ref{fig:RAW_NHI_map}.  The curve is the prediction of
eq.~\ref{fla:noise}, showing the effect of the primary beam correction applied to the VLA data, as it
appears after merging with the GBT data.  It is clear that noise from the VLA data completely dominates the noise in the
final cubes, as the points in Fig.~\ref{fig:rawIMMERGEnoise} are well described by the curve.  For
each cube we measure empirically the noise level at the field center, $\sigma_0$, using off-line channels, and
construct a noise function $\sigma(r)$ similar to that shown in Fig.~\ref{fig:rawIMMERGEnoise}.  These are
given in Table~\ref{tab:sigma_0} in Kelvins and the equivalent error in \NHI\ for a 25 \kms\ wide velocity interval and the channel width of 0.64~\kms.

\subsection{Noise Amplitude}

Because of the angular resolution difference, noise from the GBT has little influence on the noise in the final cube,
but rather appears as a systematic error in flux measurement  over areas of a size comparable to the GBT beam.
Rms noise values in the GBT cubes are 0.08 -- 0.14~K, a factor 3-4 times smaller
than the  noise
at the very center of any of the final \HI maps. Thus the dominant noise in the final data
comes from the VLA, and errors due to the GBT noise can be neglected.

We have processed the data in a number of non-trivial ways so it is important to
check that the final noise level is reasonable, and is consistent with the noise in the
VLA data  at earlier reduction stages.  Examination of the noise in line-free channels in the
 dirty continuum-subtracted cubes and in the clean cubes before PBCOR
for four clouds is given in Table~\ref{tab:noise_prop}.
Comparing  $\lbr\sigma_{\mathrm{clean}}\rbr$ for each cloud from Table~\ref{tab:noise_prop} with
 $\sigma_0$ in Table~\ref{tab:sigma_0}  we conclude that
 despite a significant number of processing steps following the cleaning of the VLA image,
 the rms noise value remains virtually unchanged, except by the correction
for the main beam gain of eq.~(\ref{fla:noise}).

Two conclusions can be drawn: 1) because rms noise values are good indicators of the finest scale of the
image, the fact that they do not change much from the VLA to the final data cube shows that our procedure
of recovering the short spacings from the GBT data has not distorted the small-scale structure of the interferometric data;
2) the values of $\sigma_0$ measured in \S~\ref{subsec:noise_pattern} and the noise pattern of
eq.~(\ref{fla:noise}) are indeed valid indicators of the rms noise in the final data
and can be used with confidence.


\subsection{A Noise Threshold}
\label{subsec:noise_threshold}

\begin{figure}
\includegraphics[width=\textwidth]{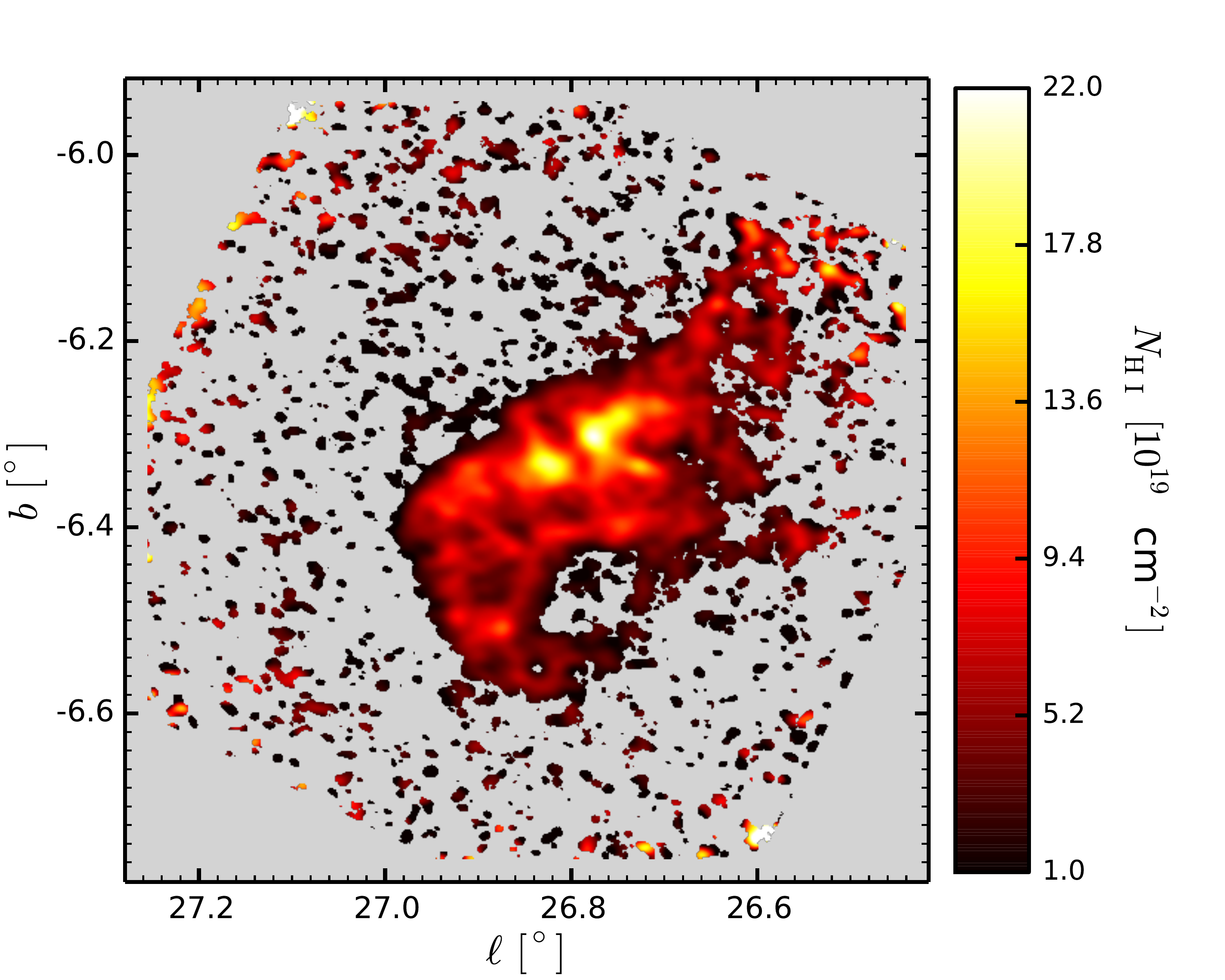}
\caption{\HI column density map of G$26.9-6.3$
displaying the same data as Fig. ~\ref{fig:RAW_NHI_map},
where spectra with values below $3\sigma(r)$ were flagged and not plotted, as explained in \S~\ref{subsec:noise_threshold}. The noise $\sigma(r)$ was calculated using eq.~(\ref{fla:noise}) with values of $\sigma_0$ and $f$ from
Table~\ref{tab:sigma_0}.
The figure shows data over the interval $1 \times 10^{19}\leq$~\NHI~$\leq2.2 \times 10^{20}$~cm$^{-2}$,
where the lower limit is three times the column density noise at the center of the field.
}
\label{fig:269filtered_NHI_map}
\end{figure}

Using eq.~(\ref{fla:noise}) we can
establish a noise threshold for every pixel in a cube.
If there is no emission $>3\sigma(r)$ over the channel range of interest the spectrum is flagged
and not used for making column density maps or other types of analysis. Figure~\ref{fig:269filtered_NHI_map} shows the same data as Figure~\ref{fig:RAW_NHI_map},
only with pixels blanked below the $3\sigma(r)$ level, leaving only emission detected significantly above the noise.

\section{Results}

\subsection{Column Density Maps, Mass Profiles, Spectra}
\label{subsec:column_dens_maps}

\begin{figure}
\centering
    \vspace{-2cm}

    \captionsetup[subfigure]{labelformat=empty}
    \subfloat[][]{
        \centering
        \hspace{-1.0cm}
        \includegraphics[width=0.35\textwidth]{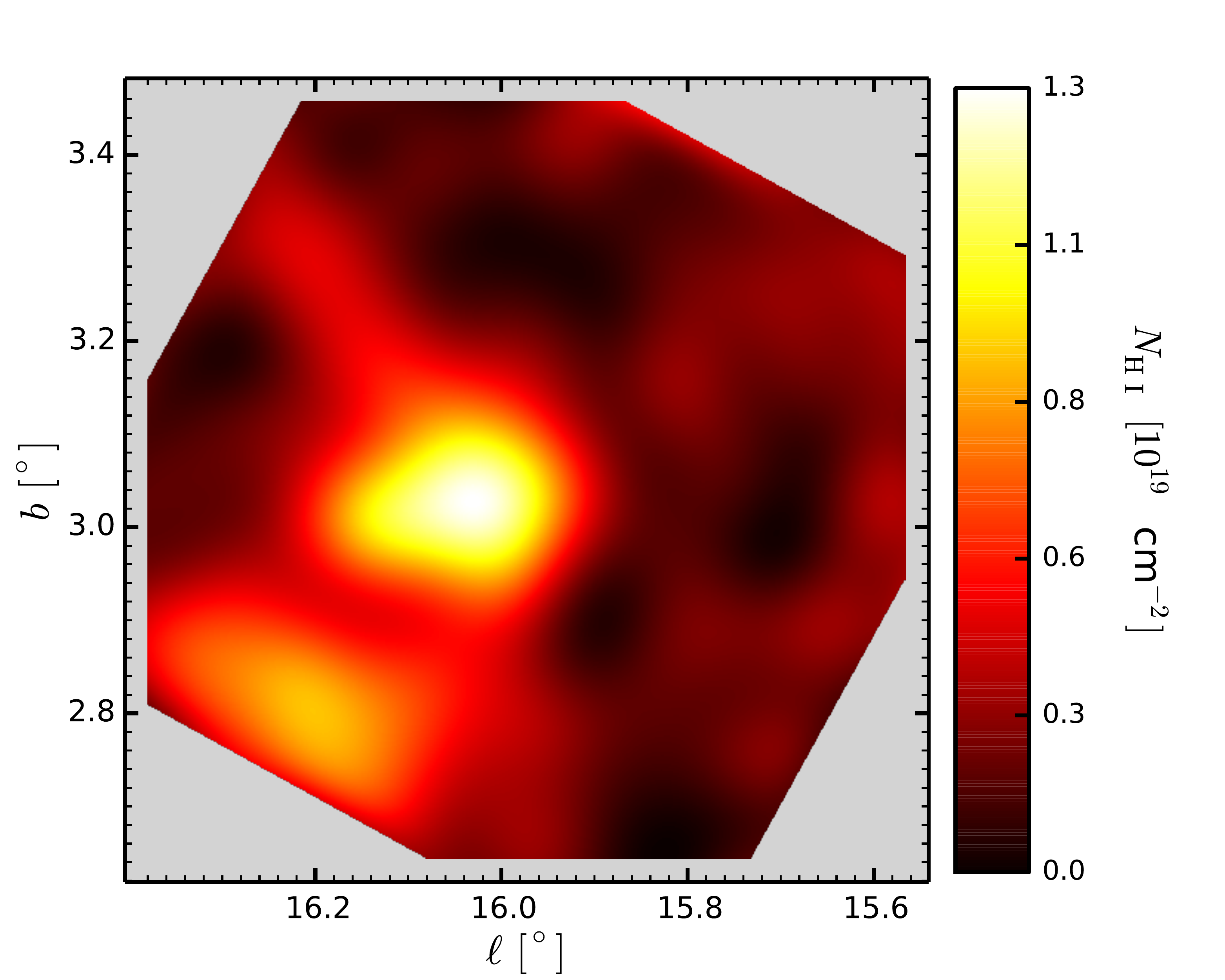}
        \hspace{-0.1cm}}
    \subfloat[][]{
        \centering
        \includegraphics[width=0.35\textwidth]{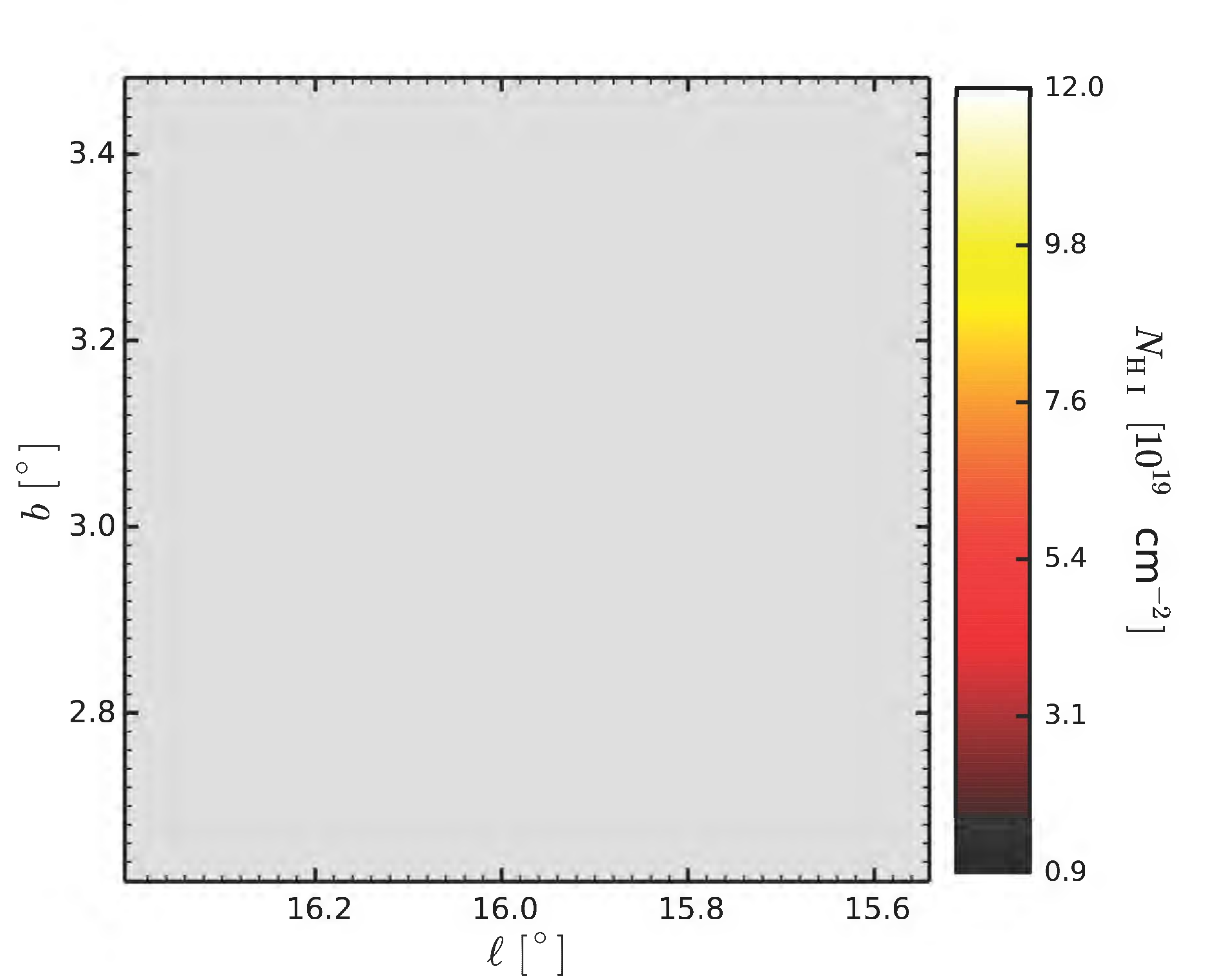}
        }

    \vspace{-1.3cm}

    \subfloat[][]{
        \centering
        \hspace{-1.0cm}
        \includegraphics[width=0.35\textwidth]{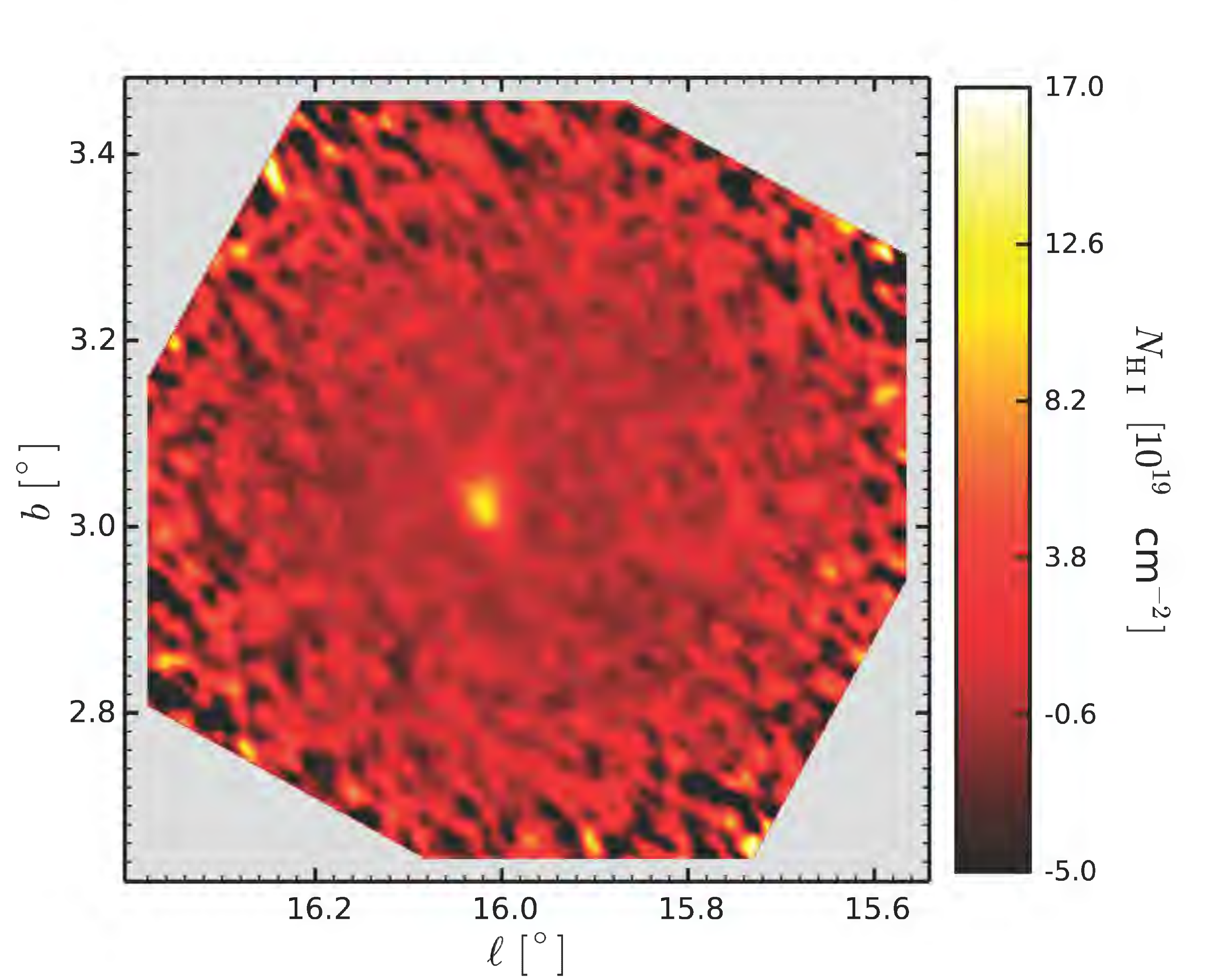}
        \hspace{-0.1cm}}
    \subfloat[][]{
        \centering
        \includegraphics[width=0.35\textwidth]{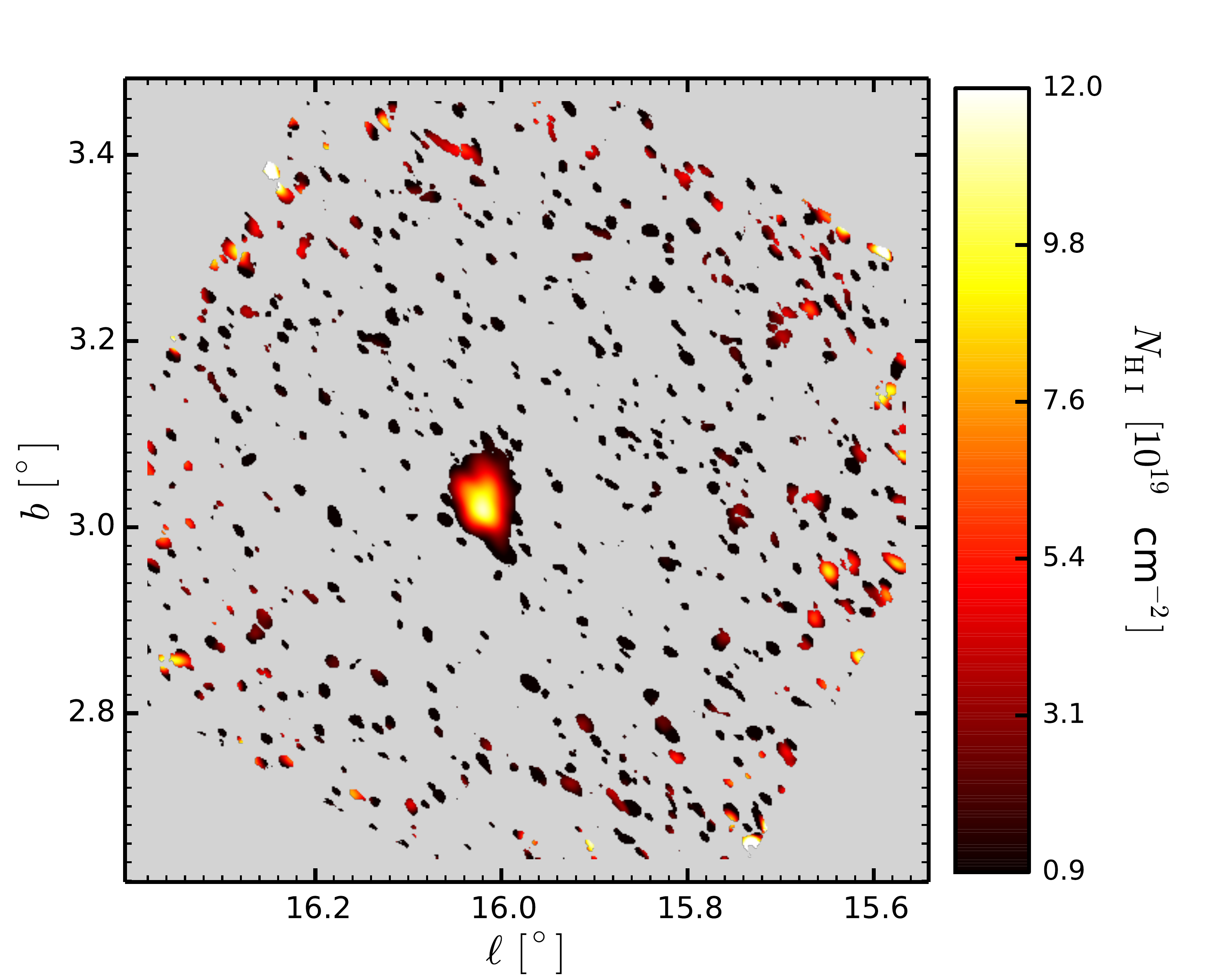}
        }

     \vspace{-1.3cm}

     \subfloat[][]{
        \centering
        \hspace{-1.0cm}
        \includegraphics[width=0.35\textwidth]{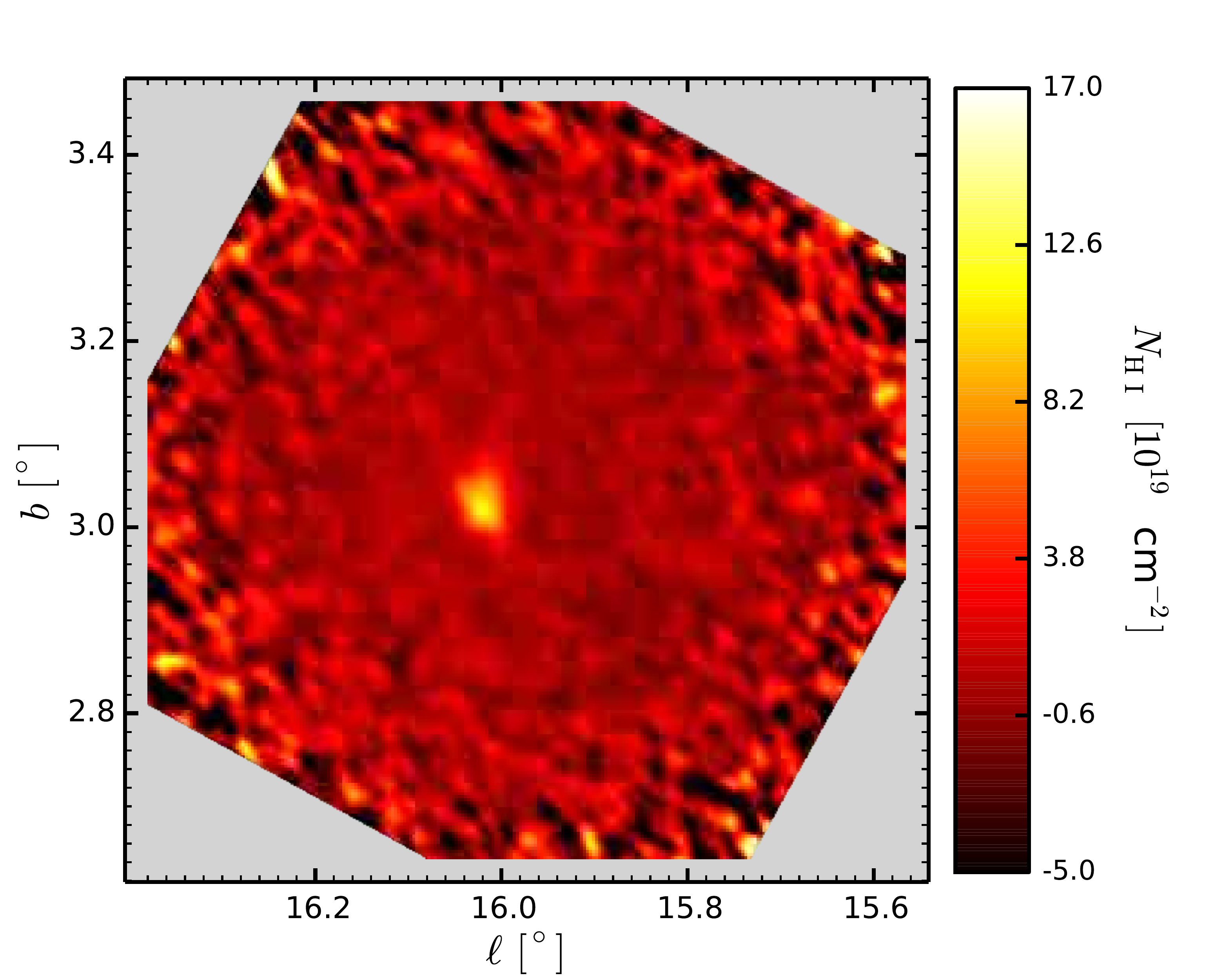}
        \hspace{-0.1cm}}
    \subfloat[][]{
        \centering
        \includegraphics[width=0.35\textwidth]{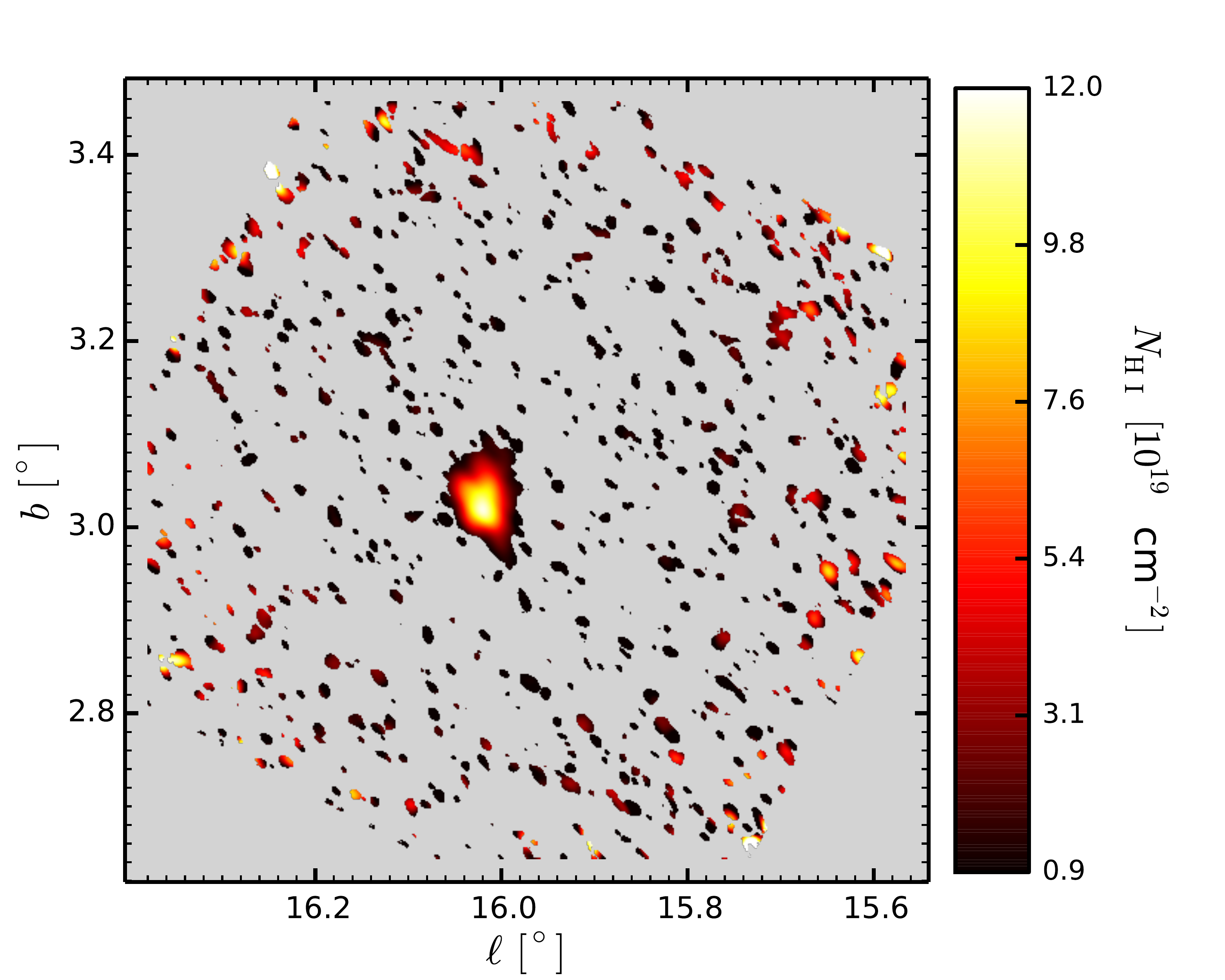}
        }
\vspace{-0.9cm}
\caption{Comparison of \HI column density maps of G$16.0-3.0$, integrated over 36 spectral channels in the interval $132.3 \leq \VLSR \leq 154.8$~\kms. 3 channels at 139.4 -- 140.6~\kms were flagged due to RFI and replaced with a linear interpolation.
Upper row: GBT data regridded to match the VLA.
Middle row: VLA data.
Bottom row: The result of IMMERGE, VLA+GBT.
Left column: full data.
The column density range of the full VLA and VLA+GBT maps (left middle and left bottom correspondingly)
of this and all similar figures
is fixed at $-5 \times 10^{19}\leq$~\NHI~$\leq1.7 \times 10^{20}$~cm$^{-2}$.
Right column: The same maps as in left column, only thresholded by 3$\sigma(r)$ of the VLA noise level, similar to Fig.~\ref{fig:269filtered_NHI_map}. All three are shown over the same interval of column densities, $9 \times 10^{18}\leq$~\NHI~$\leq1.2 \times 10^{20}$~cm$^{-2}$, where the lower limit is three times the column density noise at the center of the field.
This shows the GBT data (upper right) compared to both VLA signal and noise values.
In this particular case, alone among the ten clouds under consideration, the GBT signal is never greater than the VLA 3$\sigma(r)$ level.}
\label{fig:160_HImaps}
\end{figure}

\begin{figure}
    \centering
    \vspace{-1.0cm}

    \captionsetup[subfigure]{labelformat=empty}
    \subfloat[][]{
        \centering
        \hspace{-0.5cm}
        \includegraphics[width=0.7\textwidth]{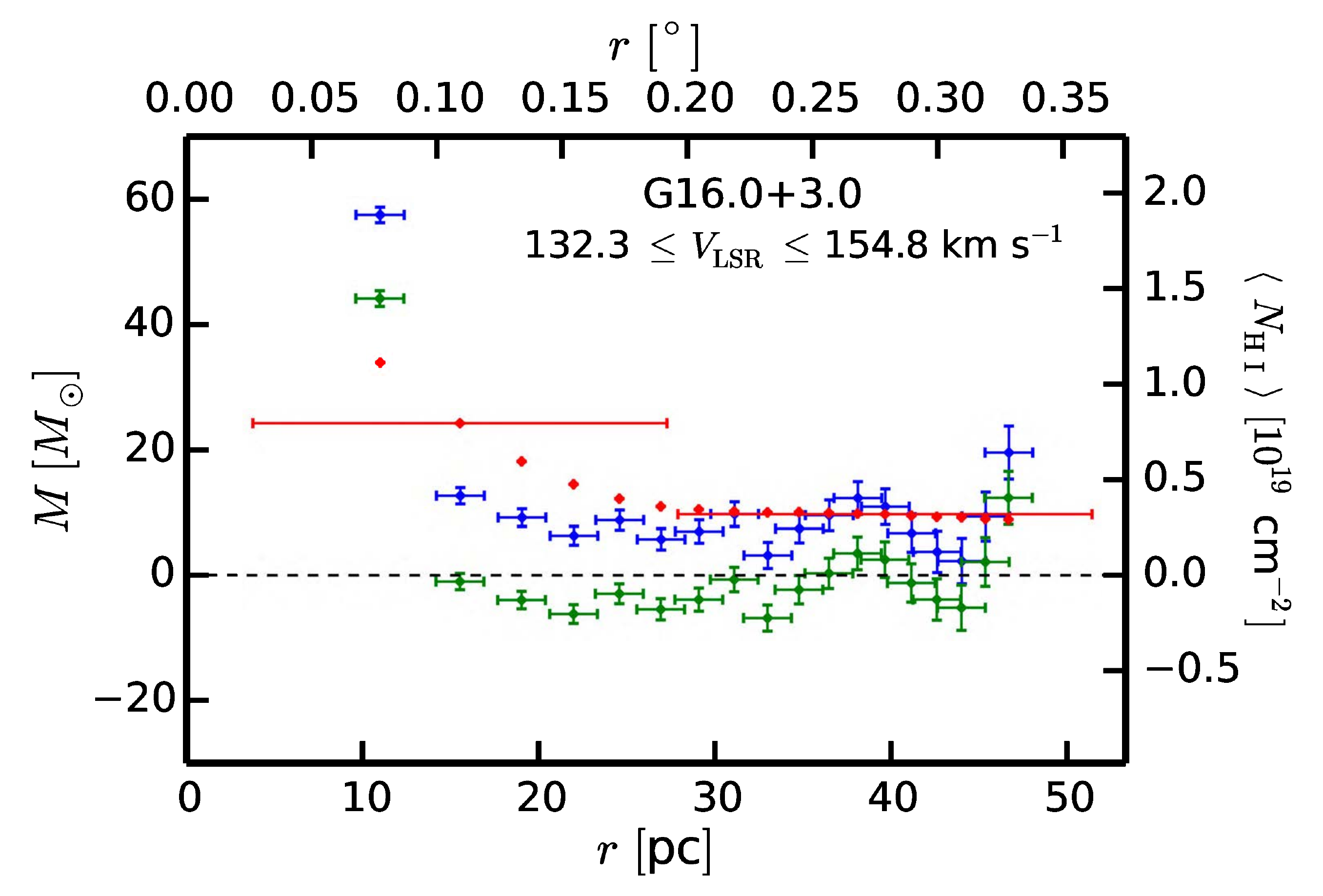}
        }

    \vspace{-1.0cm}

    \subfloat[][]{
        \centering
        \hspace{-1.8cm}
        \includegraphics[width=0.6\textwidth]{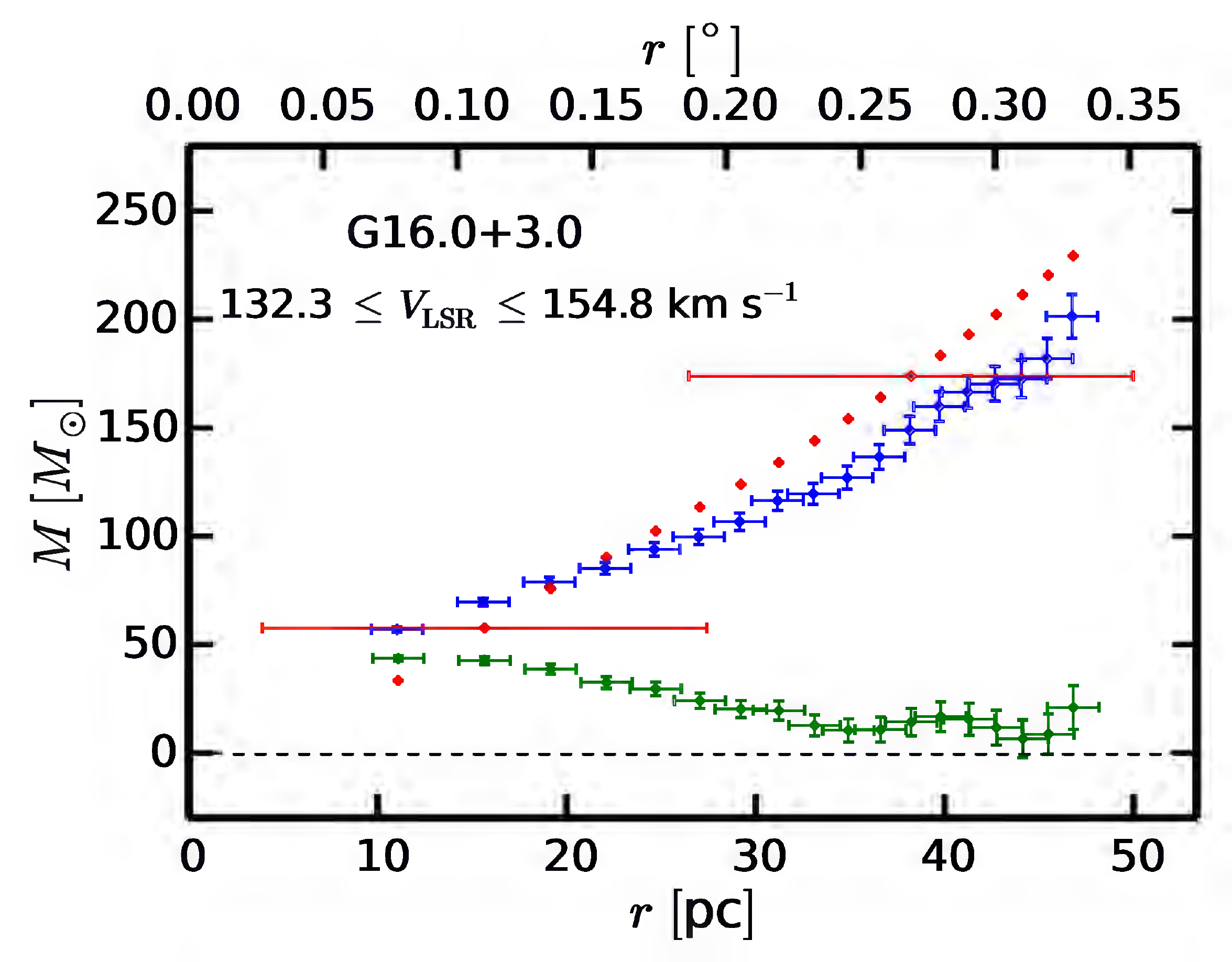}
        }

\caption{Radial mass distribution for G$16.0+3.0$. Top: mass and average column density per annulus, with sequence designed to ensure equal areas in each bin, as explained in \S~\ref{subsec:column_dens_maps}. Bottom: integral mass within circles of each radius. Red -- GBT data, corresponding to Fig.~\ref{fig:160_HImaps} top left (only two of the error bars are shown to avoid clutter). Green -- VLA data, corresponding to Fig.~\ref{fig:160_HImaps} middle left. Blue -- VLA+GBT data, corresponding to Fig.~\ref{fig:160_HImaps} bottom left.}\label{fig:160_mass_graphs}
\end{figure}

\begin{figure}
\vspace{-2.0cm}
\centering
\captionsetup[subfigure]{labelformat=empty}
    \subfloat[][]{
        \centering
        \hspace{-1cm}
        \includegraphics[width=1.0\textwidth]{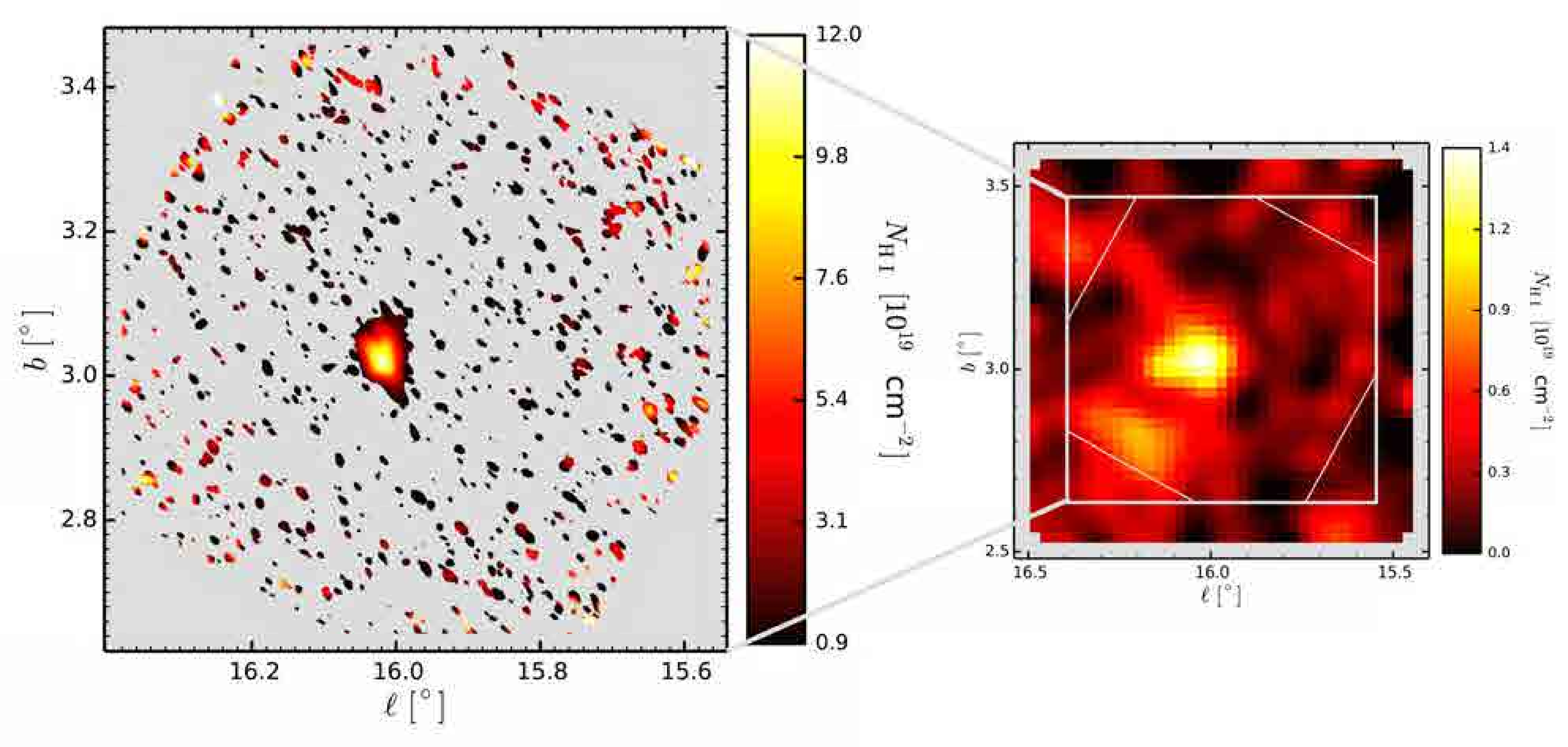}
        \hspace{-0cm}
        }

    \vspace{-1.0cm}

    \subfloat[][]{
        \centering
        \hspace{-1cm}
        \includegraphics[width=0.5\textwidth]{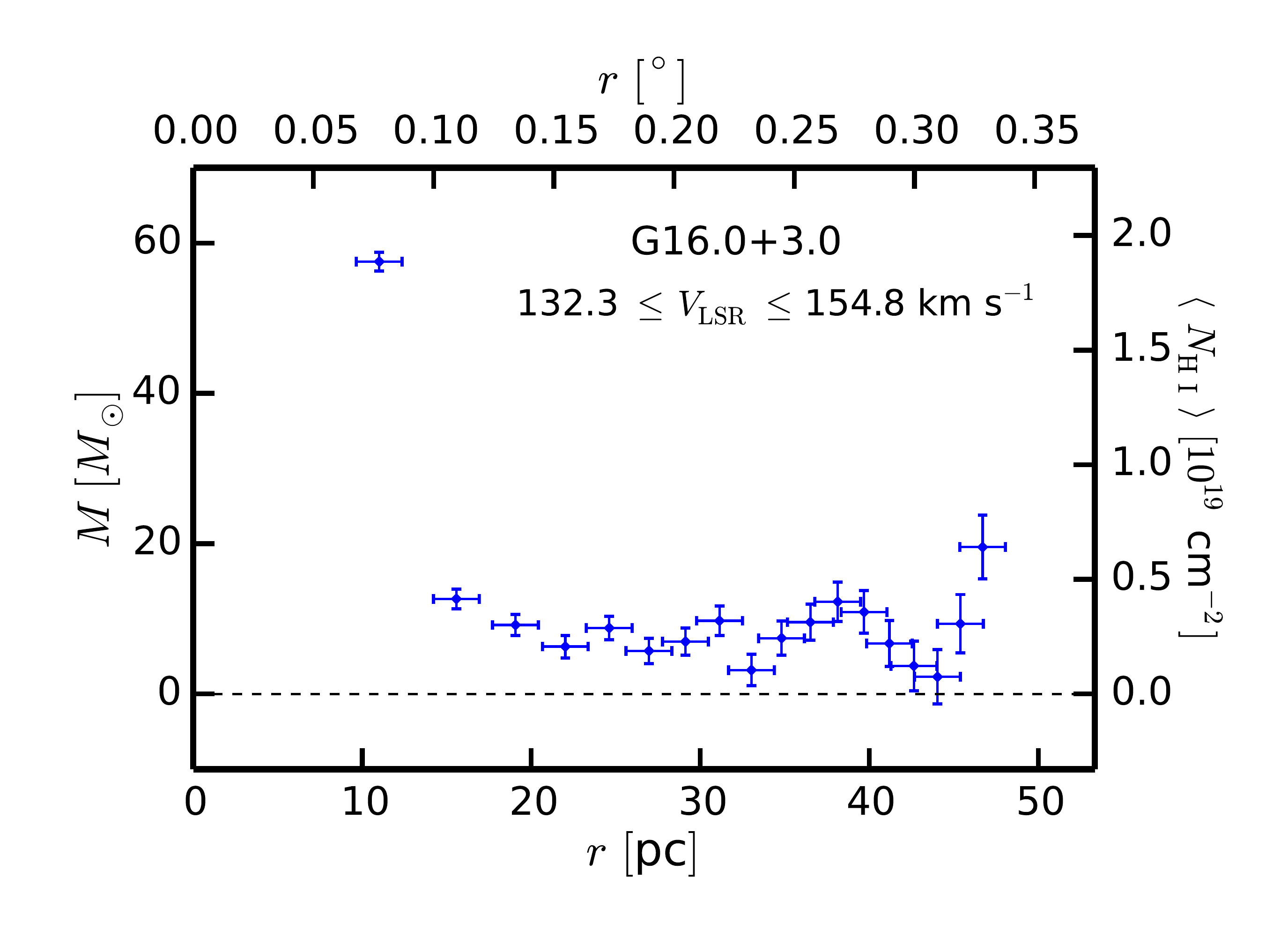}
        \hspace{-0.7cm}
        }
    \subfloat[][]{
        \centering
        \includegraphics[viewport = 0 -30 1150 650, width=0.59\textwidth]{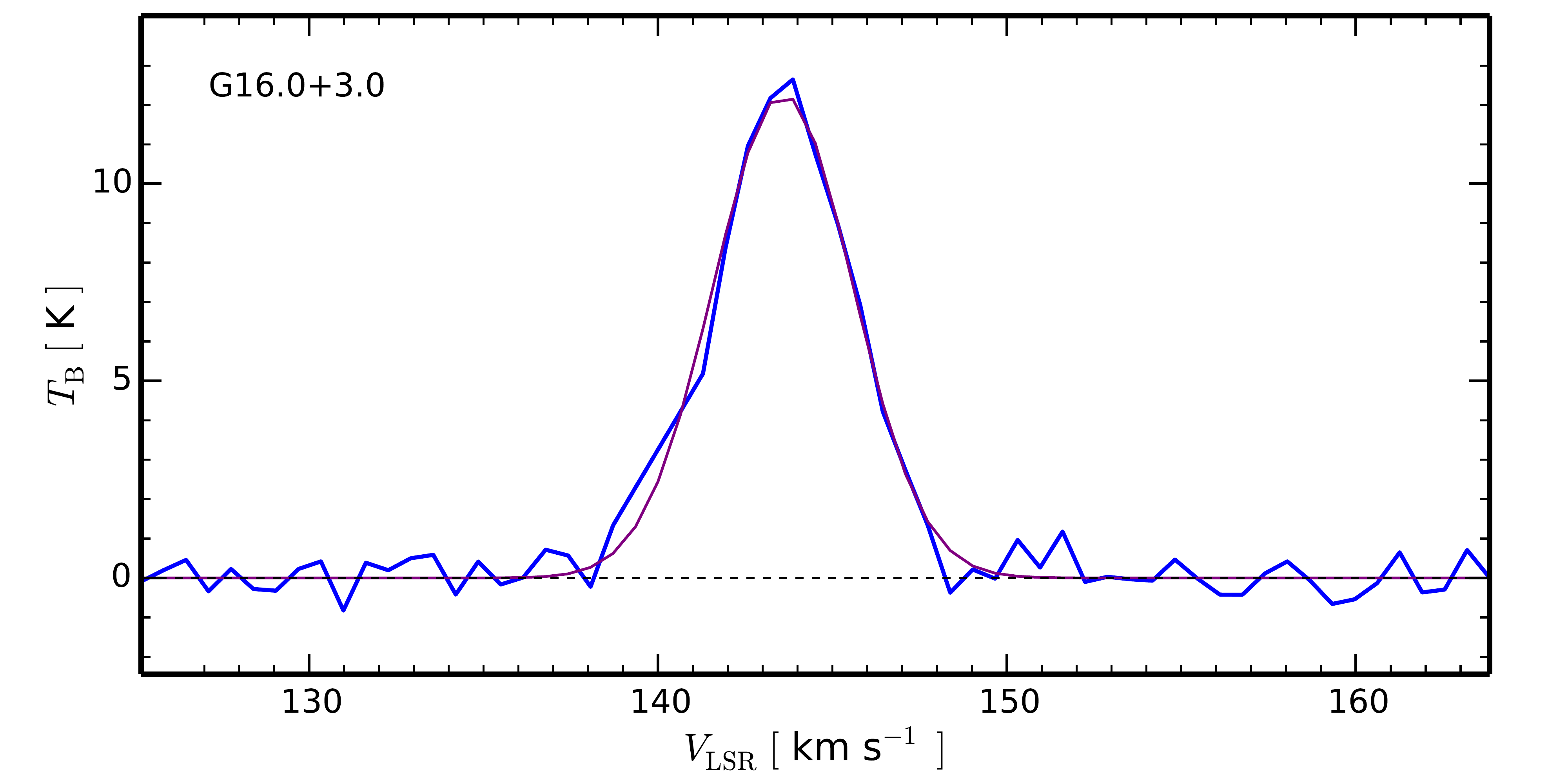}
        }
\vspace{-1.1cm}
\caption{Summary of G$16.0+3.0$. Top left: 3$\sigma(r)$ thresholded VLA+GBT map, same as Fig.~\ref{fig:160_HImaps} bottom right. Top right: GBT column density map integrated over the same velocity range, with the box marking the extent of the VLA map. Lower left: radial mass distribution, same as blue points in Fig.~\ref{fig:160_mass_graphs} top. Lower right: final spectrum toward the position of the peak \NHI with the Gaussian whose parameters are given in Table~\ref{tab:observed-props}. This is a very little cloud with no structure, unlike any other cloud in the group.  It is the smallest cloud by far, has one of the narrowest line widths implying a kinetic temperature $<500$ K, and has the largest mean density by a factor $\approx 5$  (see \S\ref{subsec:discussion}). About one half of the cloud mass is confined to the dense and very compact core. It may be in the region of the Galaxy excavated by a hot wind from the Galactic nucleus as discussed in \S\ref{subsec:discussion}. This isolated cloud is similar to the dense cores seen in other clouds, e.g., G$26.9-6.3$.}
\label{fig:G160-4plots}
\end{figure}

Fig.~\ref{fig:160_HImaps} shows in comparison both full and thresholded column density maps of G$16.0+3.0$ based separately on the GBT and VLA data, and final VLA+GBT results. Fig.~\ref{fig:160_mass_graphs} gives the corresponding mass profiles. Finally, Fig.~\ref{fig:G160-4plots} presents a summary of G$16.0+3.0$: thresholded VLA+GBT maps in relation to full GBT images, mass profile of the VLA+GBT image, and spectra toward the \NHI\ peak of the clouds. Caption to the latter figure contains comments about the structure of the cloud.

\begin{figure}
\centering
    \vspace{-2cm}

    \captionsetup[subfigure]{labelformat=empty}
    \subfloat[][]{
        \centering
        \hspace{0.0cm}
        \includegraphics[width=0.5\textwidth]{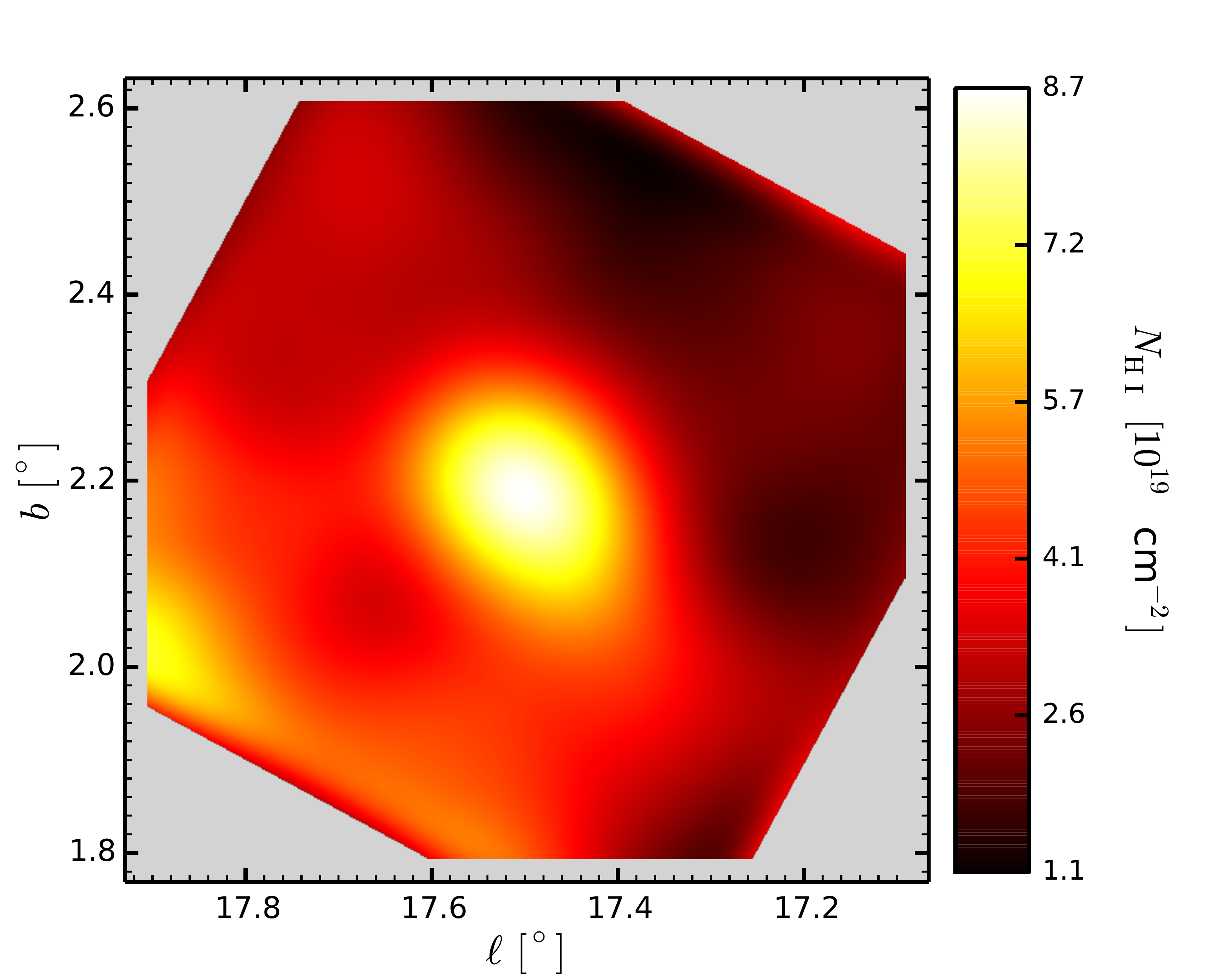}
        \hspace{-0.1cm}}
    \subfloat[][]{
        \centering
        \includegraphics[width=0.5\textwidth]{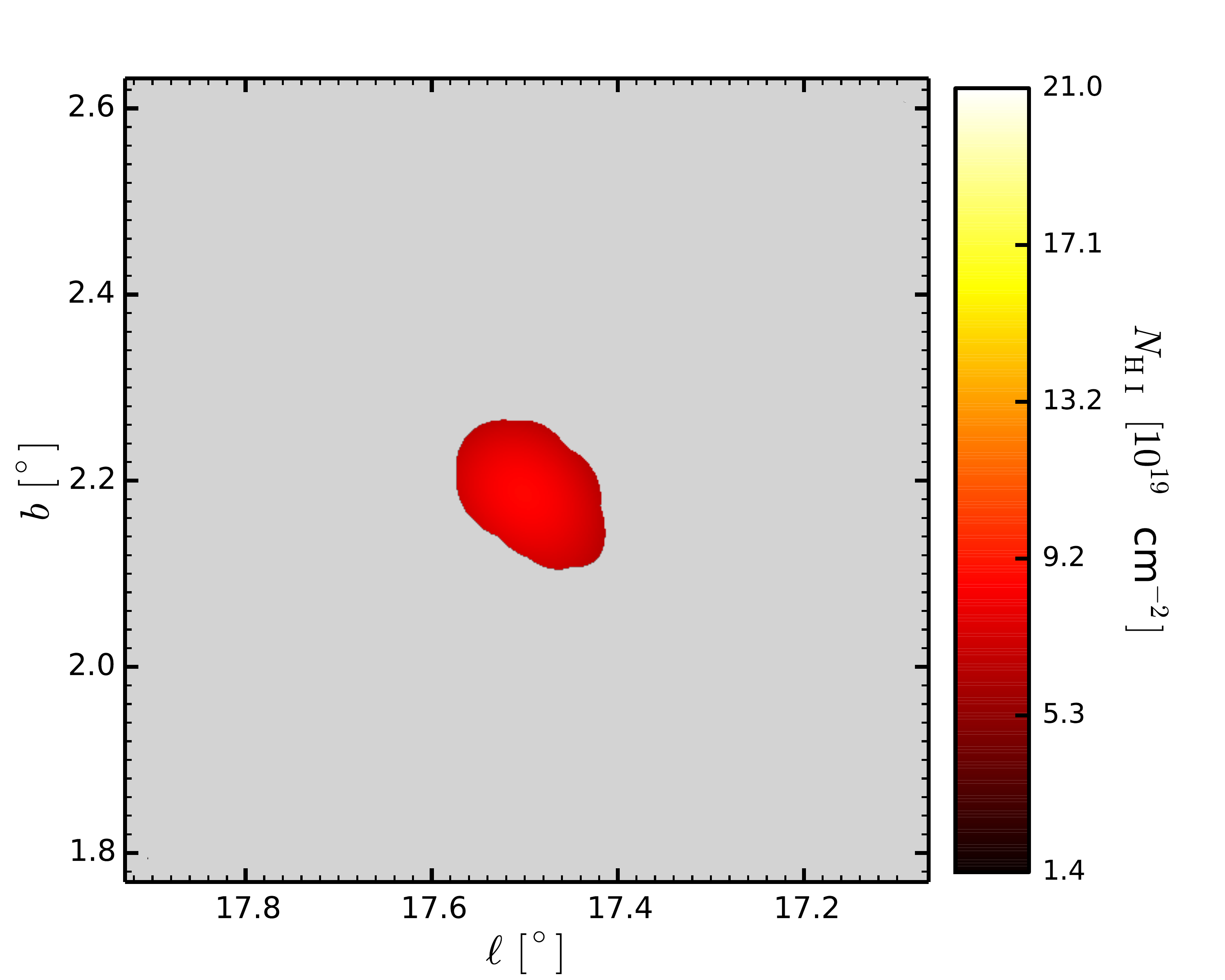}
        }

    \vspace{-1.3cm}

    \subfloat[][]{
        \centering
        \hspace{0.0cm}
        \includegraphics[width=0.5\textwidth]{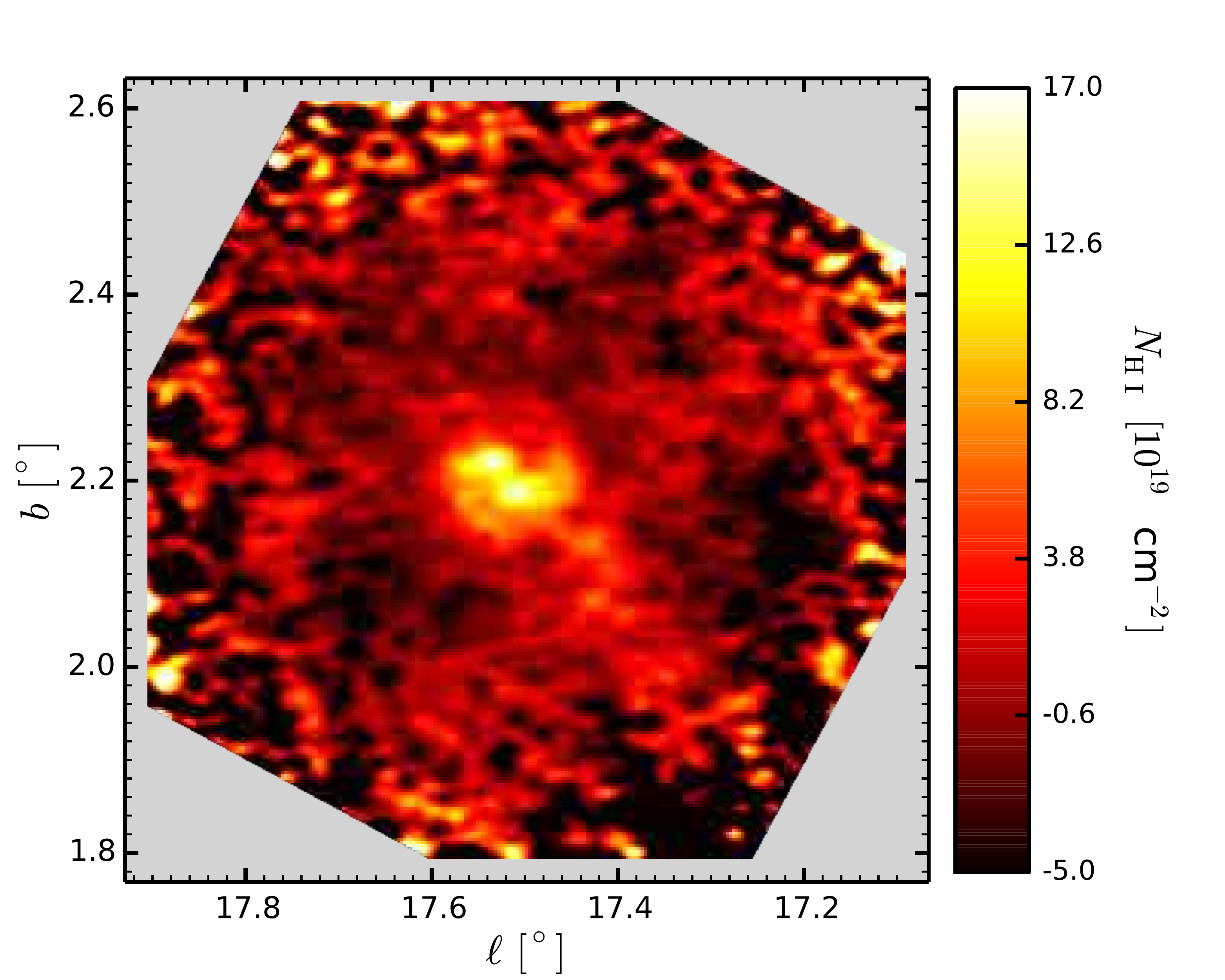}
        \hspace{-0.1cm}}
    \subfloat[][]{
        \centering
        \includegraphics[width=0.5\textwidth]{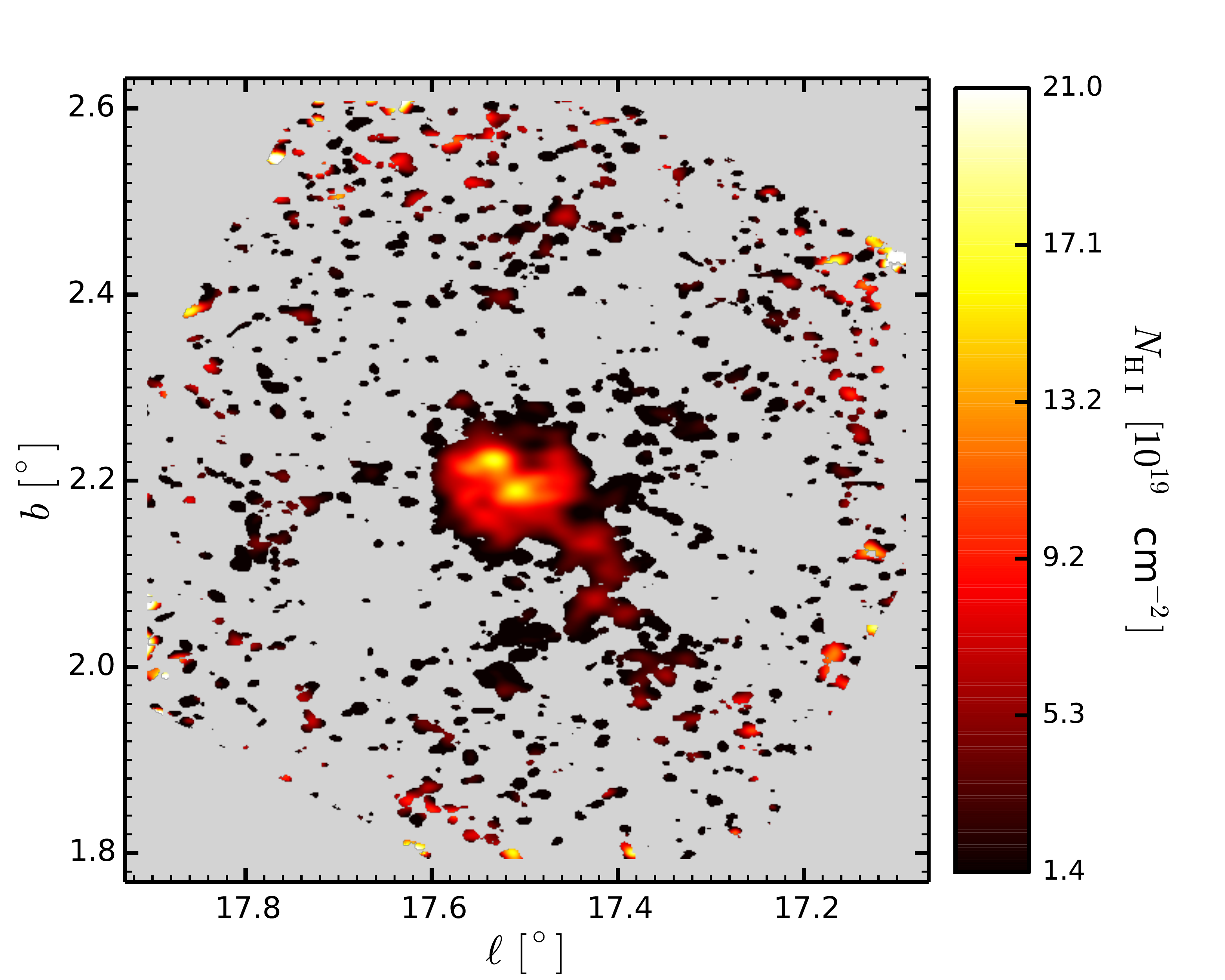}
        }

     \vspace{-1.3cm}

     \subfloat[][]{
        \centering
        \hspace{0.0cm}
        \includegraphics[width=0.5\textwidth]{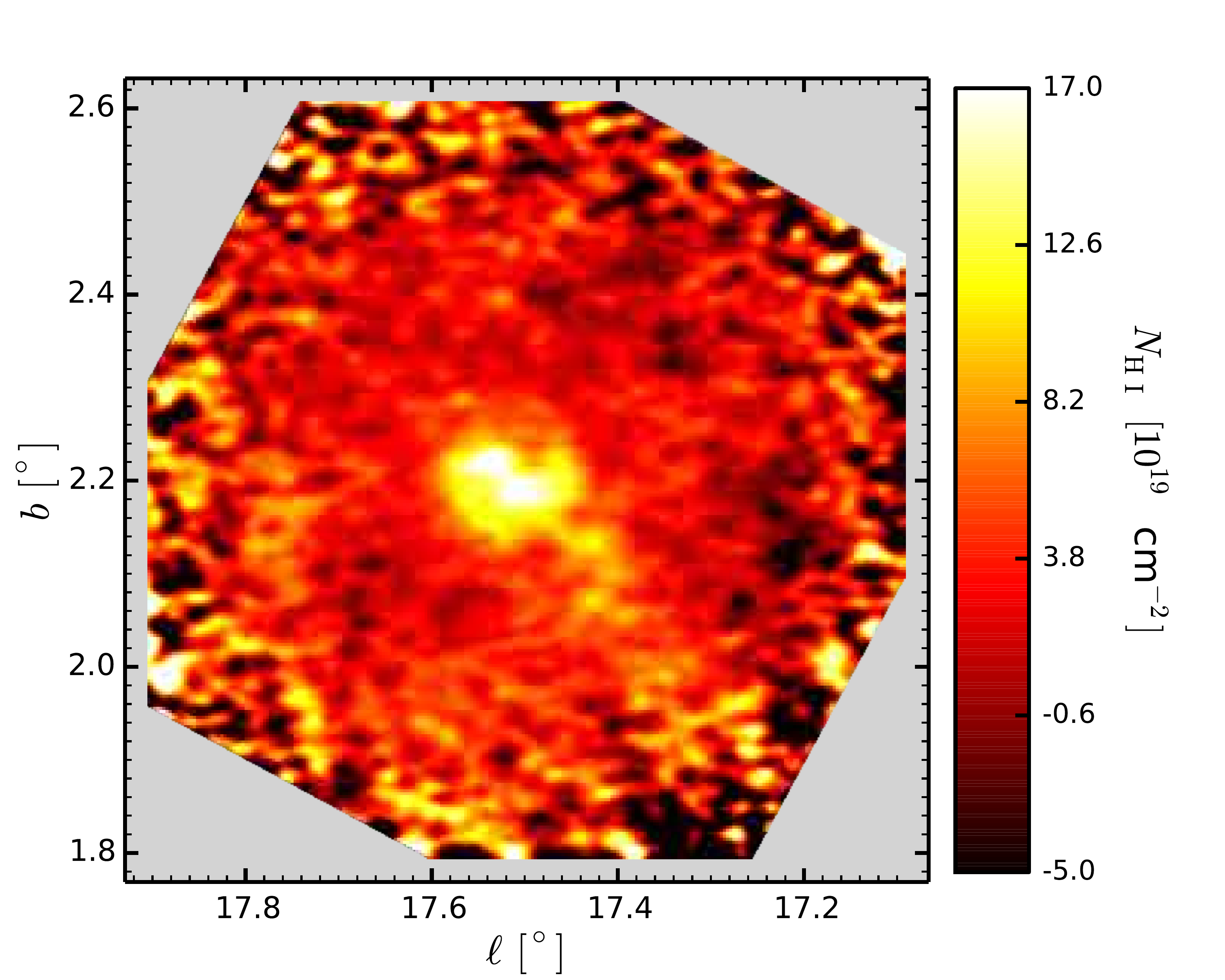}
        \hspace{-0.1cm}}
    \subfloat[][]{
        \centering
        \includegraphics[width=0.5\textwidth]{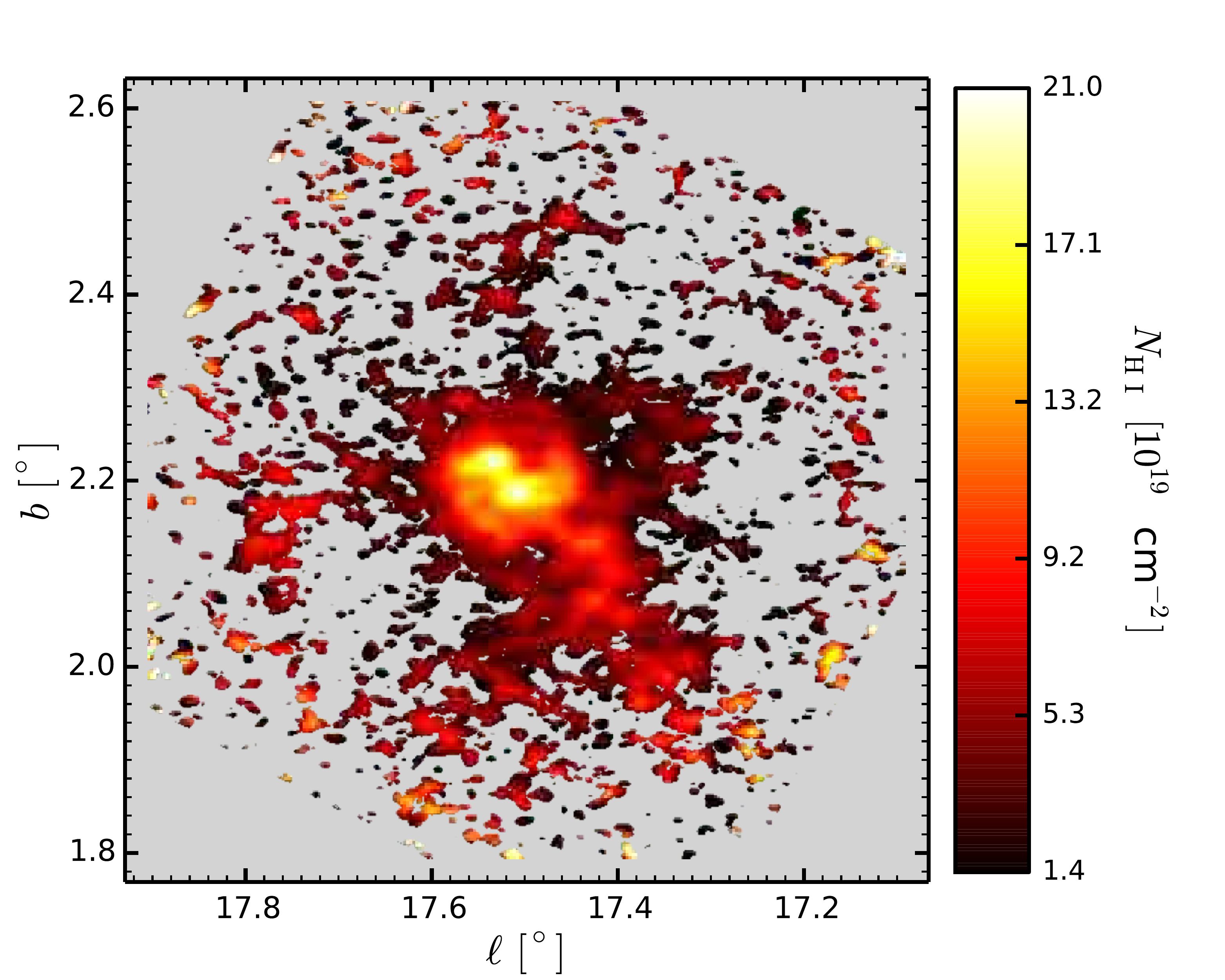}
        }
\vspace{-0.5cm}
\caption{\HI column density maps for G$17.5+2.2$, integrated over 48 spectral channels in the interval $122.8 \leq \VLSR \leq 153.0$~\kms, as described in the caption to Fig.~\ref{fig:160_HImaps}
}
\label{fig:175_HImaps}
\end{figure}

\begin{figure}
    \centering
    \vspace{-1.0cm}

    \captionsetup[subfigure]{labelformat=empty}
    \subfloat[][]{
        \centering
        \hspace{-0.5cm}
        \includegraphics[width=0.75\textwidth]{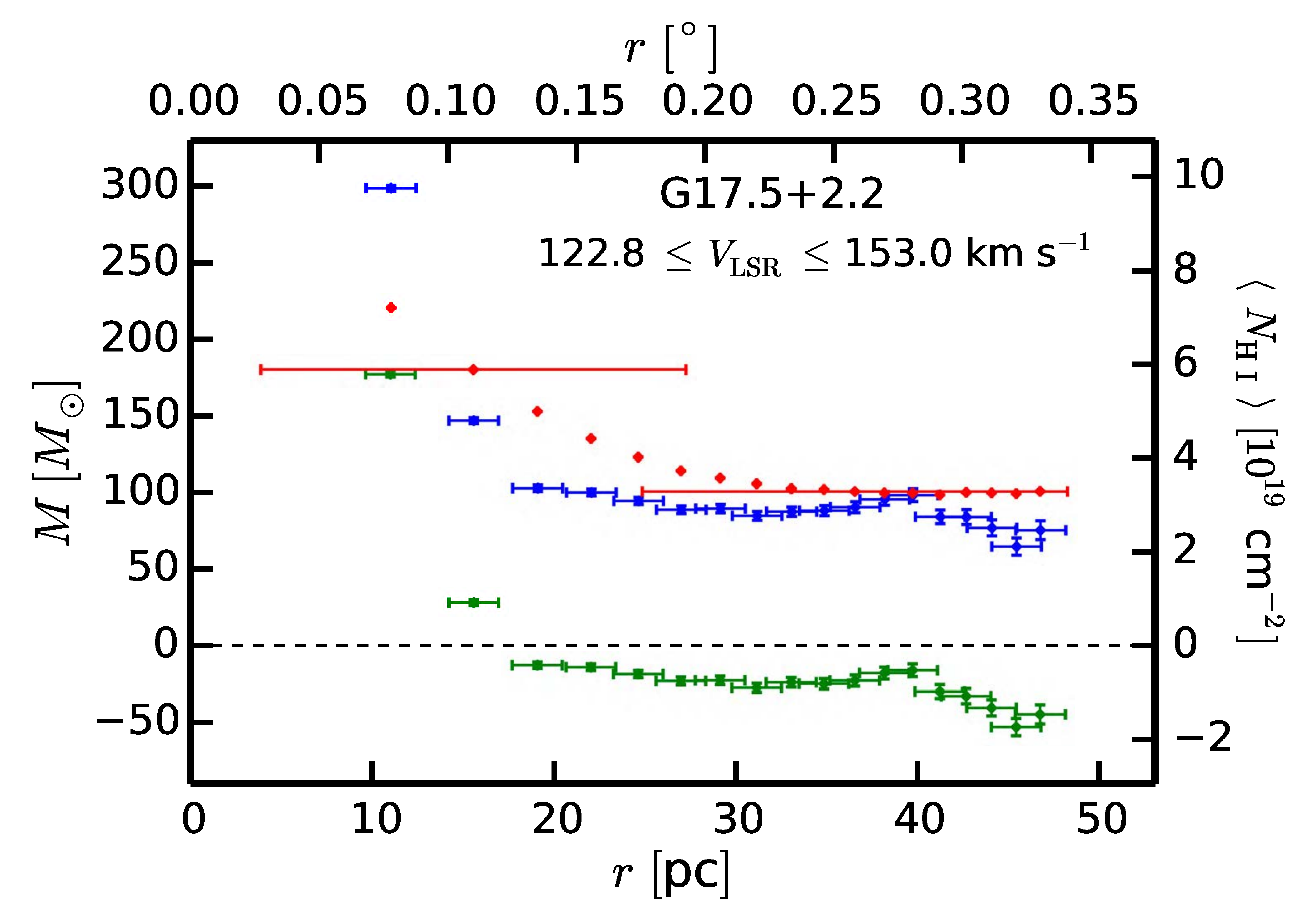}
        }

    \vspace{-1.0cm}

    \subfloat[][]{
        \centering
        \hspace{-1.8cm}
        \includegraphics[width=0.7\textwidth]{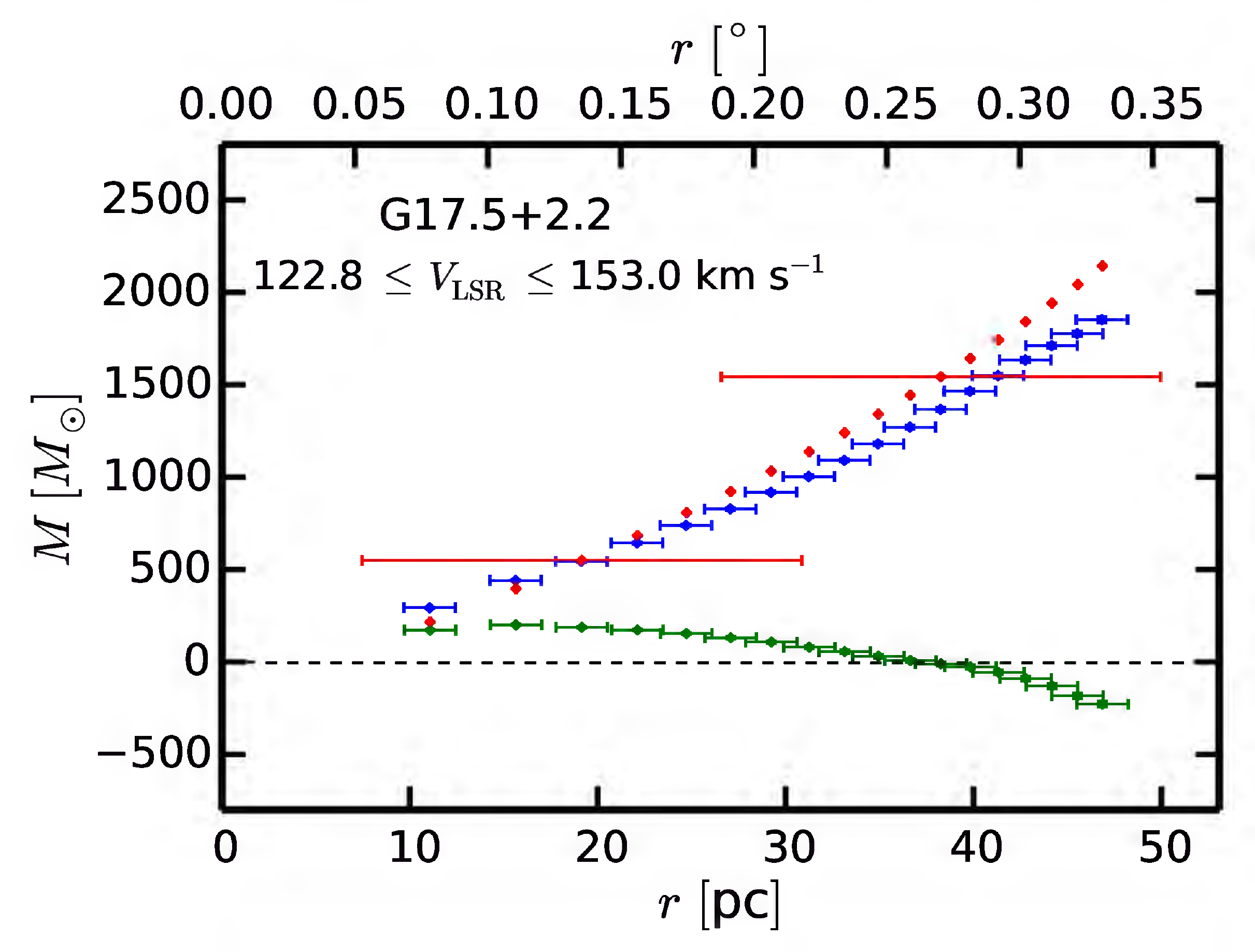}
        }

\caption{Radial mass profiles for G$17.5+2.2$ as described in the caption to Fig.~\ref{fig:160_mass_graphs}.}\label{fig:175_mass_graphs}
\end{figure}

\begin{figure}
\vspace{-2.0cm}
\centering
\captionsetup[subfigure]{labelformat=empty}
    \subfloat[][]{
        \centering
        \hspace{-1cm}
        \includegraphics[width=1.0\textwidth]{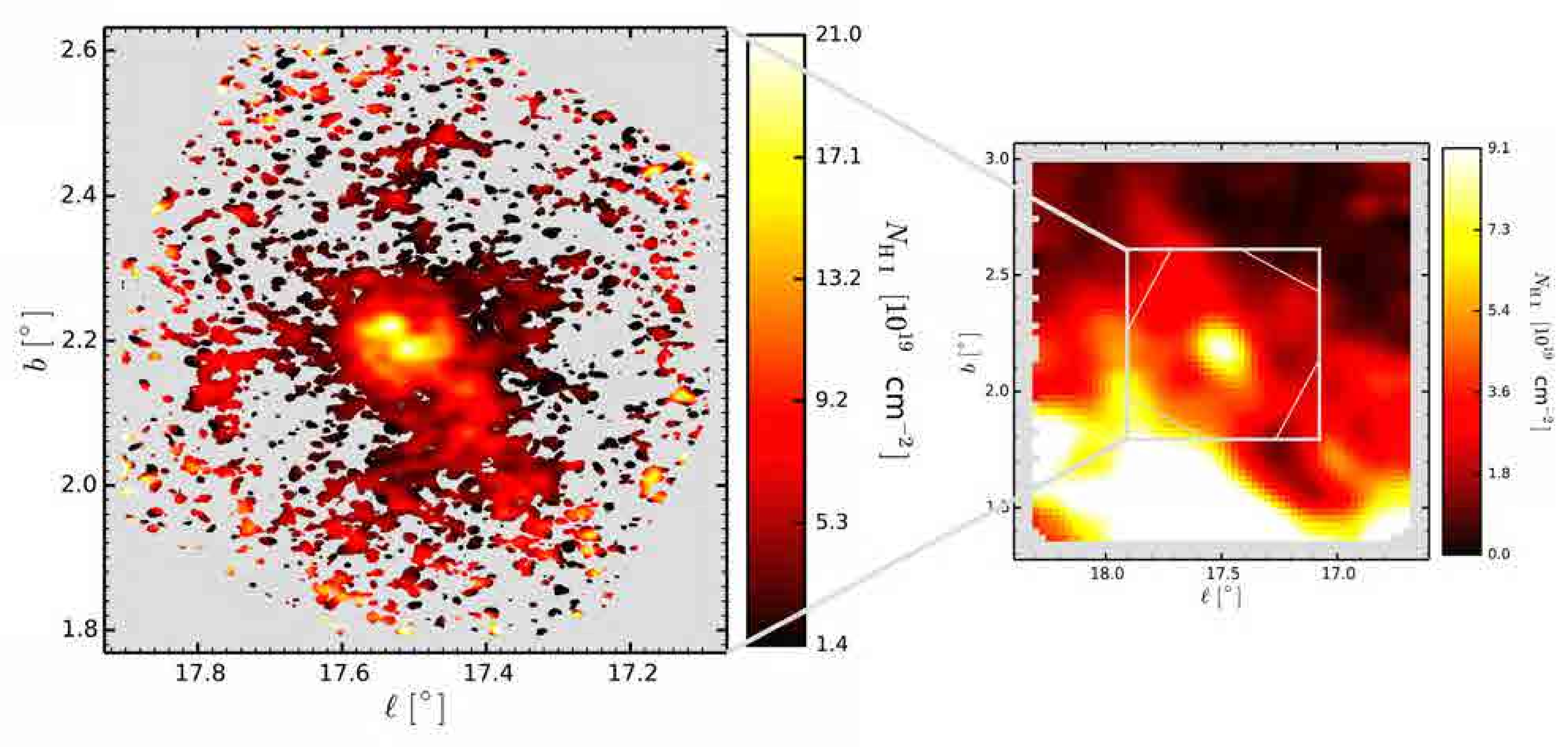}
        \hspace{-0cm}
        }

    \vspace{-1.0cm}

    \subfloat[][]{
        \centering
        \hspace{-1cm}
        \includegraphics[width=0.5\textwidth]{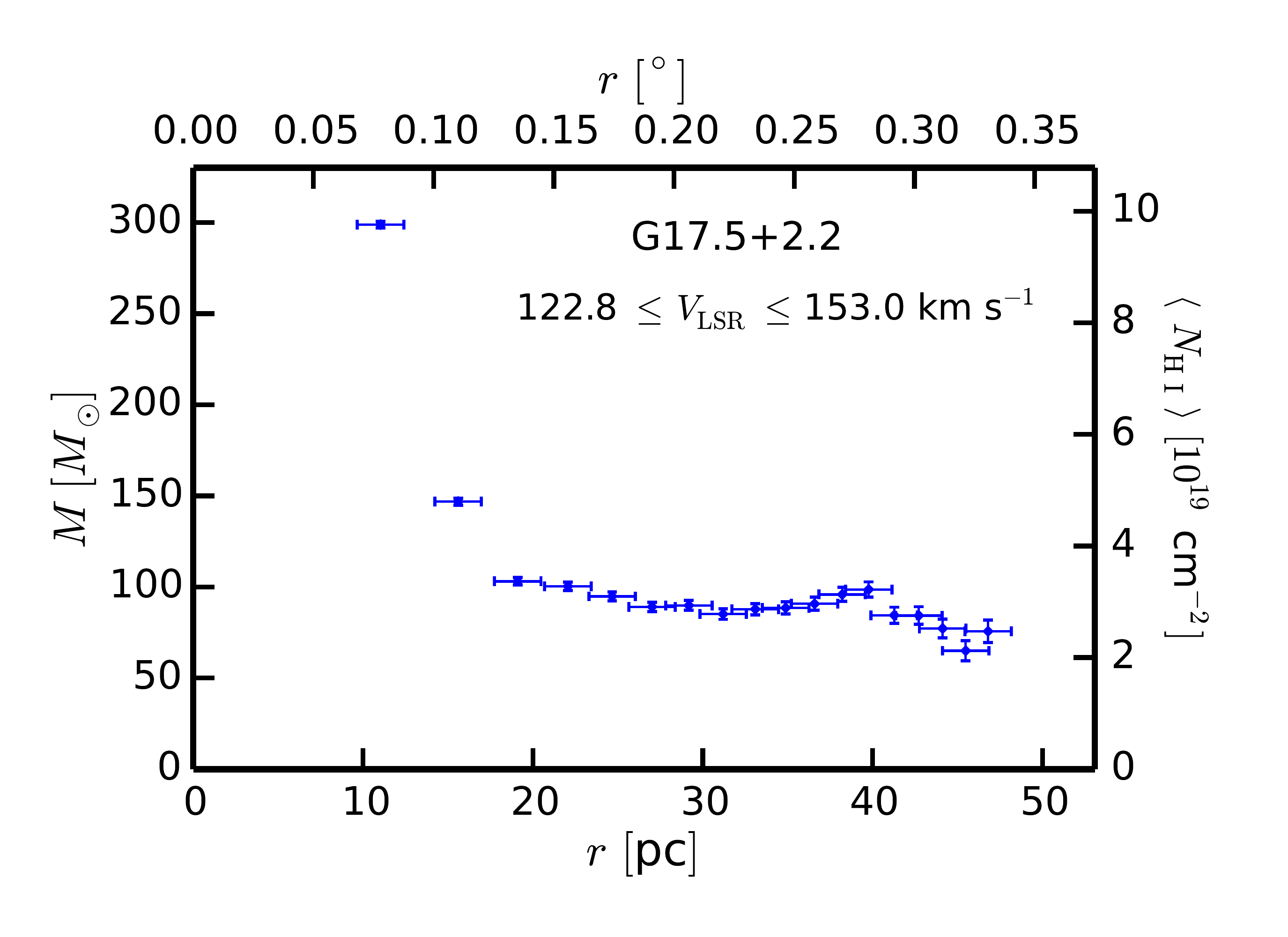}
        \hspace{-0.7cm}
        }
    \subfloat[][]{
        \centering
        \includegraphics[viewport = 0 -30 1150 650, width=0.59\textwidth]{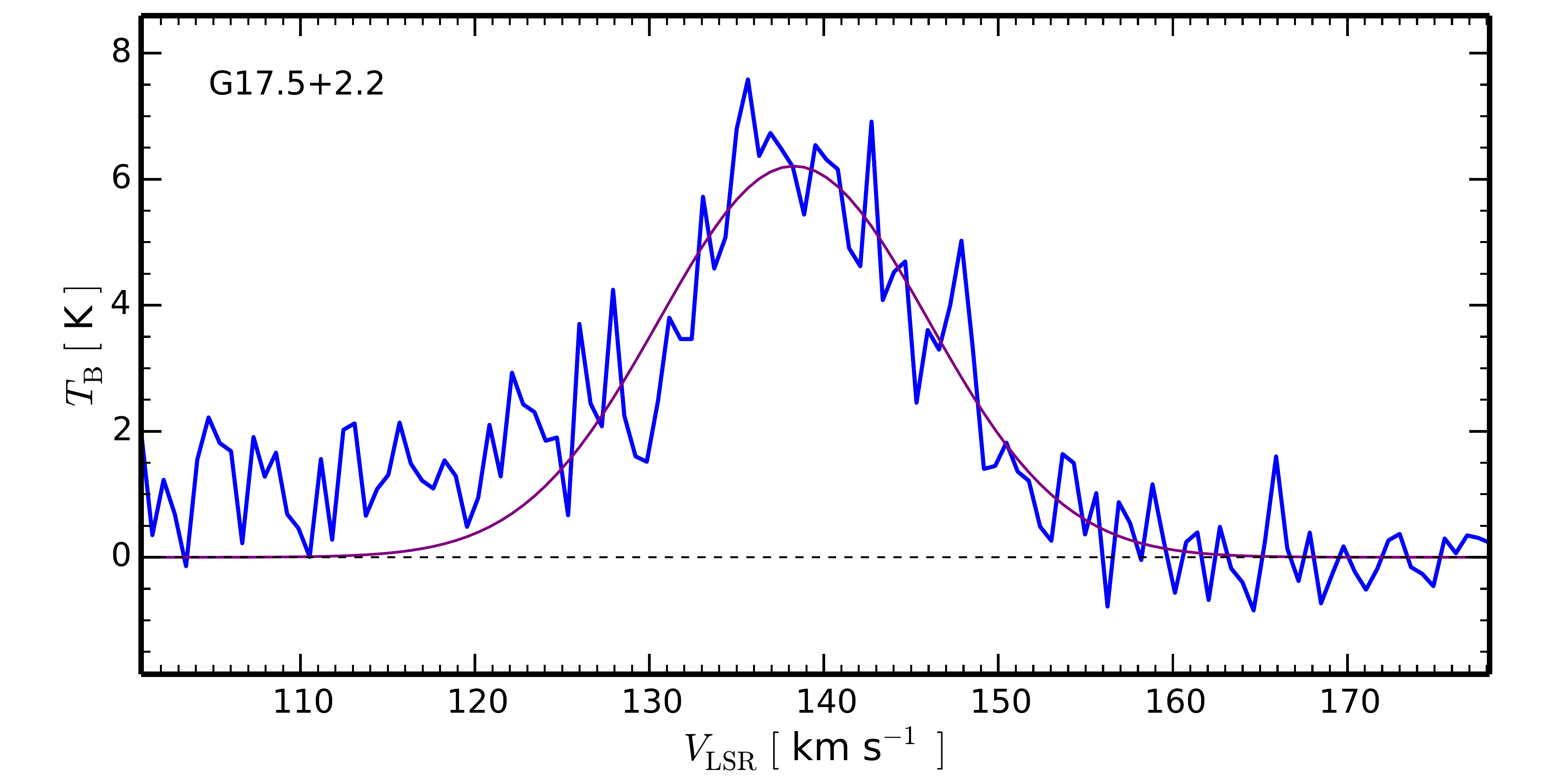}
        }
\vspace{-0.5cm}
\caption{Summary of G$17.5+2.2$ as described in the caption to Fig.~\ref{fig:G160-4plots}. A cloud with an almost circular core $\approx 30$~pc in diameter that contains two peaks of similar \NHI\ and about half of the cloud's mass.  Surrounding the core are irregular structures of low \NHI, the densest of which has a cometary appearance stretching to lower longitude and latitude.}
\label{fig:G175-4plots}
\end{figure}

\begin{figure}
\centering
    \vspace{-2cm}

    \captionsetup[subfigure]{labelformat=empty}
    \subfloat[][]{
        \centering
        \hspace{0.0cm}
        \includegraphics[width=0.5\textwidth]{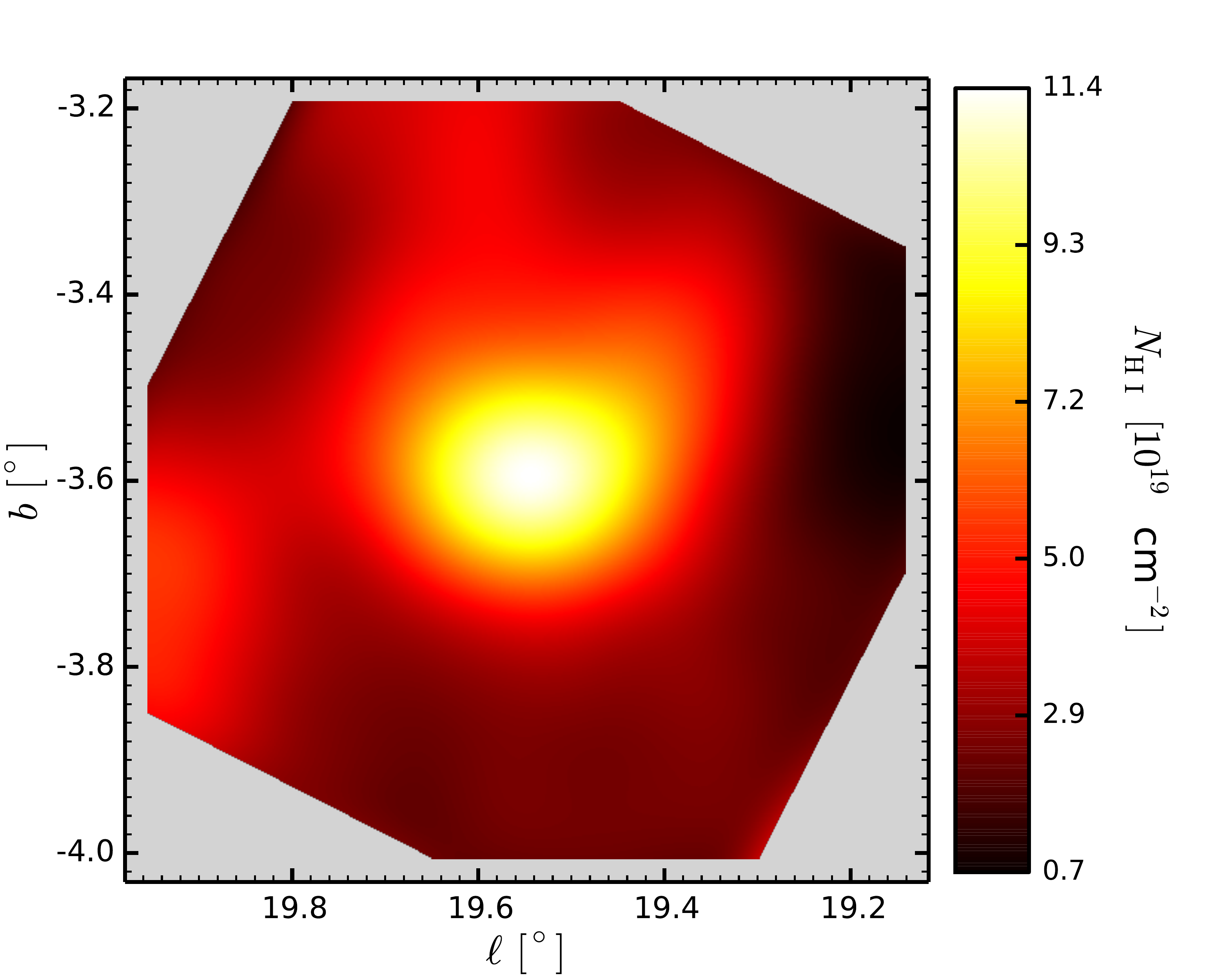}
        \hspace{-0.1cm}}
    \subfloat[][]{
        \centering
        \includegraphics[width=0.5\textwidth]{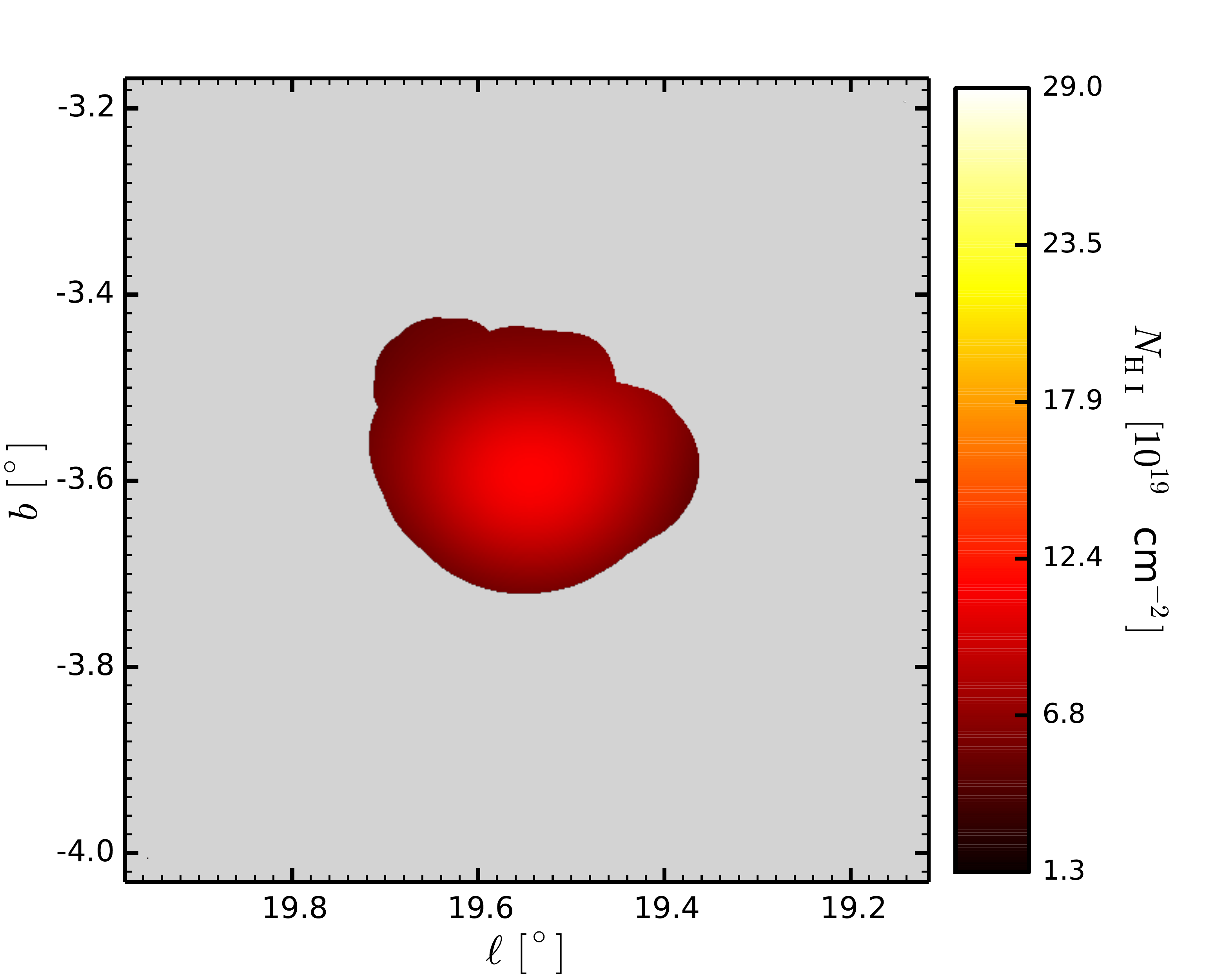}
        }

    \vspace{-1.3cm}

    \subfloat[][]{
        \centering
        \hspace{0.0cm}
        \includegraphics[width=0.5\textwidth]{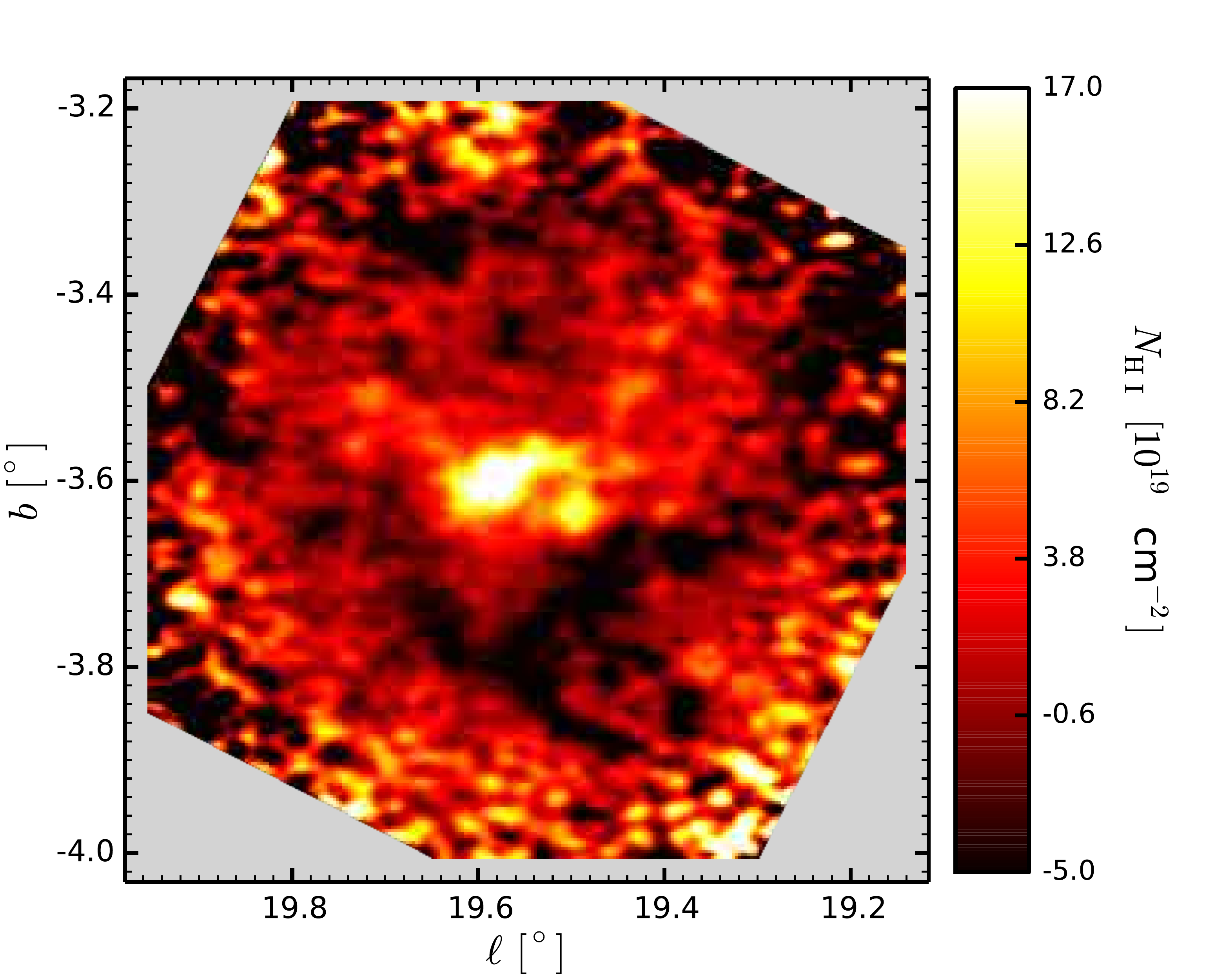}
        \hspace{-0.1cm}}
    \subfloat[][]{
        \centering
        \includegraphics[width=0.5\textwidth]{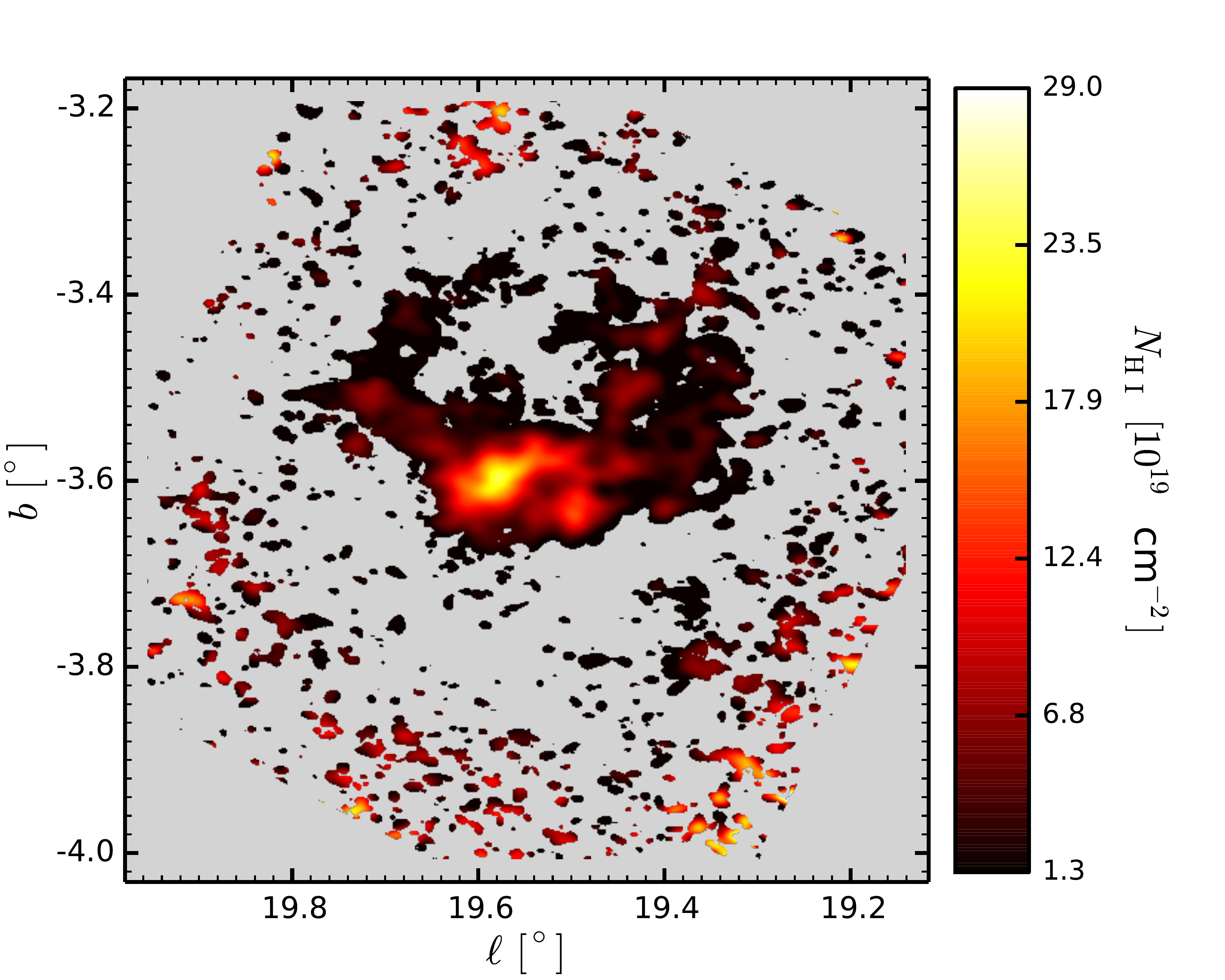}
        }

     \vspace{-1.3cm}

     \subfloat[][]{
        \centering
        \hspace{0.0cm}
        \includegraphics[width=0.5\textwidth]{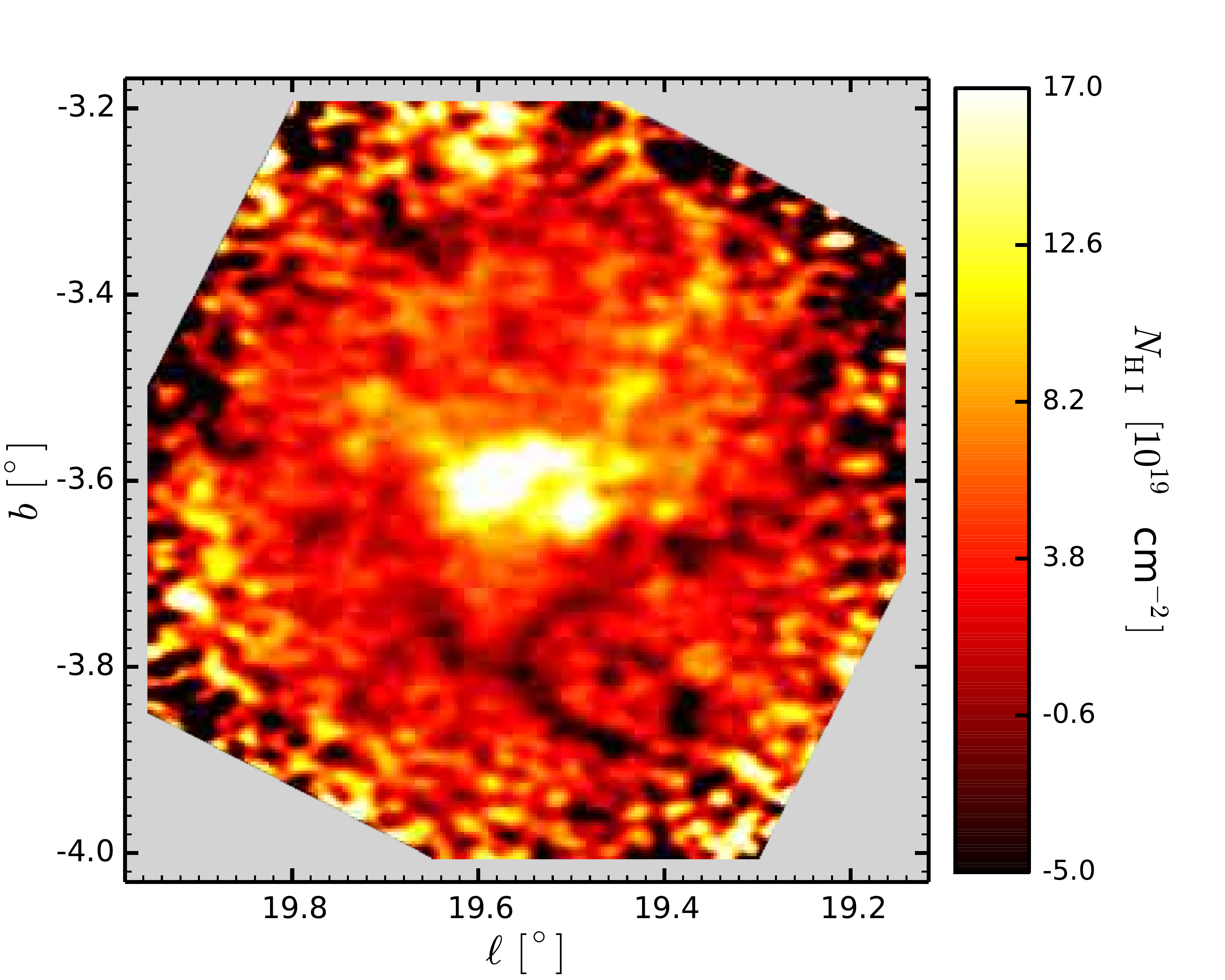}
        \hspace{-0.1cm}}
    \subfloat[][]{
        \centering
        \includegraphics[width=0.5\textwidth]{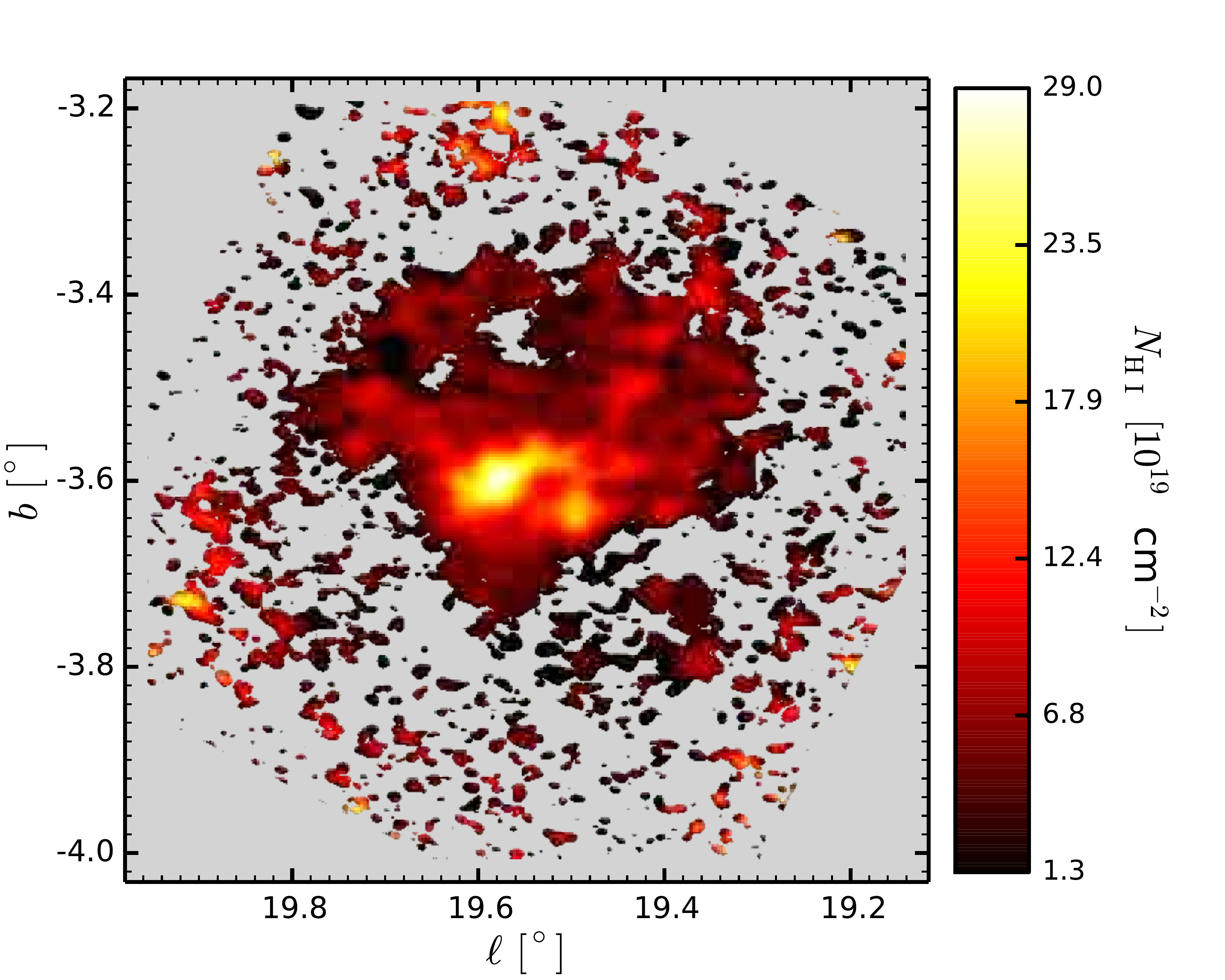}
        }
\vspace{-0.5cm}
\caption{\HI column density maps for G$19.5-3.6$, integrated over 56 spectral channels in the interval $107.8 \leq \VLSR \leq 143.2$~\kms, as described in the caption to Fig.~\ref{fig:160_HImaps}.
}
\label{fig:195_HImaps}
\end{figure}

\begin{figure}
    \centering
    \vspace{-1.0cm}

    \captionsetup[subfigure]{labelformat=empty}
    \subfloat[][]{
        \centering
        \hspace{-0.5cm}
        \includegraphics[width=0.75\textwidth]{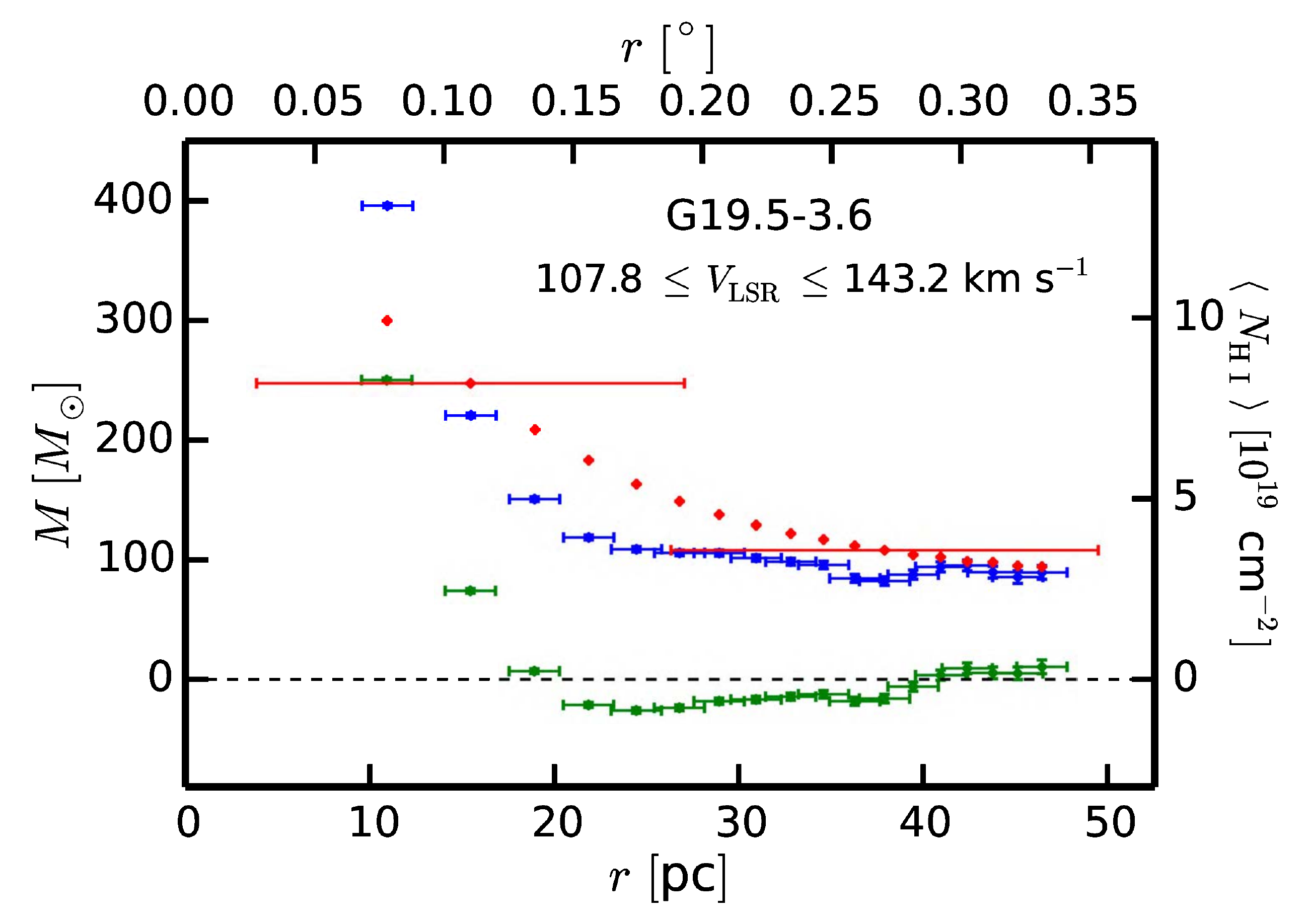}
        }

    \vspace{-1.0cm}

    \subfloat[][]{
        \centering
        \hspace{-1.8cm}
        \includegraphics[width=0.7\textwidth]{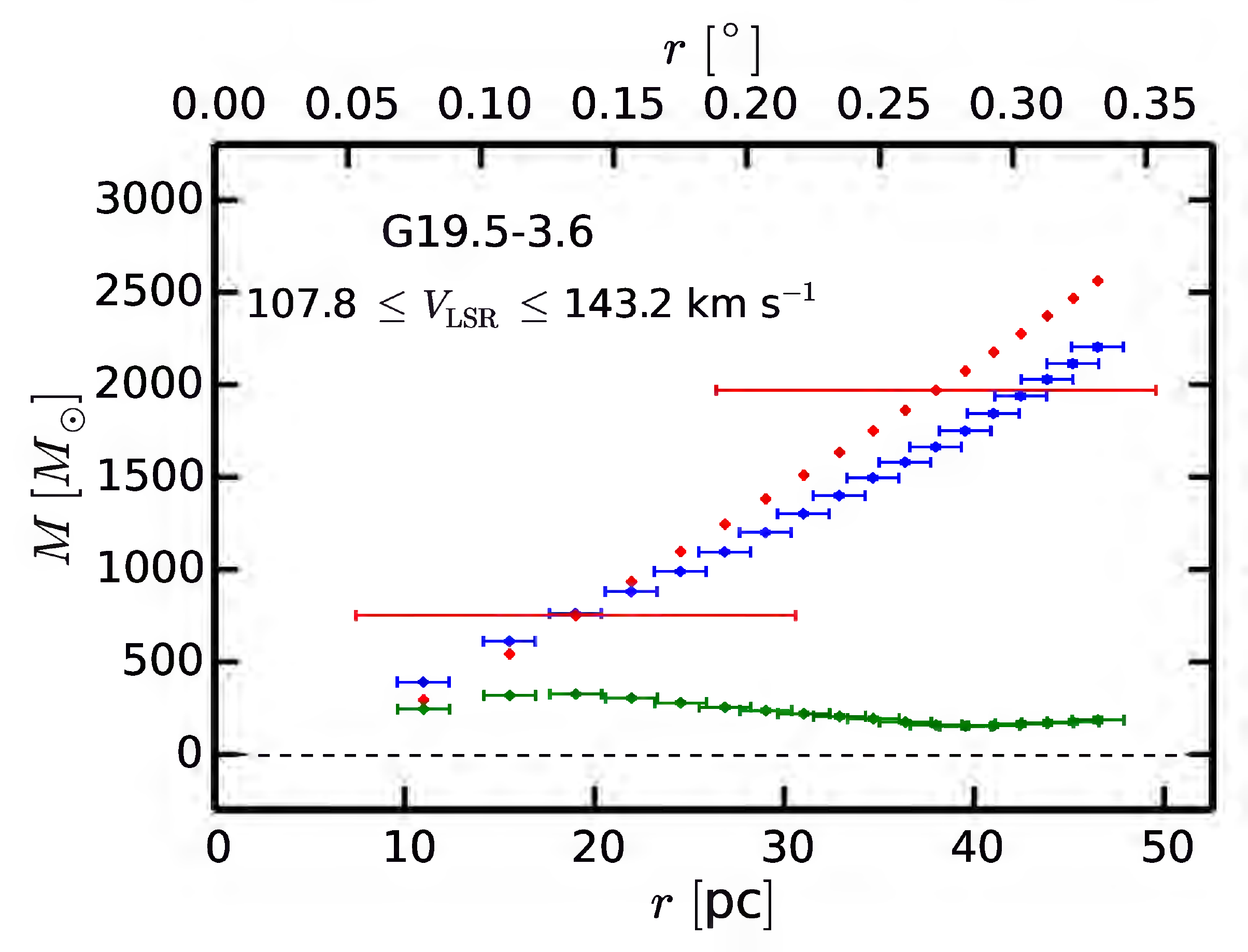}
        }

\caption{Radial mass profiles for G$19.5-3.6$ as described in the caption to Fig.~\ref{fig:160_mass_graphs}.}\label{fig:195_mass_graphs}
\end{figure}

\begin{figure}
\vspace{-2.0cm}
\centering
\captionsetup[subfigure]{labelformat=empty}
    \subfloat[][]{
        \centering
        \hspace{-1cm}
        \includegraphics[width=1.0\textwidth]{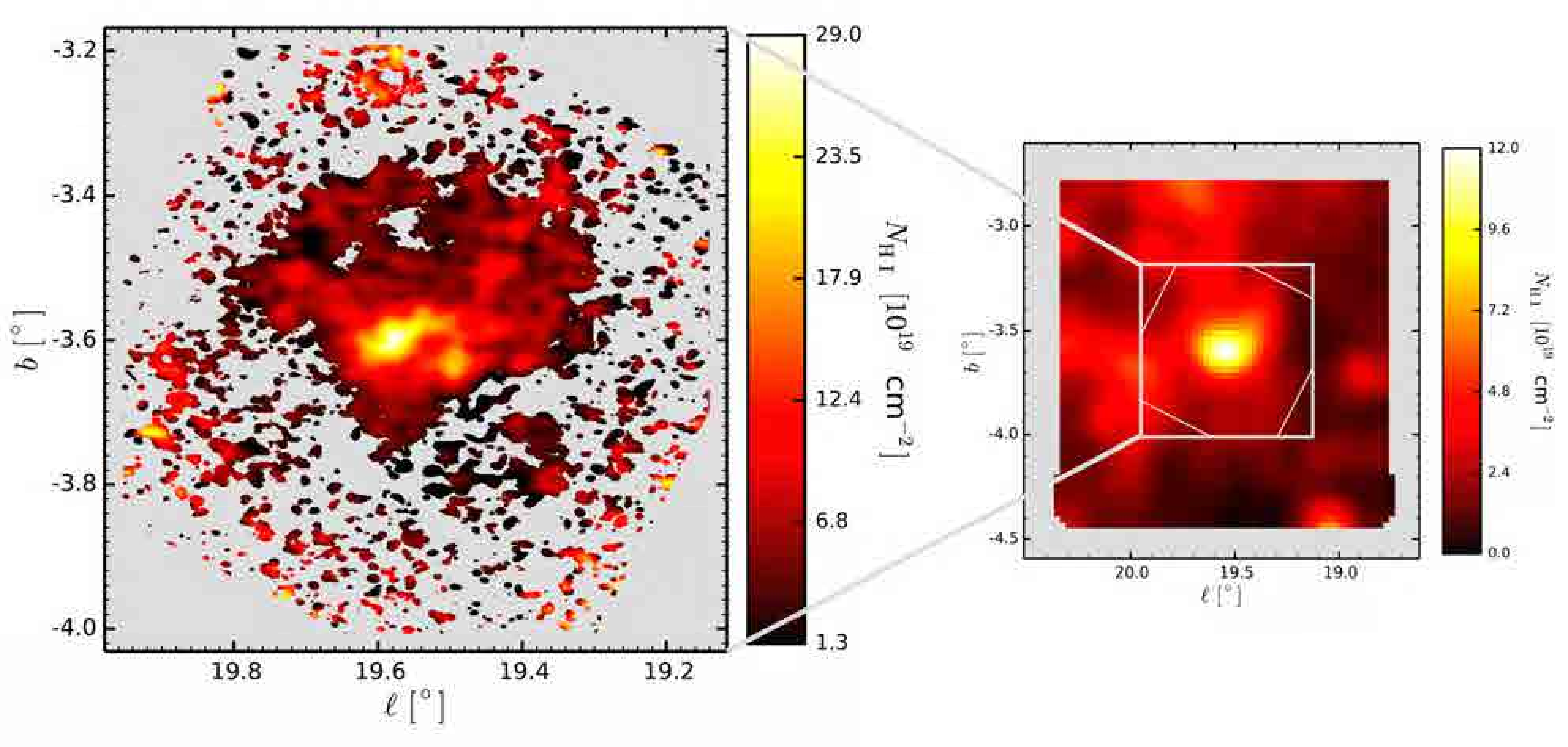}
        \hspace{-0cm}
        }

    \vspace{-1.0cm}

    \subfloat[][]{
        \centering
        \hspace{-1cm}
        \includegraphics[width=0.5\textwidth]{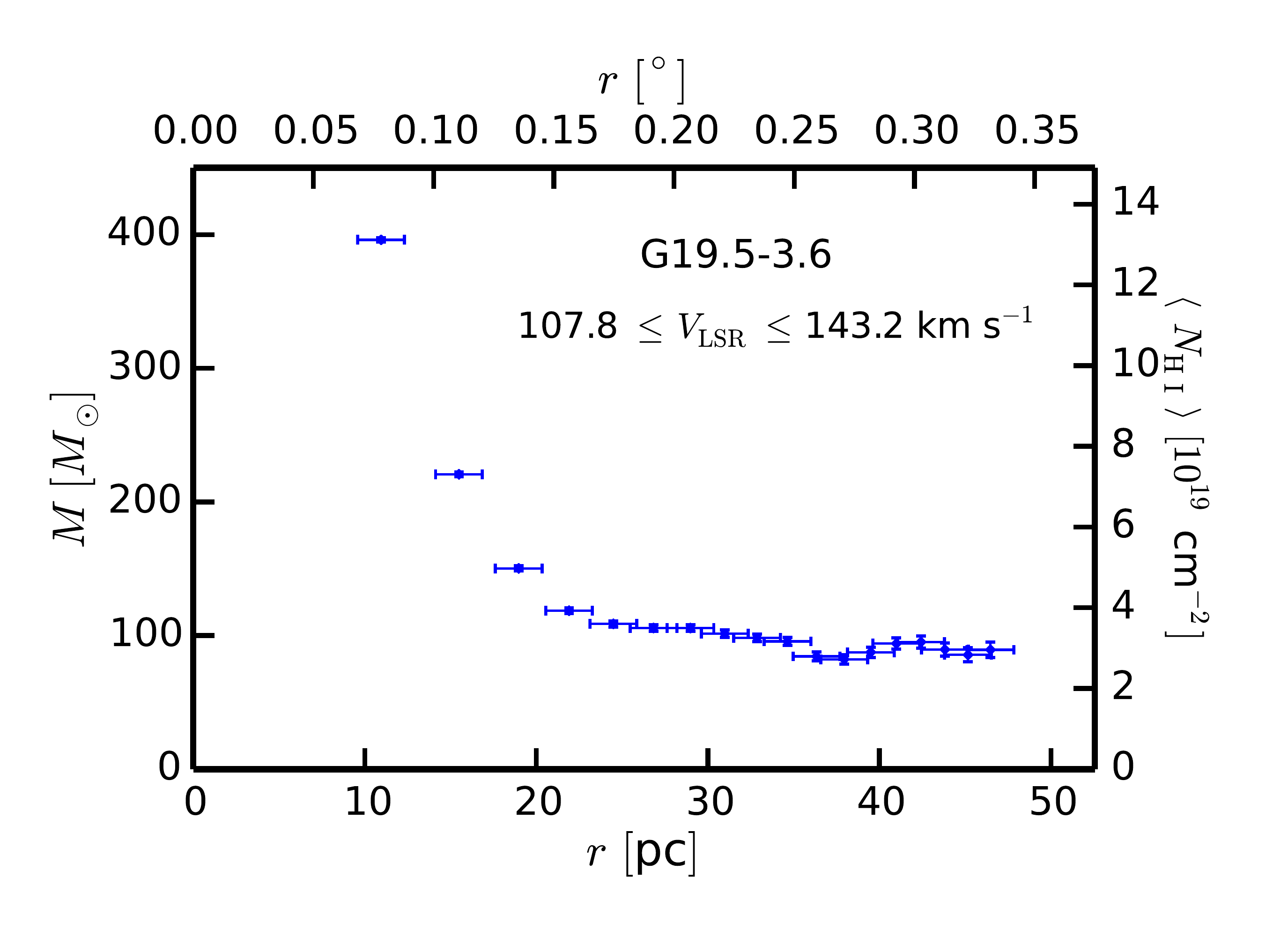}
        \hspace{-0.7cm}
        }
    \subfloat[][]{
        \centering
        \includegraphics[viewport = 0 -30 1150 650, width=0.59\textwidth]{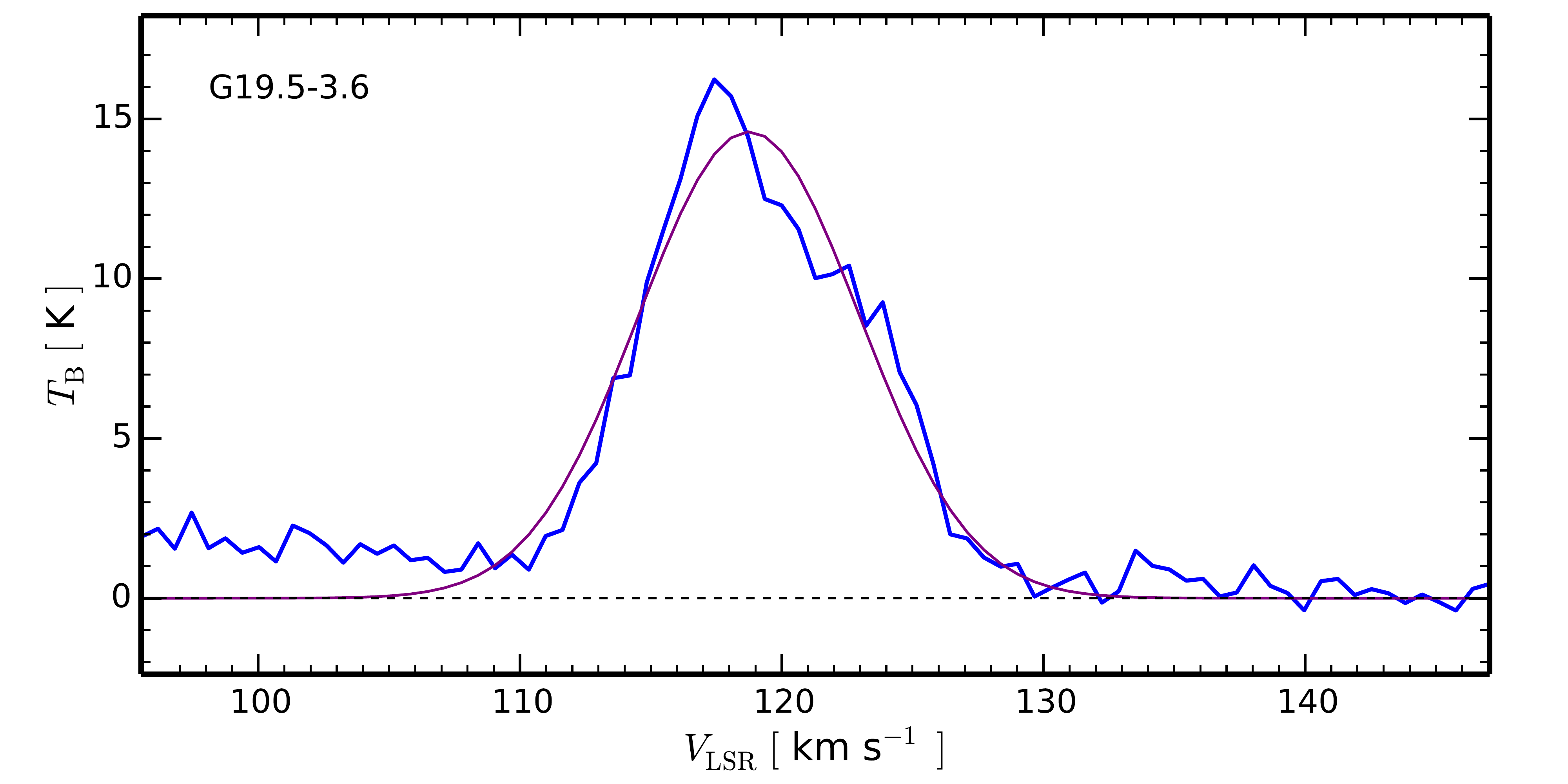}
        }
\vspace{-0.5cm}
\caption{Summary of G$19.5-3.6$ as described in the caption to Fig.~\ref{fig:G160-4plots}. One of the two largest clouds in the sample with a diameter $>60$ pc and a mass around $800$~\Msun in the central region and more than $1000$~\Msun in its extended envelope. The cloud's boundary is sharpest away from the Galactic plane and there are two ``tails'' of less bright emission stretching back toward the plane.}
\label{fig:G195-4plots}
\end{figure}

\begin{figure}
\centering
    \vspace{-2cm}

    \captionsetup[subfigure]{labelformat=empty}
    \subfloat[][]{
        \centering
        \hspace{0.0cm}
        \includegraphics[width=0.5\textwidth]{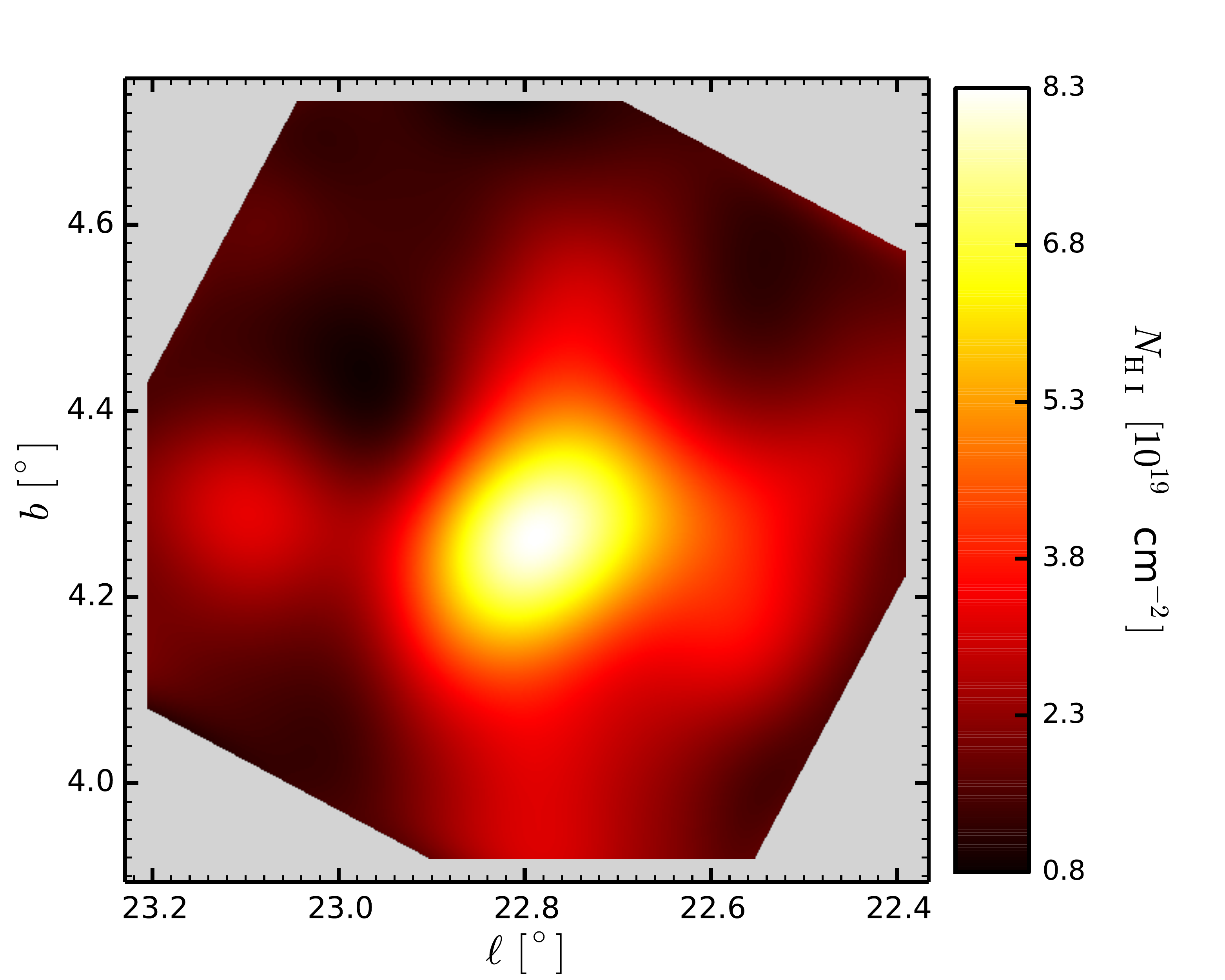}
        \hspace{-0.1cm}}
    \subfloat[][]{
        \centering
        \includegraphics[width=0.5\textwidth]{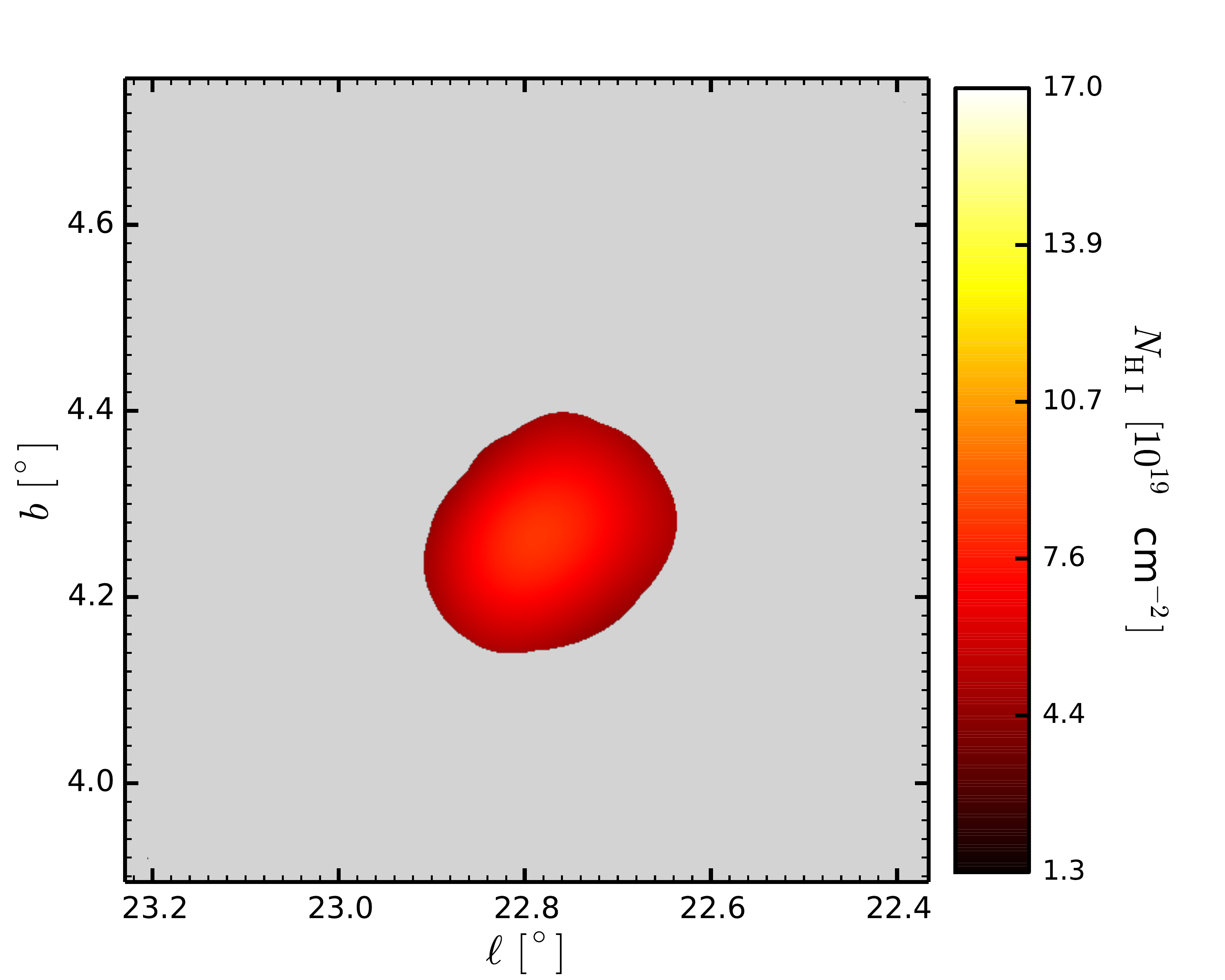}
        }

    \vspace{-1.3cm}

    \subfloat[][]{
        \centering
        \hspace{0.0cm}
        \includegraphics[width=0.5\textwidth]{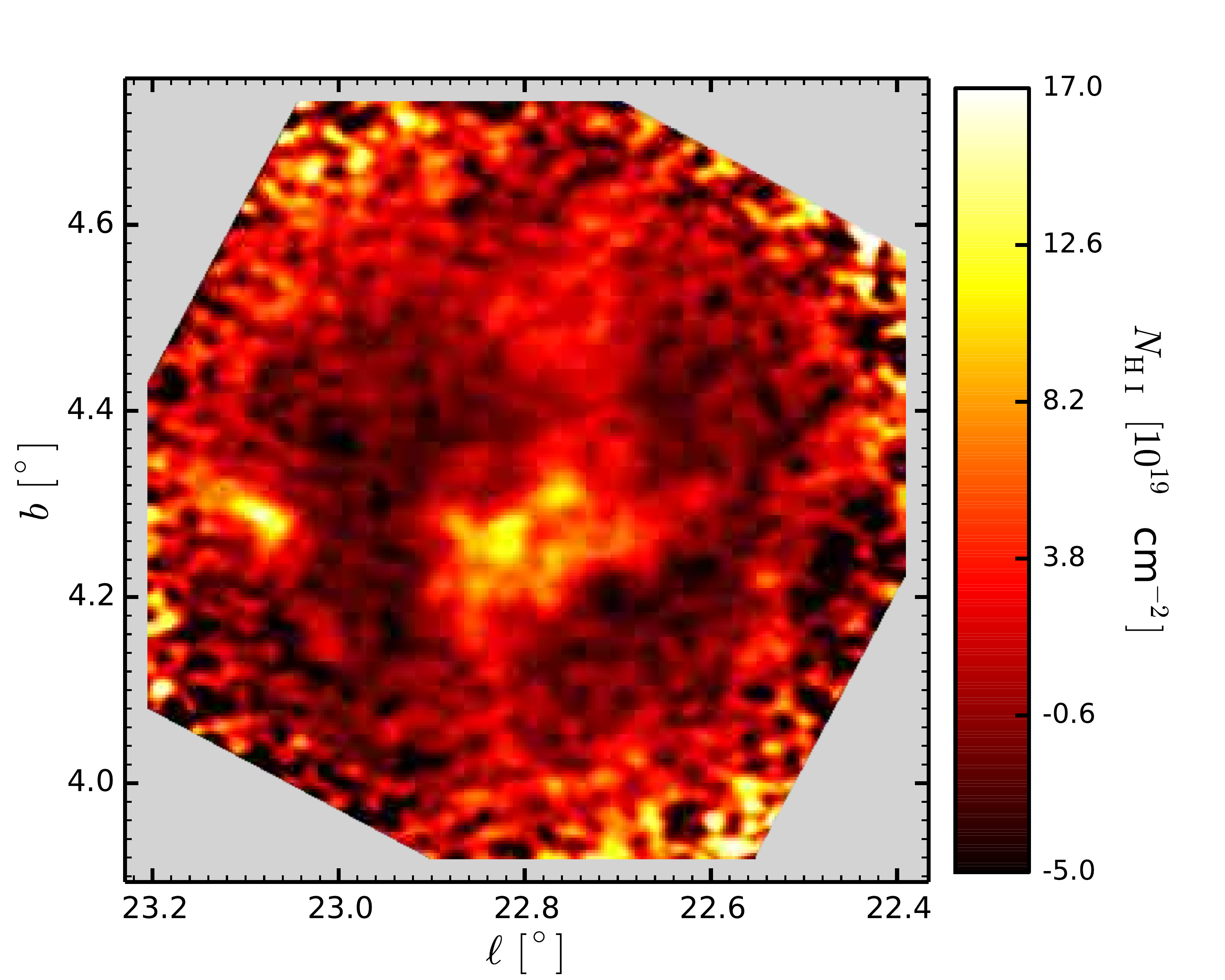}
        \hspace{-0.1cm}}
    \subfloat[][]{
        \centering
        \includegraphics[width=0.5\textwidth]{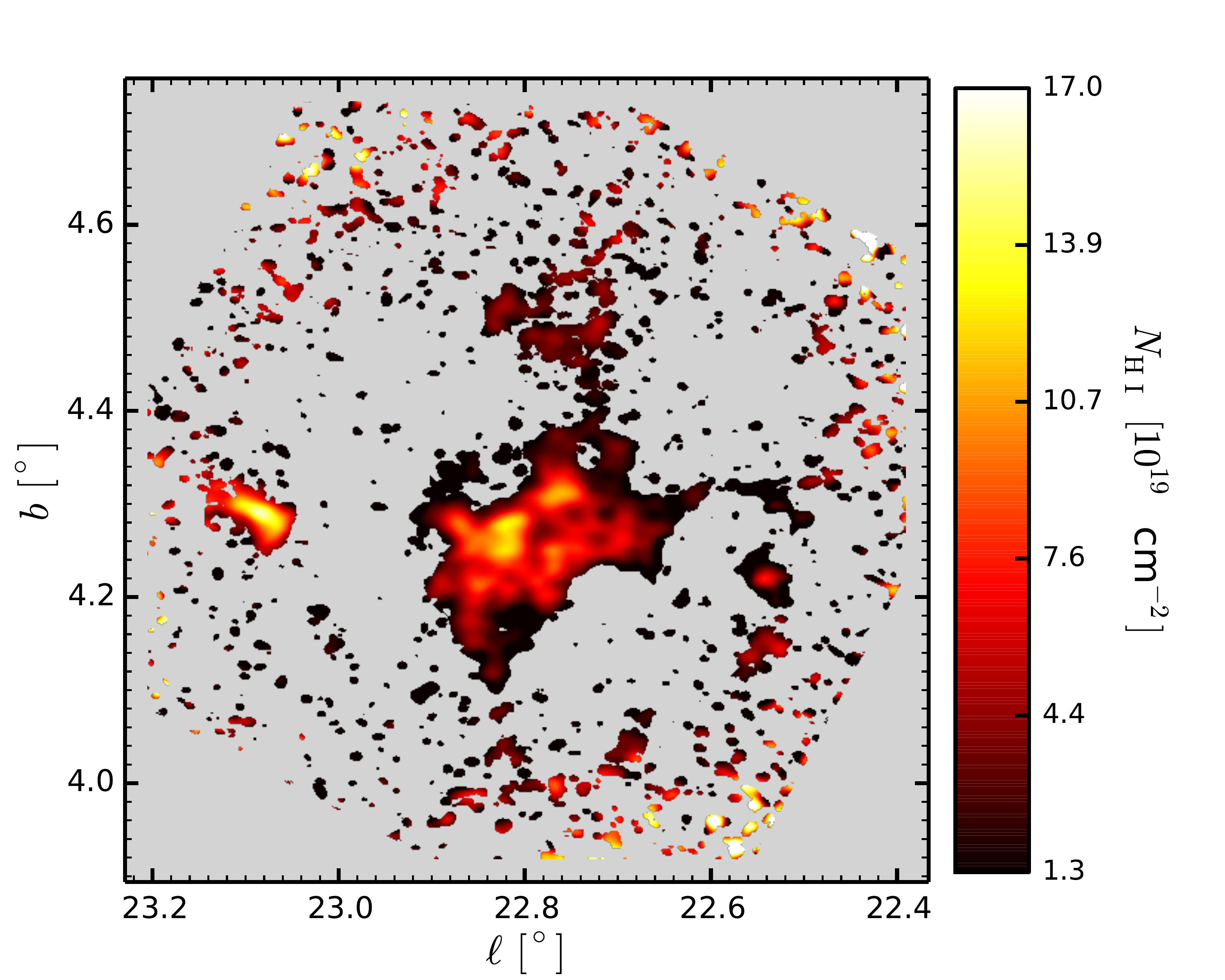}
        }

     \vspace{-1.3cm}

     \subfloat[][]{
        \centering
        \hspace{0.0cm}
        \includegraphics[width=0.5\textwidth]{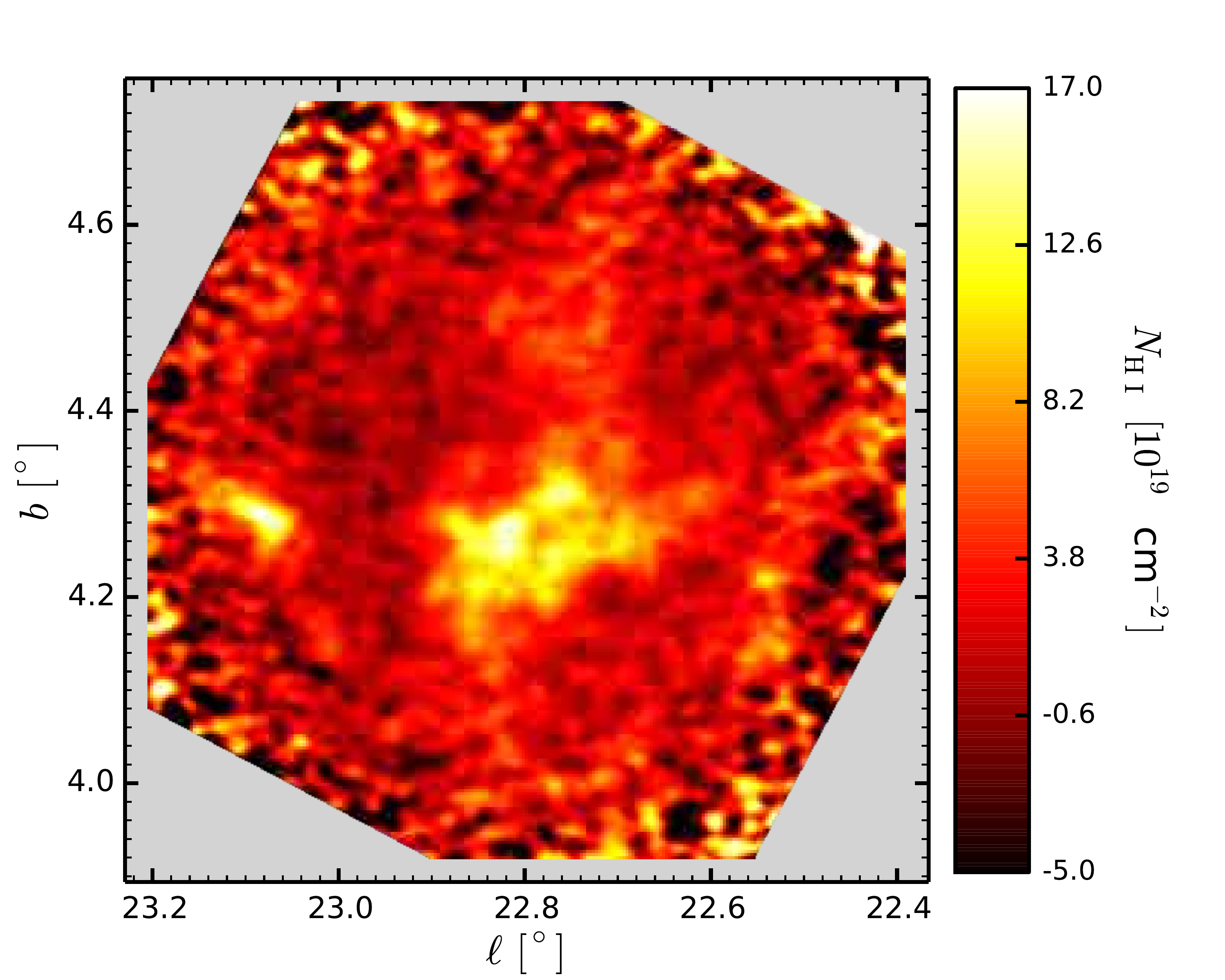}
        \hspace{-0.1cm}}
    \subfloat[][]{
        \centering
        \includegraphics[width=0.5\textwidth]{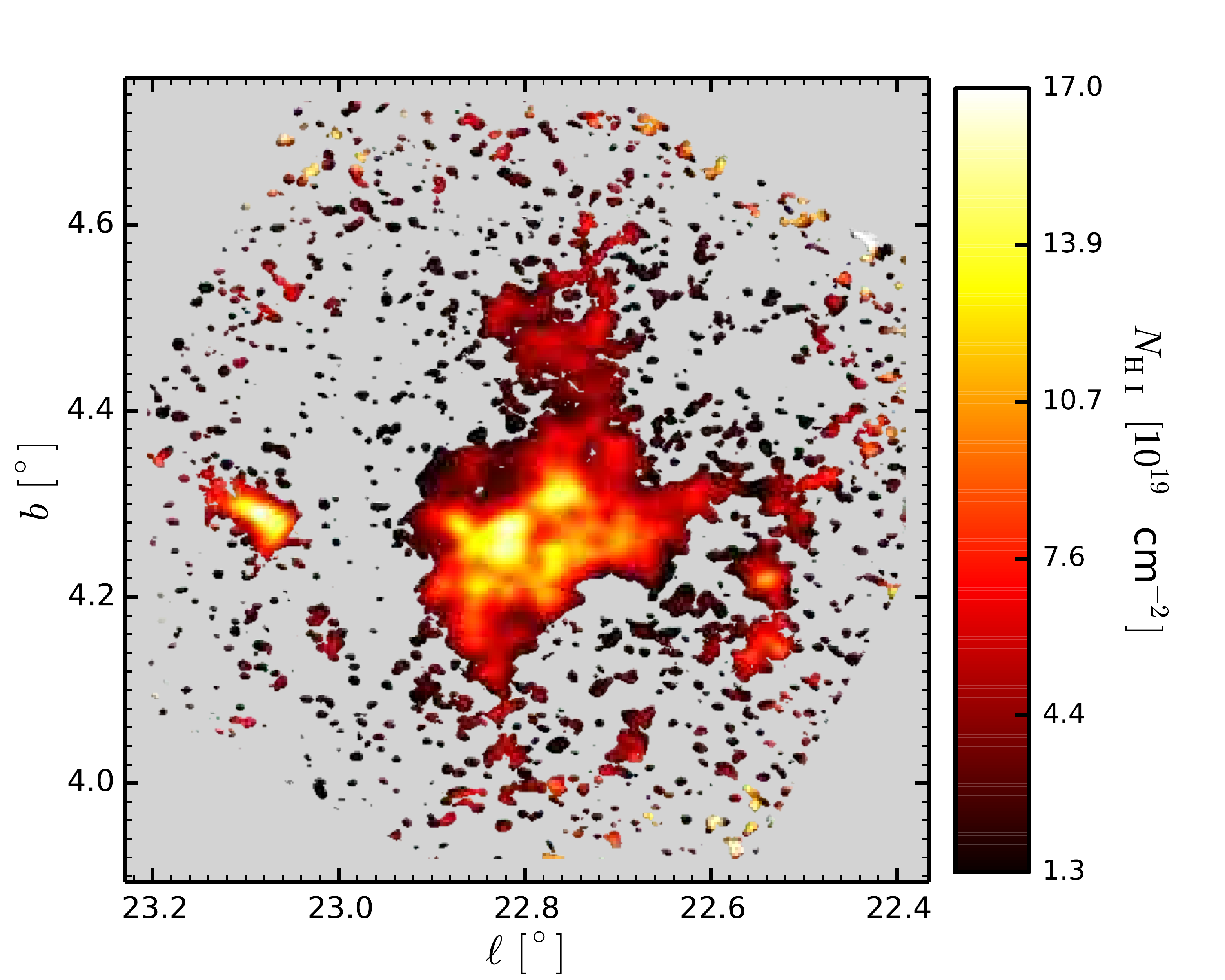}
        }
\vspace{-0.5cm}
\caption{\HI column density maps for G$22.8+4.3$, integrated over 46 spectral channels in the interval $125.2 \leq \VLSR \leq 154.1$~\kms, as described in the caption to Fig.~\ref{fig:160_HImaps}.
}
\label{fig:228_HImaps}
\end{figure}

\begin{figure}
    \centering
    \vspace{-1.0cm}

    \captionsetup[subfigure]{labelformat=empty}
    \subfloat[][]{
        \centering
        \hspace{-0.5cm}
        \includegraphics[width=0.75\textwidth]{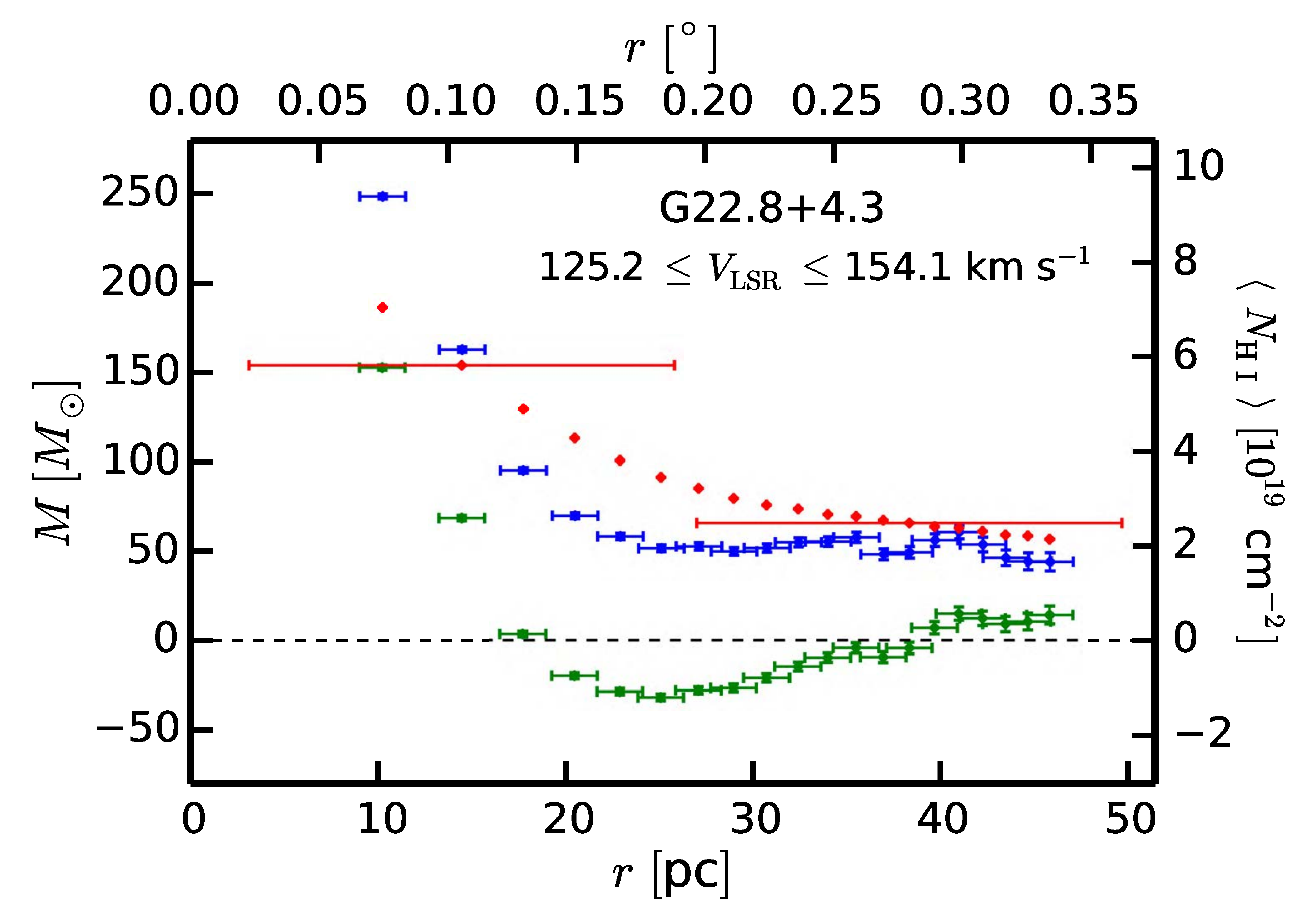}
        }

    \vspace{-1.0cm}

    \subfloat[][]{
        \centering
        \hspace{-1.8cm}
        \includegraphics[width=0.7\textwidth]{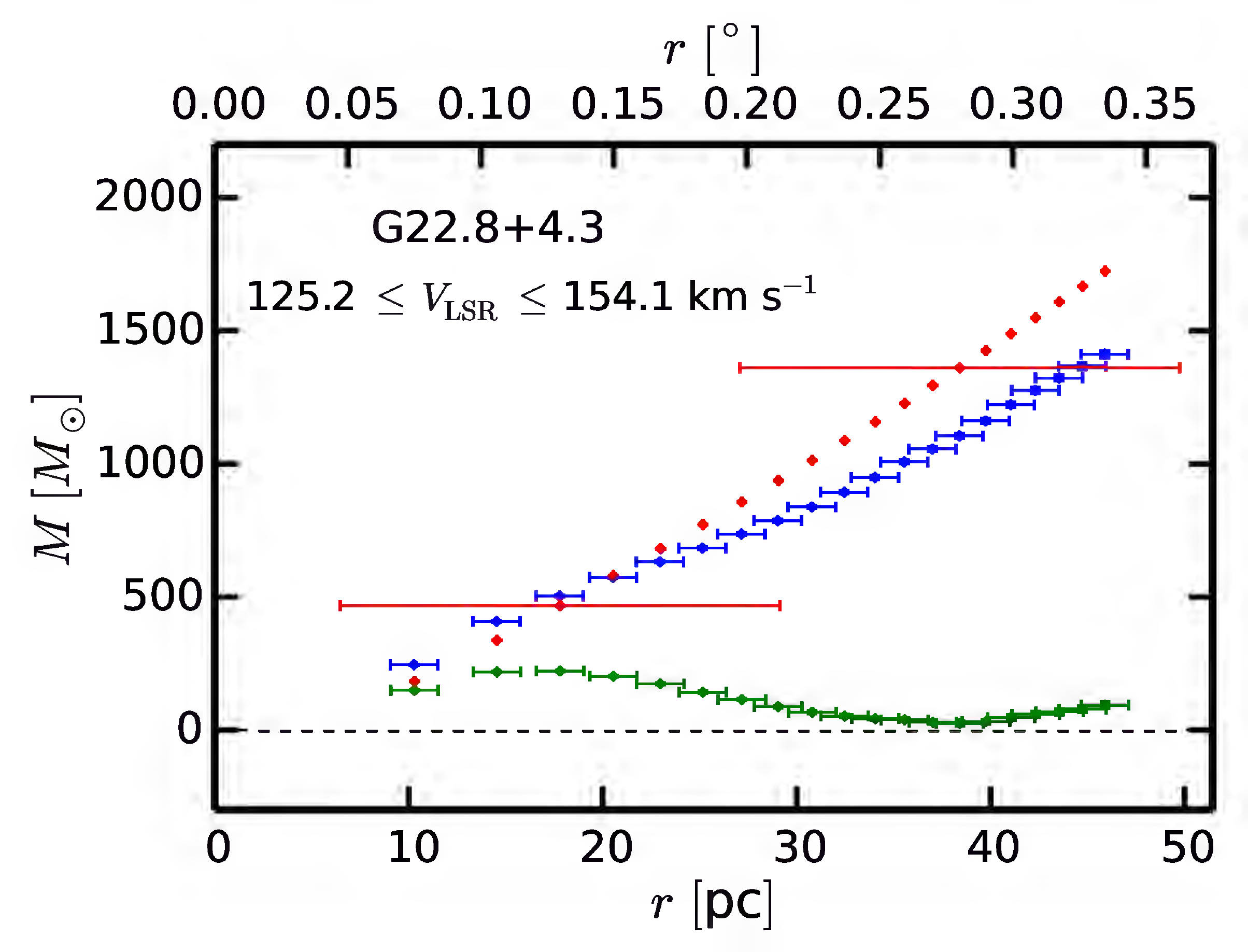}
        }

\caption{Radial mass profiles for G$22.8+4.3$ as described in the caption to Fig.~\ref{fig:160_mass_graphs}.}\label{fig:228_mass_graphs}
\end{figure}

\begin{figure}
\vspace{-2.0cm}
\centering
\captionsetup[subfigure]{labelformat=empty}
    \subfloat[][]{
        \centering
        \hspace{-1cm}
        \includegraphics[width=1.0\textwidth]{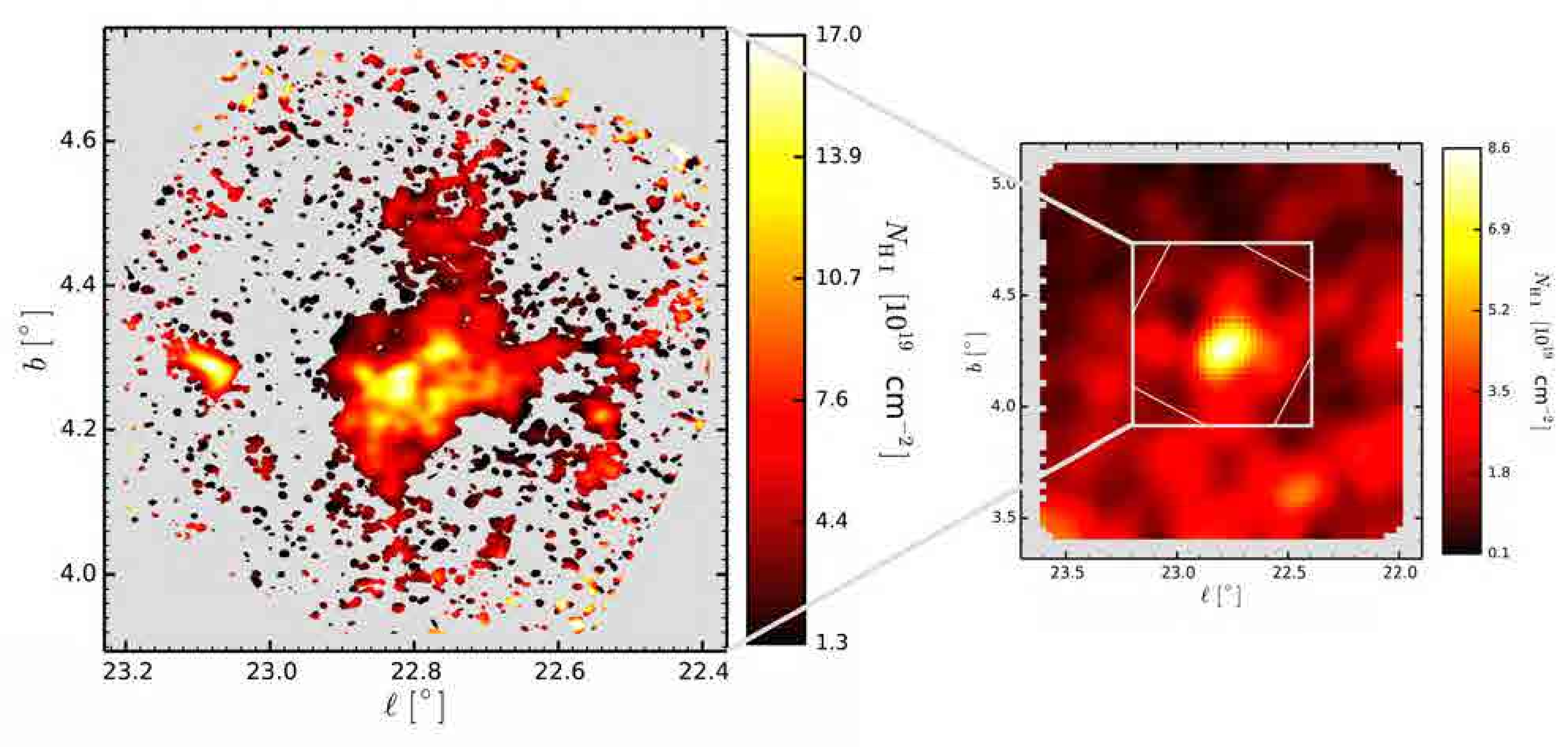}
        \hspace{-0cm}
        }

    \vspace{-1.0cm}

    \subfloat[][]{
        \centering
        \hspace{-1cm}
        \includegraphics[width=0.5\textwidth]{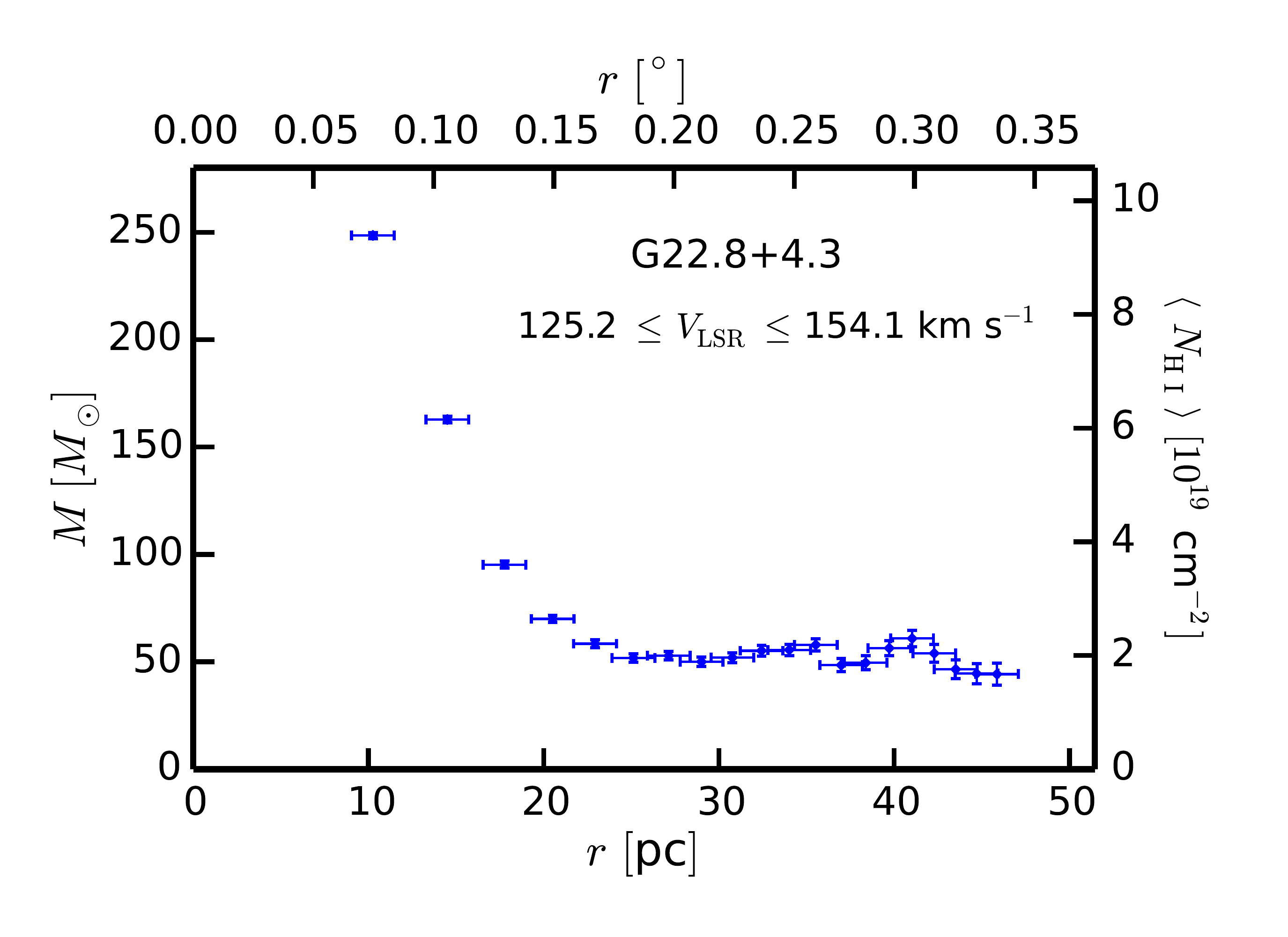}
        \hspace{-0.7cm}
        }
    \subfloat[][]{
        \centering
        \includegraphics[viewport = 0 -30 1150 650, width=0.59\textwidth]{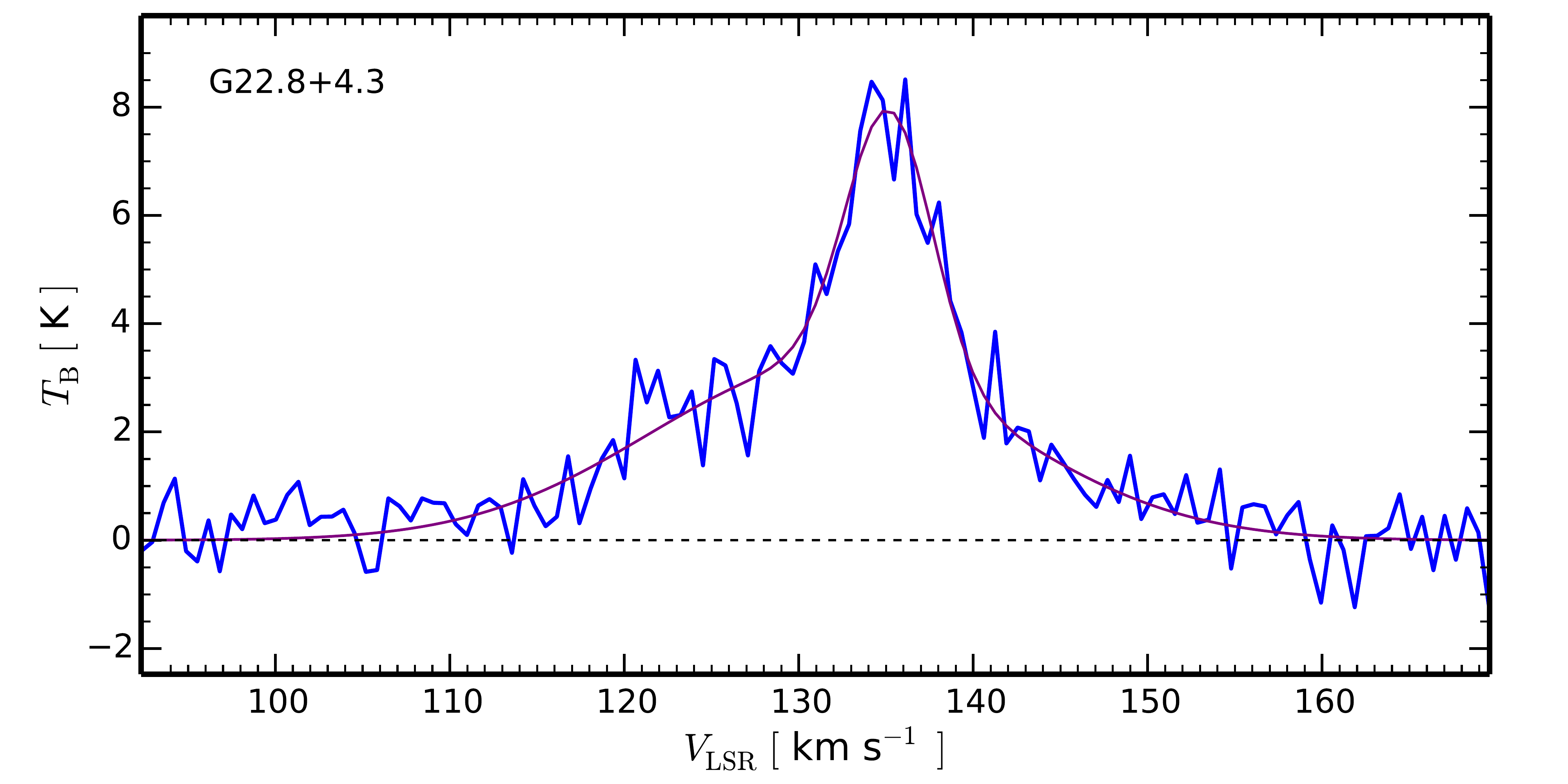}
        }
\vspace{-0.5cm}
\caption{Summary of G$22.8+4.3$ as described in the caption to Fig.~\ref{fig:G160-4plots}. This cloud consists of several fragments, one of which (G$23.1+4.3$) has a size of 20 pc and a mass of 70 \Msun, similar to the small isolated cloud  G$16.0+3.0$. The line at the peak \NHI\ consists of two components, one broad and one narrow.}
\label{fig:G228-4plots}
\end{figure}

\begin{figure}
\centering
    \vspace{-2cm}

    \captionsetup[subfigure]{labelformat=empty}
    \subfloat[][]{
        \centering
        \hspace{0.0cm}
        \includegraphics[width=0.5\textwidth]{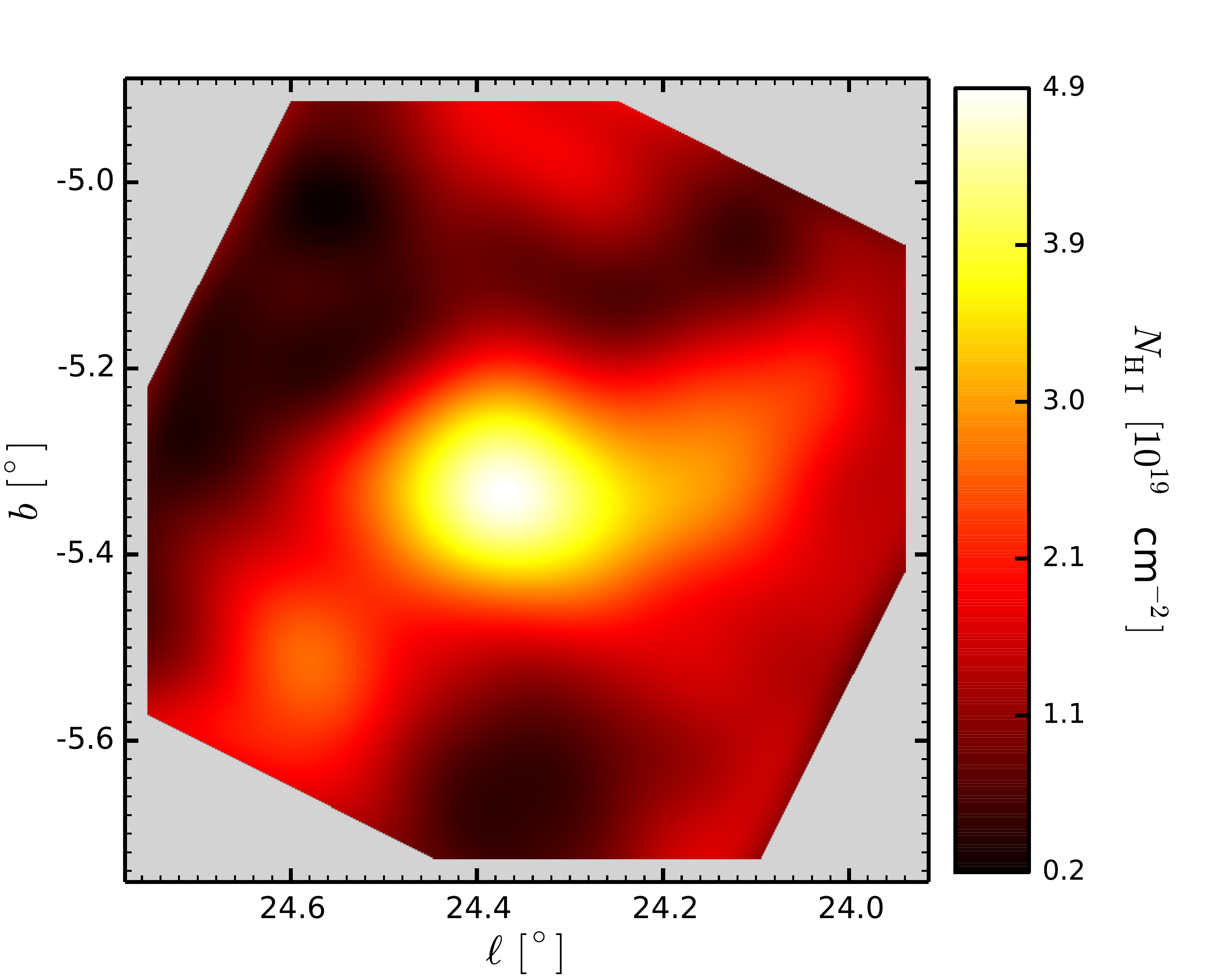}
        \hspace{-0.1cm}}
    \subfloat[][]{
        \centering
        \includegraphics[width=0.5\textwidth]{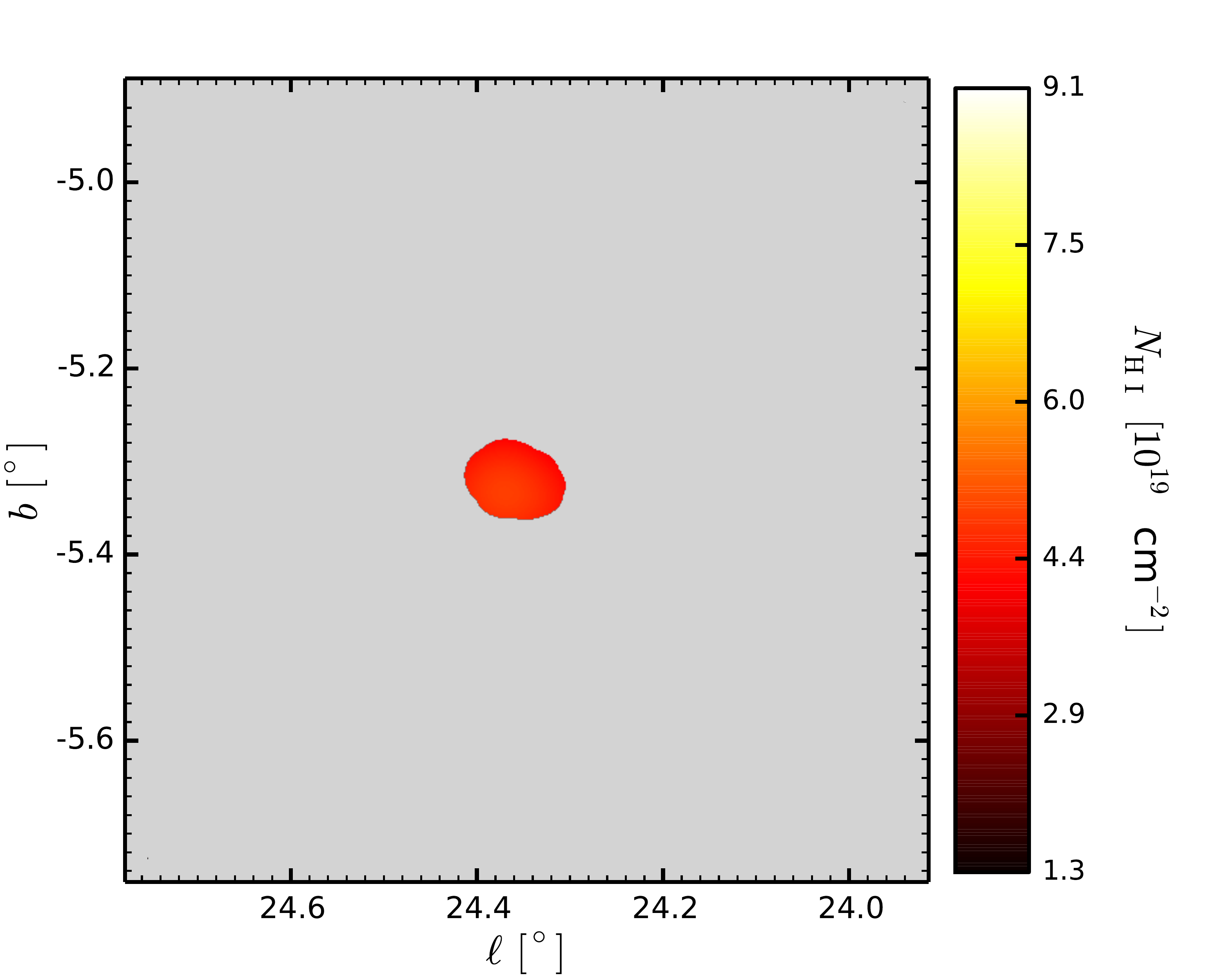}
        }

    \vspace{-1.3cm}

    \subfloat[][]{
        \centering
        \hspace{0.0cm}
        \includegraphics[width=0.5\textwidth]{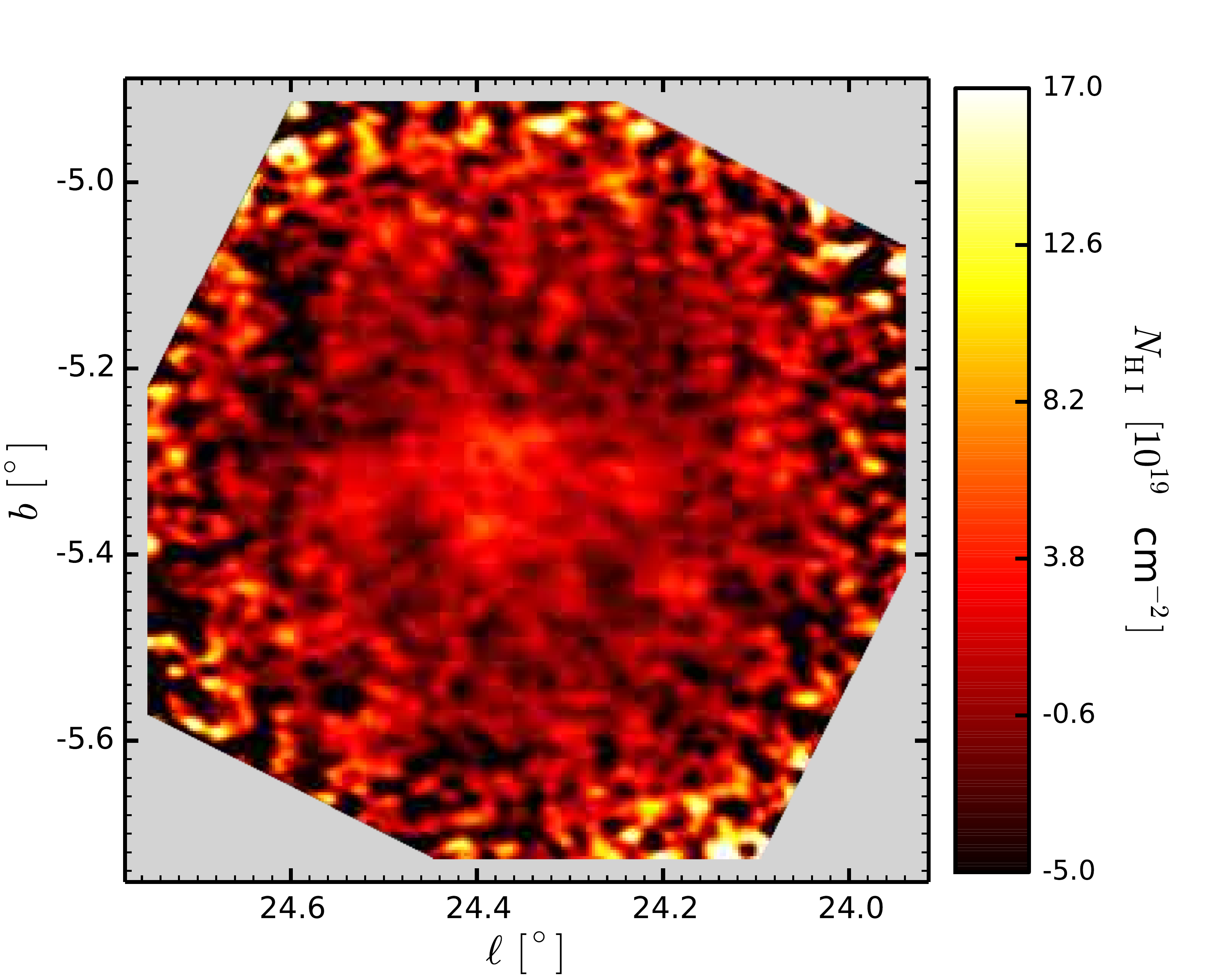}
        \hspace{-0.1cm}}
    \subfloat[][]{
        \centering
        \includegraphics[width=0.5\textwidth]{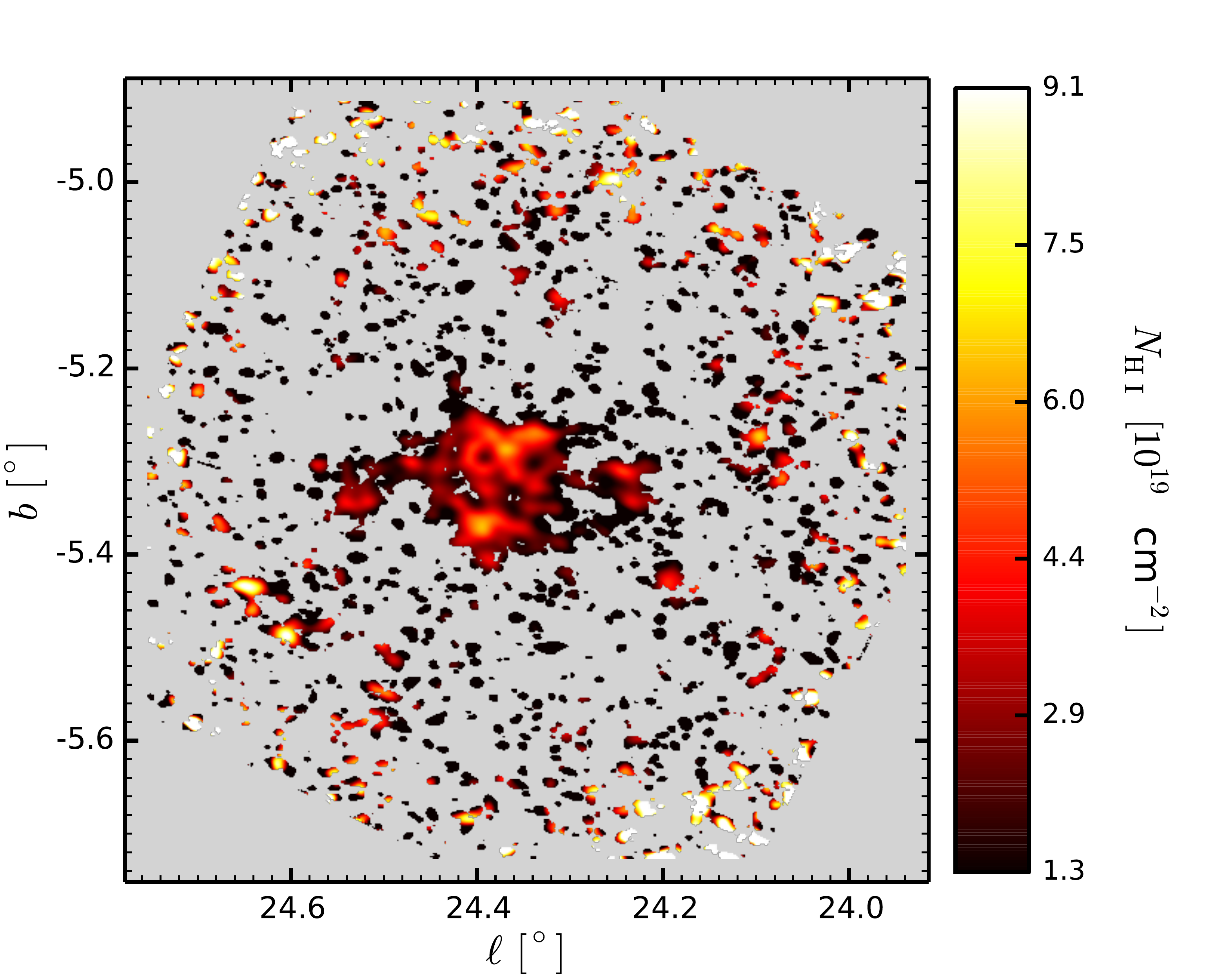}
        }

     \vspace{-1.3cm}

     \subfloat[][]{
        \centering
        \hspace{0.0cm}
        \includegraphics[width=0.5\textwidth]{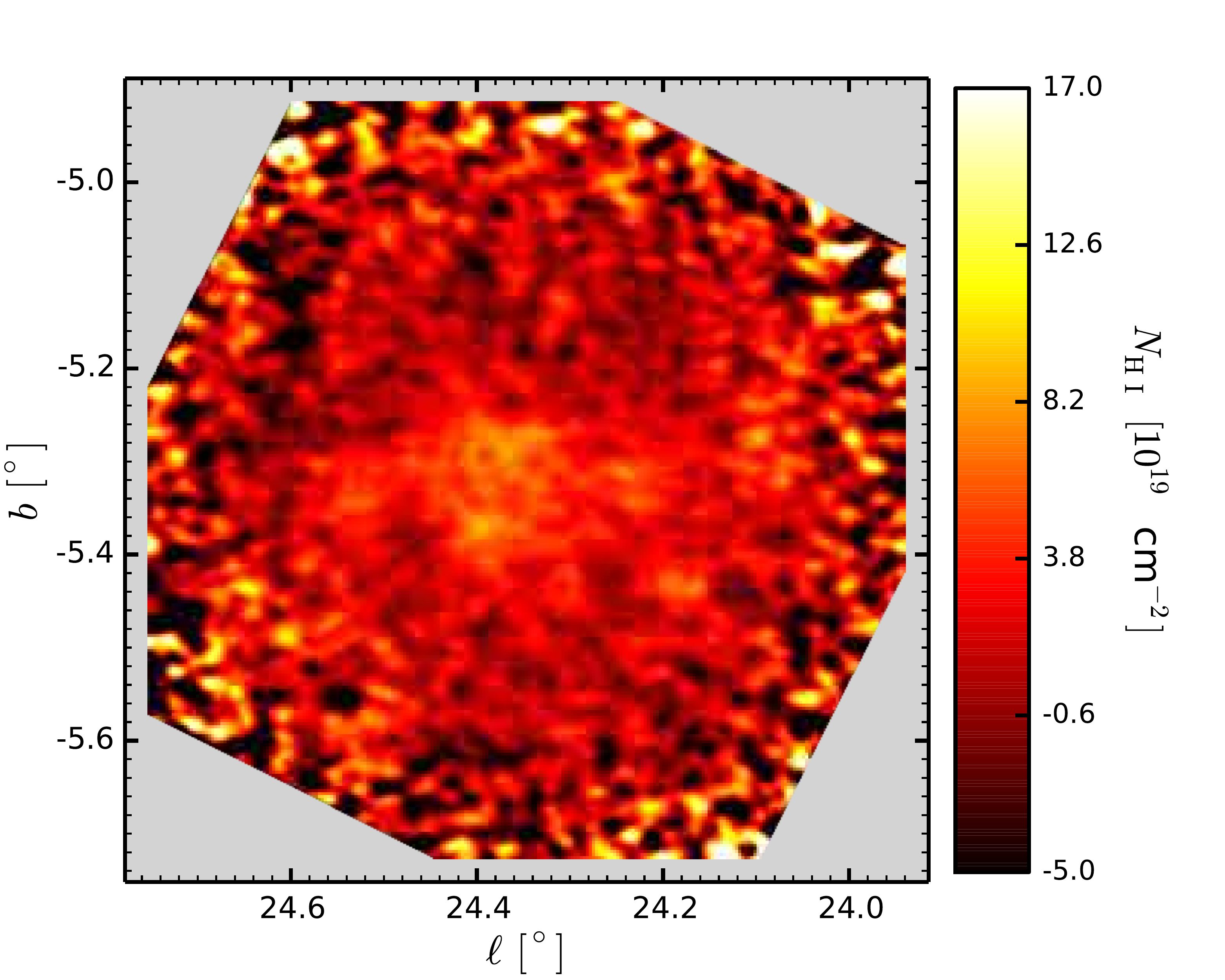}
        \hspace{-0.1cm}}
    \subfloat[][]{
        \centering
        \includegraphics[width=0.5\textwidth]{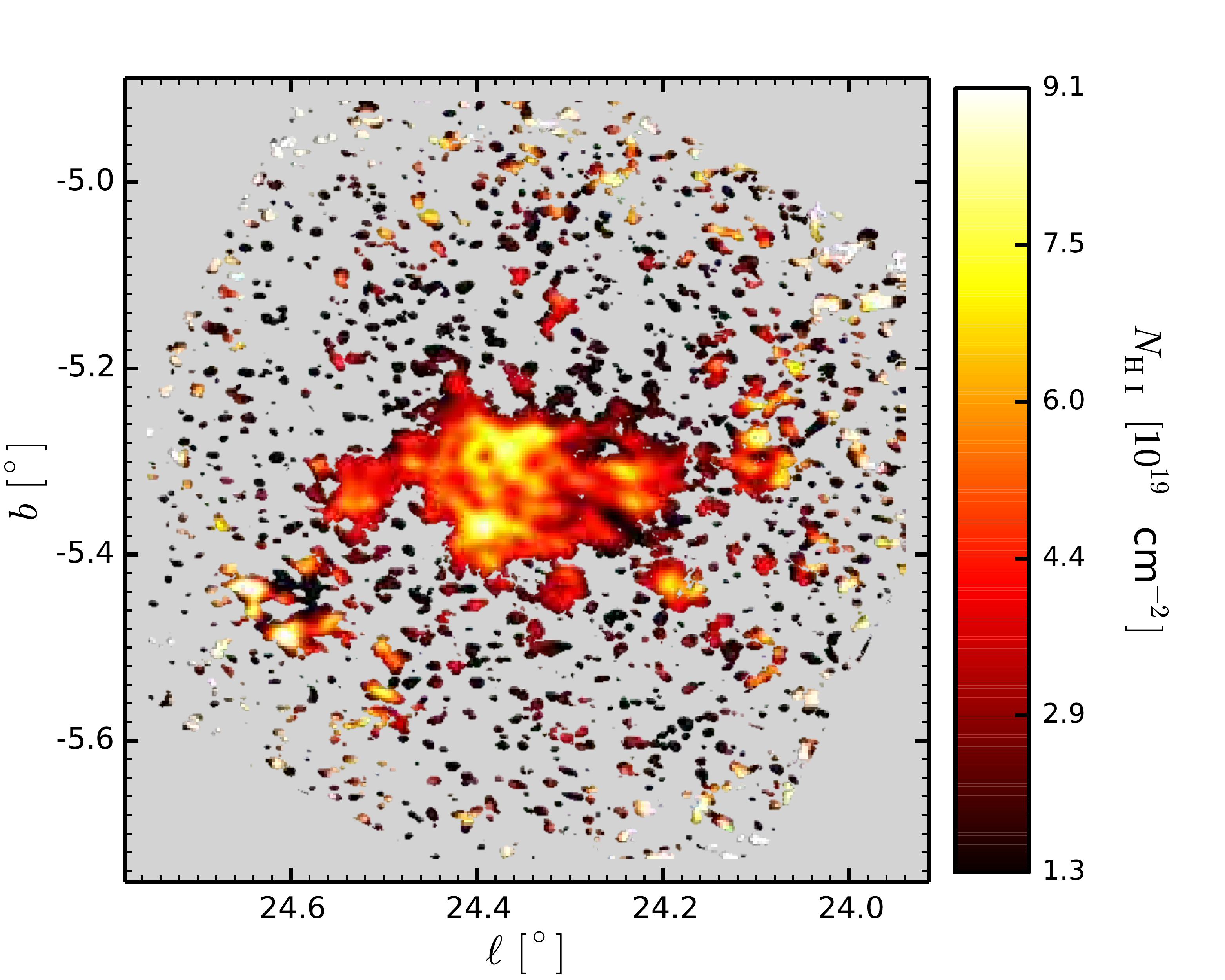}
        }
\vspace{-0.5cm}
\caption{\HI column density maps for G$24.3-5.3$, integrated over 70 spectral channels in the interval $107.0 \leq \VLSR \leq 151.4$~\kms, as described in the caption to Fig.~\ref{fig:160_HImaps}.
}
\label{fig:243_HImaps}
\end{figure}

\begin{figure}
    \centering
    \vspace{-1.0cm}

    \captionsetup[subfigure]{labelformat=empty}
    \subfloat[][]{
        \centering
        \hspace{-0.5cm}
        \includegraphics[width=0.75\textwidth]{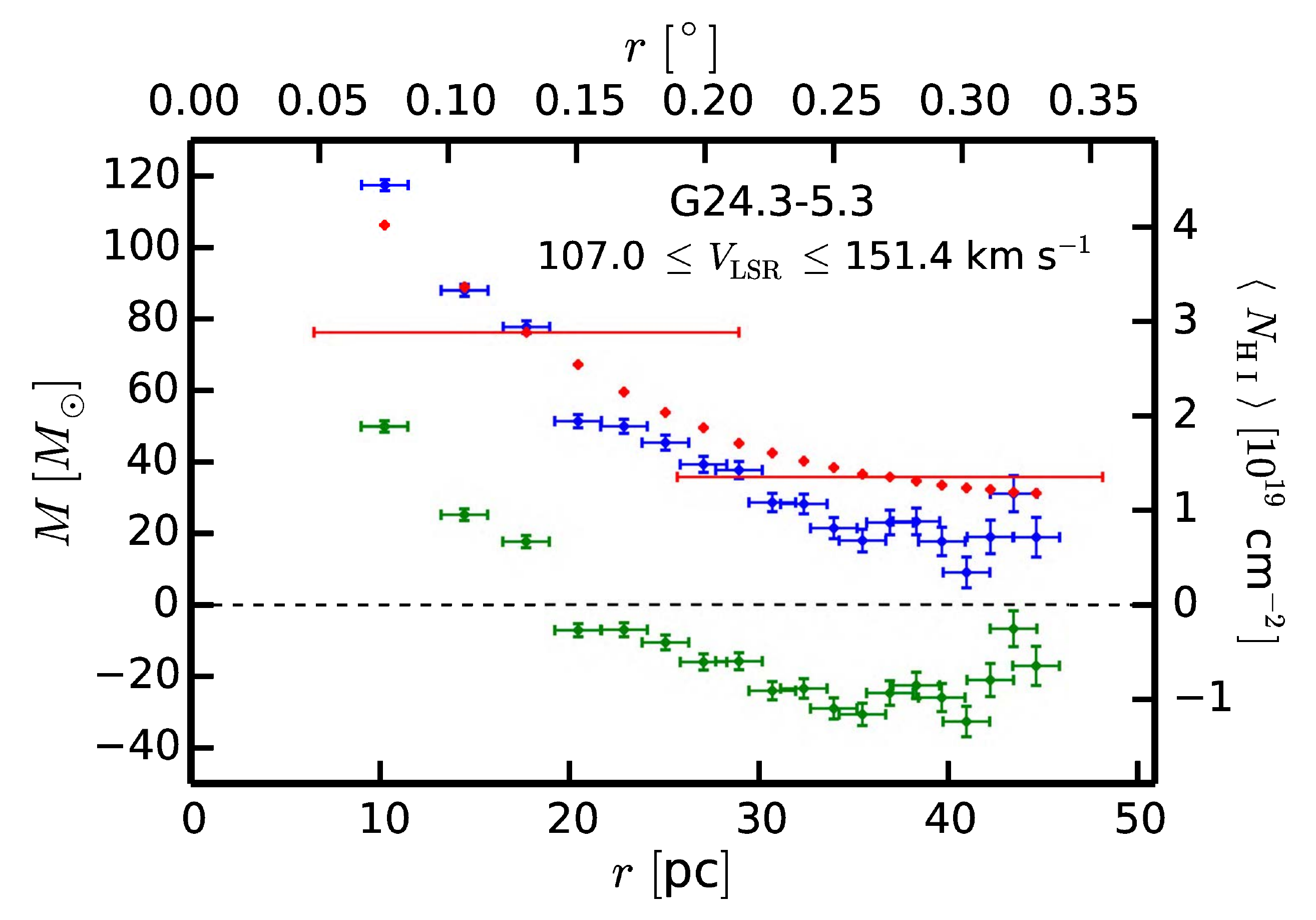}
        }

    \vspace{-1.0cm}

    \subfloat[][]{
        \centering
        \hspace{-1.8cm}
        \includegraphics[width=0.7\textwidth]{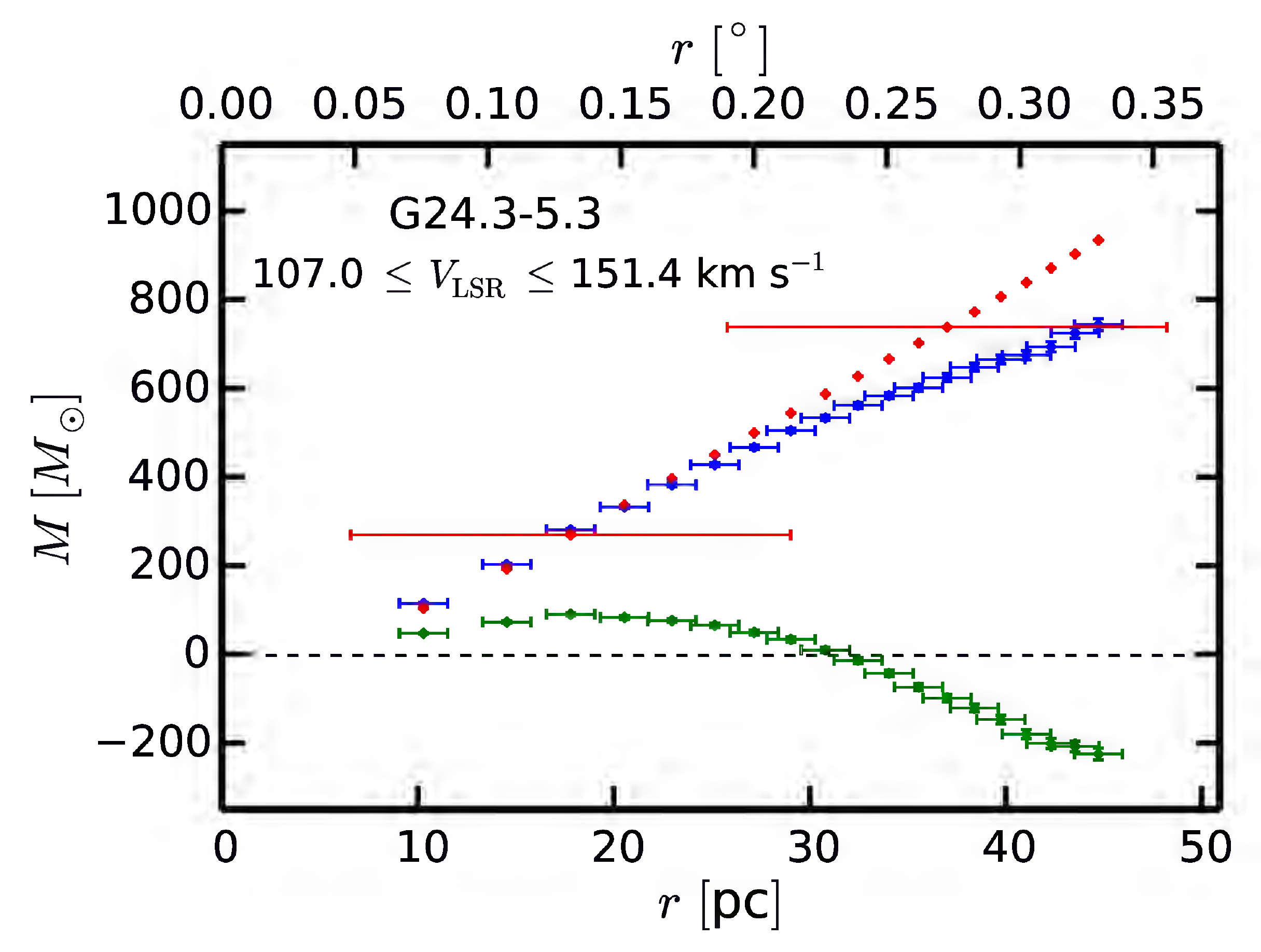}
        }

\caption{Radial mass profiles for G$24.3-5.3$ as described in the caption to Fig.~\ref{fig:160_mass_graphs}.}\label{fig:243_mass_graphs}
\end{figure}

\begin{figure}
\vspace{-2.0cm}
\centering
\captionsetup[subfigure]{labelformat=empty}
    \subfloat[][]{
        \centering
        \hspace{-1cm}
        \includegraphics[width=1.0\textwidth]{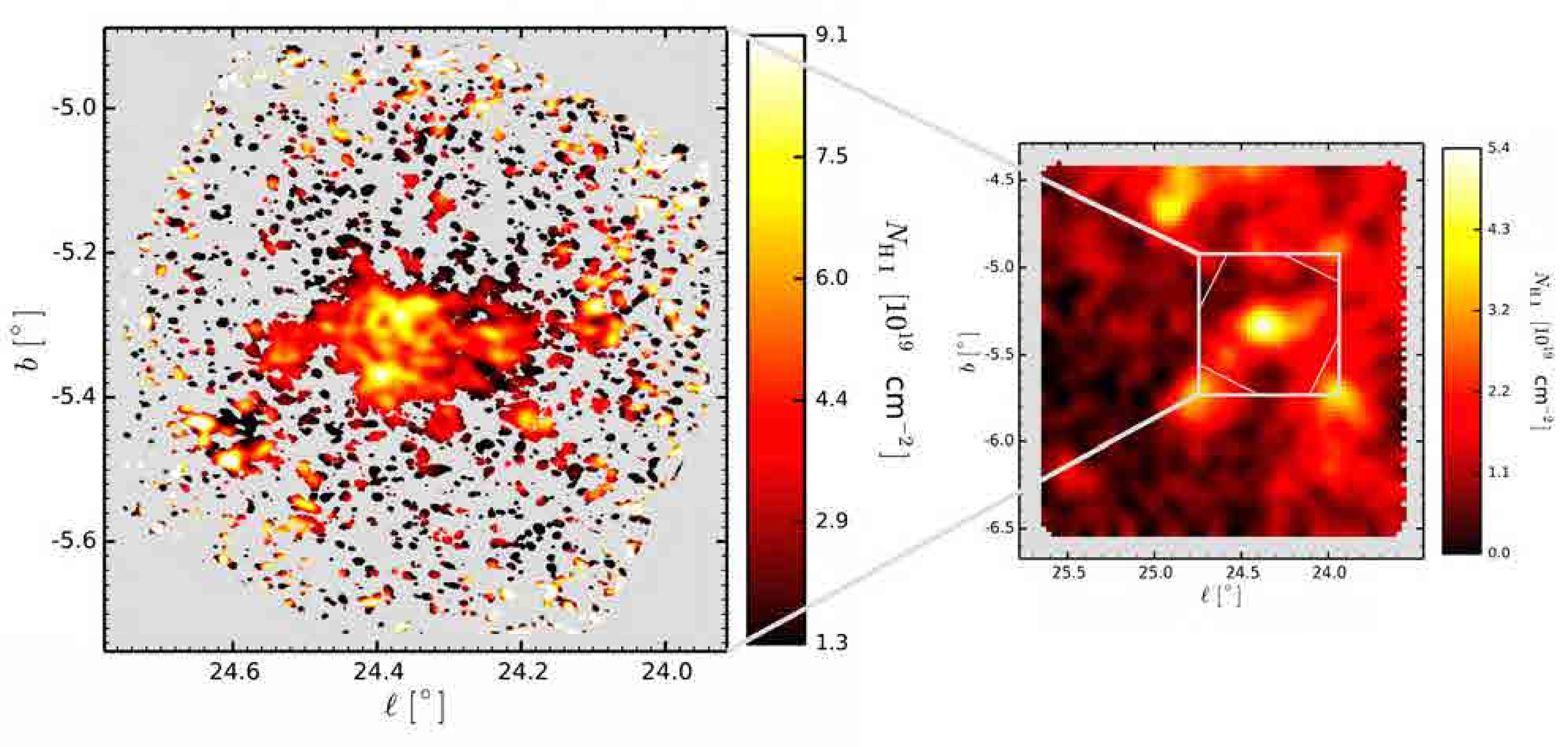}
        \hspace{-0cm}
        }

    \vspace{-1.0cm}

    \subfloat[][]{
        \centering
        \hspace{-1cm}
        \includegraphics[width=0.5\textwidth]{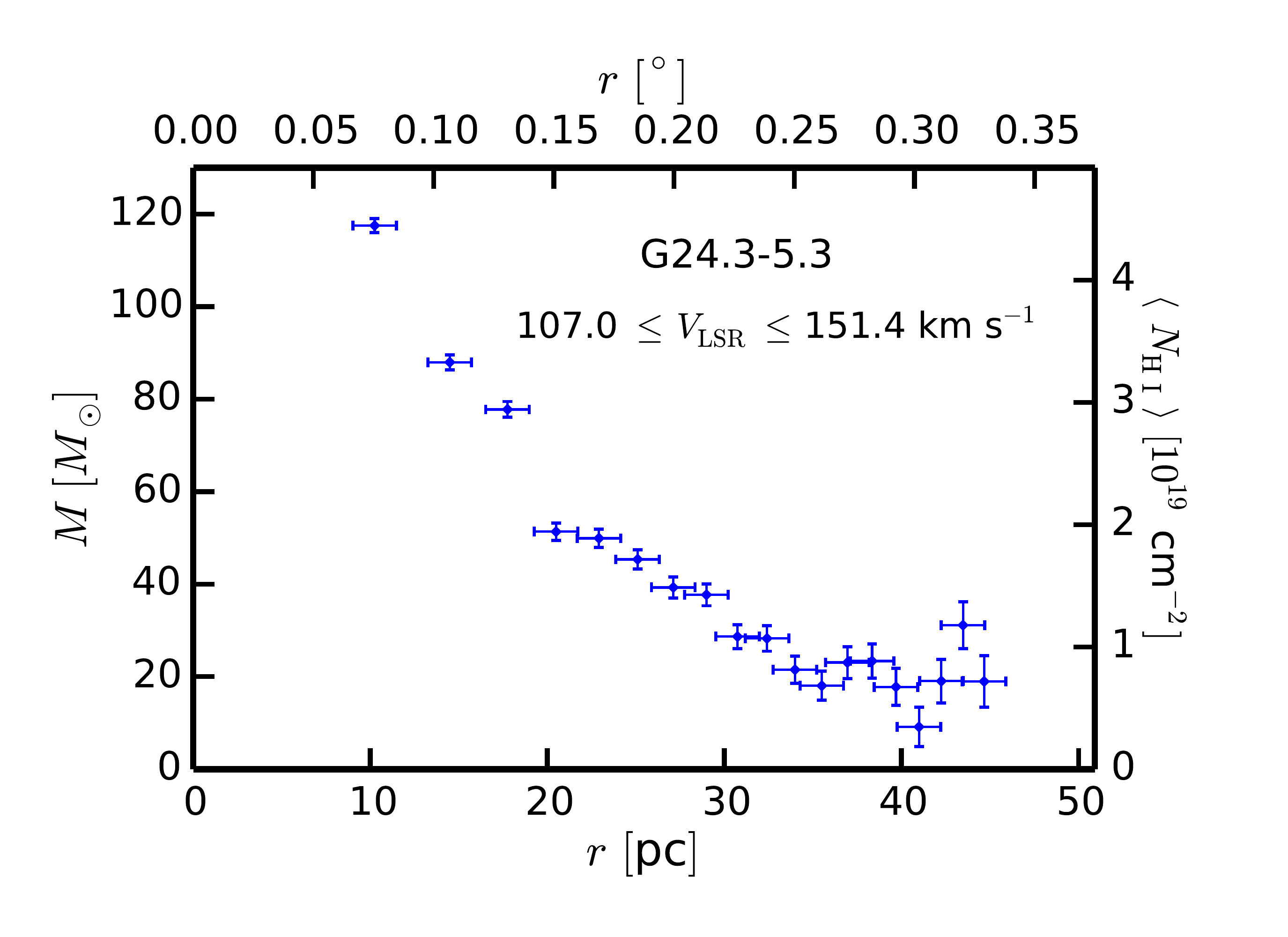}
        \hspace{-0.7cm}
        }
    \subfloat[][]{
        \centering
        \includegraphics[viewport = 0 -30 1150 650, width=0.59\textwidth]{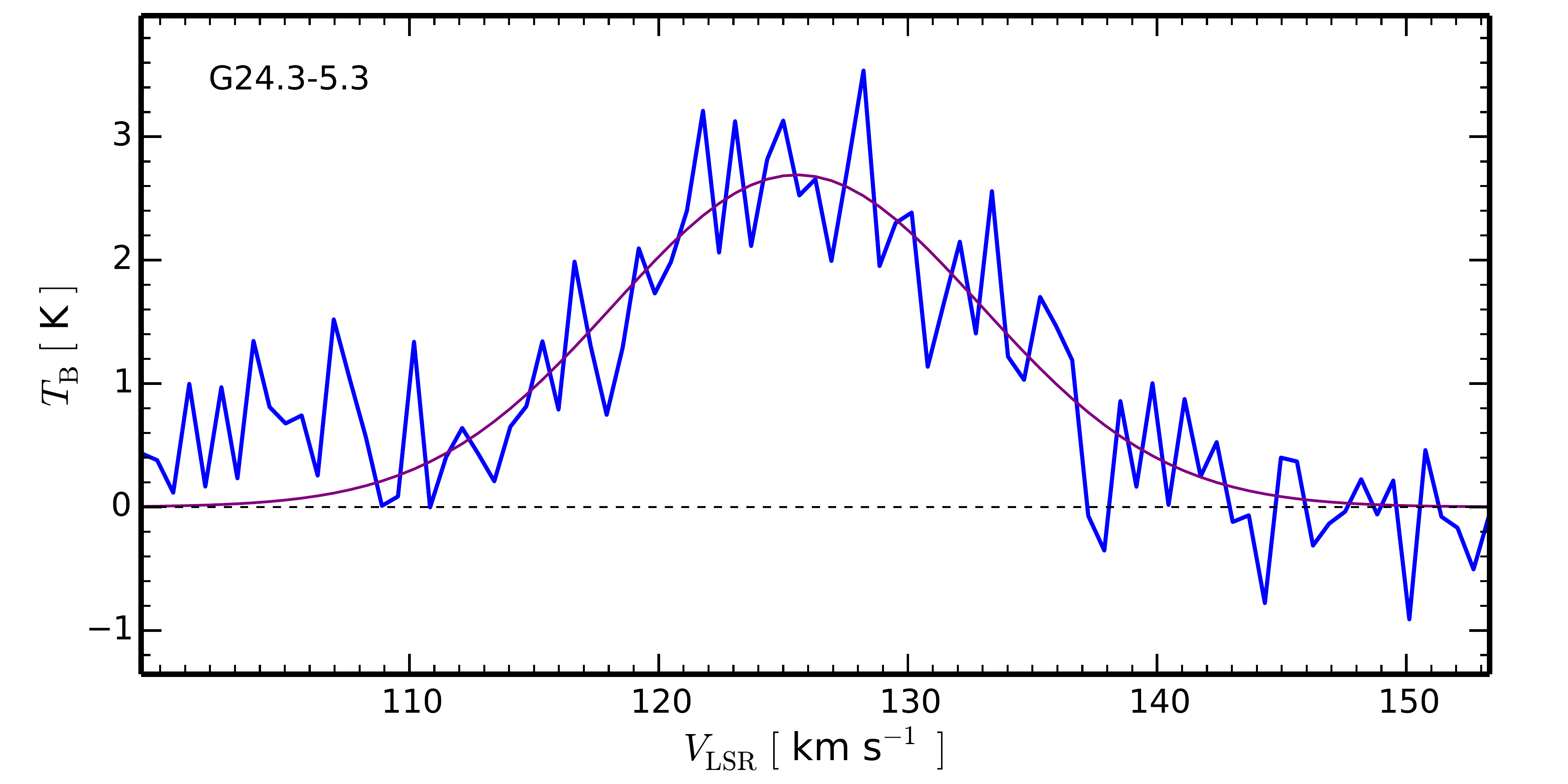}
        }
\vspace{-0.5cm}
\caption{Summary of G$24.3-5.3$ as described in the caption to Fig.~\ref{fig:G160-4plots}. This cloud, together with G$24.7-5.7$ and a smaller one that lies between them, form a chain of three clouds connected by diffuse \HI emission.  G$24.3-5.3$ has no clear core, but rather several column density peaks of similar values spread over the body of the cloud.  Its radial mass profile shows the lack of central concentration. In contrast to G$16.0+3.0$, which is a core without an envelope, this cloud is an envelope without a core. With a FWHM of $17.4\pm1.1$~\kms, its line is one of the broadest in the sample.}
\label{fig:G243-4plots}
\end{figure}

\begin{figure}
\centering
    \vspace{-2cm}

    \captionsetup[subfigure]{labelformat=empty}
    \subfloat[][]{
        \centering
        \hspace{0.0cm}
        \includegraphics[width=0.5\textwidth]{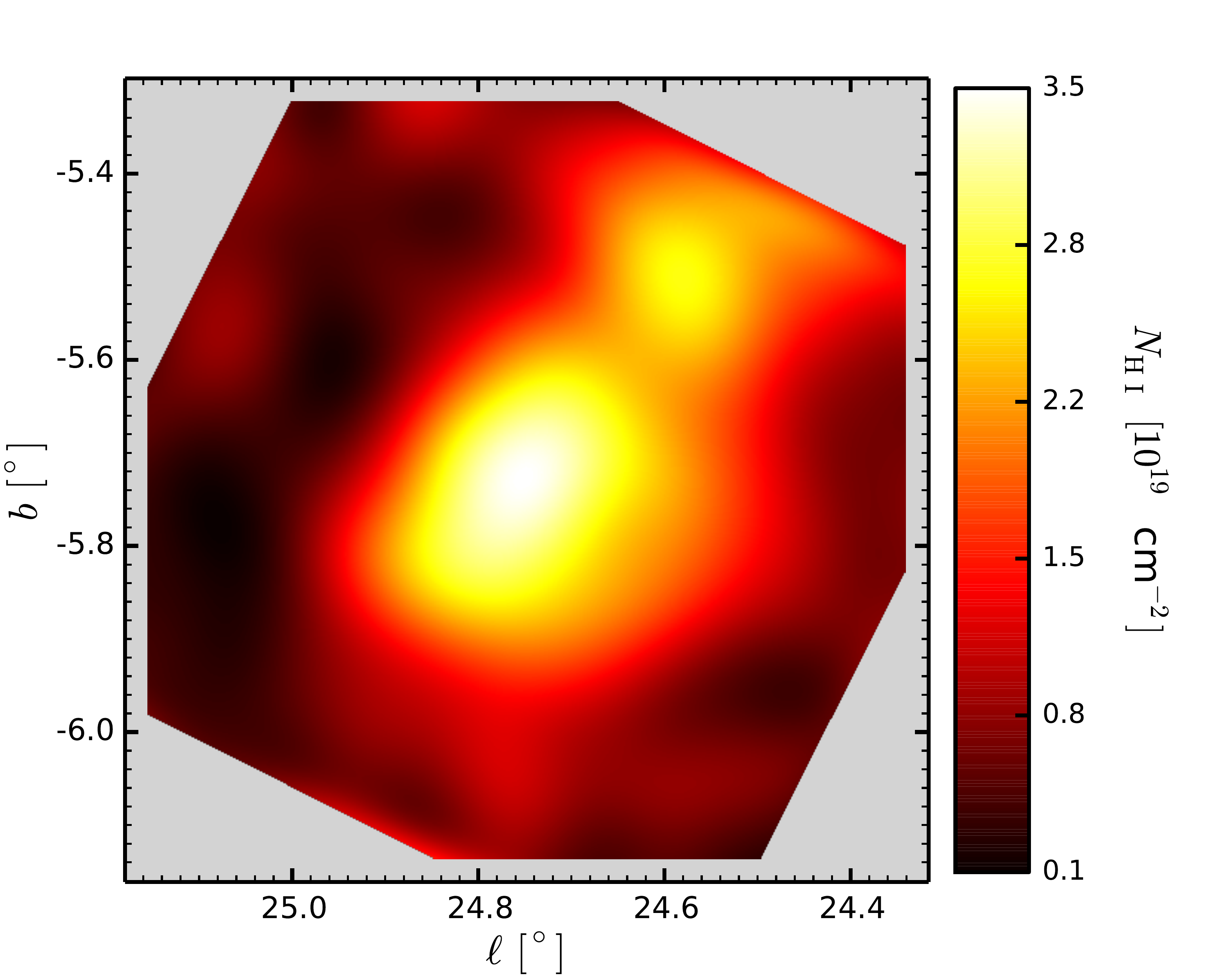}
        \hspace{-0.1cm}}
    \subfloat[][]{
        \centering
        \includegraphics[width=0.5\textwidth]{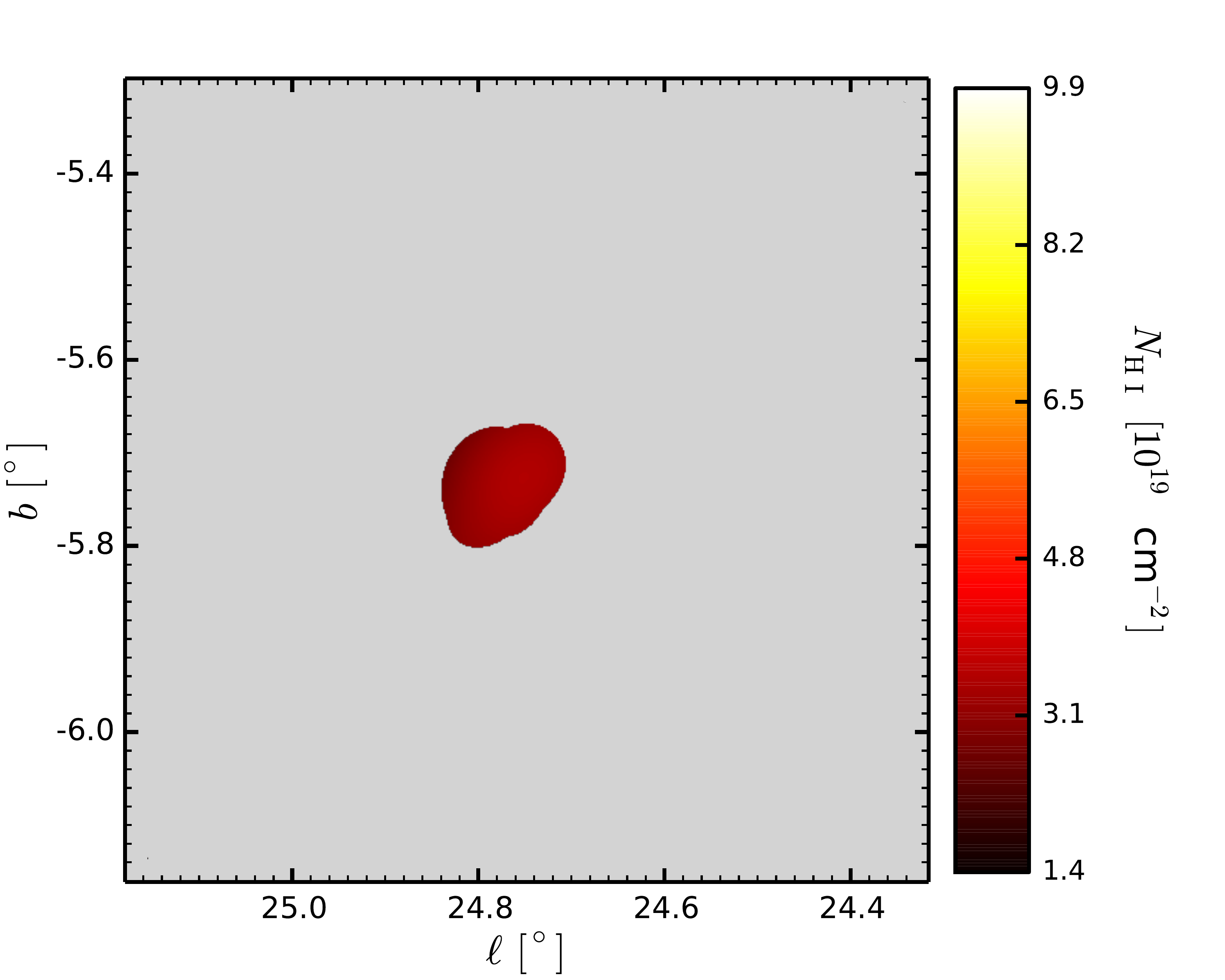}
        }

    \vspace{-1.3cm}

    \subfloat[][]{
        \centering
        \hspace{0.0cm}
        \includegraphics[width=0.5\textwidth]{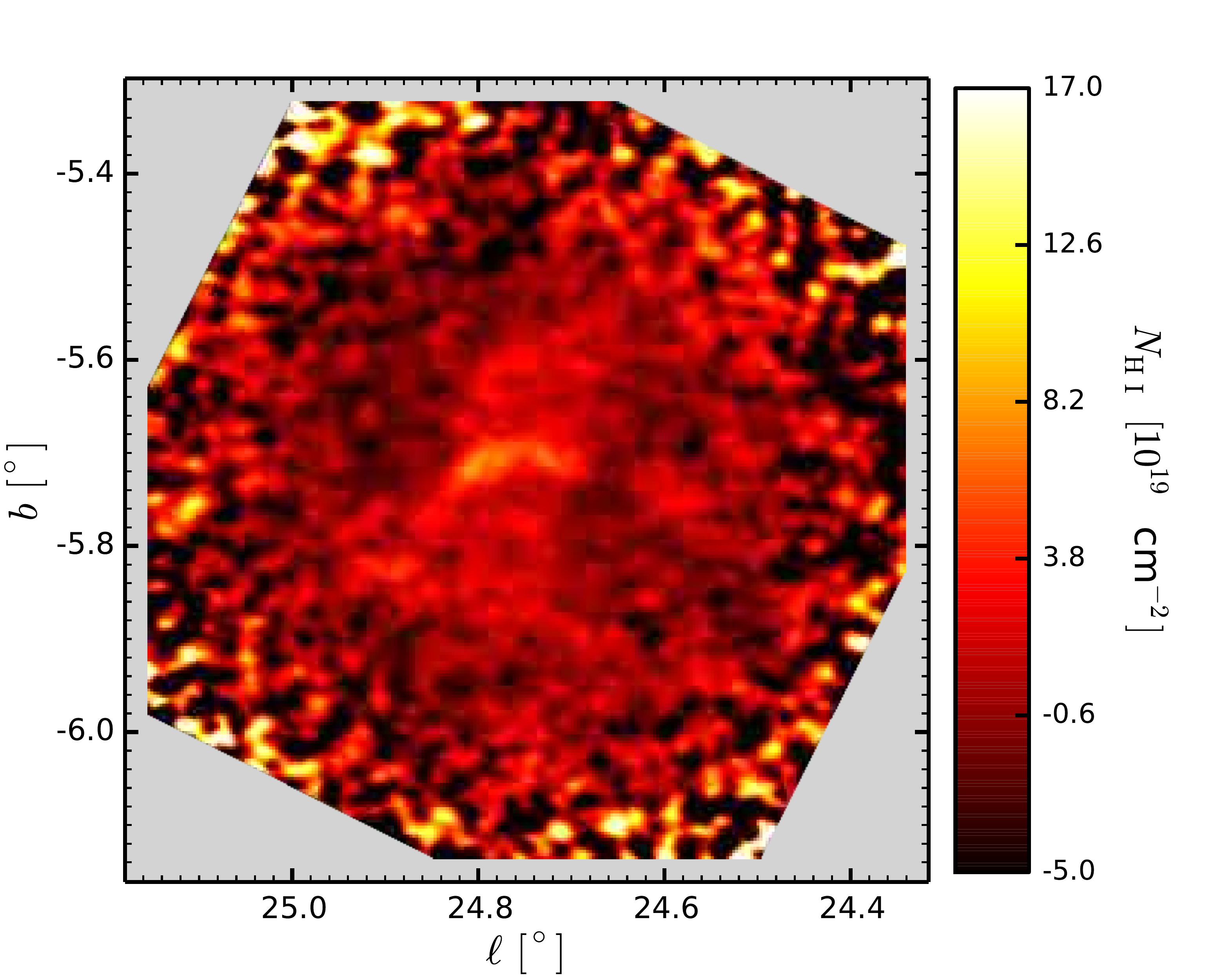}
        \hspace{-0.1cm}}
    \subfloat[][]{
        \centering
        \includegraphics[width=0.5\textwidth]{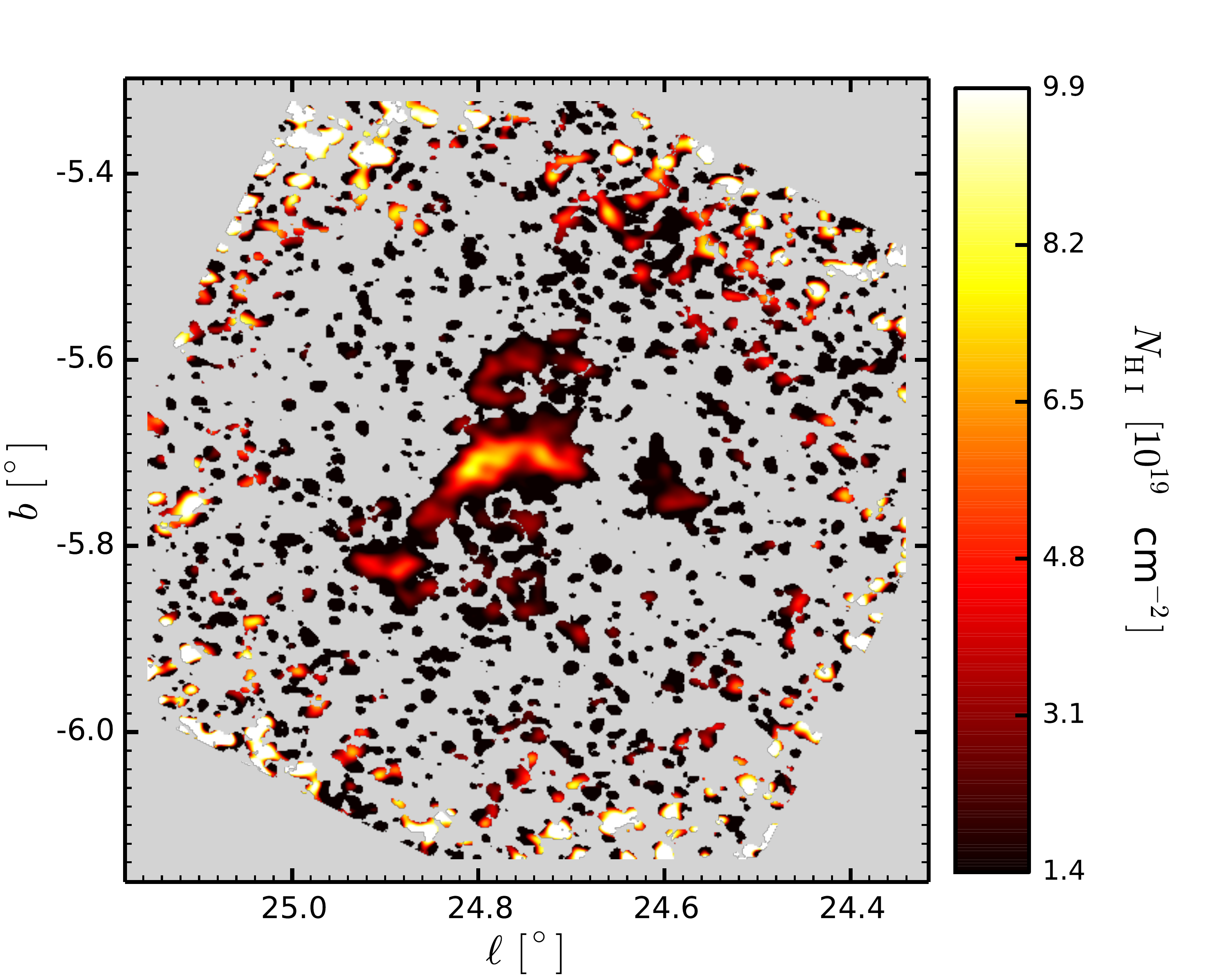}
        }

     \vspace{-1.3cm}

     \subfloat[][]{
        \centering
        \hspace{0.0cm}
        \includegraphics[width=0.5\textwidth]{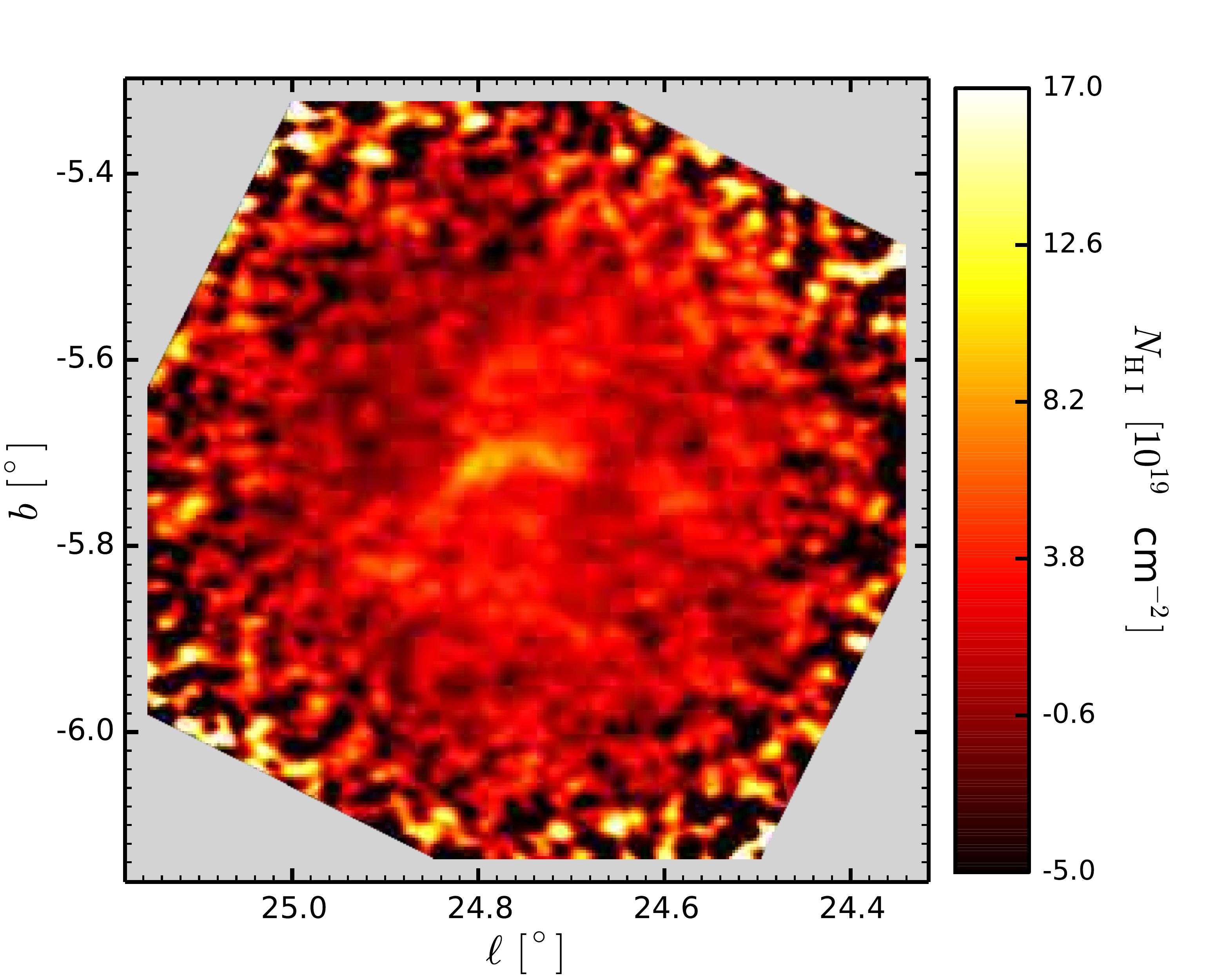}
        \hspace{-0.1cm}}
    \subfloat[][]{
        \centering
        \includegraphics[width=0.5\textwidth]{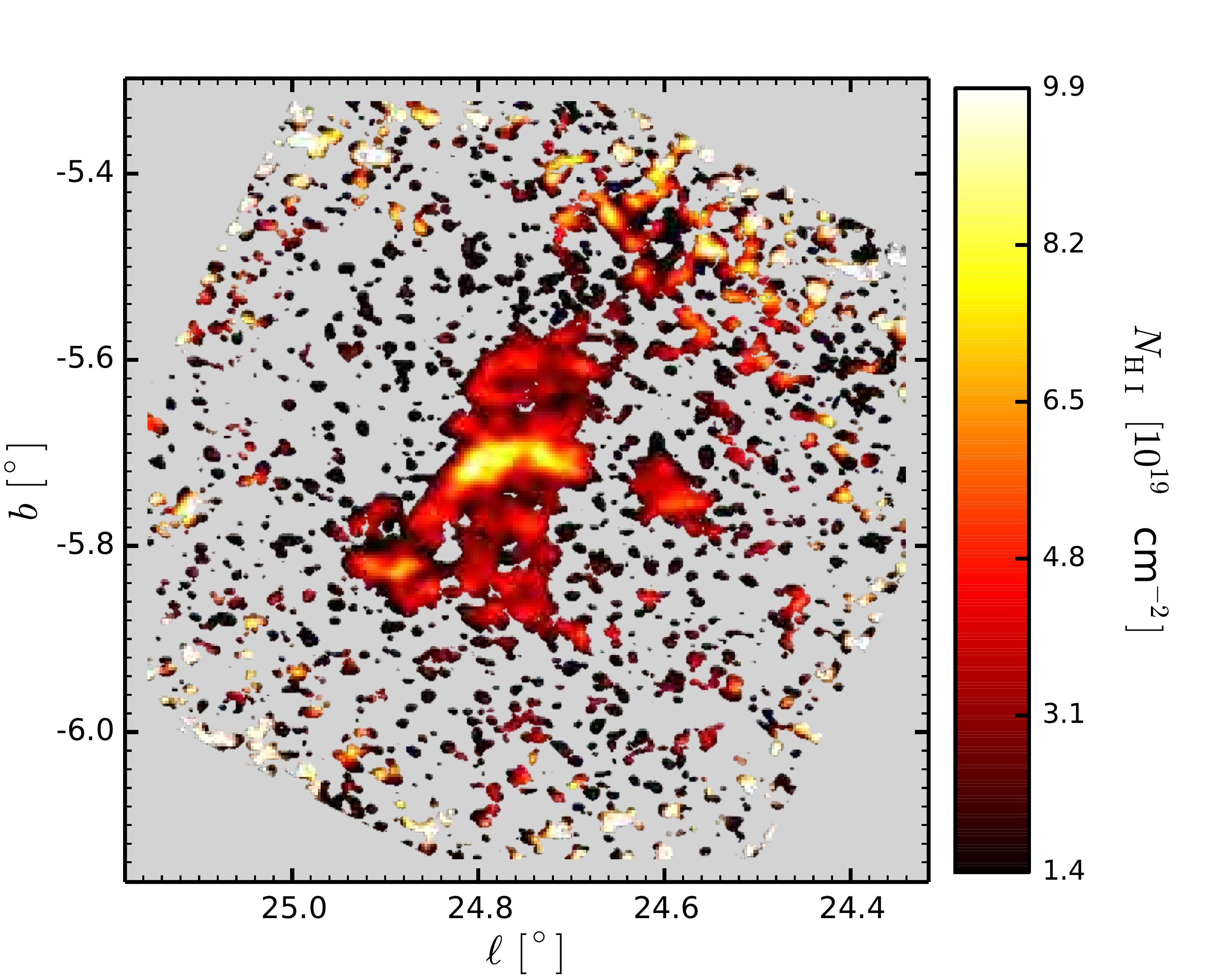}
        }
\vspace{-0.5cm}
\caption{\HI column density maps for G$24.7-5.7$, integrated over 66 spectral channels in the interval $103.7 \leq \VLSR \leq 145.6$~\kms, as described in the caption to Fig.~\ref{fig:160_HImaps}.
}
\label{fig:247_HImaps}
\end{figure}

\begin{figure}
    \centering
    \vspace{-1.0cm}

    \captionsetup[subfigure]{labelformat=empty}
    \subfloat[][]{
        \centering
        \hspace{-0.5cm}
        \includegraphics[width=0.75\textwidth]{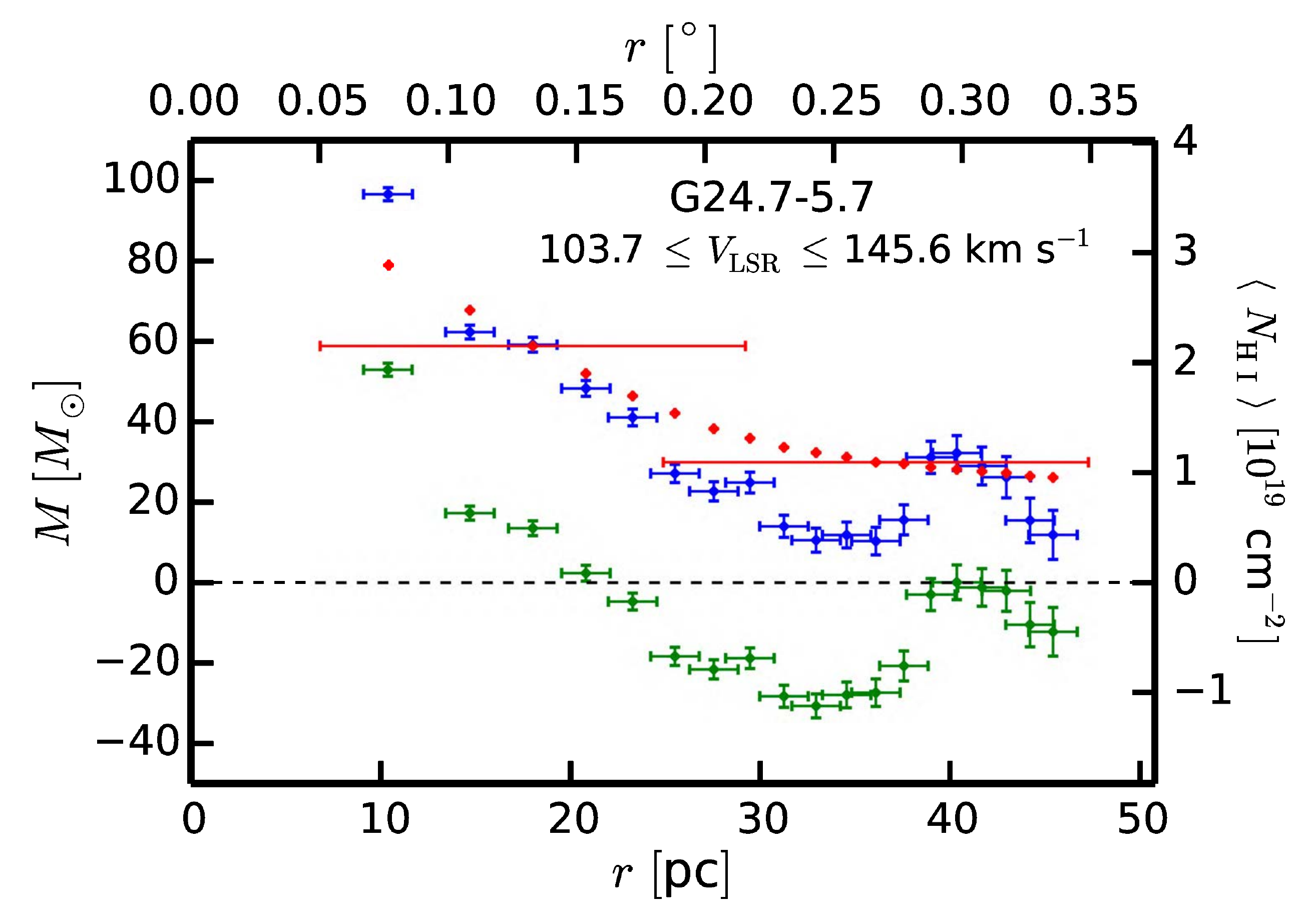}
        }

    \vspace{-1.0cm}

    \subfloat[][]{
        \centering
        \hspace{-1.8cm}
        \includegraphics[width=0.7\textwidth]{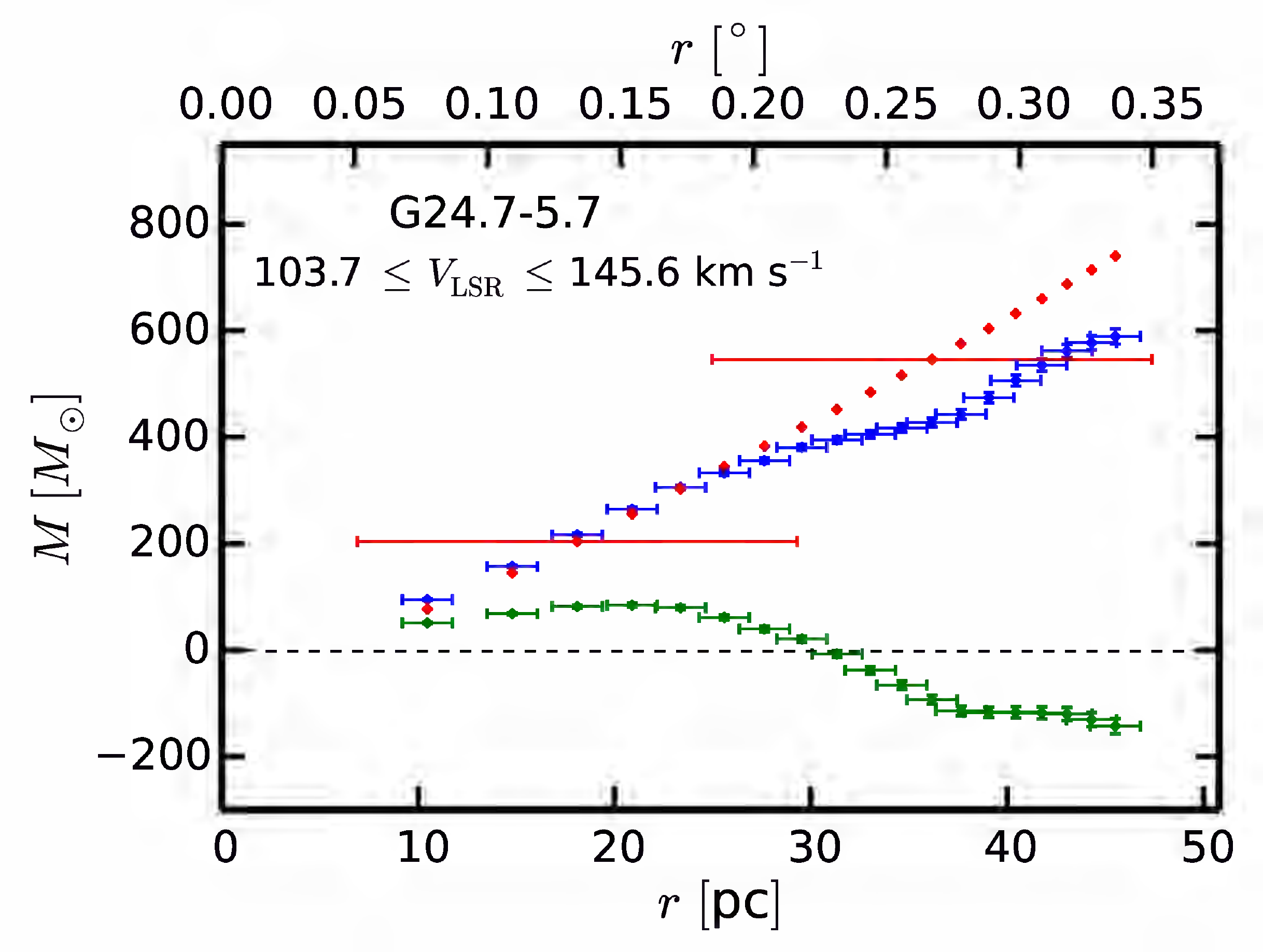}
        }

\caption{Radial mass profiles for G$24.7-5.7$ as described in the caption to Fig.~\ref{fig:160_mass_graphs}.
}\label{fig:247_mass_graphs}
\end{figure}

\begin{figure}
\vspace{-2.0cm}
\centering
\captionsetup[subfigure]{labelformat=empty}
    \subfloat[][]{
        \centering
        \hspace{-1cm}
        \includegraphics[width=1.0\textwidth]{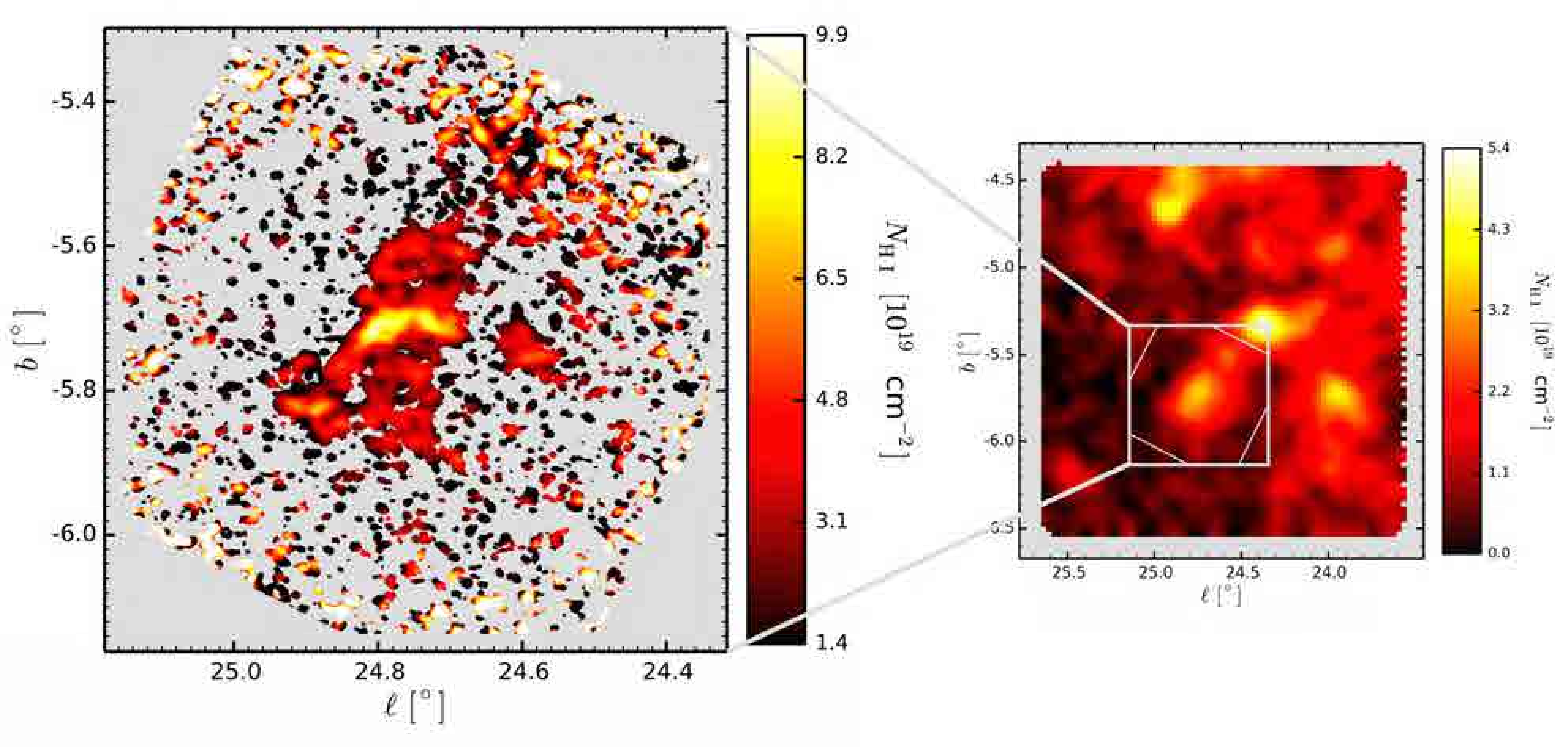}
        \hspace{-0cm}
        }

    \vspace{-1.0cm}

    \subfloat[][]{
        \centering
        \hspace{-1cm}
        \includegraphics[width=0.5\textwidth]{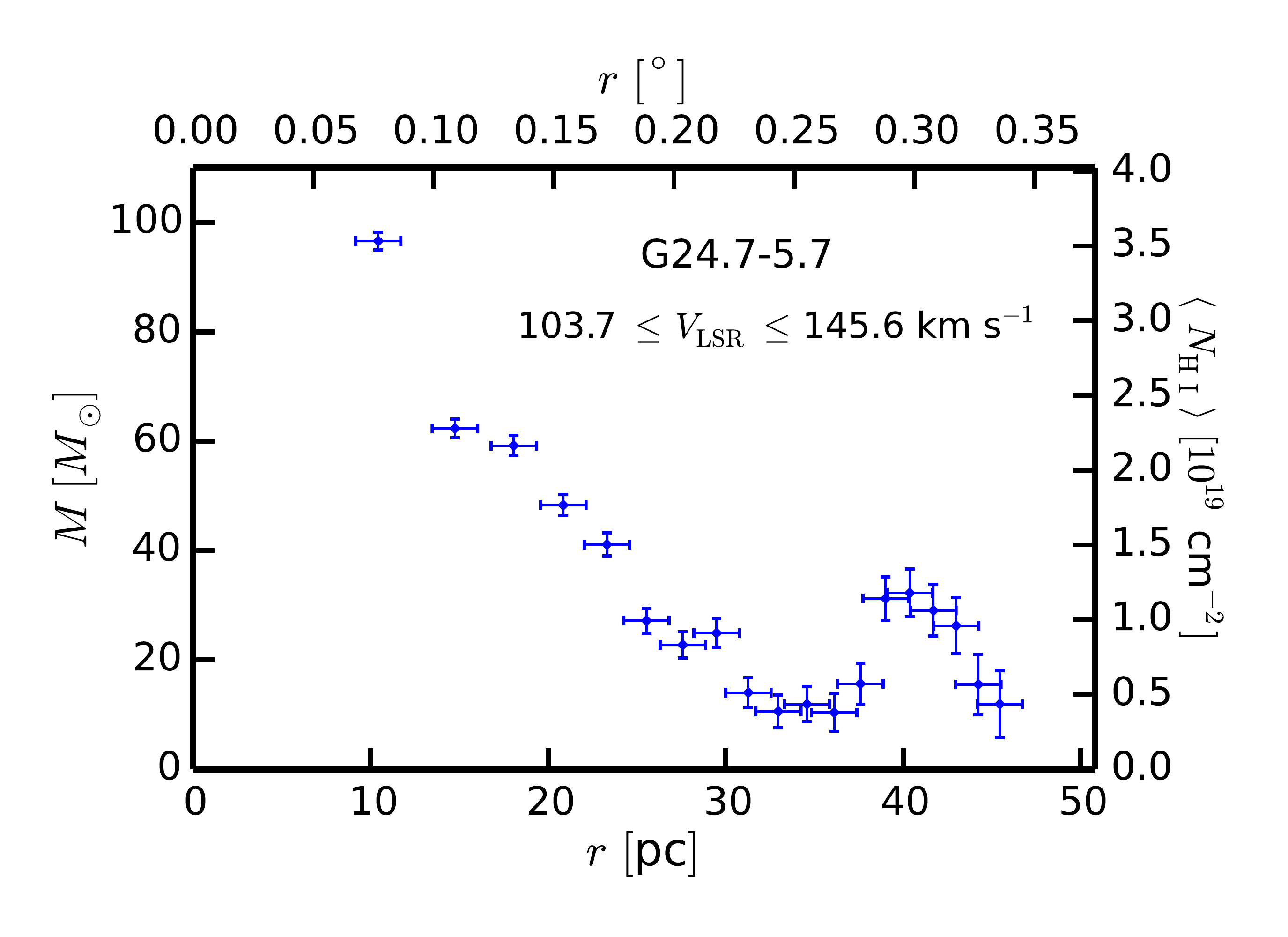}
        \hspace{-0.7cm}
        }
    \subfloat[][]{
        \centering
        \includegraphics[viewport = 0 -30 1150 650, width=0.59\textwidth]{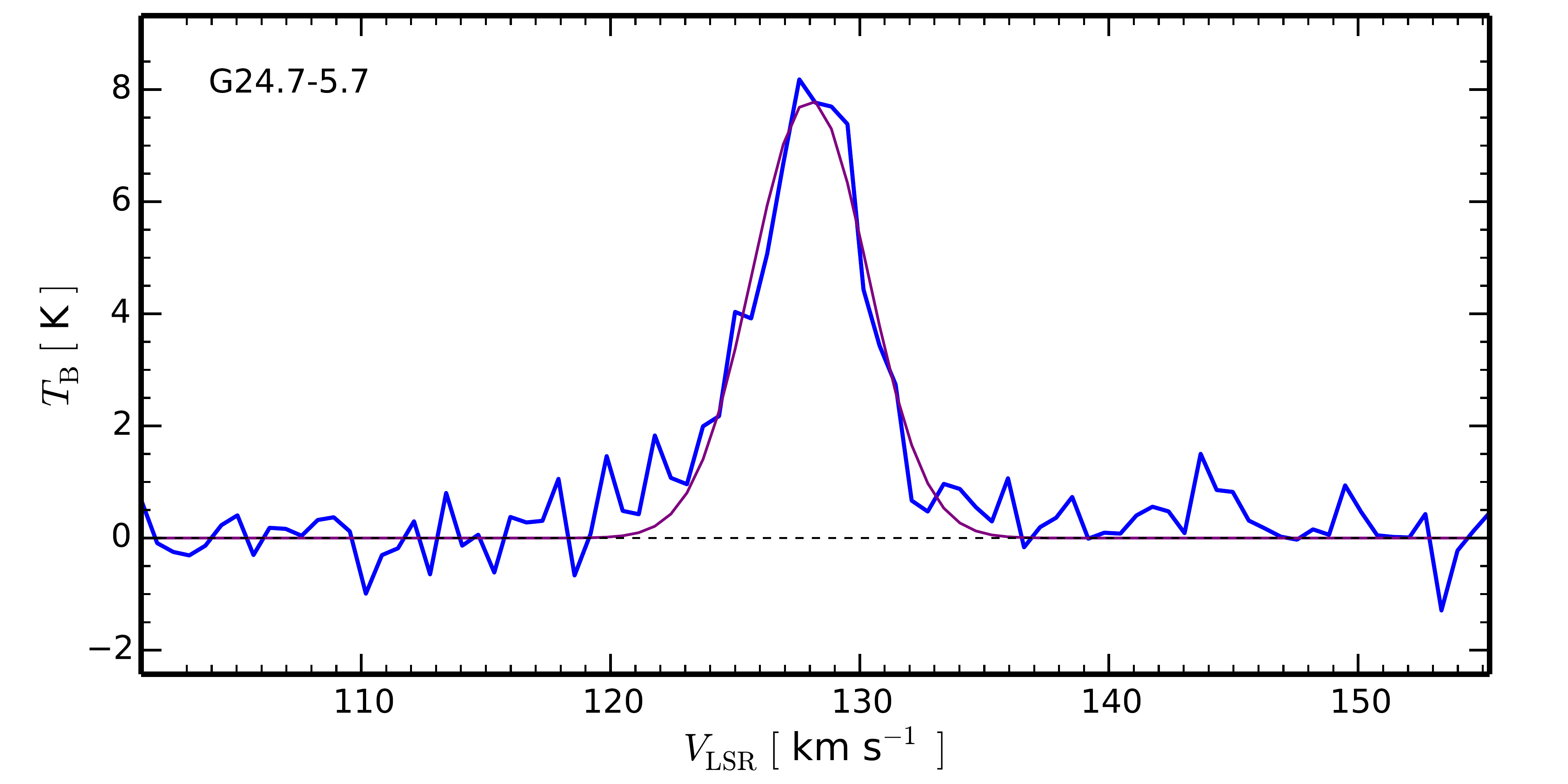}
        }
\vspace{-0.5cm}
\caption{Summary of G$24.7-5.7$ as described in the caption to Fig.~\ref{fig:G160-4plots}. A relatively small cloud that together with G$24.3-5.3$ and a cloud between them forms a chain of three clouds connected spatially and kinematically. Unlike its companion G$24.3-5.3$, it has a clear core with a narrow line of FWHM=5.5 \kms, but still has a similar range of column densities.  The edge of the chain of clouds is quite sharp at higher longitudes.}
\label{fig:G247-4plots}
\end{figure}

\begin{figure}
\centering
    \vspace{-2cm}

    \captionsetup[subfigure]{labelformat=empty}
    \subfloat[][]{
        \centering
        \hspace{0.0cm}
        \includegraphics[width=0.5\textwidth]{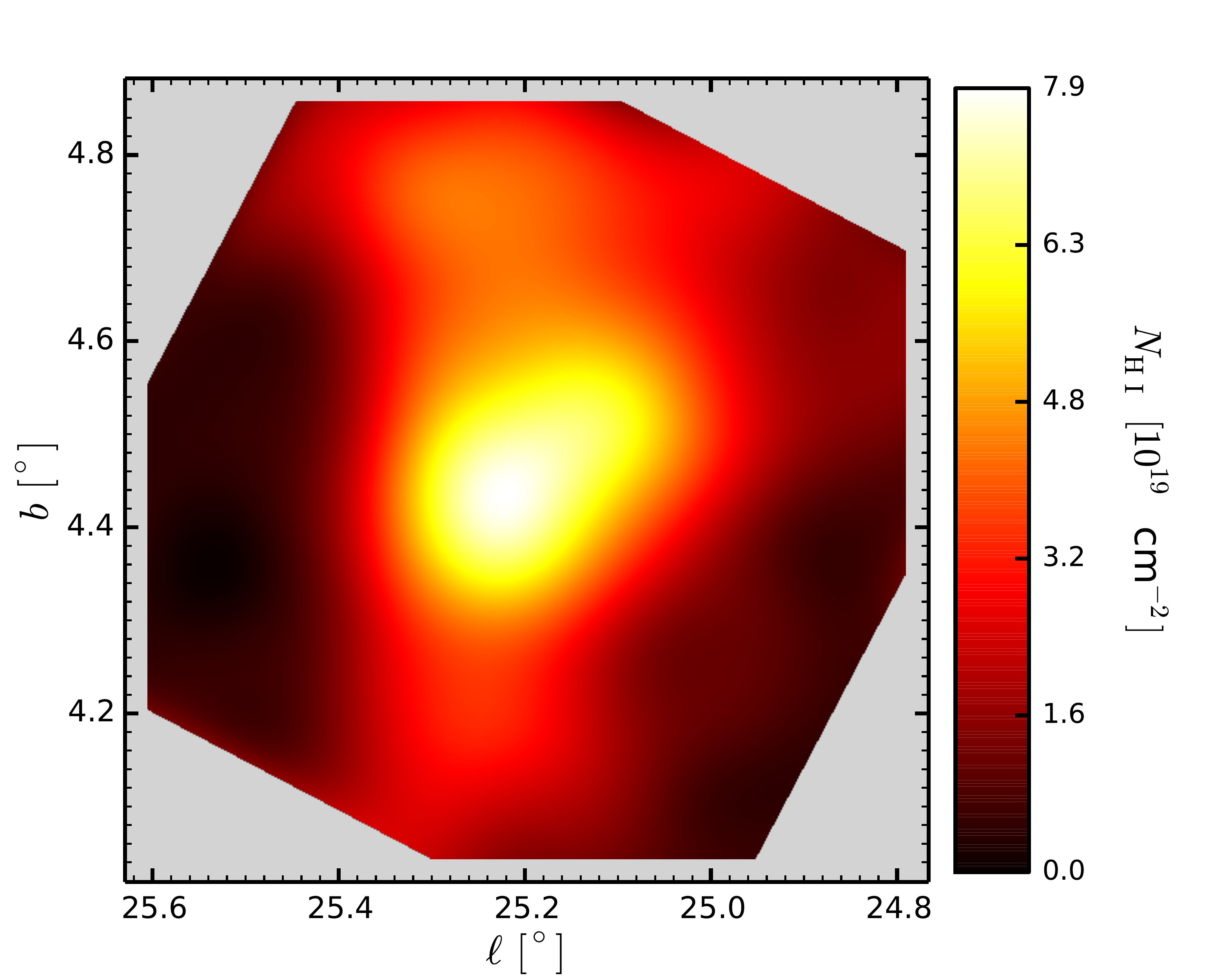}
        \hspace{-0.1cm}}
    \subfloat[][]{
        \centering
        \includegraphics[width=0.5\textwidth]{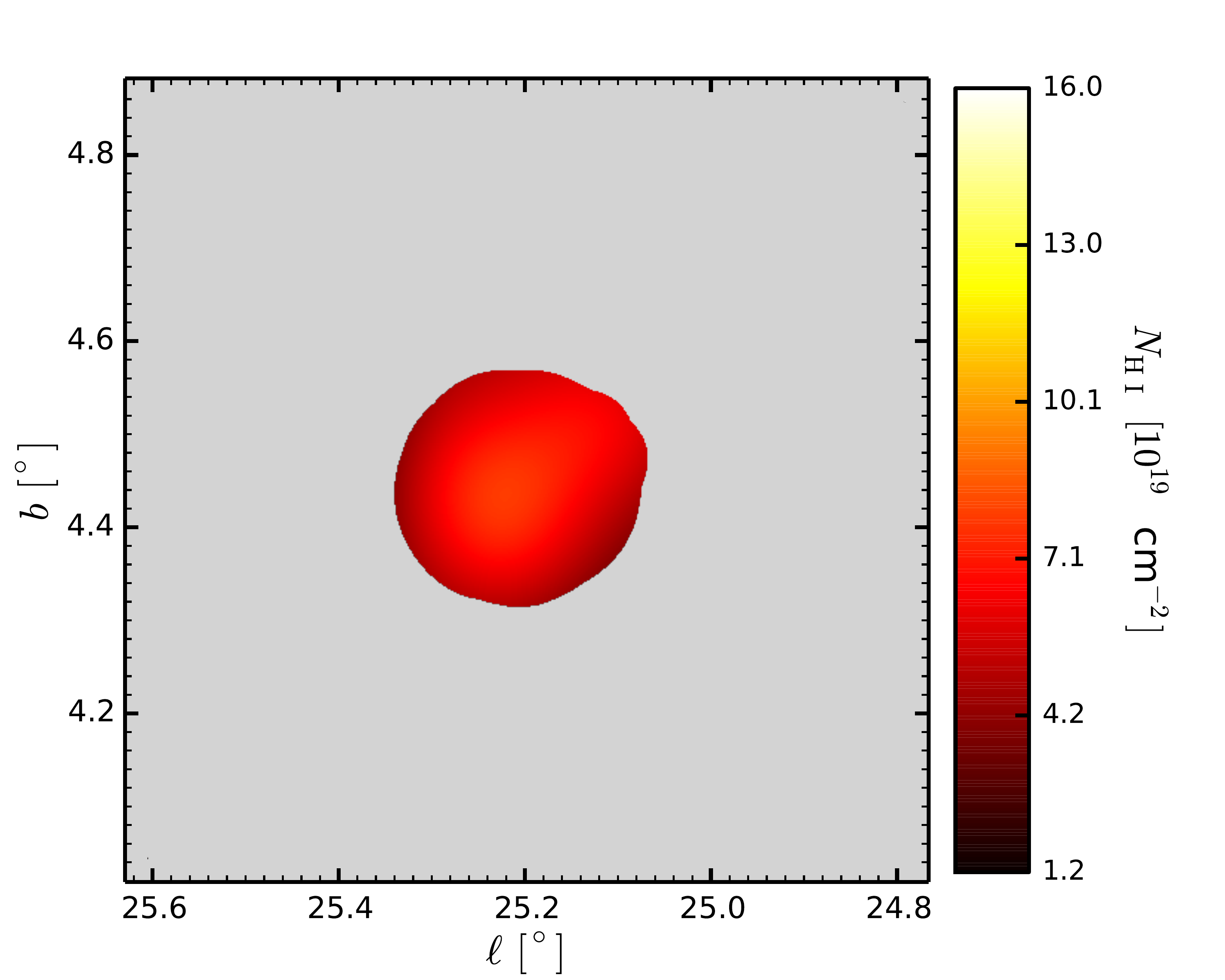}
        }

    \vspace{-1.3cm}

    \subfloat[][]{
        \centering
        \hspace{0.0cm}
        \includegraphics[width=0.5\textwidth]{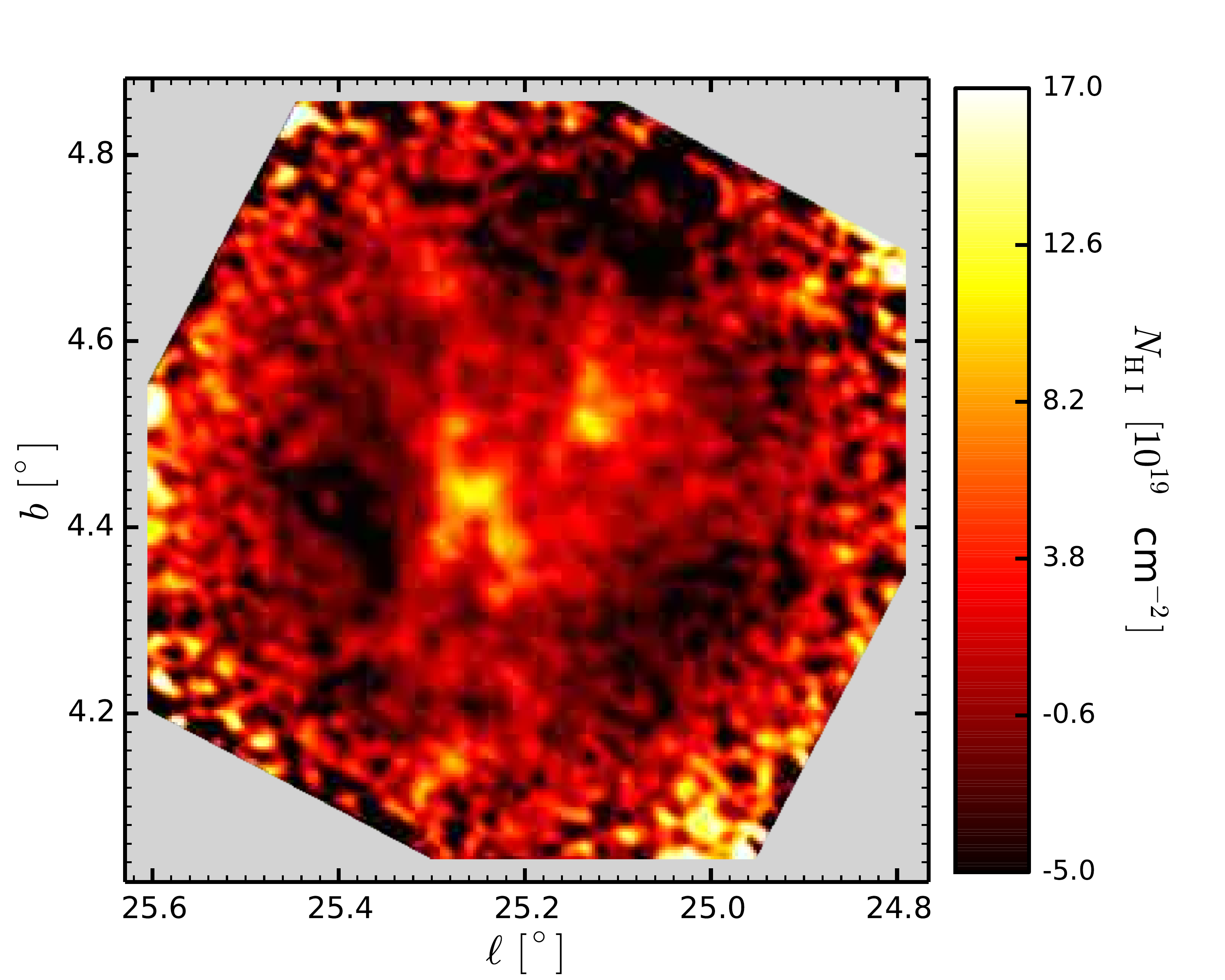}
        \hspace{-0.1cm}}
    \subfloat[][]{
        \centering
        \includegraphics[width=0.5\textwidth]{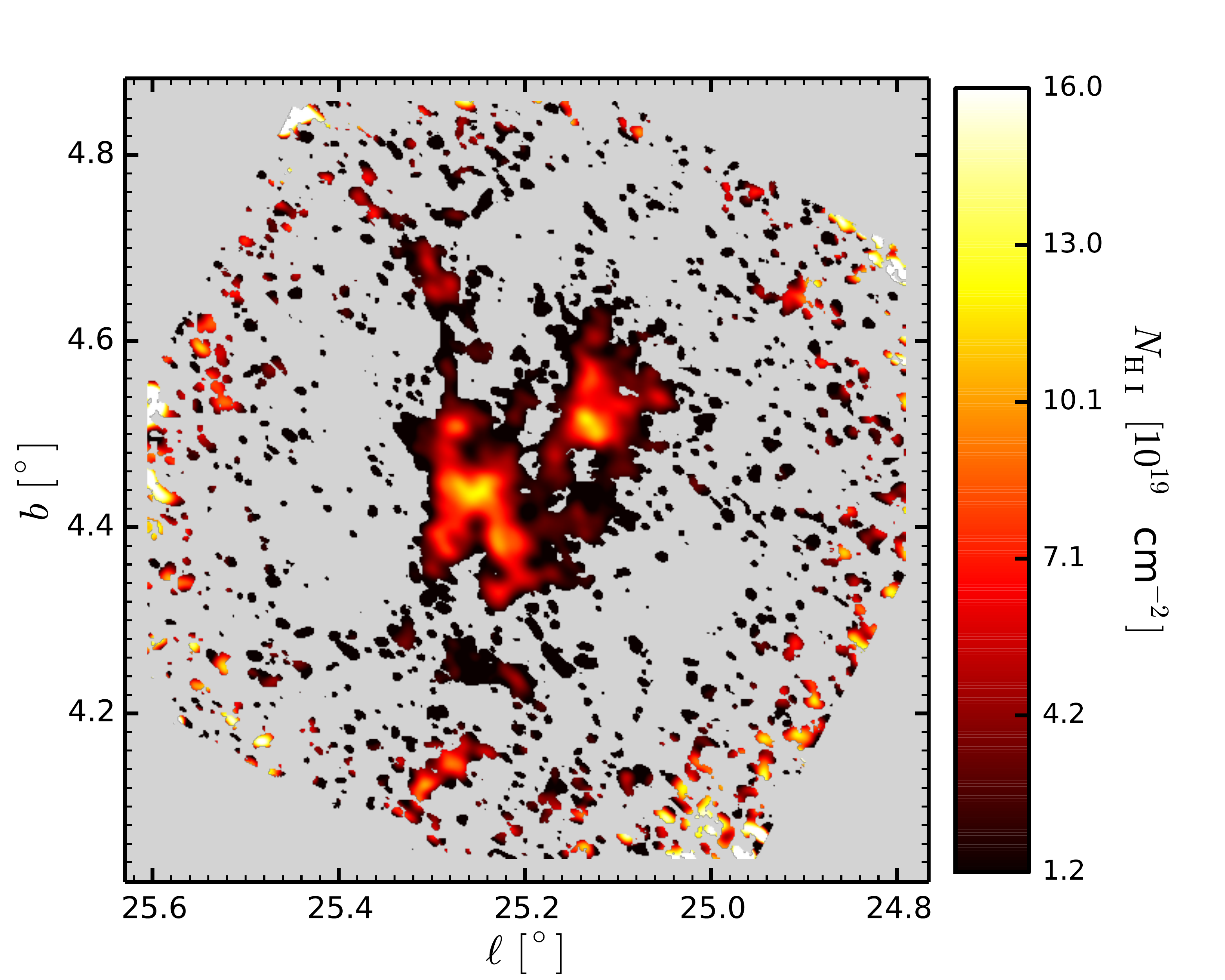}
        }

     \vspace{-1.3cm}

     \subfloat[][]{
        \centering
        \hspace{0.0cm}
        \includegraphics[width=0.5\textwidth]{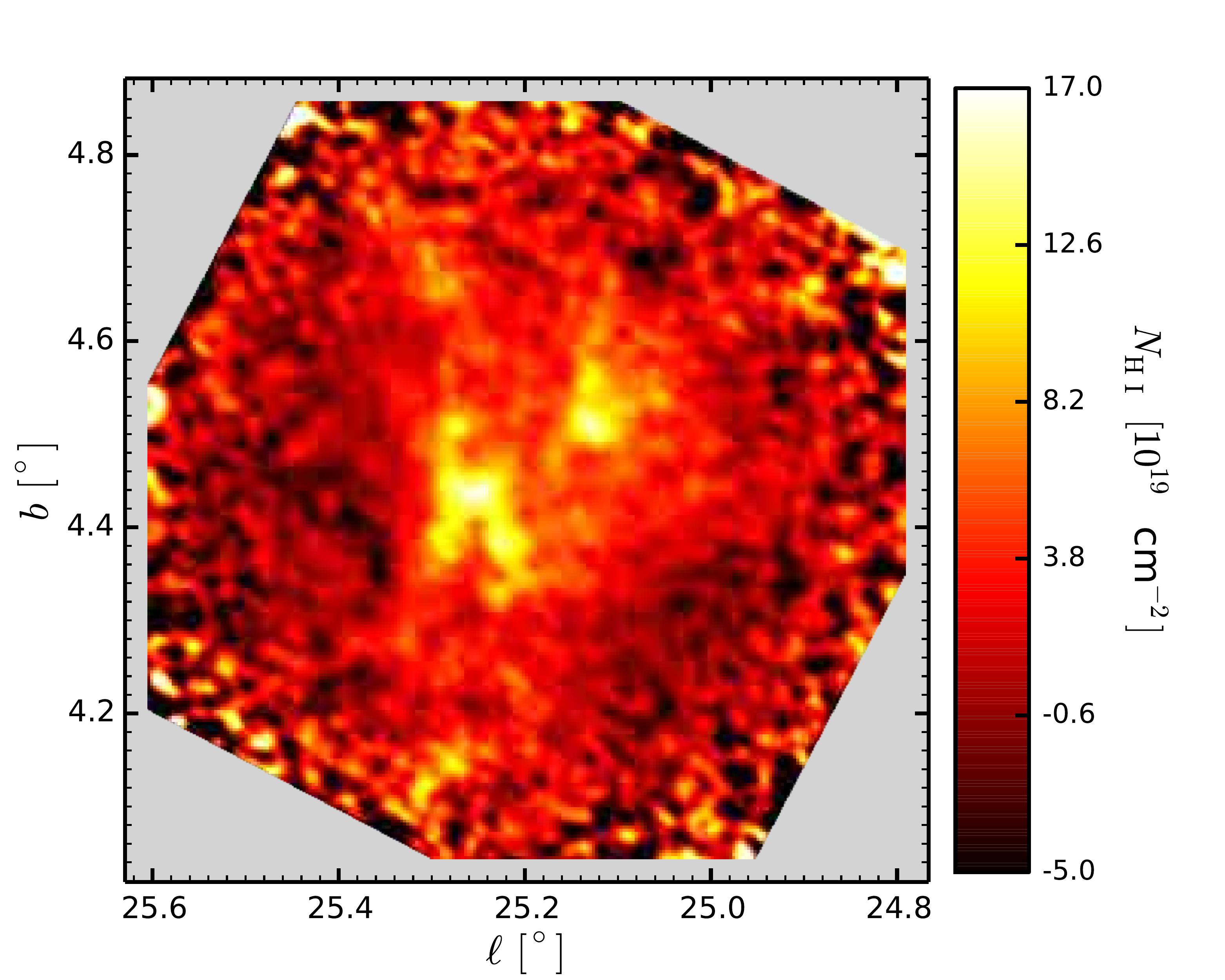}
        \hspace{-0.1cm}}
    \subfloat[][]{
        \centering
        \includegraphics[width=0.5\textwidth]{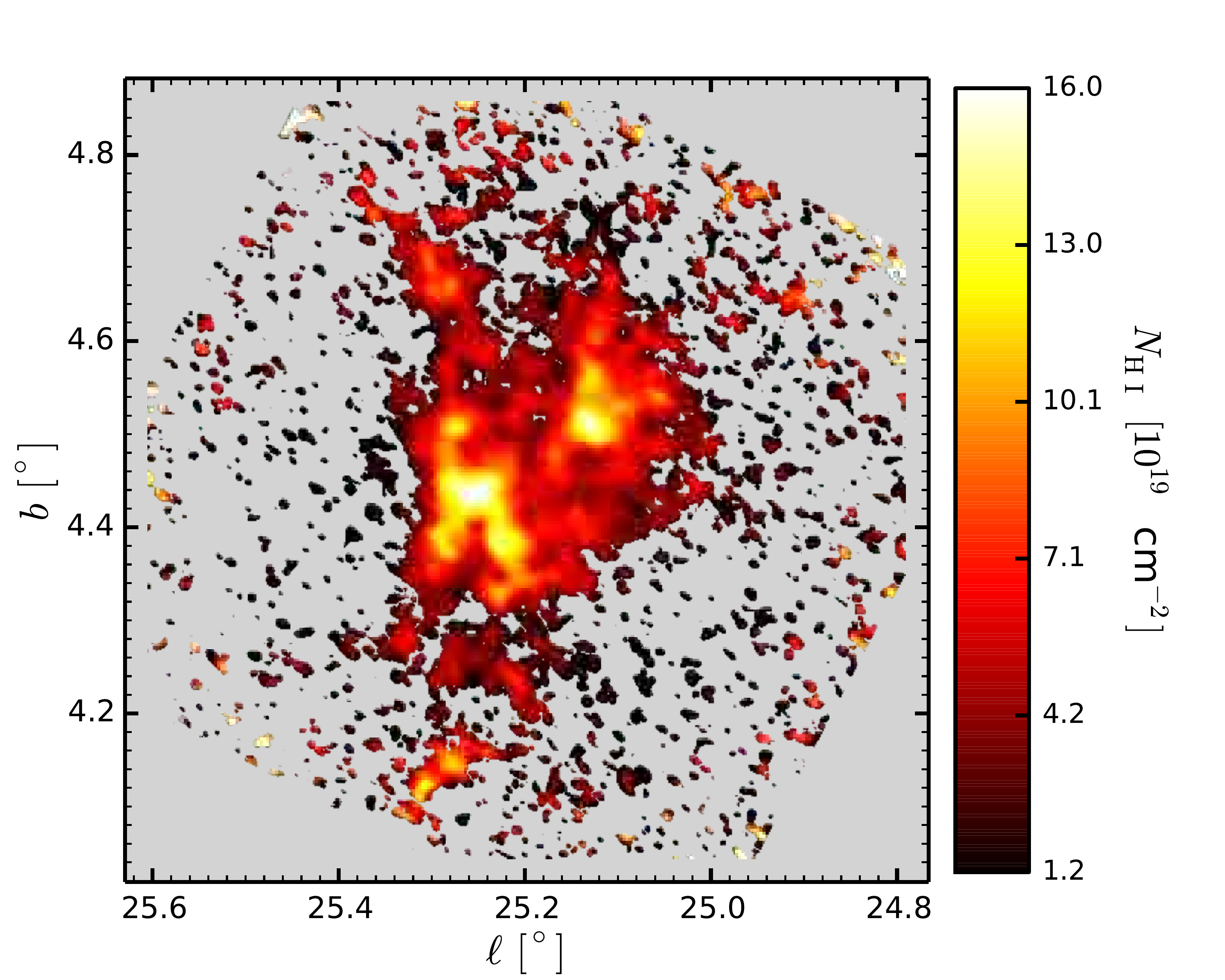}
        }
\vspace{-0.5cm}
\caption{\HI column density maps for G$25.2+4.5$, integrated over 64 spectral channels in the interval $125.7 \leq \VLSR \leq 166.3$~\kms, as described in the caption to Fig.~\ref{fig:160_HImaps}.
}
\label{fig:252_HImaps}
\end{figure}

\begin{figure}
    \centering
    \vspace{-1.0cm}

    \captionsetup[subfigure]{labelformat=empty}
    \subfloat[][]{
        \centering
        \hspace{-0.5cm}
        \includegraphics[width=0.75\textwidth]{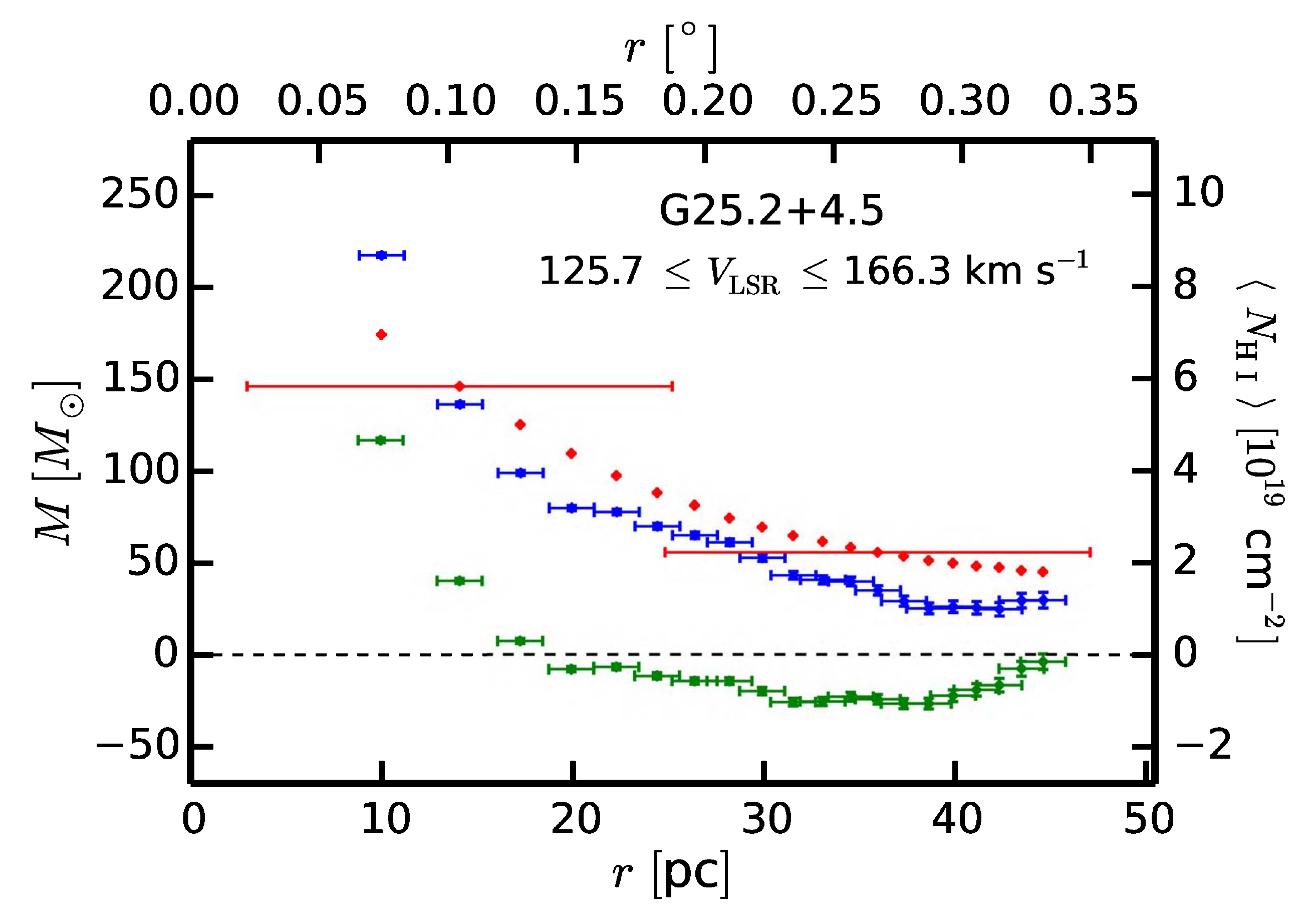}
        }

    \vspace{-1.0cm}

    \subfloat[][]{
        \centering
        \hspace{-1.8cm}
        \includegraphics[width=0.7\textwidth]{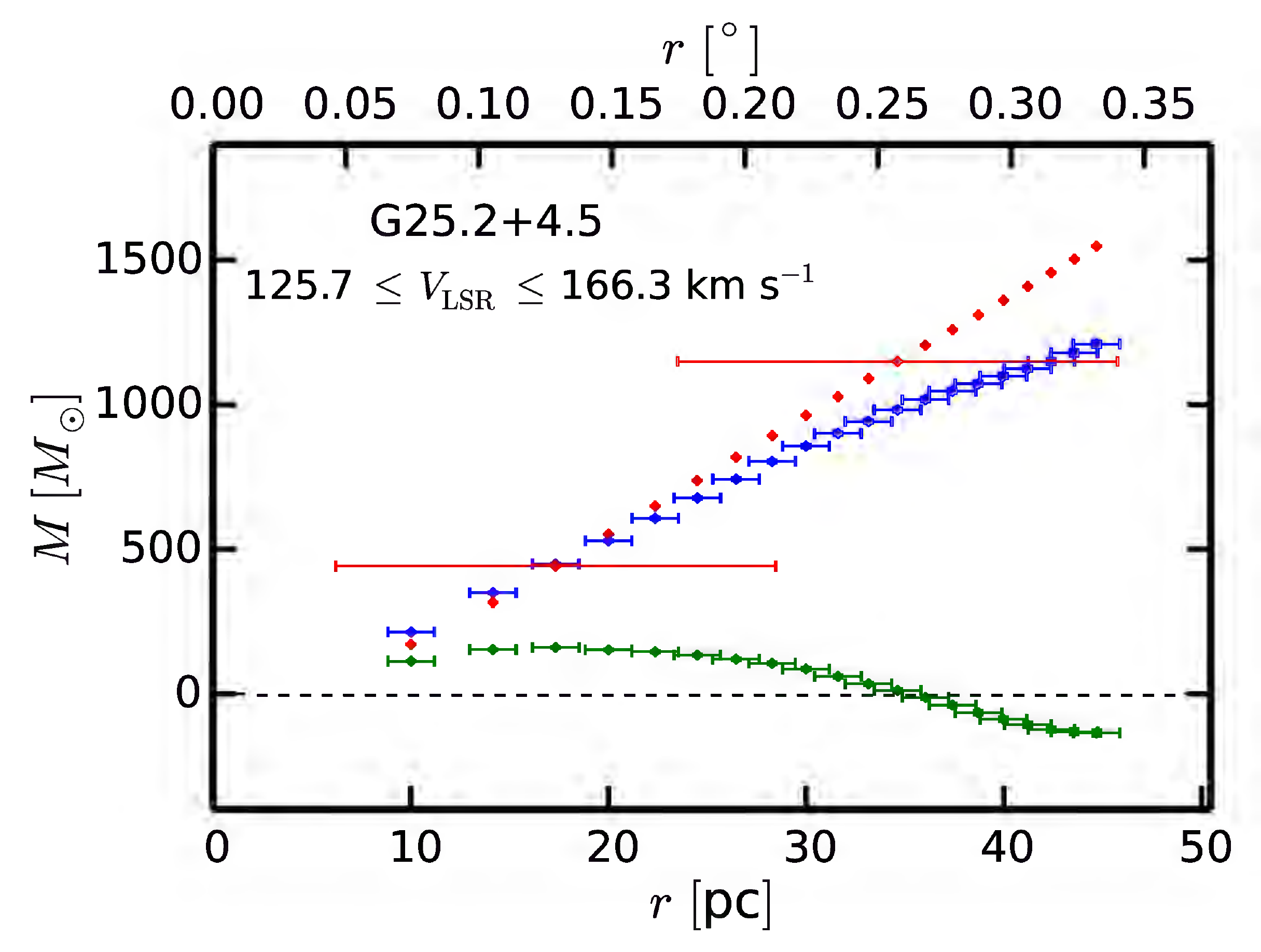}
        }

\caption{Radial mass profiles for G$25.2+4.5$ as described in the caption to Fig.~\ref{fig:160_mass_graphs}.}\label{fig:252_mass_graphs}
\end{figure}

\begin{figure}
\vspace{-2.0cm}
\centering
\captionsetup[subfigure]{labelformat=empty}
    \subfloat[][]{
        \centering
        \hspace{-1cm}
        \includegraphics[width=1.0\textwidth]{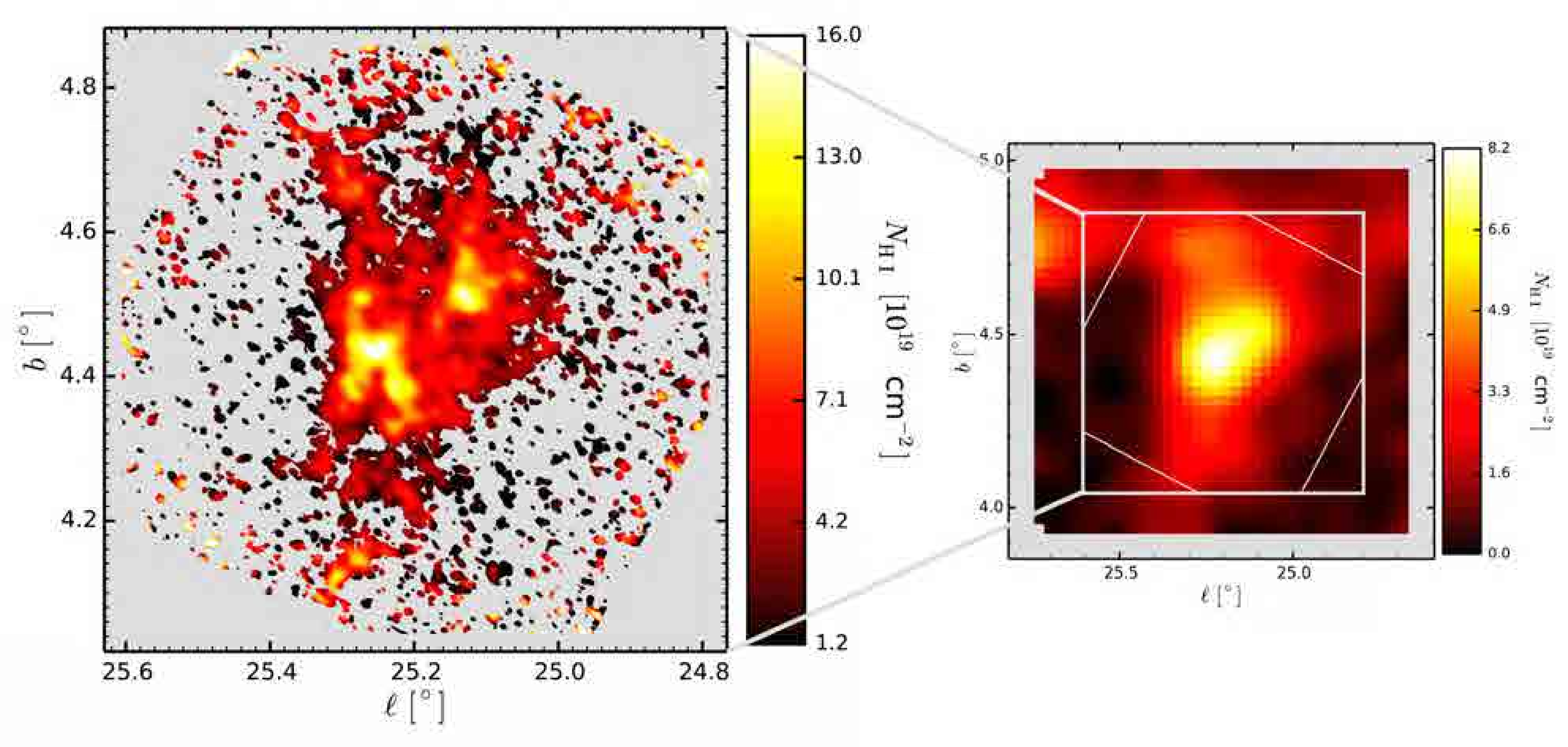}
        \hspace{-0cm}
        }

    \vspace{-1.0cm}

    \subfloat[][]{
        \centering
        \hspace{-1cm}
        \includegraphics[width=0.5\textwidth]{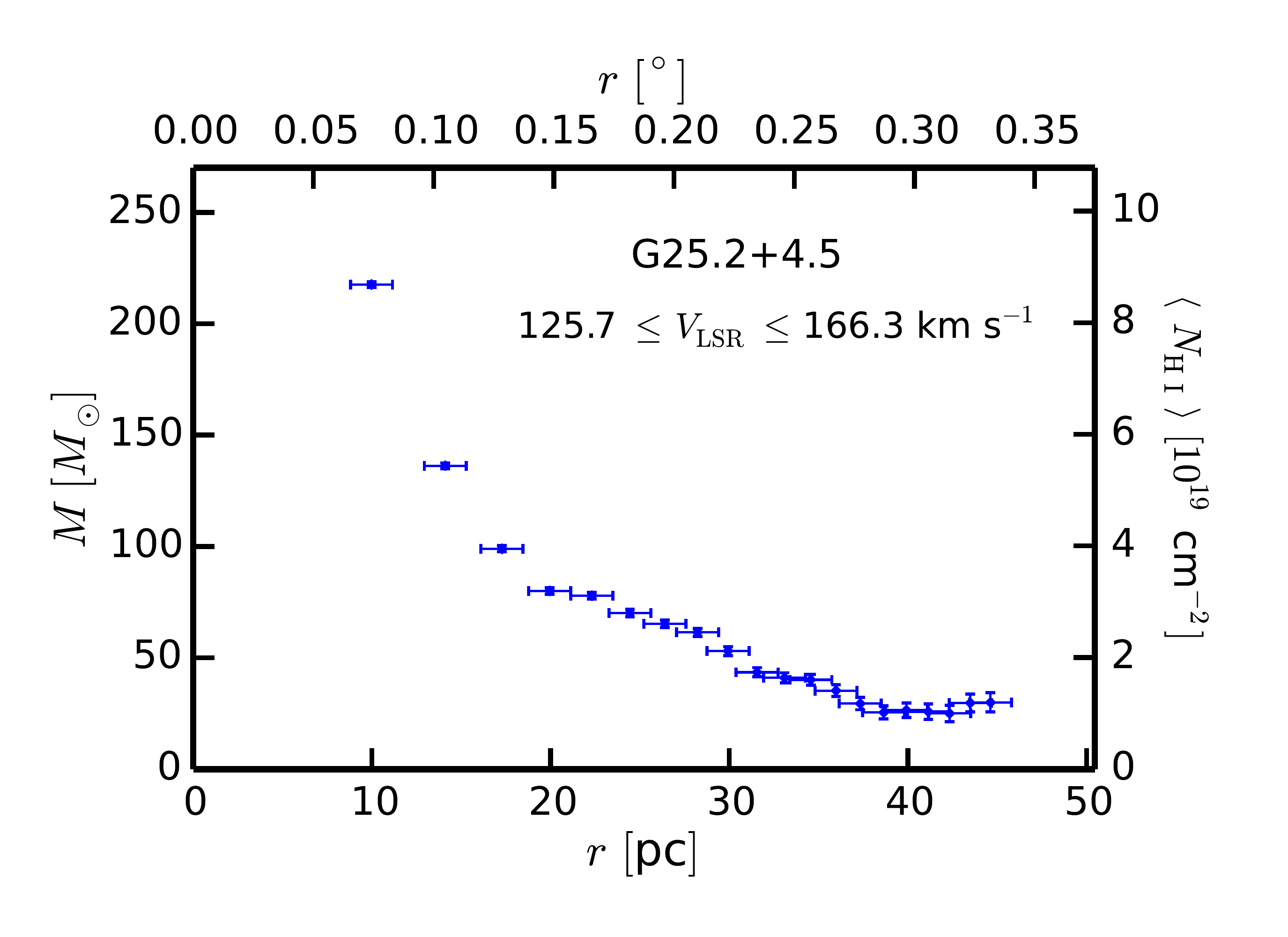}
        \hspace{-0.7cm}
        }
    \subfloat[][]{
        \centering
        \includegraphics[viewport = 0 -30 1150 650, width=0.59\textwidth]{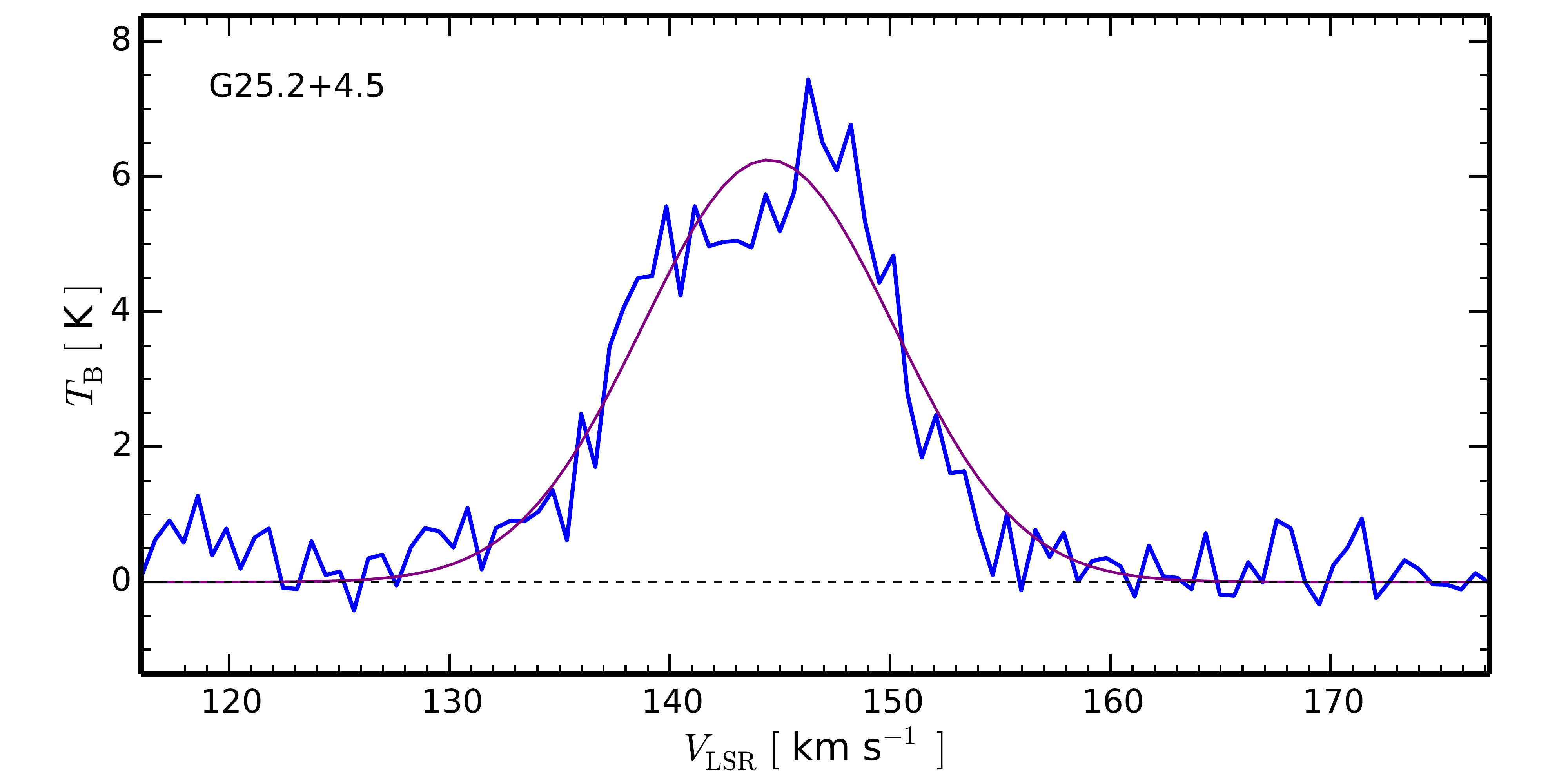}
        }
\vspace{-0.5cm}
\caption{Summary of G$25.2+4.5$ as described in the caption to Fig.~\ref{fig:G160-4plots}. The VLA data reveal the considerable complexity of this cloud.  There seems to be no single core, but instead many dense regions.  The broad line with FWHM=13.4~\kms\ may indicate the blending of several kinematically distinct components within the 1\farcm05 angular resolution of the map, equivalent to 2.35 pc at the adopted distance of the cloud.  The cloud has a relatively sharp edge to higher longitudes with diffuse material spreading out to lower longitudes.  The relatively broad mass profile and the GBT data suggest that the diffuse components of the cloud may extend well beyond the VLA field of view.}
\label{fig:G252-4plots}
\end{figure}

\begin{figure}
\centering
    \vspace{-2cm}

    \captionsetup[subfigure]{labelformat=empty}
    \subfloat[][]{
        \centering
        \hspace{0.0cm}
        \includegraphics[width=0.5\textwidth]{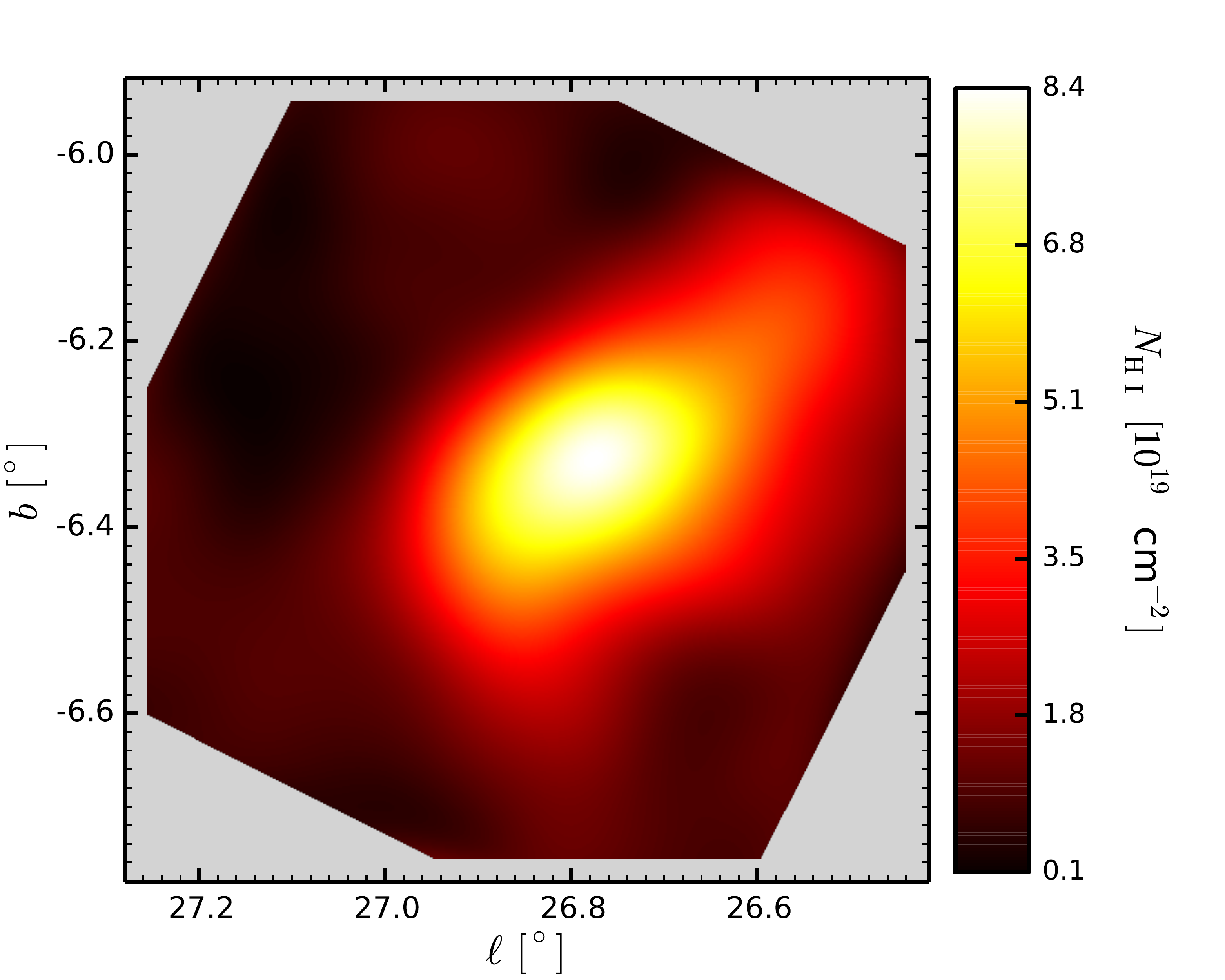}
        \hspace{-0.1cm}}
    \subfloat[][]{
        \centering
        \includegraphics[width=0.5\textwidth]{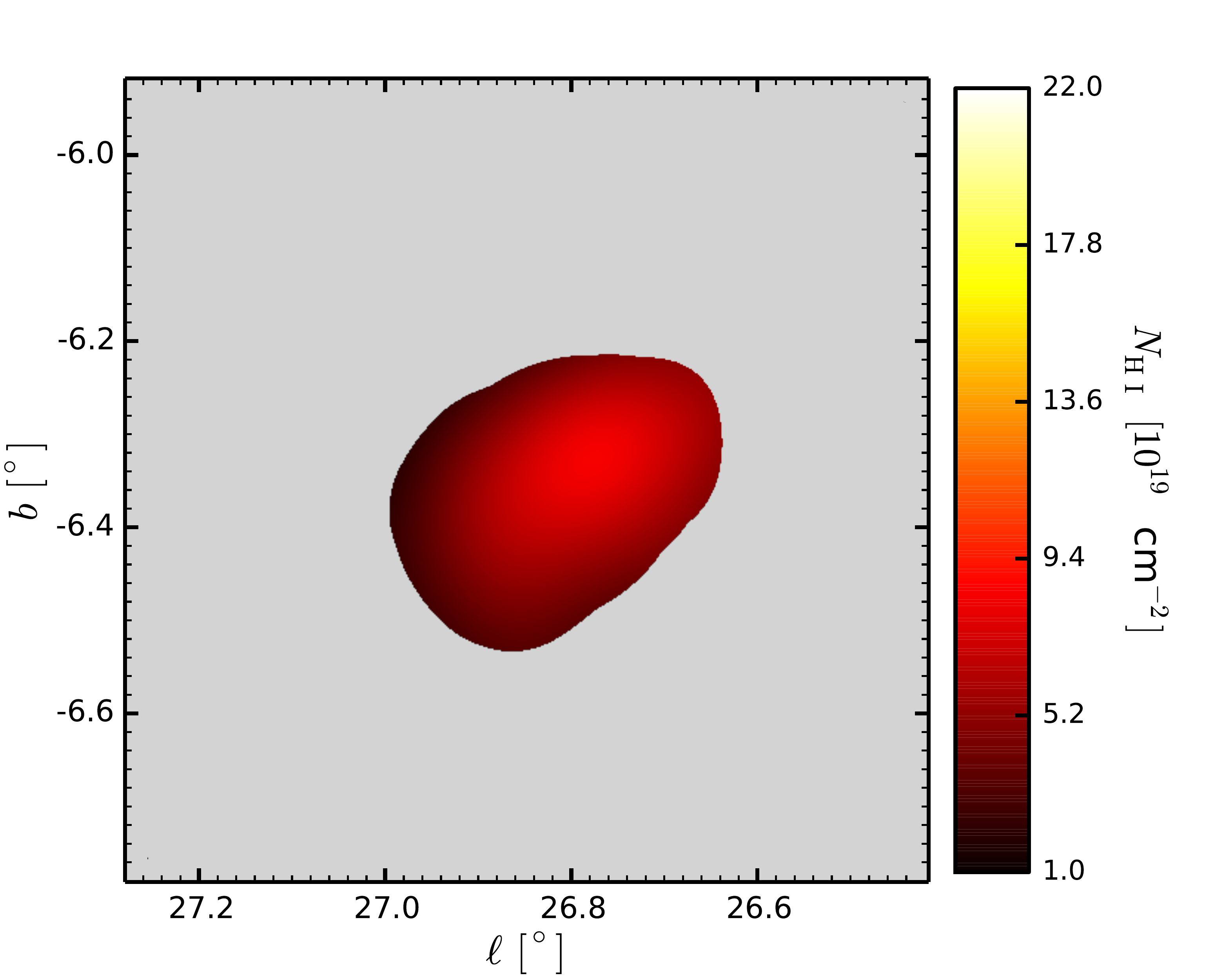}
        }

    \vspace{-1.3cm}

    \subfloat[][]{
        \centering
        \hspace{0.0cm}
        \includegraphics[width=0.5\textwidth]{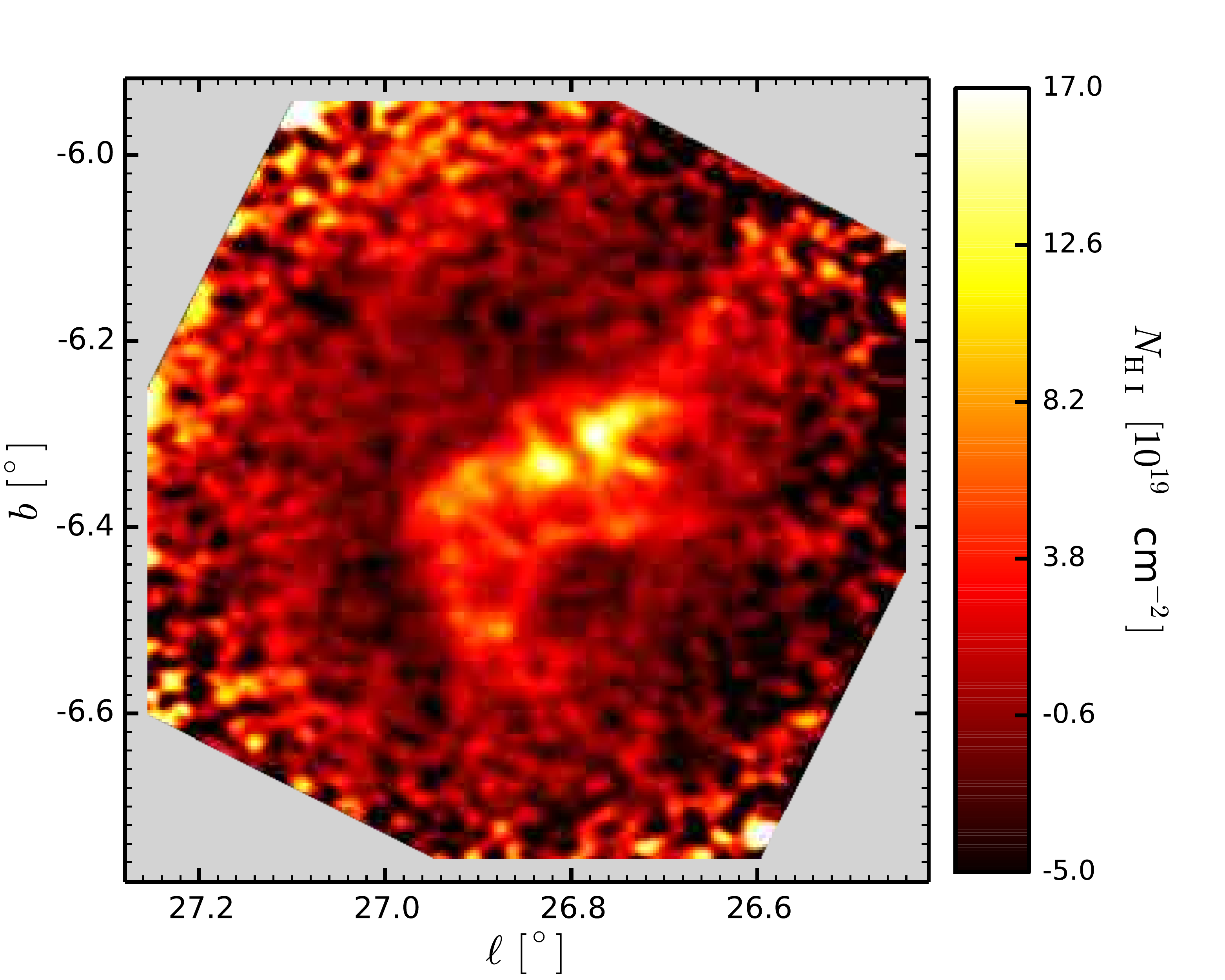}
        \hspace{-0.1cm}}
    \subfloat[][]{
        \centering
        \includegraphics[width=0.5\textwidth]{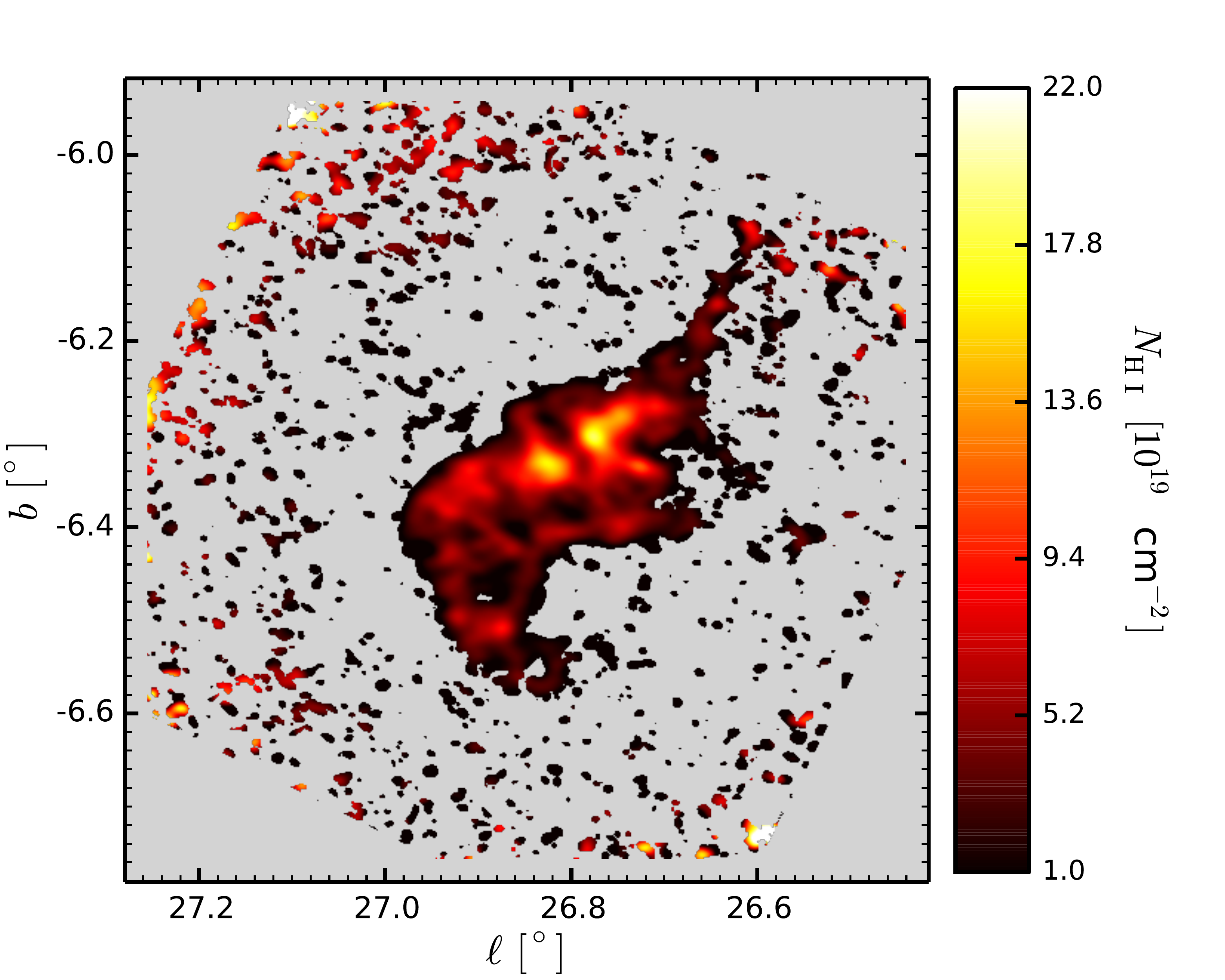}
        }

     \vspace{-1.3cm}

     \subfloat[][]{
        \centering
        \hspace{0.0cm}
        \includegraphics[width=0.5\textwidth]{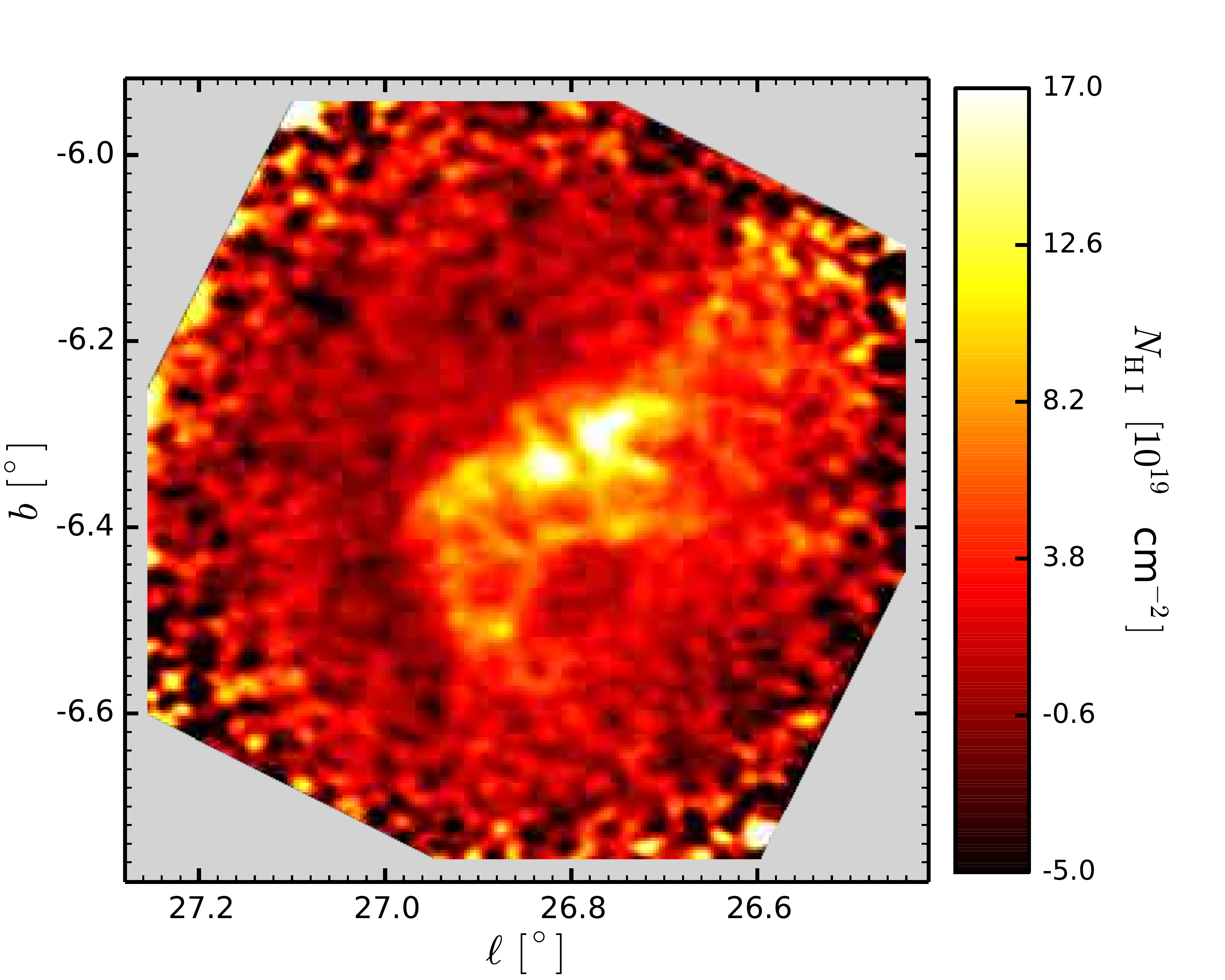}
        \hspace{-0.1cm}}
    \subfloat[][]{
        \centering
        \includegraphics[width=0.5\textwidth]{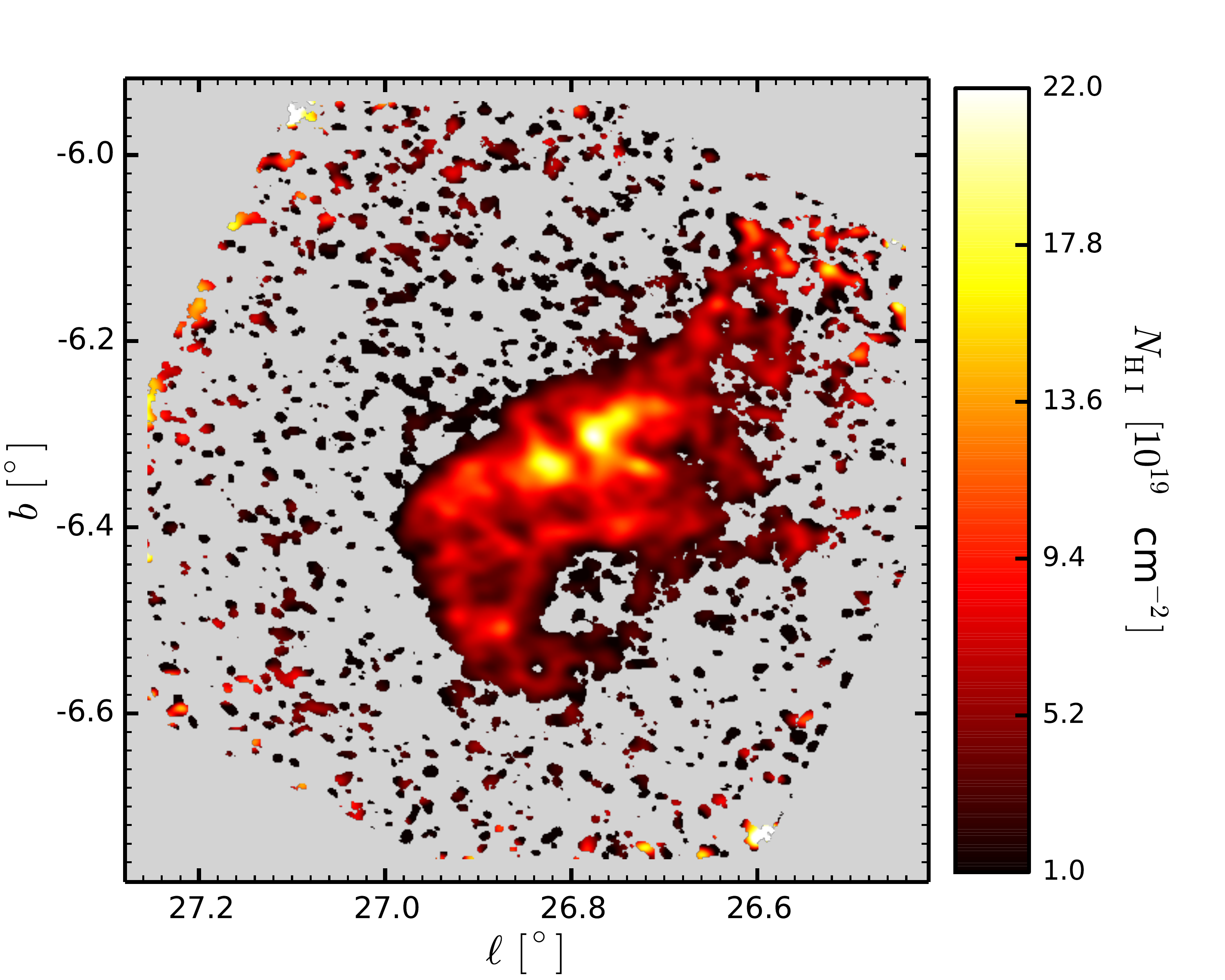}
        }
\vspace{-0.5cm}
\caption{\HI column density maps for G$26.9-6.3$, integrated over 51 spectral channels in the interval $108.5 \leq \VLSR \leq 140.7$~\kms, as described in the caption to Fig.~\ref{fig:160_HImaps}.}
\label{fig:269_HImaps}
\end{figure}

\begin{figure}
    \centering
    \vspace{-1.0cm}

    \captionsetup[subfigure]{labelformat=empty}
    \subfloat[][]{
        \centering
        \hspace{-0.5cm}
        \includegraphics[width=0.75\textwidth]{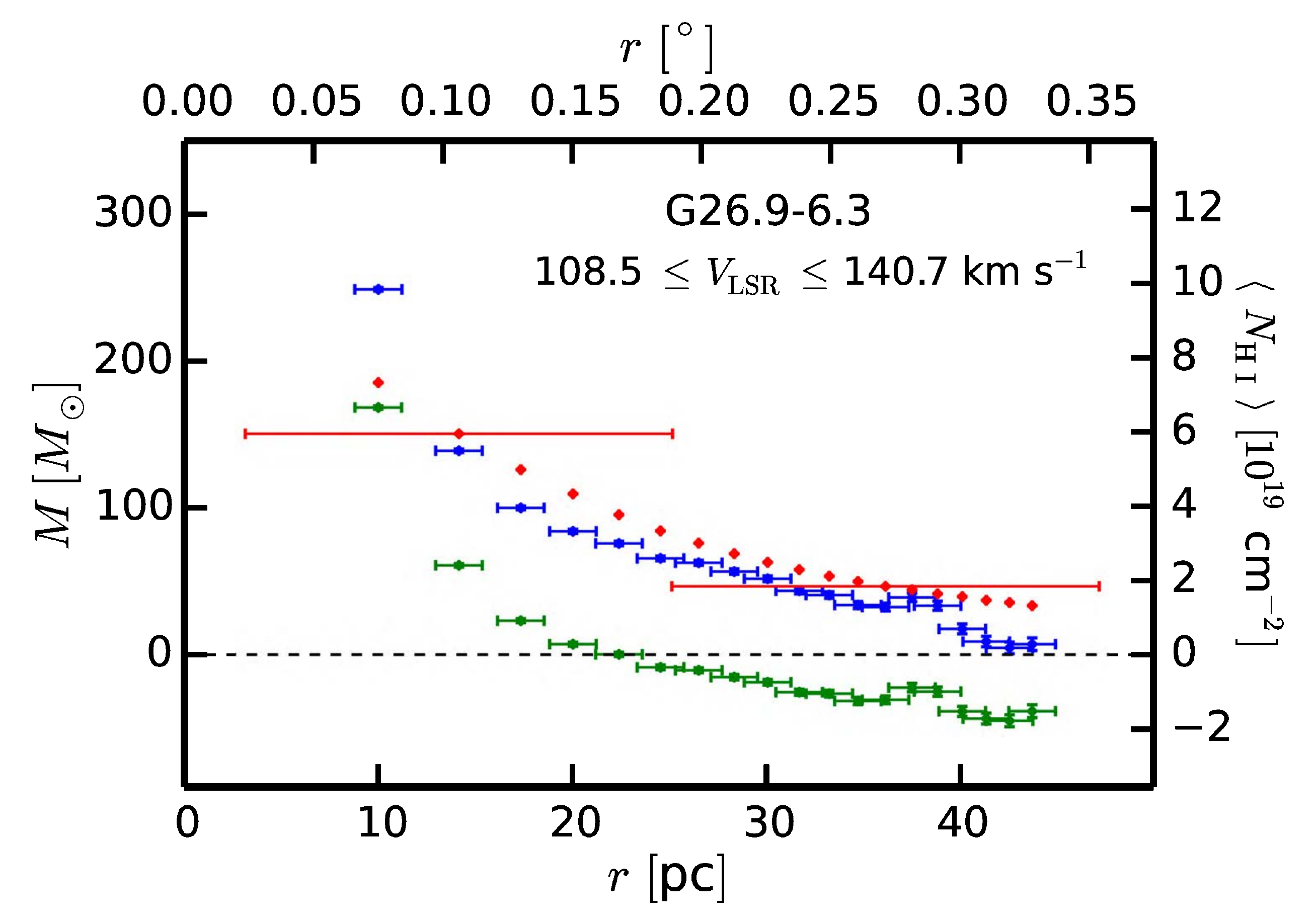}
        }

    \vspace{-1.0cm}

    \subfloat[][]{
        \centering
        \hspace{-1.8cm}
        \includegraphics[width=0.7\textwidth]{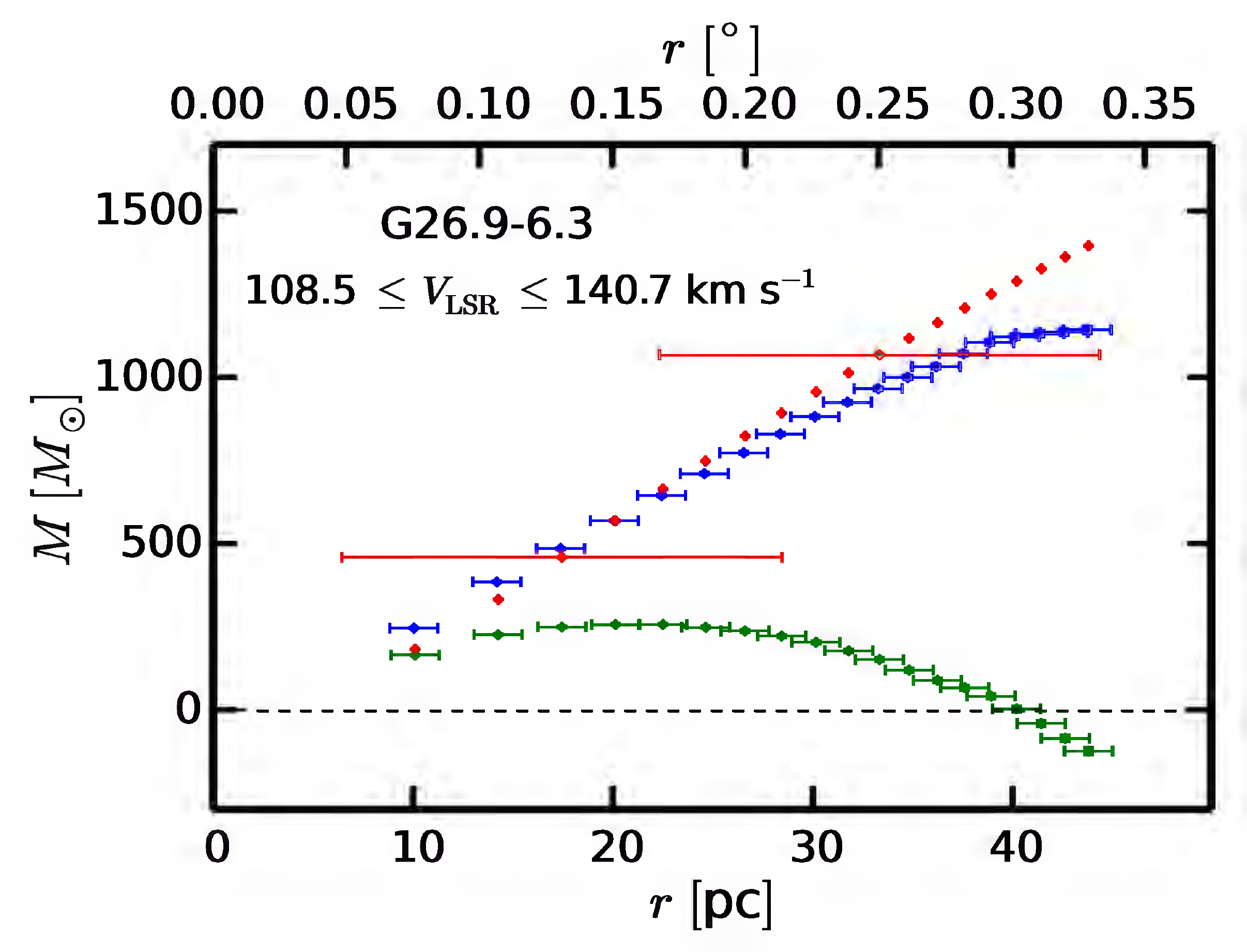}
        }

\caption{Radial mass profiles for G$26.9-6.3$ as described in the caption to Fig.~\ref{fig:160_mass_graphs}.}\label{fig:269_mass_graphs}
\end{figure}

\begin{figure}
\vspace{-2.0cm}
\centering
\captionsetup[subfigure]{labelformat=empty}
    \subfloat[][]{
        \centering
        \hspace{-1cm}
        \includegraphics[width=1.0\textwidth]{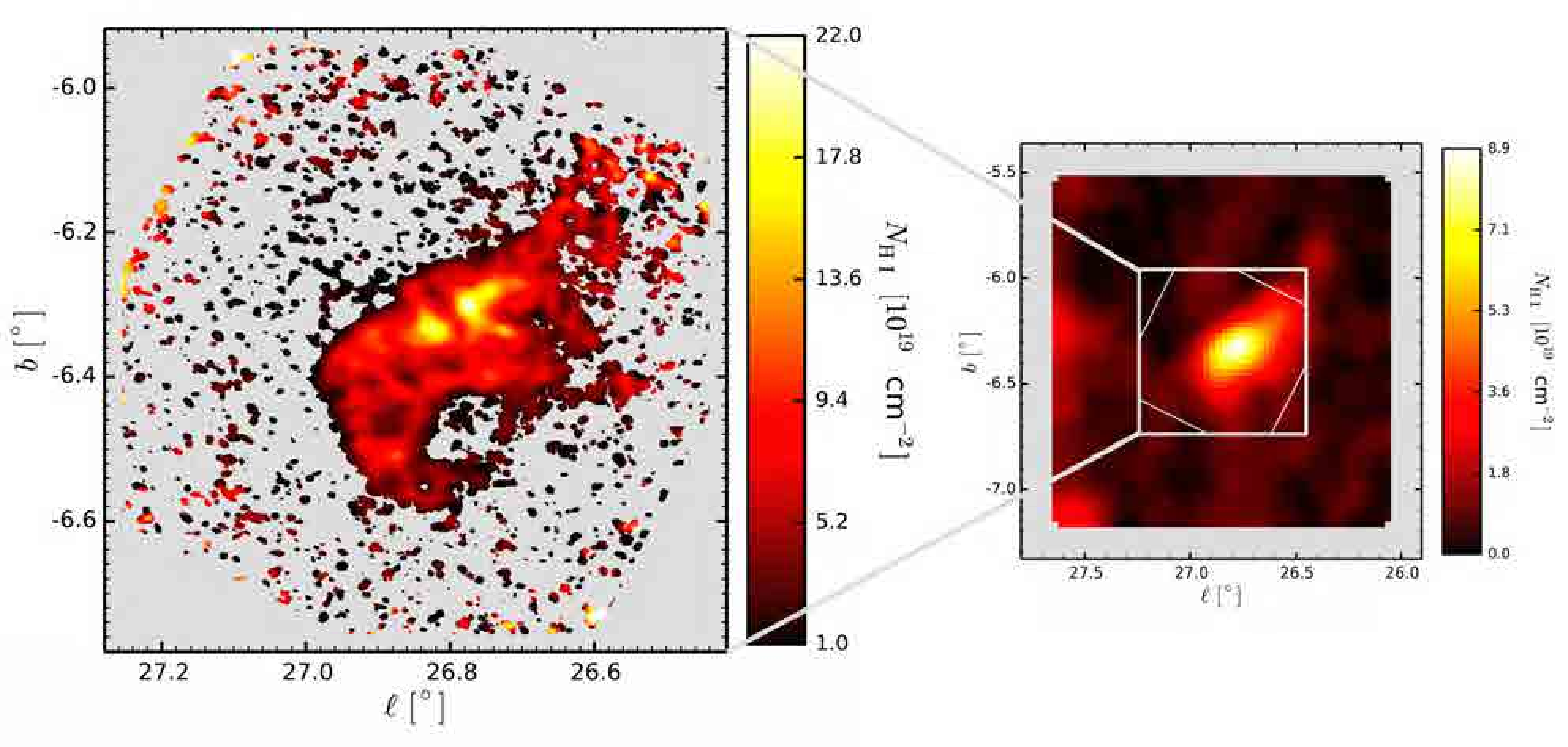}
        \hspace{-0cm}
        }

    \vspace{-1.0cm}

    \subfloat[][]{
        \centering
        \hspace{-1cm}
        \includegraphics[width=0.5\textwidth]{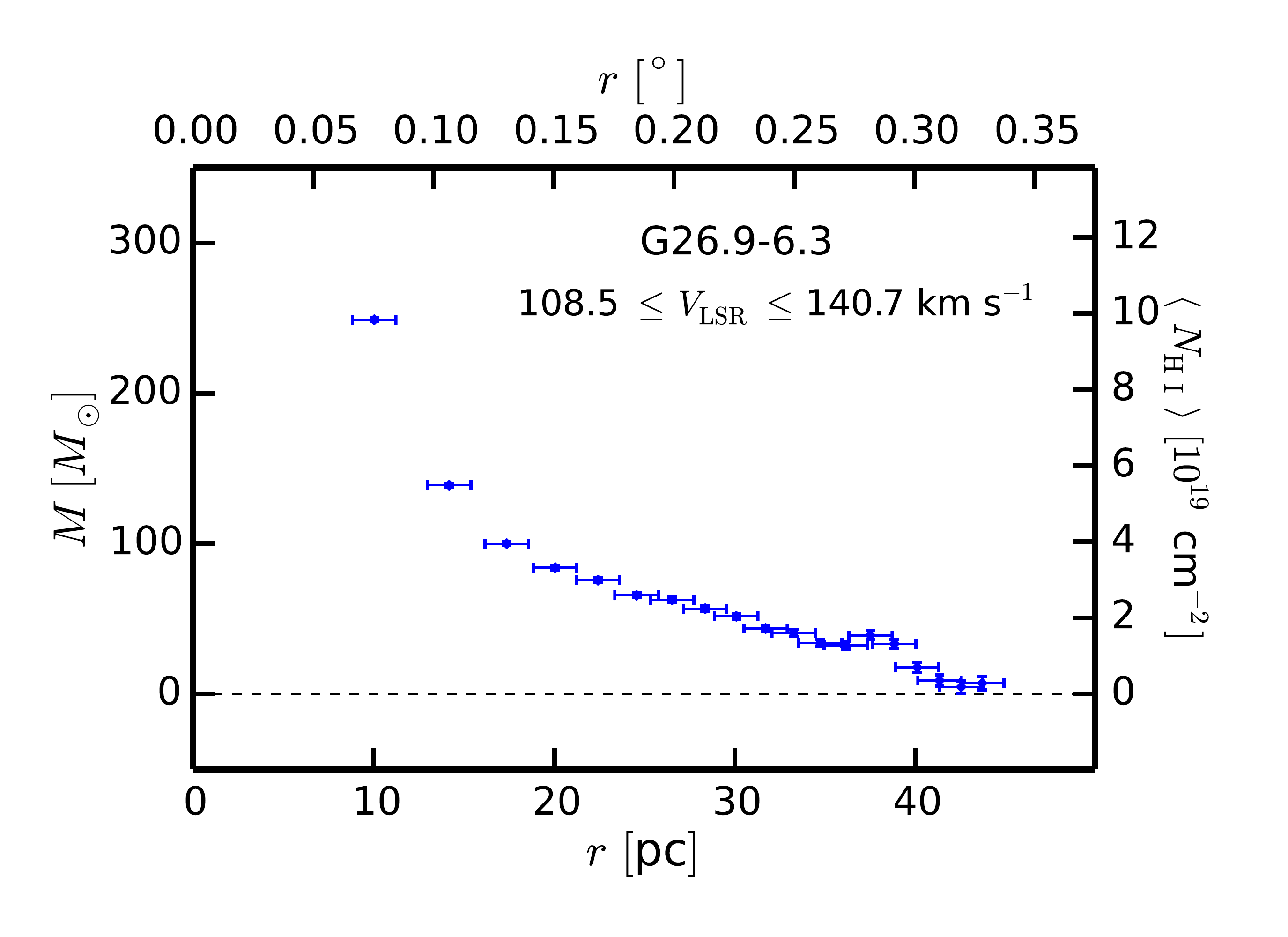}
        \hspace{-0.7cm}
        }
    \subfloat[][]{
        \centering
        \includegraphics[viewport = 0 -30 1150 650, width=0.59\textwidth]{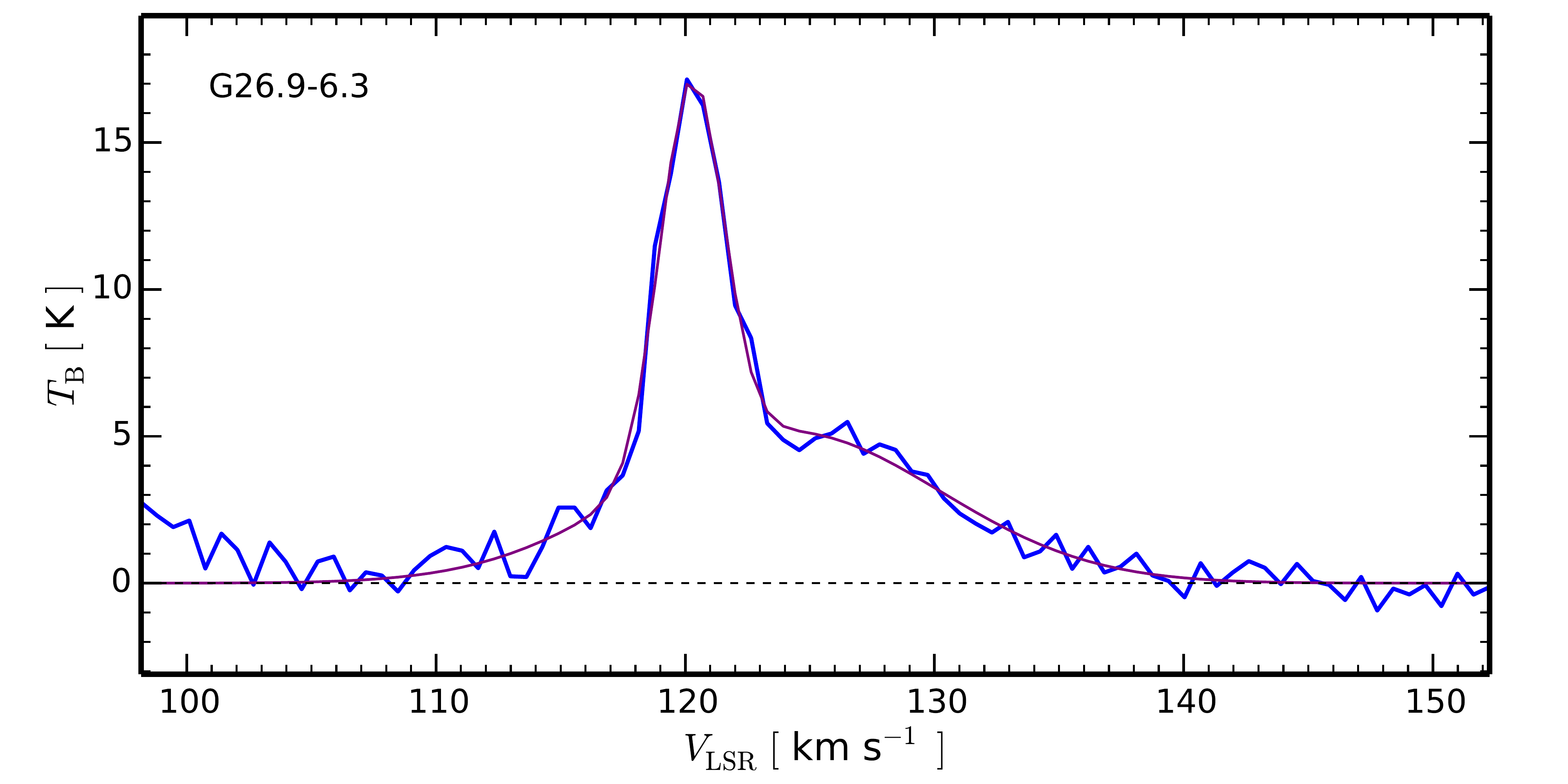}
        }
\vspace{-0.5cm}
\caption{Summary of G$26.9-6.3$ as described in the caption to Fig.~\ref{fig:G160-4plots}. A cloud with a sharp boundary toward high latitude and toward the plane, and diffuse material spreading out on the opposite side.  More than half of the mass is within 20 pc of the brightest point, but the diffuse material is responsible for the broad mass profile. Its spectrum at peak \NHI\ shows two components with broad and narrow FWHM of 14.5 and 3.0 \kms, suggestive of a cloud with a two-phase structure. The line widths of the two components place an upper limit on their temperatures of 4600 K and 200~K.}
\label{fig:G269-4plots}
\end{figure}

\begin{figure}
\centering
    \vspace{-2cm}

    \captionsetup[subfigure]{labelformat=empty}
    \subfloat[][]{
        \centering
        \hspace{0.0cm}
        \includegraphics[width=0.5\textwidth]{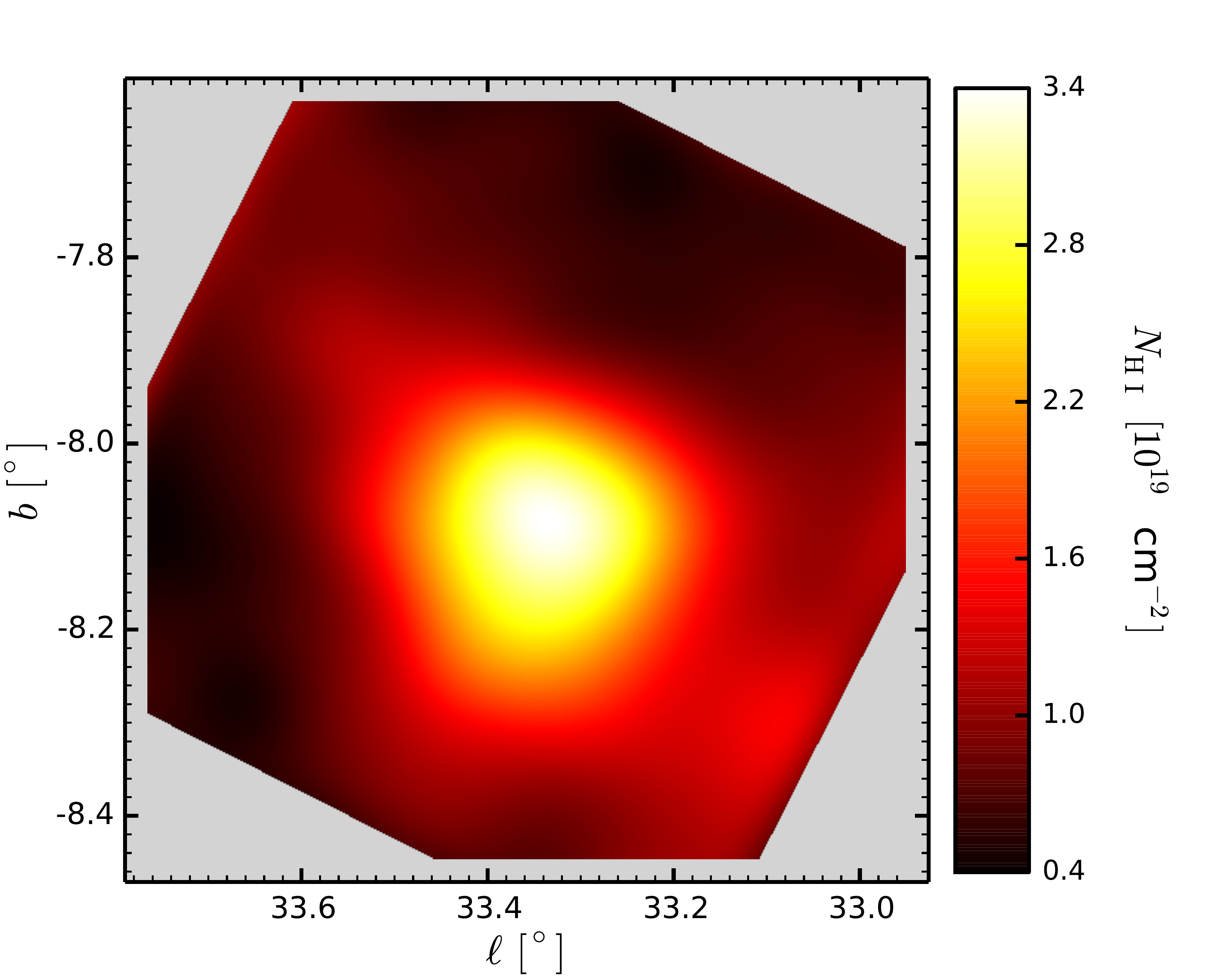}
        \hspace{-0.1cm}}
    \subfloat[][]{
        \centering
        \includegraphics[width=0.5\textwidth]{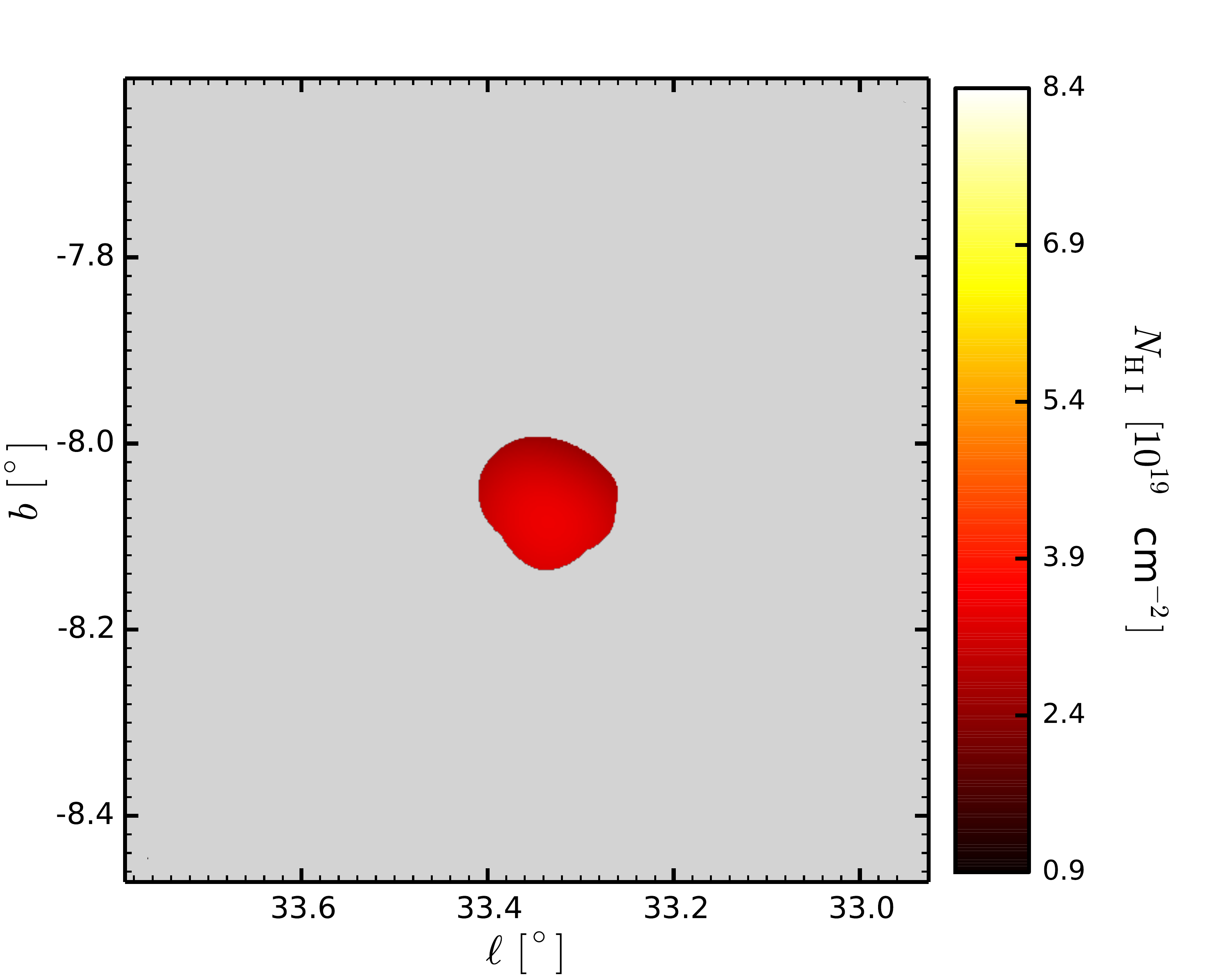}
        }

    \vspace{-1.3cm}

    \subfloat[][]{
        \centering
        \hspace{0.0cm}
        \includegraphics[width=0.5\textwidth]{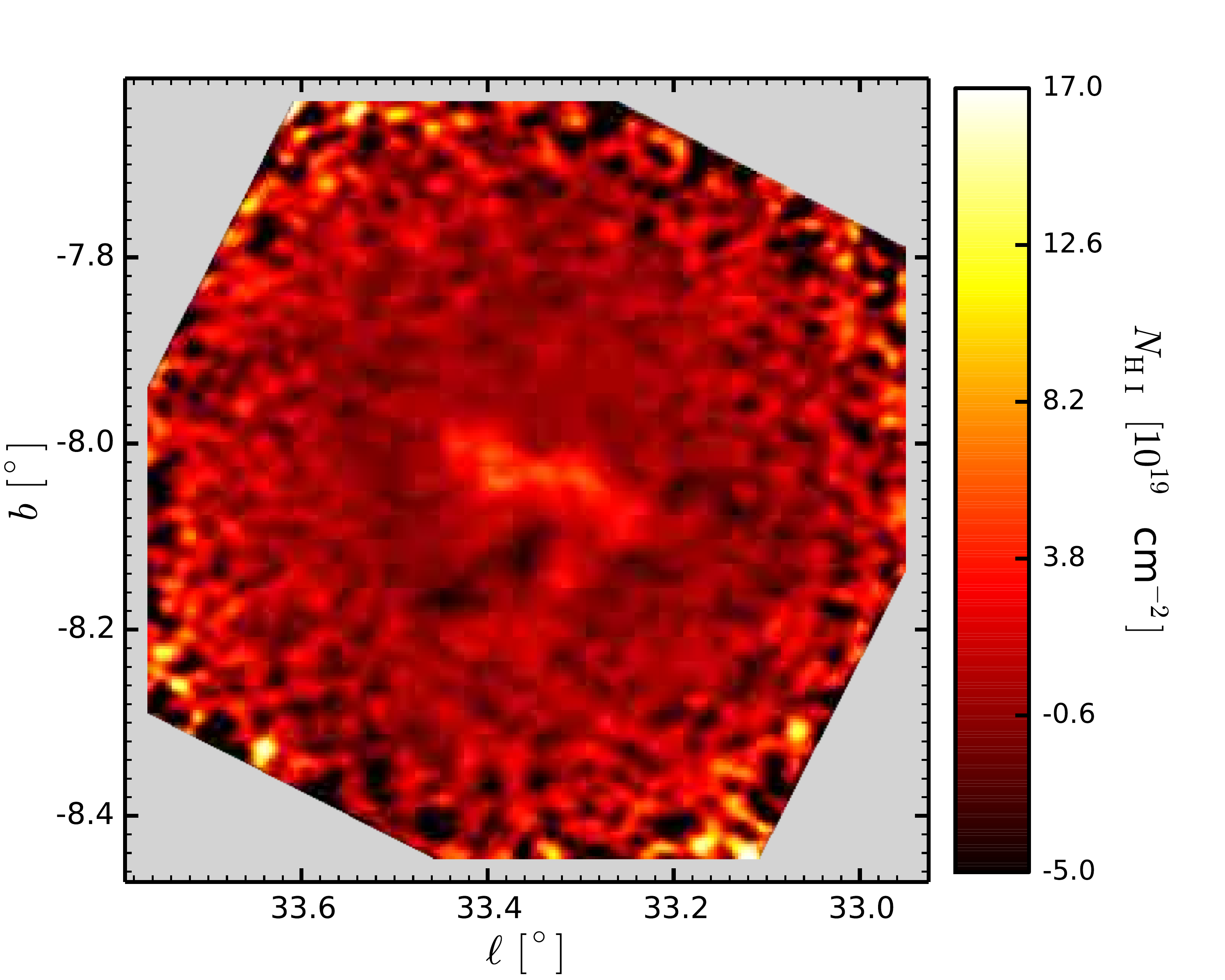}
        \hspace{-0.1cm}}
    \subfloat[][]{
        \centering
        \includegraphics[width=0.5\textwidth]{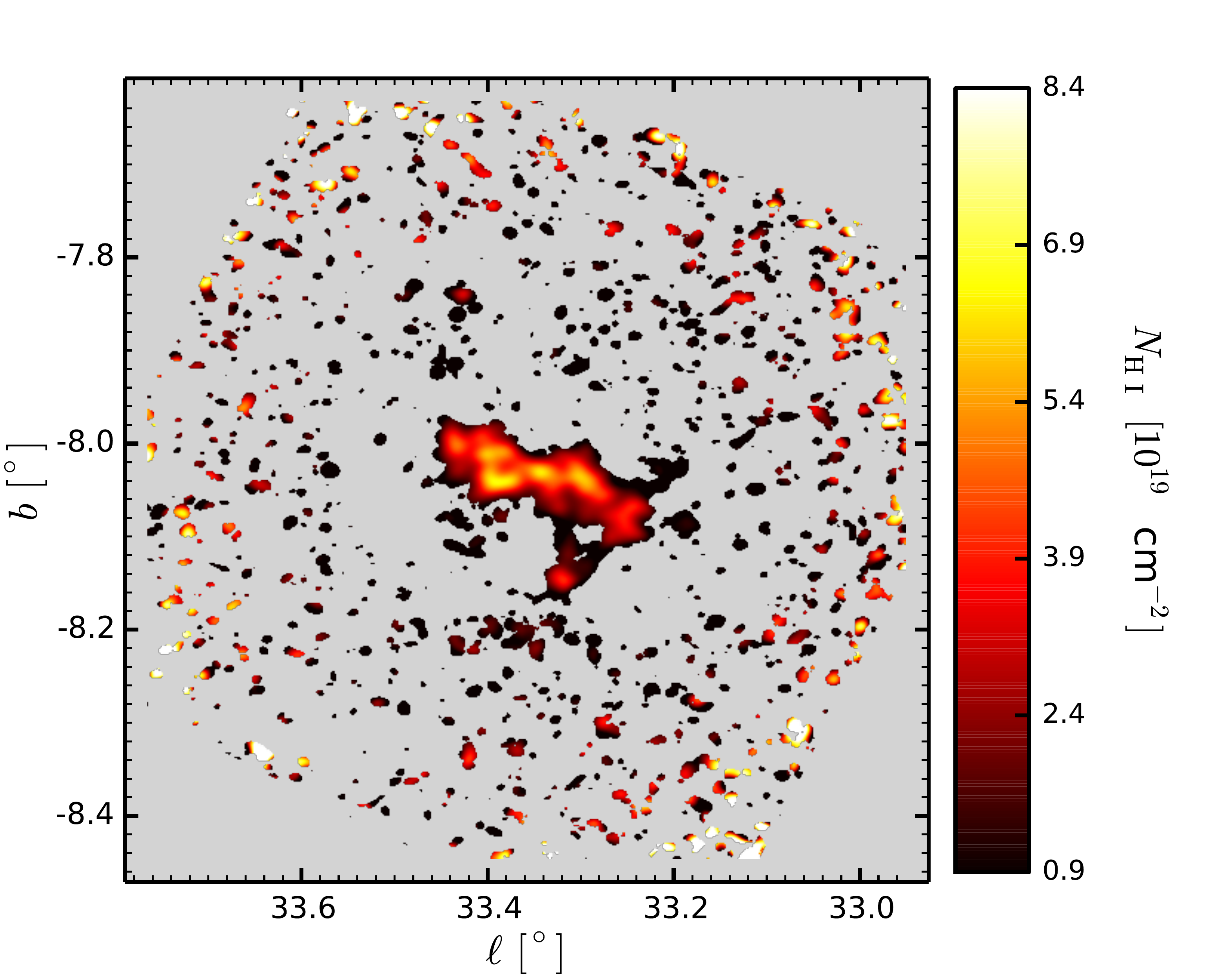}
        }

     \vspace{-1.3cm}

     \subfloat[][]{
        \centering
        \hspace{0.0cm}
        \includegraphics[width=0.5\textwidth]{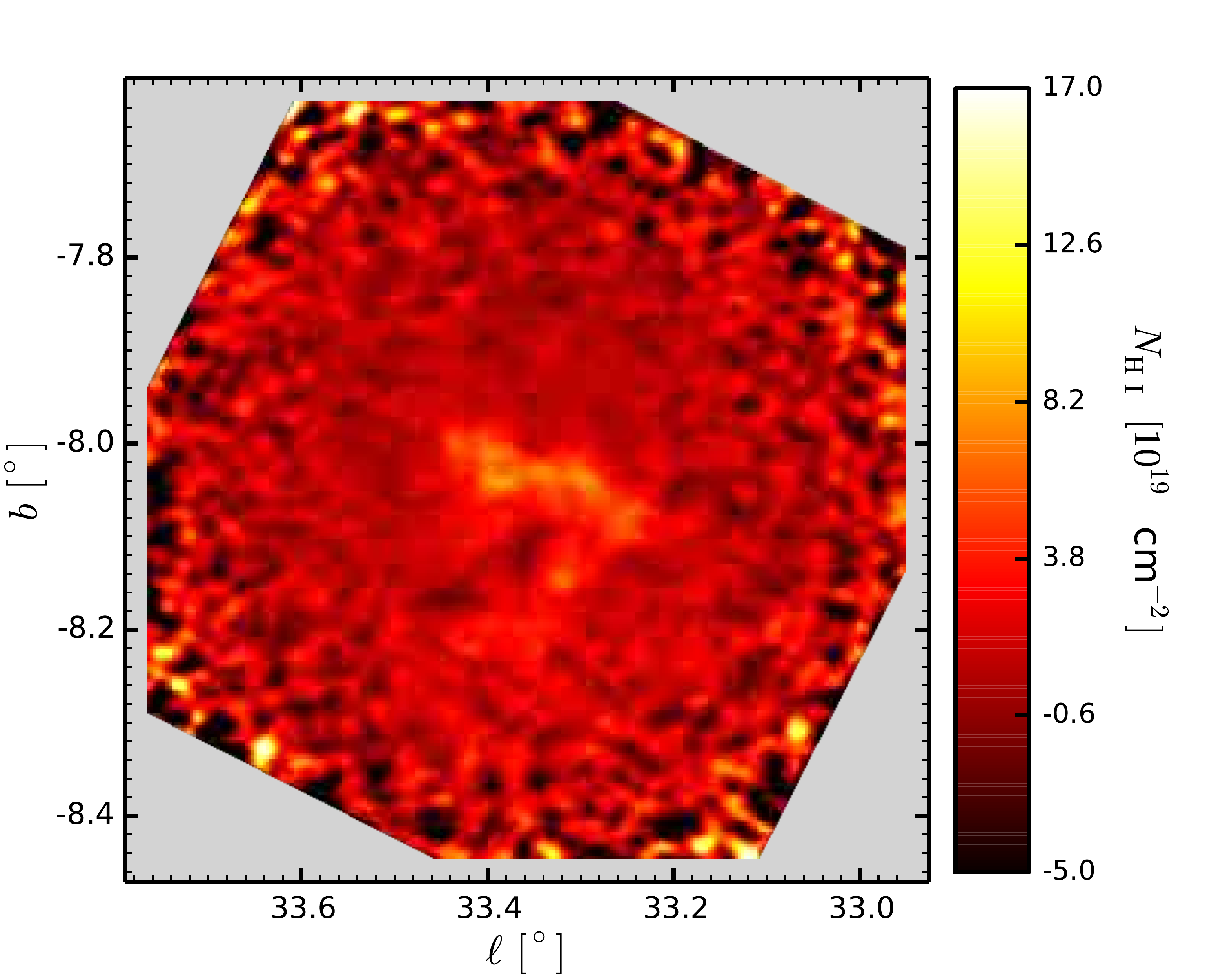}
        \hspace{-0.1cm}}
    \subfloat[][]{
        \centering
        \includegraphics[width=0.5\textwidth]{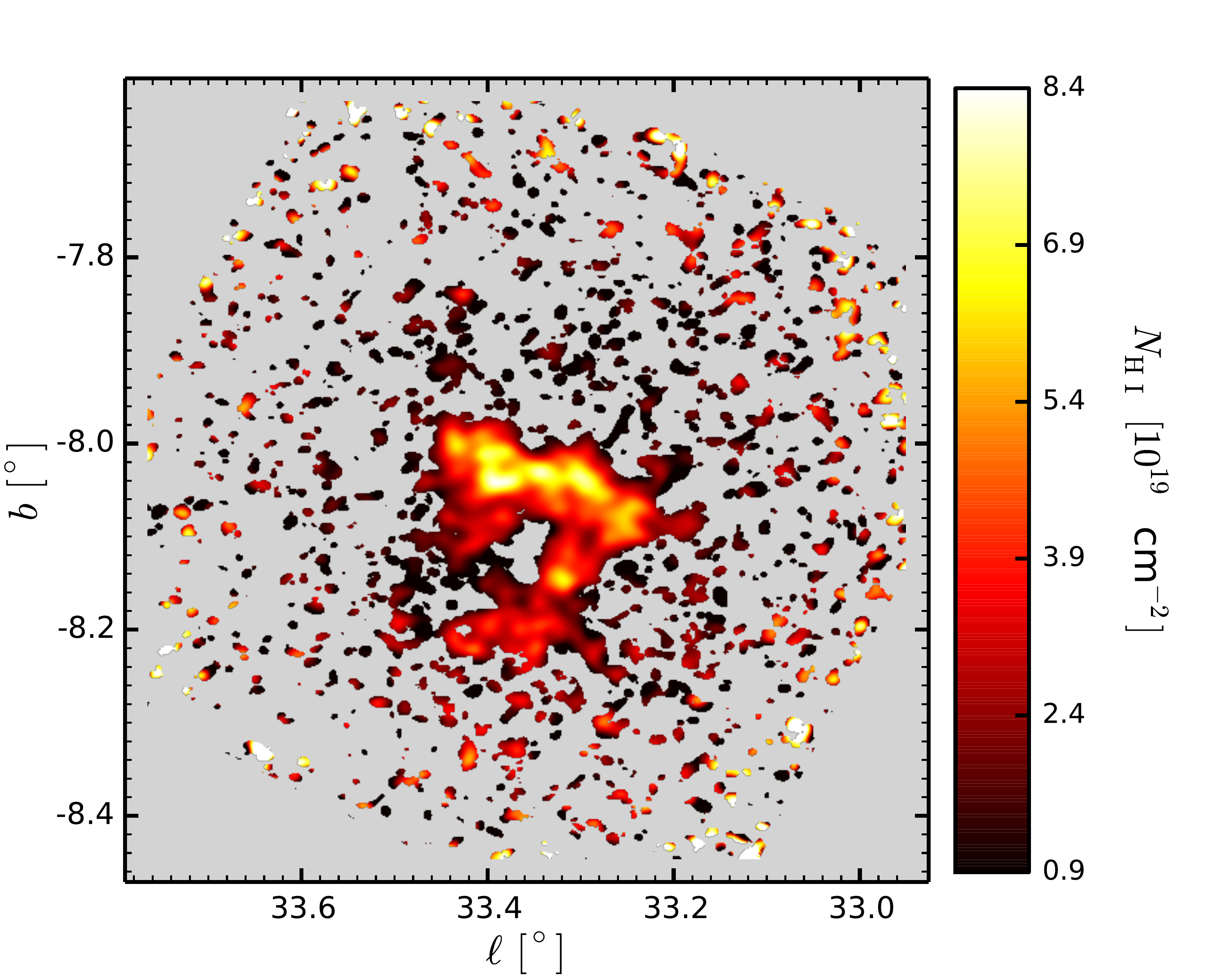}
        }
\vspace{-0.5cm}
\caption{\HI column density maps for G$33.4-8.0$, integrated over 37 spectral channels in the interval $92.3 \leq \VLSR \leq 115.5$~\kms, as described in the caption to Fig.~\ref{fig:160_HImaps}.}
\label{fig:334_HImaps}
\end{figure}

\begin{figure}
    \centering
    \vspace{-1.0cm}

    \captionsetup[subfigure]{labelformat=empty}
    \subfloat[][]{
        \centering
        \hspace{-0.5cm}
        \includegraphics[width=0.75\textwidth]{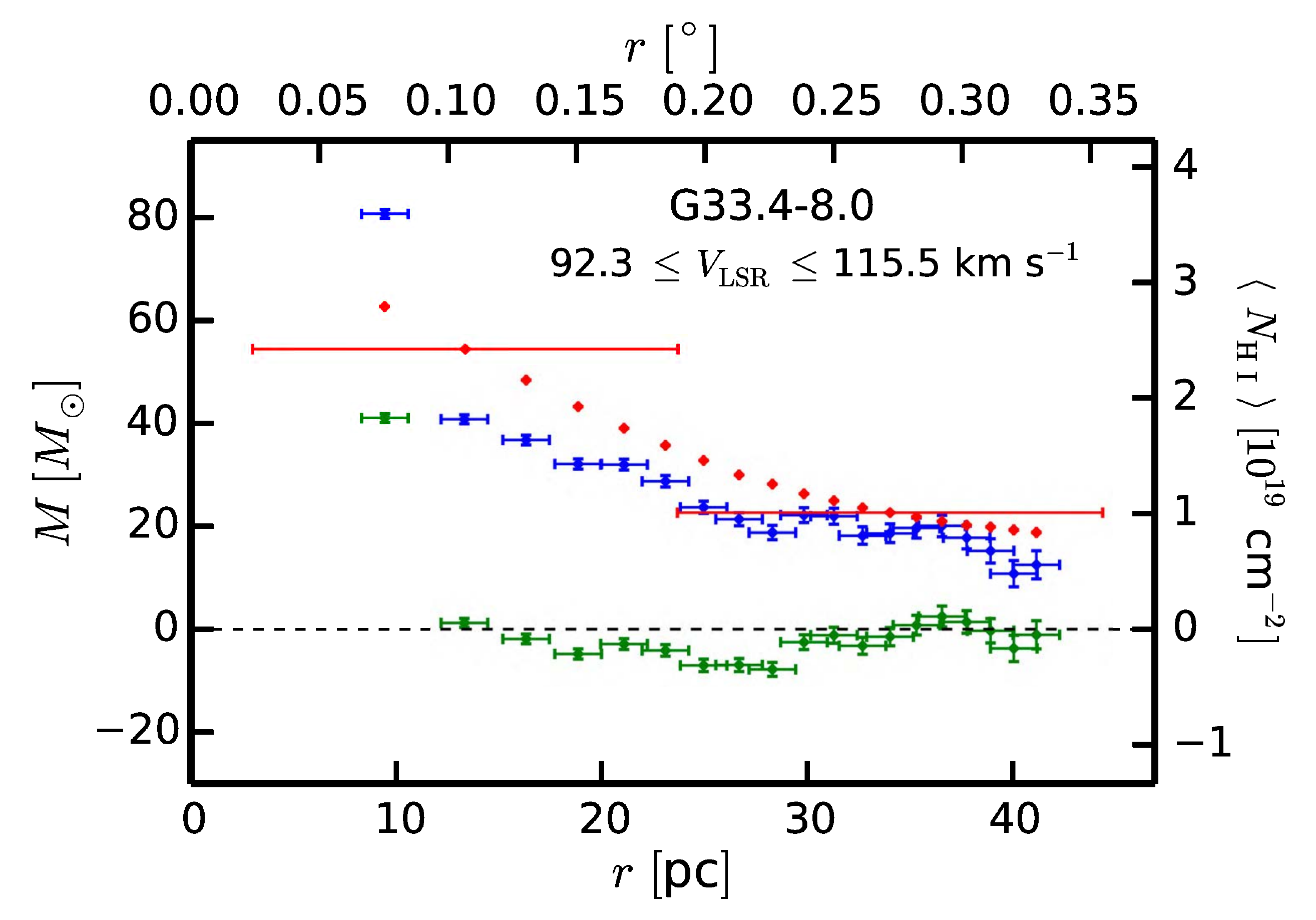}
        }

    \vspace{-1.0cm}

    \subfloat[][]{
        \centering
        \hspace{-1.8cm}
        \includegraphics[width=0.7\textwidth]{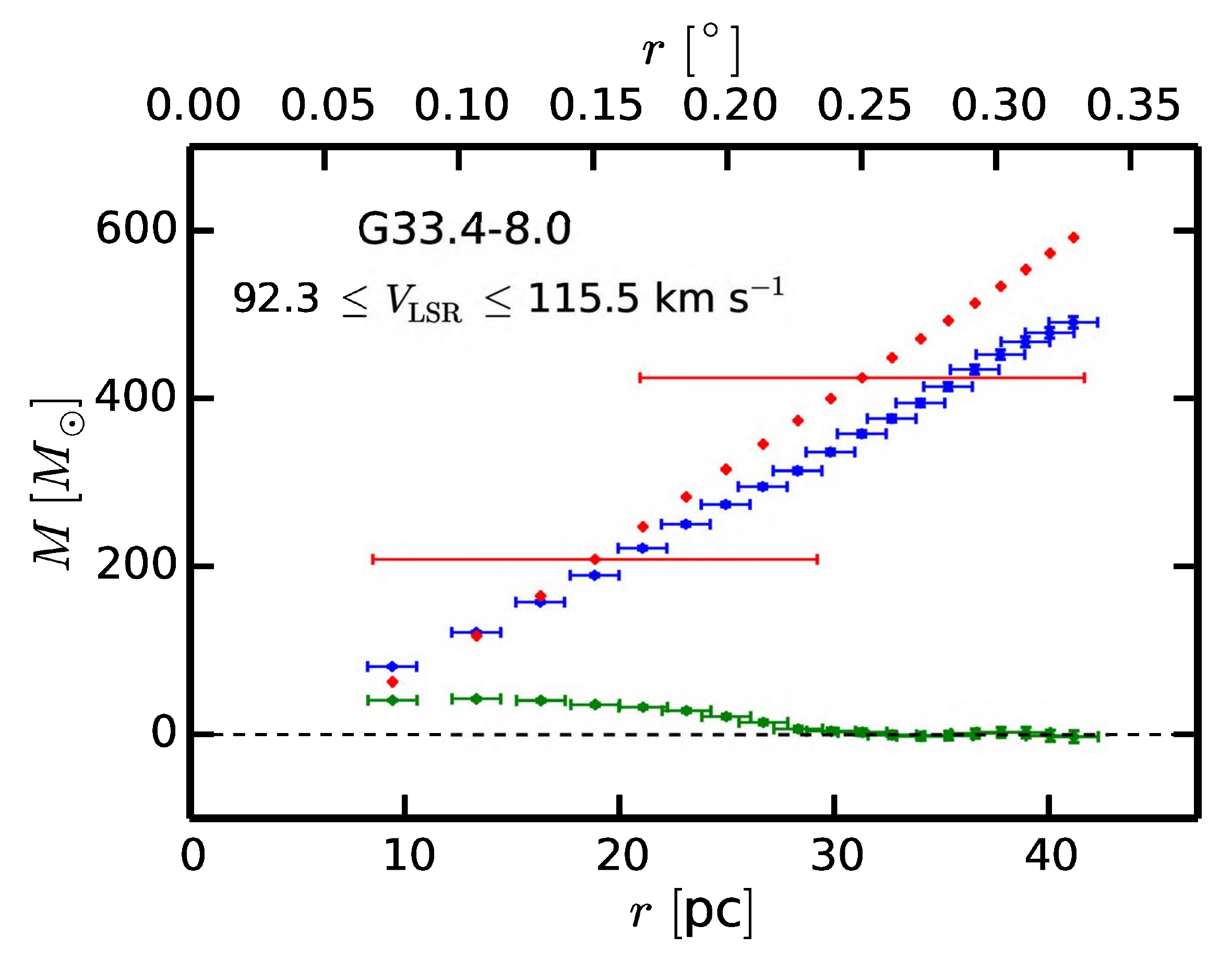}
        }

\caption{Radial mass profiles for G$33.4-8.0$ as described in the caption to Fig.~\ref{fig:160_mass_graphs}.}\label{fig:334_mass_graphs}
\end{figure}

\begin{figure}
\vspace{-2.0cm}
\centering
\captionsetup[subfigure]{labelformat=empty}
    \subfloat[][]{
        \centering
        \hspace{-1cm}
        \includegraphics[width=1.0\textwidth]{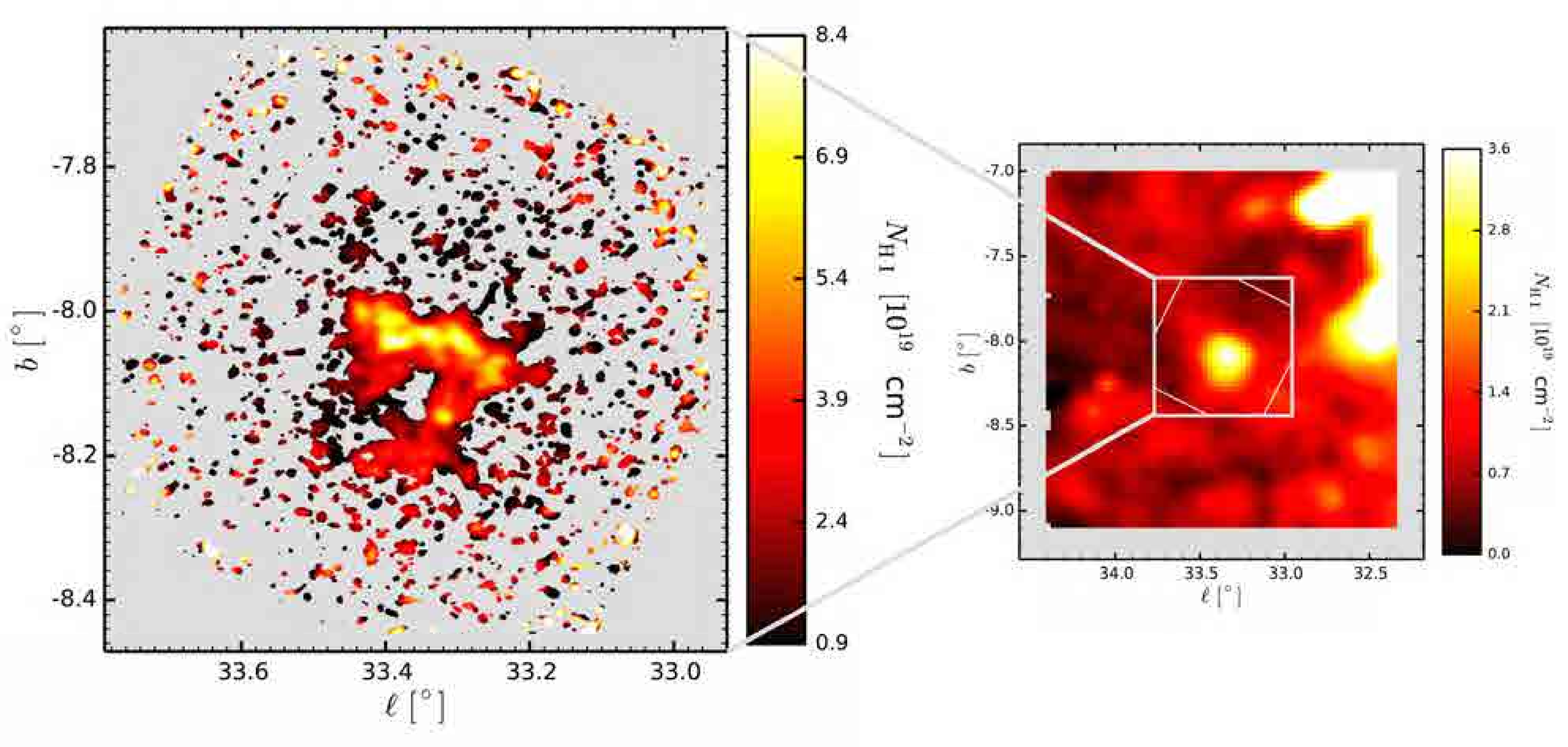}
        \hspace{-0cm}
        }

    \vspace{-1.0cm}

    \subfloat[][]{
        \centering
        \hspace{-1cm}
        \includegraphics[width=0.5\textwidth]{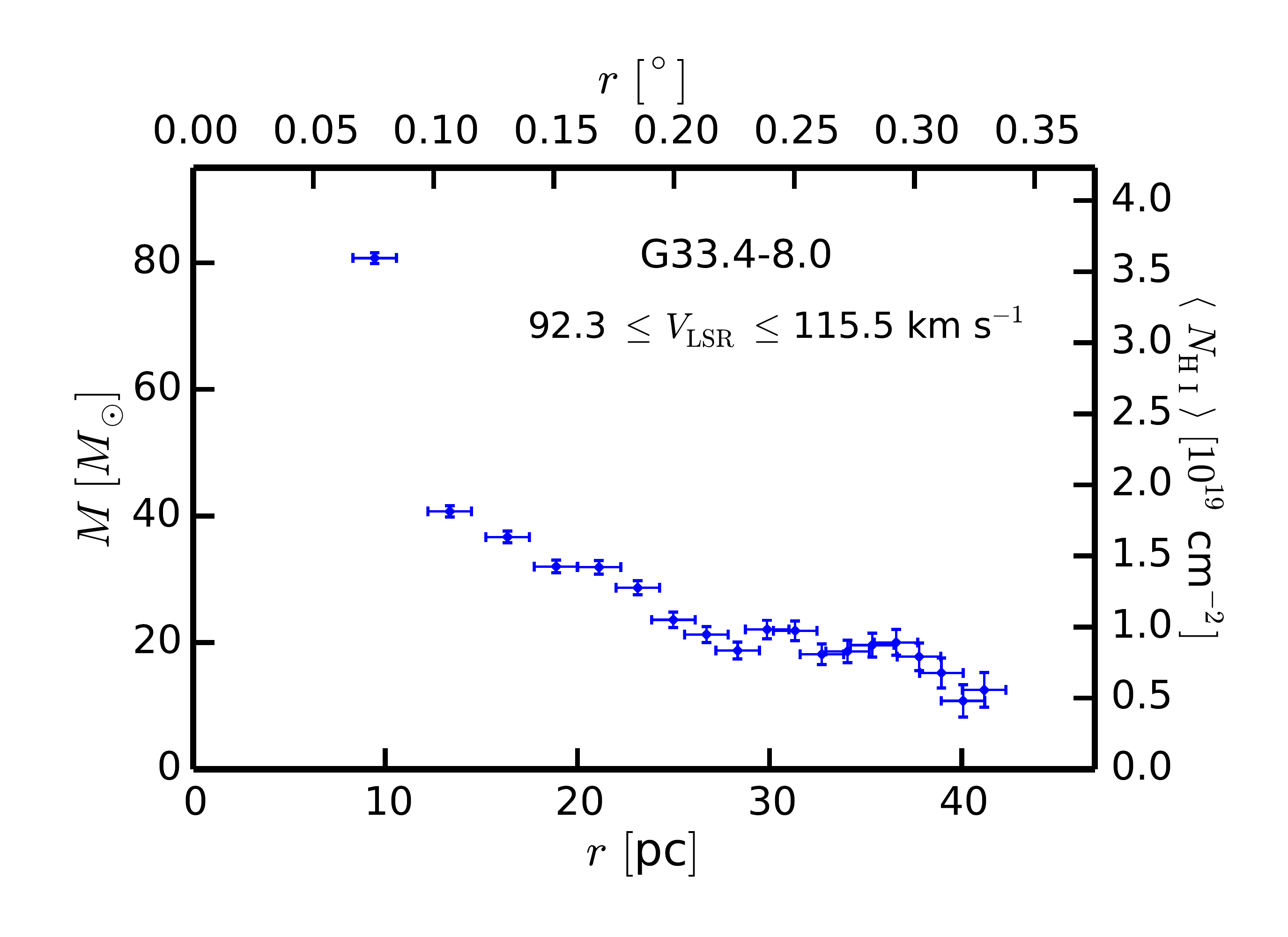}
        \hspace{-0.7cm}
        }
    \subfloat[][]{
        \centering
        \includegraphics[viewport = 0 -30 1150 650, width=0.59\textwidth]{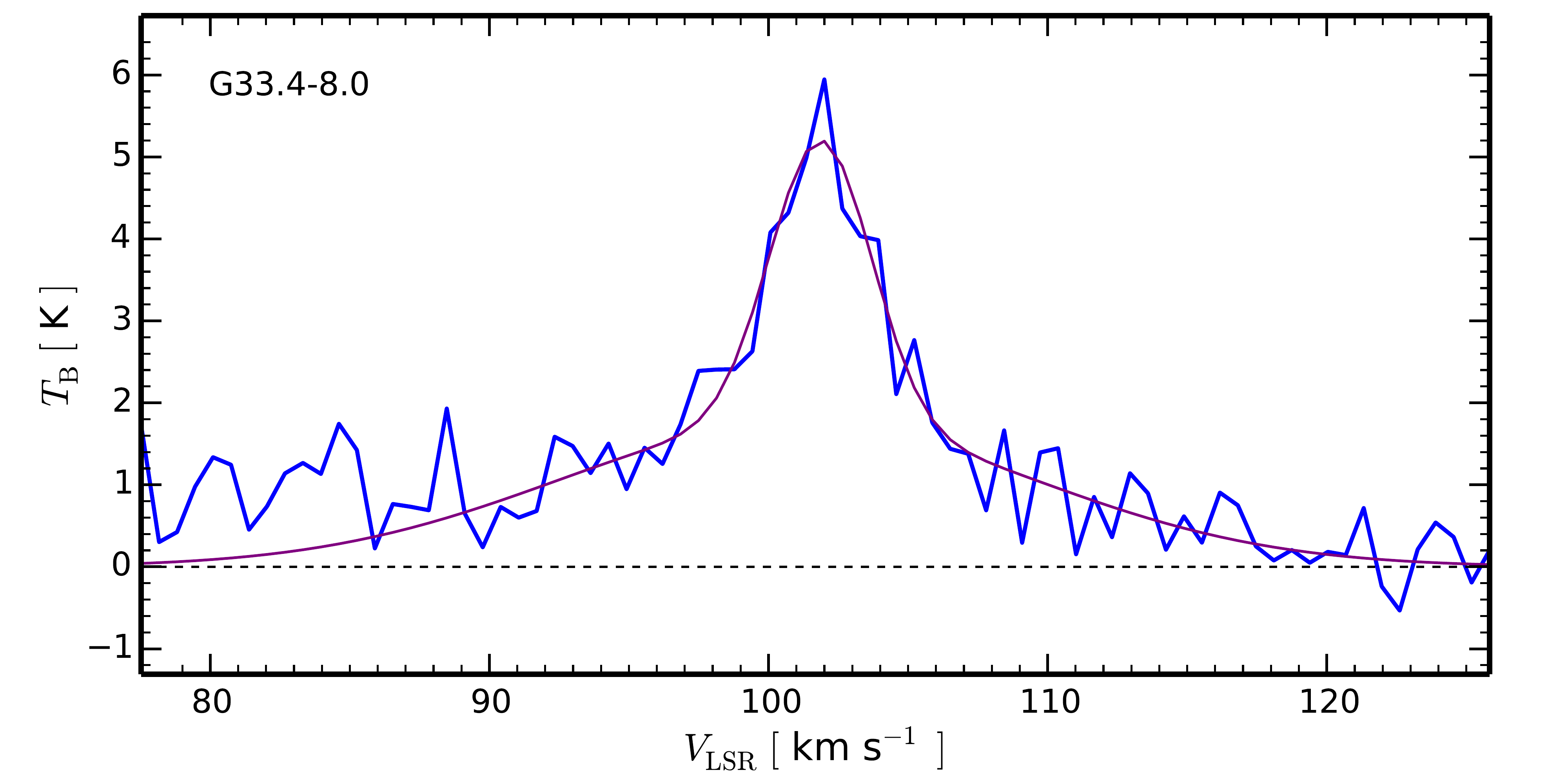}
        }
\vspace{-0.5cm}
\caption{Summary of G$33.4-8.0$ as described in the caption to Fig.~\ref{fig:G160-4plots}. This small isolated cloud of $\sim 160$ \Msun\ in its dense part is the most distant from the Galactic plane of any in the sample and has a complex structure. It has many column density maxima spread along a central ridge on the edge closest to the Galactic plane. The spectrum at the highest column density point has two components, one broad and one narrow, suggesting a two-phase thermal structure.}
\label{fig:G334-4plots}
\end{figure}

\begin{figure}
\centering
    \vspace{-2cm}

    \captionsetup[subfigure]{labelformat=empty}
    \subfloat[][]{
        \centering
        \hspace{0.0cm}
        \includegraphics[width=0.5\textwidth]{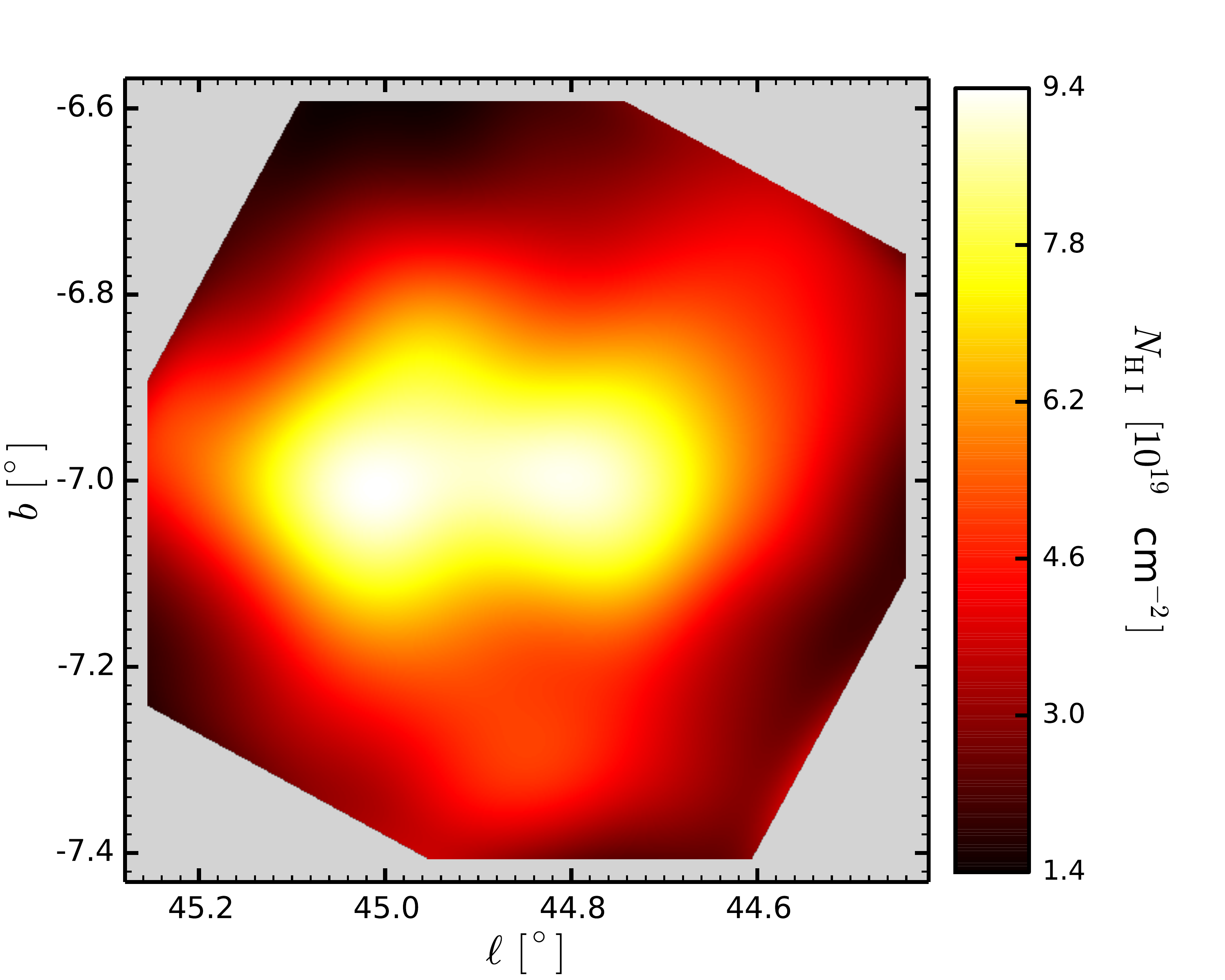}
        \hspace{-0.1cm}}
    \subfloat[][]{
        \centering
        \includegraphics[width=0.5\textwidth]{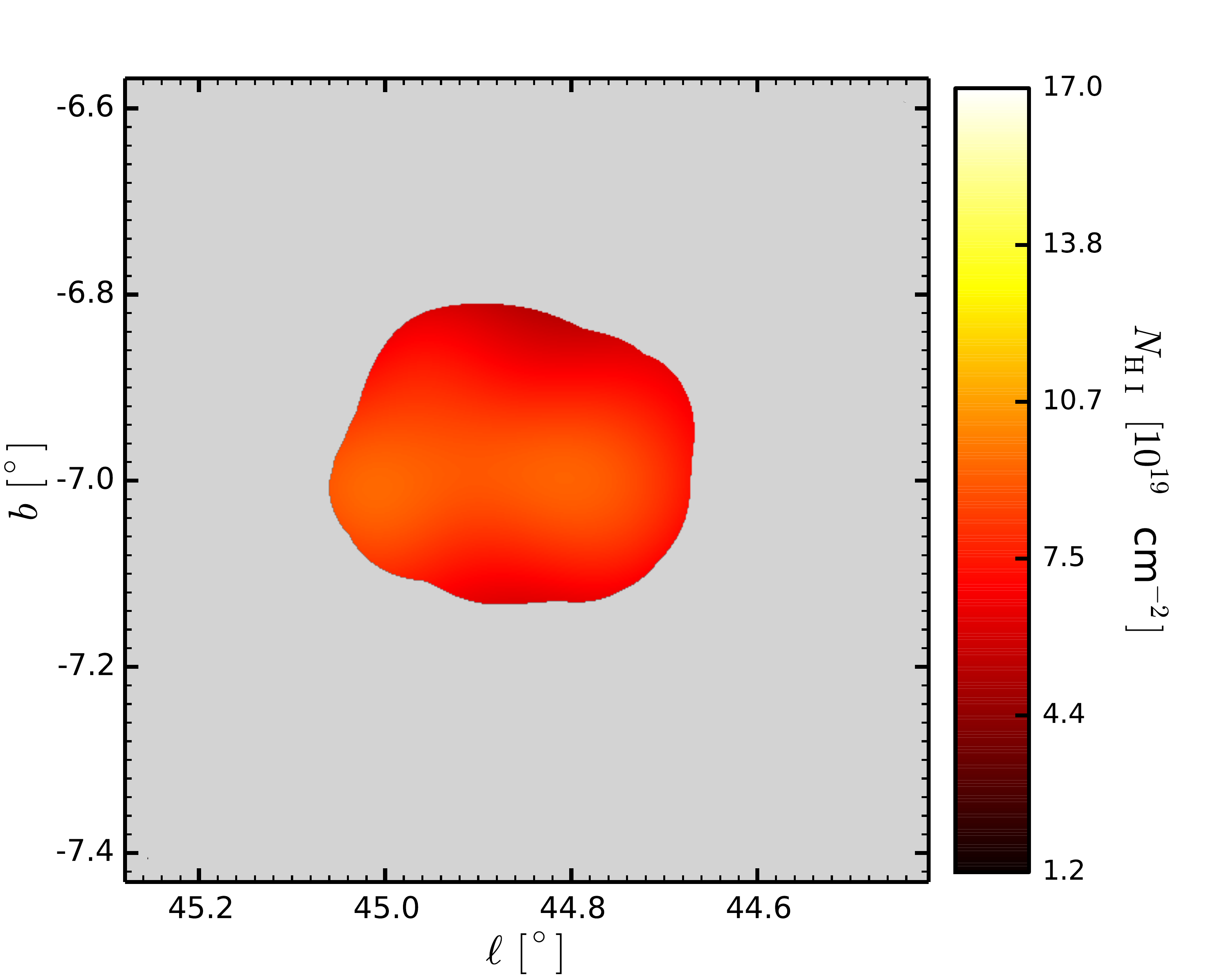}
        }

    \vspace{-1.3cm}

    \subfloat[][]{
        \centering
        \hspace{0.0cm}
        \includegraphics[width=0.5\textwidth]{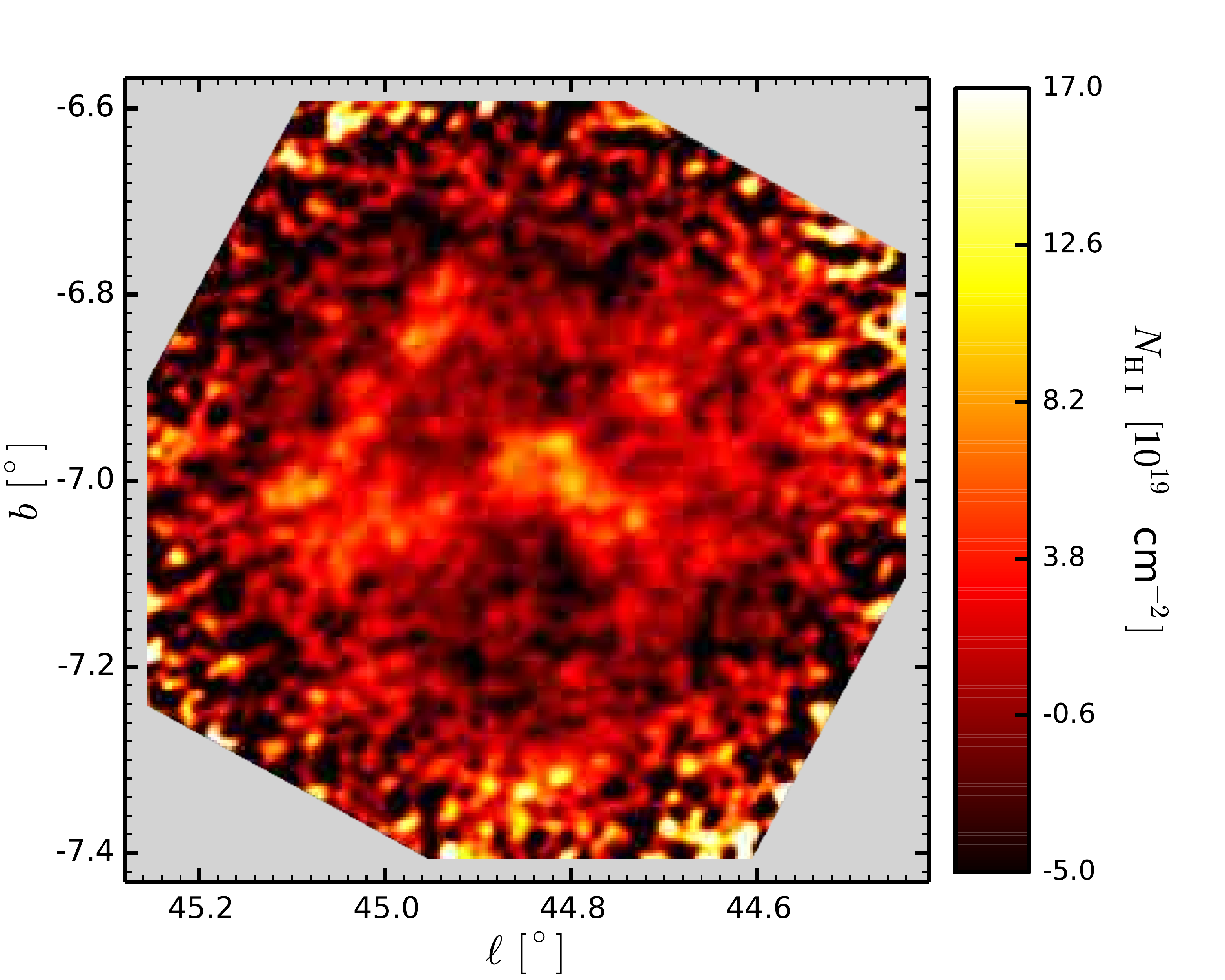}
        \hspace{-0.1cm}}
    \subfloat[][]{
        \centering
        \includegraphics[width=0.5\textwidth]{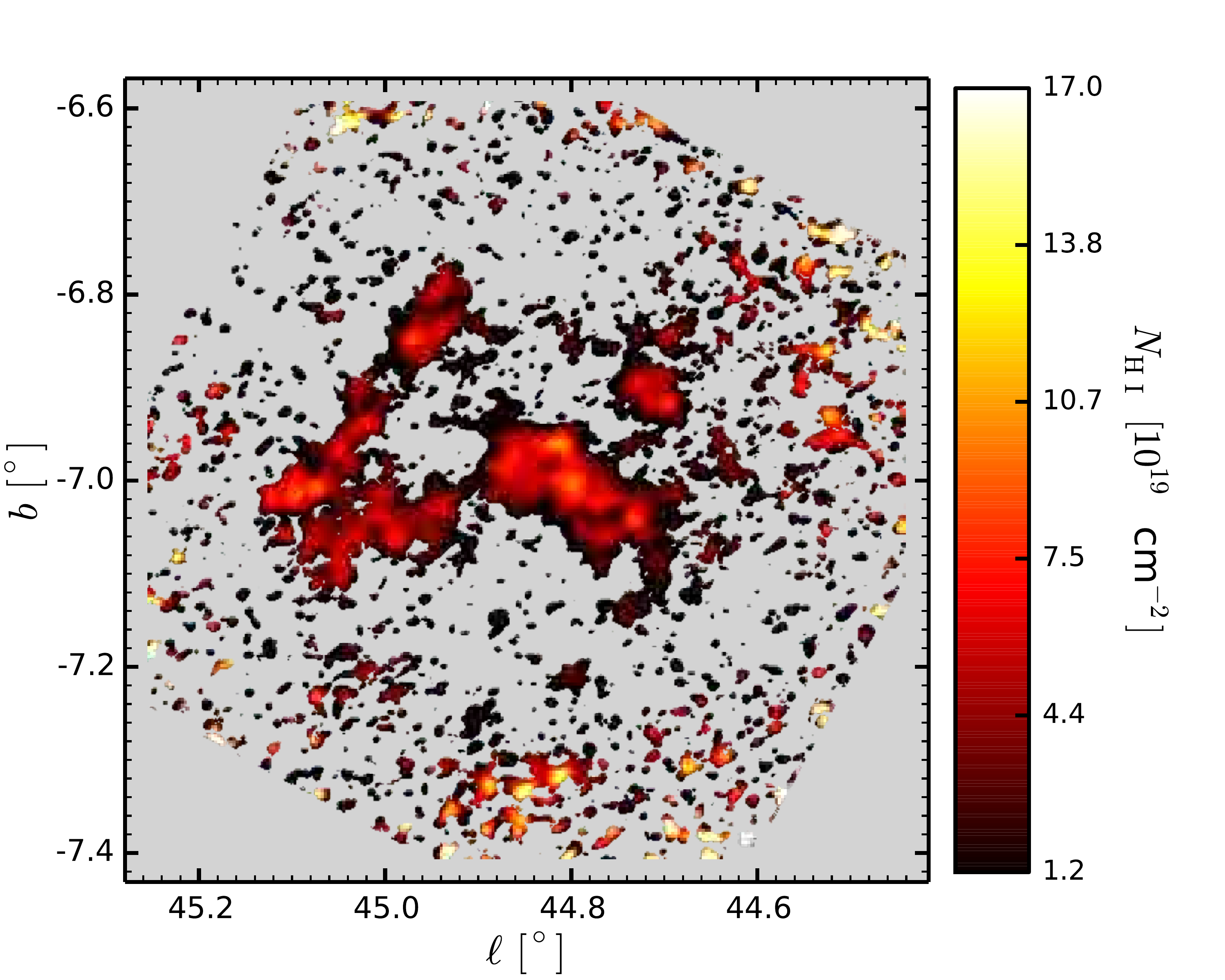}
        }

     \vspace{-1.3cm}

     \subfloat[][]{
        \centering
        \hspace{0.0cm}
        \includegraphics[width=0.5\textwidth]{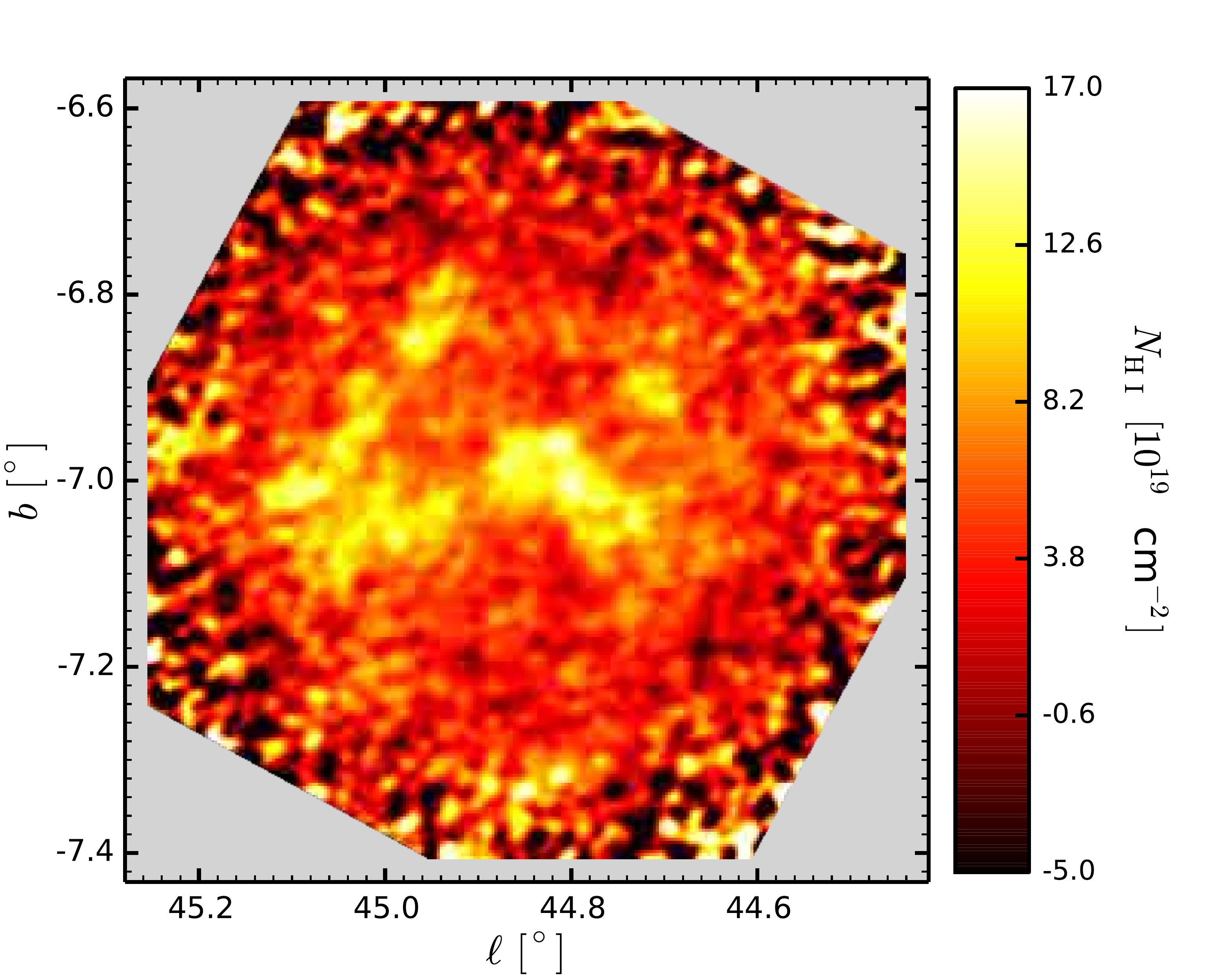}
        \hspace{-0.1cm}}
    \subfloat[][]{
        \centering
        \includegraphics[width=0.5\textwidth]{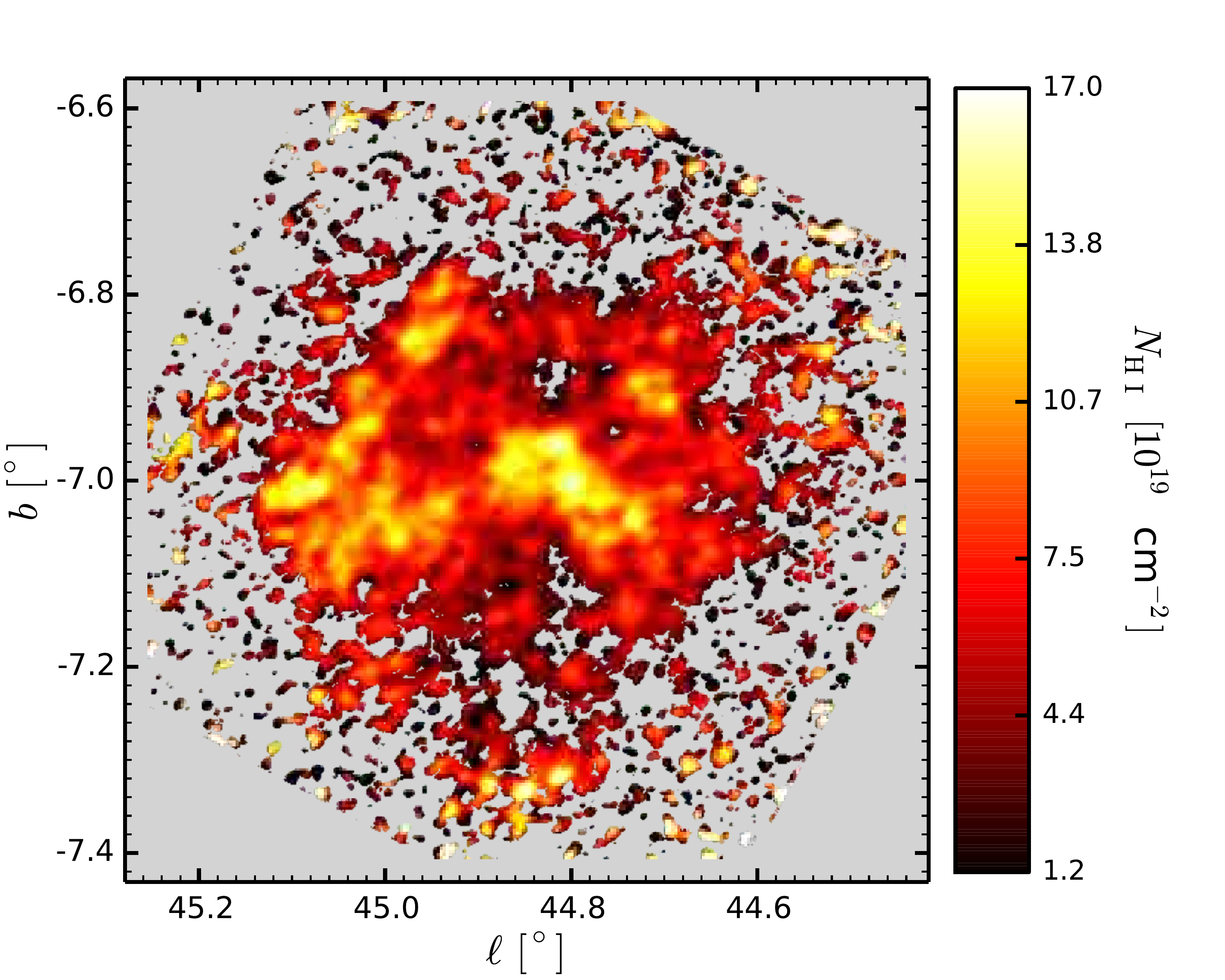}
        }
\vspace{-0.5cm}
\caption{\HI column density maps for G$44.8-7.0$, integrated over 73 spectral channels in the interval $79.8 \leq \VLSR \leq 126.2$~\kms, as described in the caption to Fig.~\ref{fig:160_HImaps}.}
\label{fig:448_HImaps}
\end{figure}

\begin{figure}
    \centering
    \vspace{-1.0cm}

    \captionsetup[subfigure]{labelformat=empty}
    \subfloat[][]{
        \centering
        \hspace{-0.5cm}
        \includegraphics[width=0.75\textwidth]{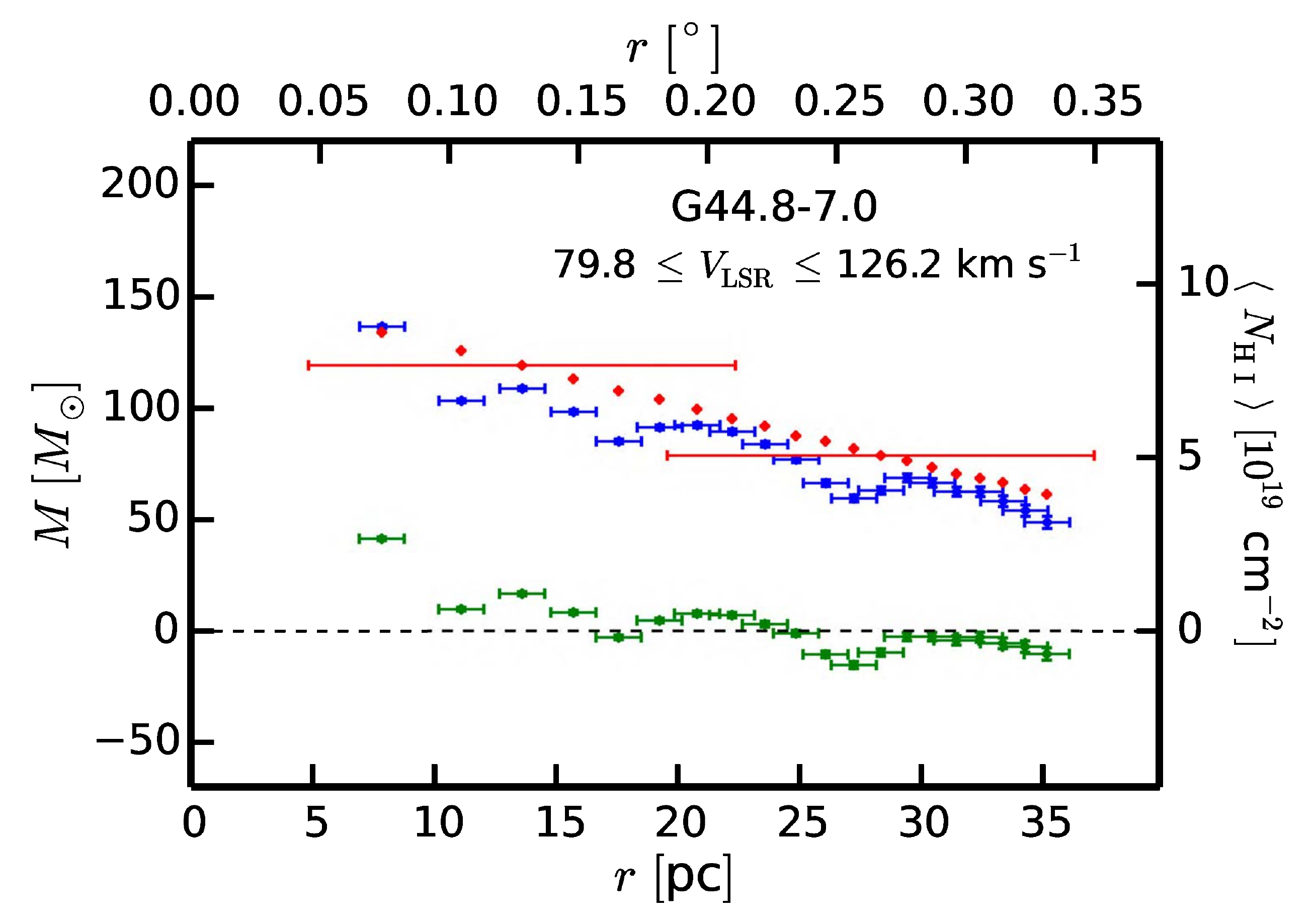}
        }

    \vspace{-1.0cm}

    \subfloat[][]{
        \centering
        \hspace{-1.8cm}
        \includegraphics[width=0.7\textwidth]{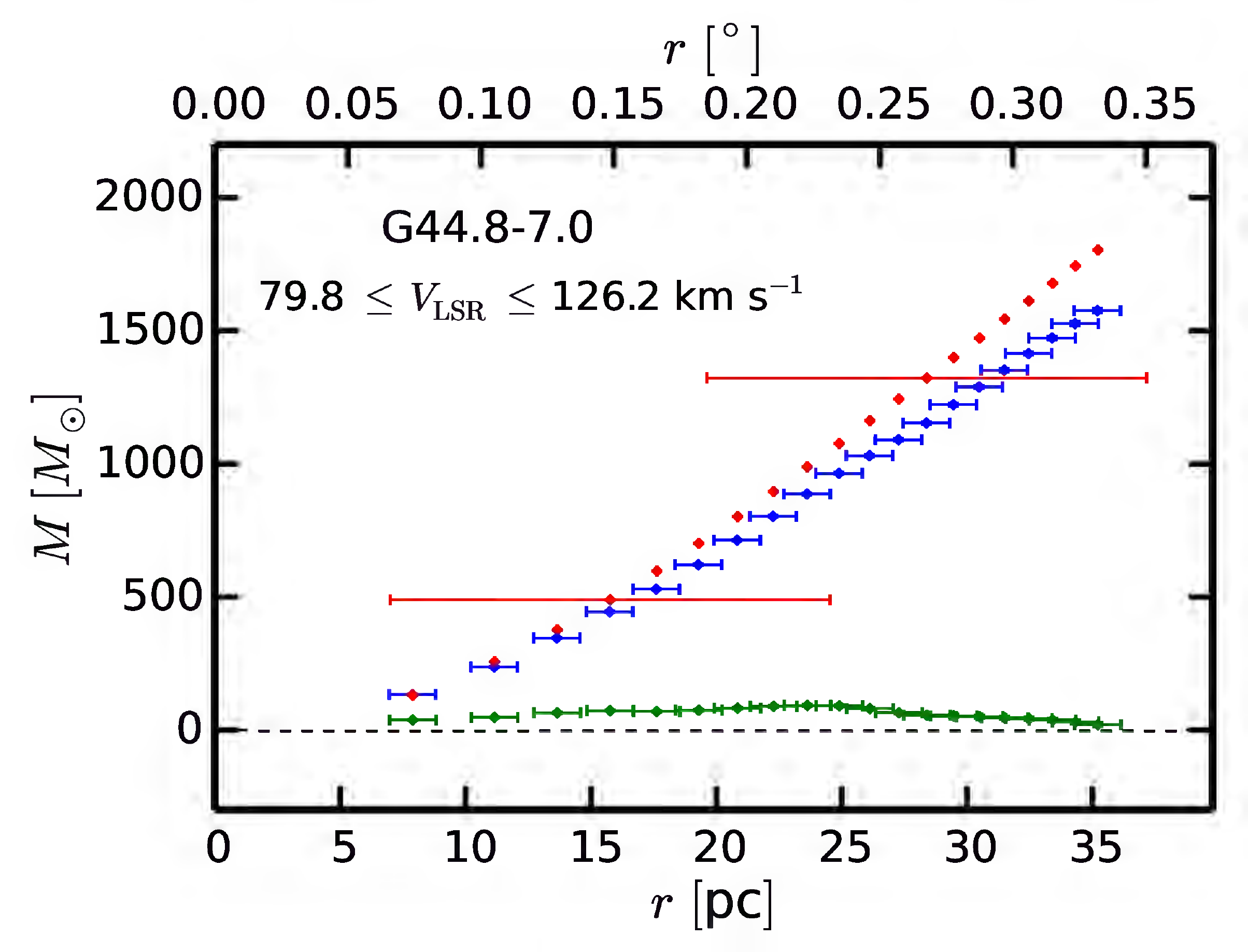}
        }

\caption{Radial mass profiles for G$44.8-7.0$ as described in the caption to Fig.~\ref{fig:160_mass_graphs}.}\label{fig:448_mass_graphs}
\end{figure}

\begin{figure}
\vspace{-2.0cm}
\centering
\captionsetup[subfigure]{labelformat=empty}
    \subfloat[][]{
        \centering
        \hspace{-1cm}
        \includegraphics[width=1.0\textwidth]{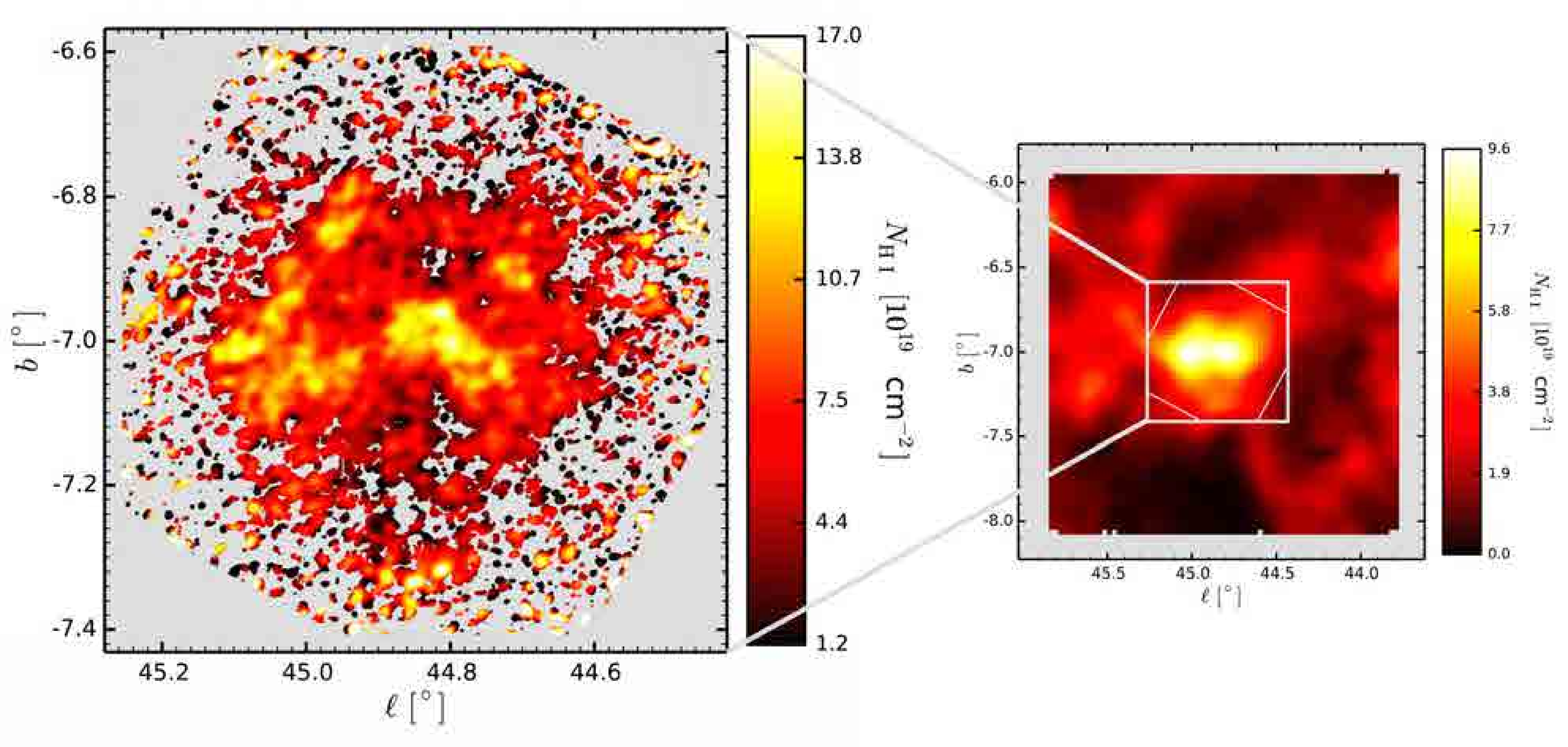}
        \hspace{-0cm}
        }

    \vspace{-1.0cm}

    \subfloat[][]{
        \centering
        \hspace{-1cm}
        \includegraphics[width=0.5\textwidth]{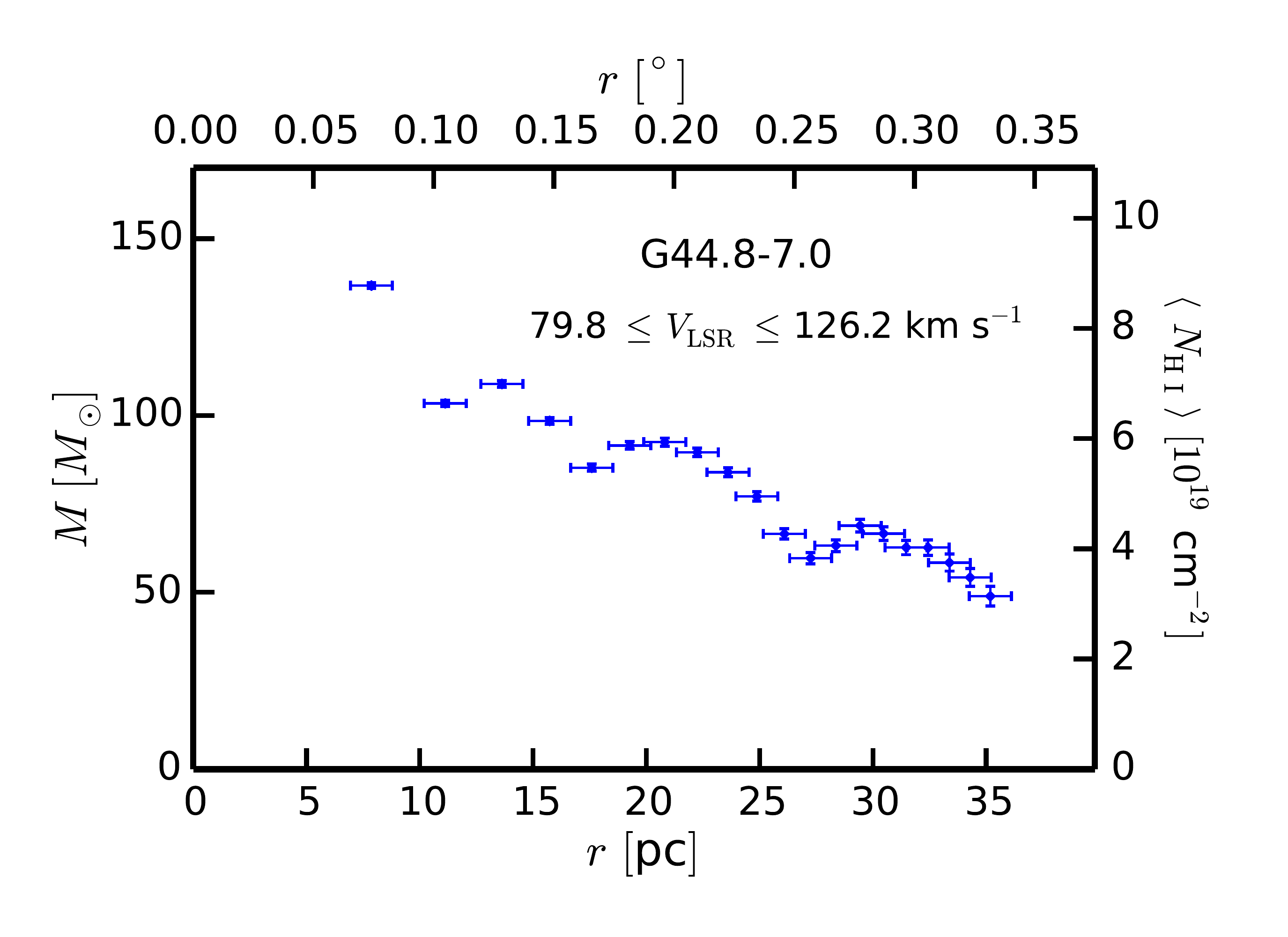}
        \hspace{-0.7cm}
        }
    \subfloat[][]{
        \centering
        \includegraphics[viewport = 0 -30 1150 650, width=0.59\textwidth]{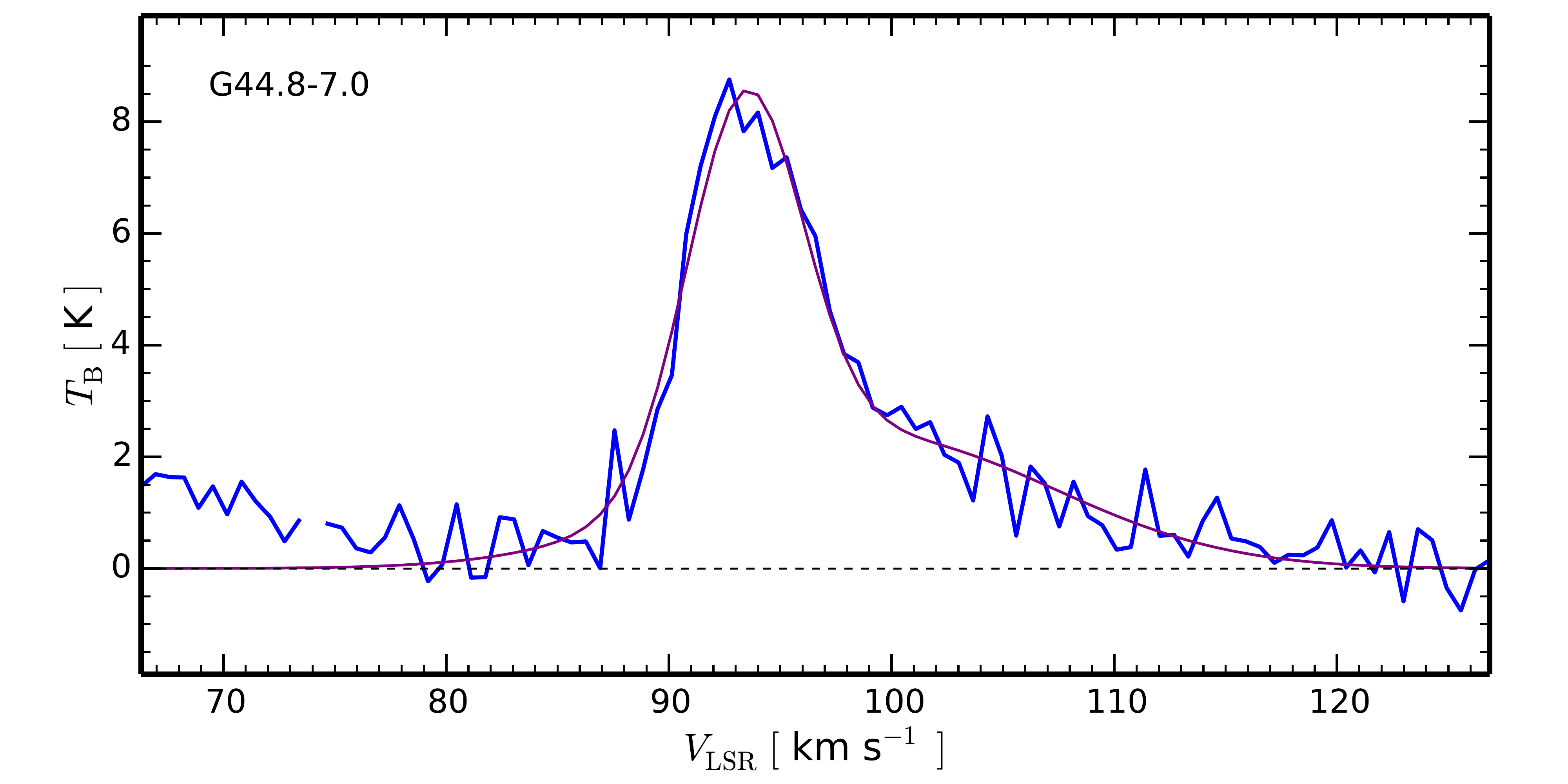}
        }
\vspace{-0.5cm}
\caption{Summary of G$44.8-7.0$ as described in the caption to Fig.~\ref{fig:G160-4plots}. In the spectrum a channel at 74.0~\kms is flagged due to an RFI spike. This large, massive cloud extends  beyond the primary beam of the VLA. This, and the very small cloud G$16.0+3.0$ are the only clouds whose shape is nearly circular, though the brightest gas is concentrated in numerous small clumps.  The spectrum at the \NHI\ peak is asymmetric with a broad wing to higher velocity.  This cloud has among the largest excess $\VLSR$ beyond the expected $V_t$ of any in the sample.}
\label{fig:G448-4plots}
\end{figure}

Figures~\ref{fig:175_HImaps}--\ref{fig:G448-4plots} present the same for the remaining 9 clouds.

Measured cloud properties are summarized in Table~\ref{tab:observed-props}.  The values of line
brightness temperature, velocity and line width are derived from the Gaussian decomposition, sampled
toward the position of the peak \NHI.  Errors are $1 \sigma$ from the Gaussian fit. Four of the clouds have spectral lines that require two Gaussians. The values of \NHI\ in col.~7 are integrals over the relevant velocity range (given in caption to each Figure) and are always close to the value from the sum of the Gaussian fits. We do not know the optical depth and so we cannot correct for self-absorption in the 21cm line, therefore the column densities and masses calculated in this paper are all lower limits.

The mass profiles were constructed in the following fashion. The mass was sampled over a set of annuli
of varying radius increment but equal area, all centered at the main column density peak of the map. With equal areas, the mass in each annulus is proportional to the average \NHI\ with the same proportionality coefficient and thus the points can be plotted simultaneously with two sets of axes: distance-dependent mass $M$ vs.~linear radius and distance-independent $\lbr \right.$\NHI$\left.\rbr$ vs.~angular radius (see Fig.~\ref{fig:160_mass_graphs} and its counterparts, upper panels). The vertical error bars show the cumulative error due to noise\footnote{Both the mass and the linear radius measurements are also subject to the distance uncertainty (see Table~\ref{tab:derived-props}) not shown in these plots.}, the horizontal ones show the average beam radius $r_{\mathrm{beam}} \equiv \onehalf (B_{maj} B_{min})^{\onehalf}$, with $B_{maj}$ and $B_{min}$ for the VLA given in column~2 of Table~\ref{tab:beams}. The GBT values are the same for all clouds: 10.2\arcmin $\times$ 9.7\arcmin, slightly increased compared to the original GBT beam due to pre-IMMERGE processing.

Since there is a significant freedom of choice for such annulus sets, we have selected a fixed maximum radius $r_n$ of the sequence, the same for all clouds and equal to 0\fdg319, which covers most of the VLA primary beam, with the exception of a small outer portion having the highest noise. Then we required the width of the largest annulus to be equal to $r_{\mathrm{beam}}$, i.e. at the resolution of the map. Designating $n$ the number of annuli and $r_1$ the smallest radius of the set, we arrive at the following equations:
\begin{align}
r_n - r_{n-1} & = r_{\mathrm{beam}}, \label{fla:annuli1} \\
r_i^2 - r_{i-1}^2 & = r_1^2, & i \in [2,n], \\
r_i^2 - r_{i-1}^2 & = \frac{r_{i-1}}{i-1}, & i \in [2,n], \\
r_i^2 & = i r_1^2, & i  \in [2,n]. \label{fla:annuli4}
\end{align}

Solving eq.~\ref{fla:annuli1} together with eq.~\ref{fla:annuli4} at $i=n$ and $i=n-1$ we get:
\be
\label{fla:annuli6}
r_1 =\left [ r_{\mathrm{beam}} ( 2r_n-r_{\mathrm{beam}}) \right ]^{\frac{1}{2}}.\
\ee
Using eqns.~\ref{fla:annuli4} and \ref{fla:annuli6} we have constructed the appropriate set of annuli for each cloud.

Comparing line parameters of the GBT-only observations of Table~\ref{tab:table0} with the
high resolution GBT+VLA
results in Table \ref{tab:observed-props} we find identical mean
velocities, with a  difference of $0 \pm 2$ \kms.  As expected
given the small angular structure revealed in the maps, the lines in the final  maps  are
brighter by factors that range from 1.6  to 11 (for the very compact G$16.0+3.0$), with a
median value around 4.  These ratios are smaller than might occur:  hydrogen clouds with sizes $<1\arcmin$
should appear $\approx 100$ times brighter to the VLA than the GBT, so it
seems that the major structures in the clouds have been resolved in the current data.
The line widths have a much smaller variation with the increased angular resolution.
For lines that appear narrow to the GBT it is typical that
they are even narrower in the combined data by around 20\%, suggesting that the higher resolution
observations are revealing colder or less turbulent material.  But the broadest lines as measured
with the GBT are sometimes even broader in the combined data, clouds G$24.3-5.3$ and G$25.2+4.5$
being examples.

\subsection{Estimating Masses, Sizes and Densities of the Clouds}
\label{subsec:MassSize}

Table~\ref{tab:derived-props} gives the derived properties of the clouds for their adopted distances. Errors on derived parameters are dominated by the the distance uncertainties, whose estimates are discussed in \S\ref{subsec:distance}.

\begin{figure}
\centering
    \vspace{-1cm}

    \captionsetup[subfigure]{labelformat=empty}
    \subfloat[][]{
        \centering
        \hspace{0.0cm}
        \includegraphics[width=0.5\textwidth]{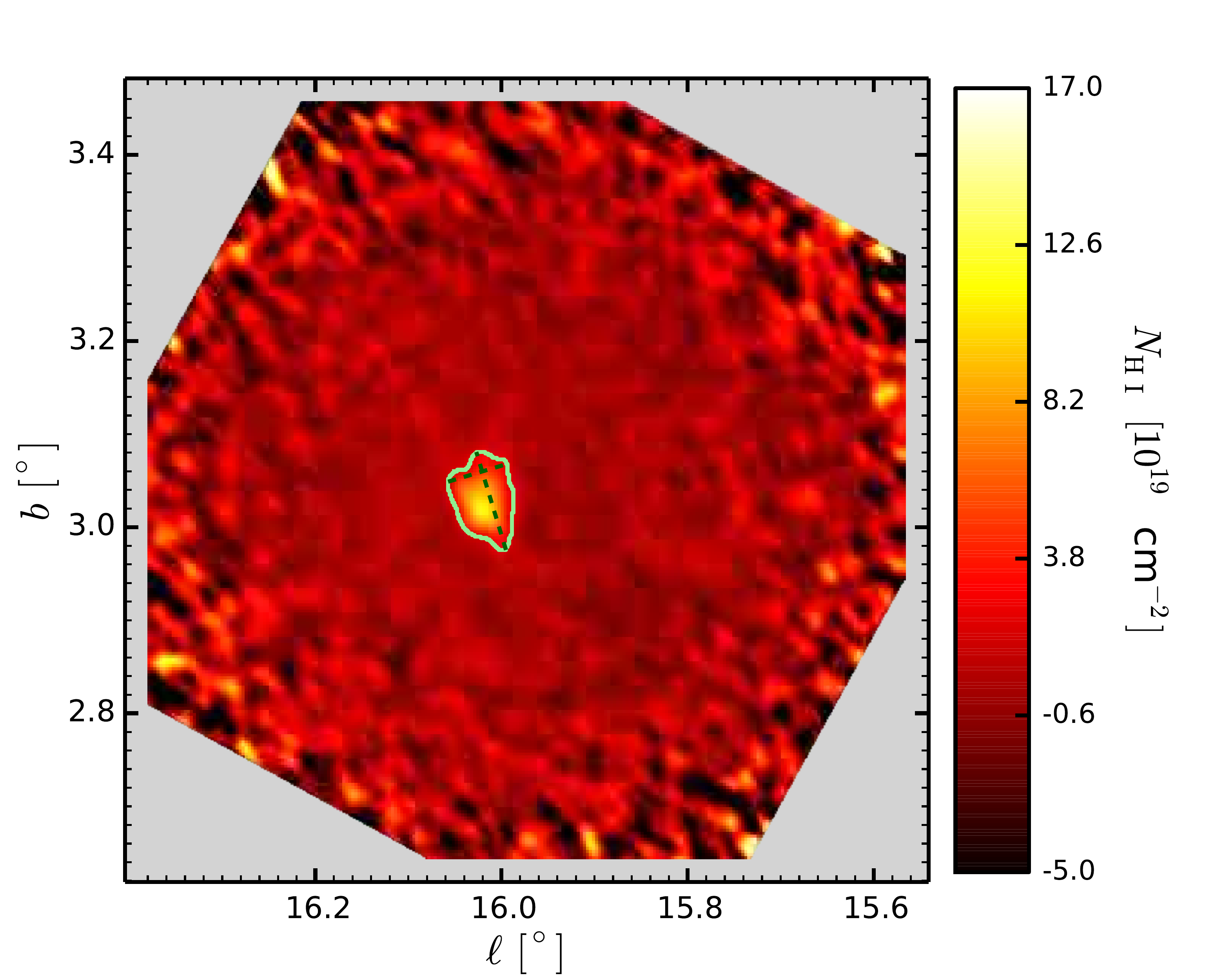}
        \hspace{-0.1cm}}
    \subfloat[][]{
        \centering
        \includegraphics[width=0.5\textwidth]{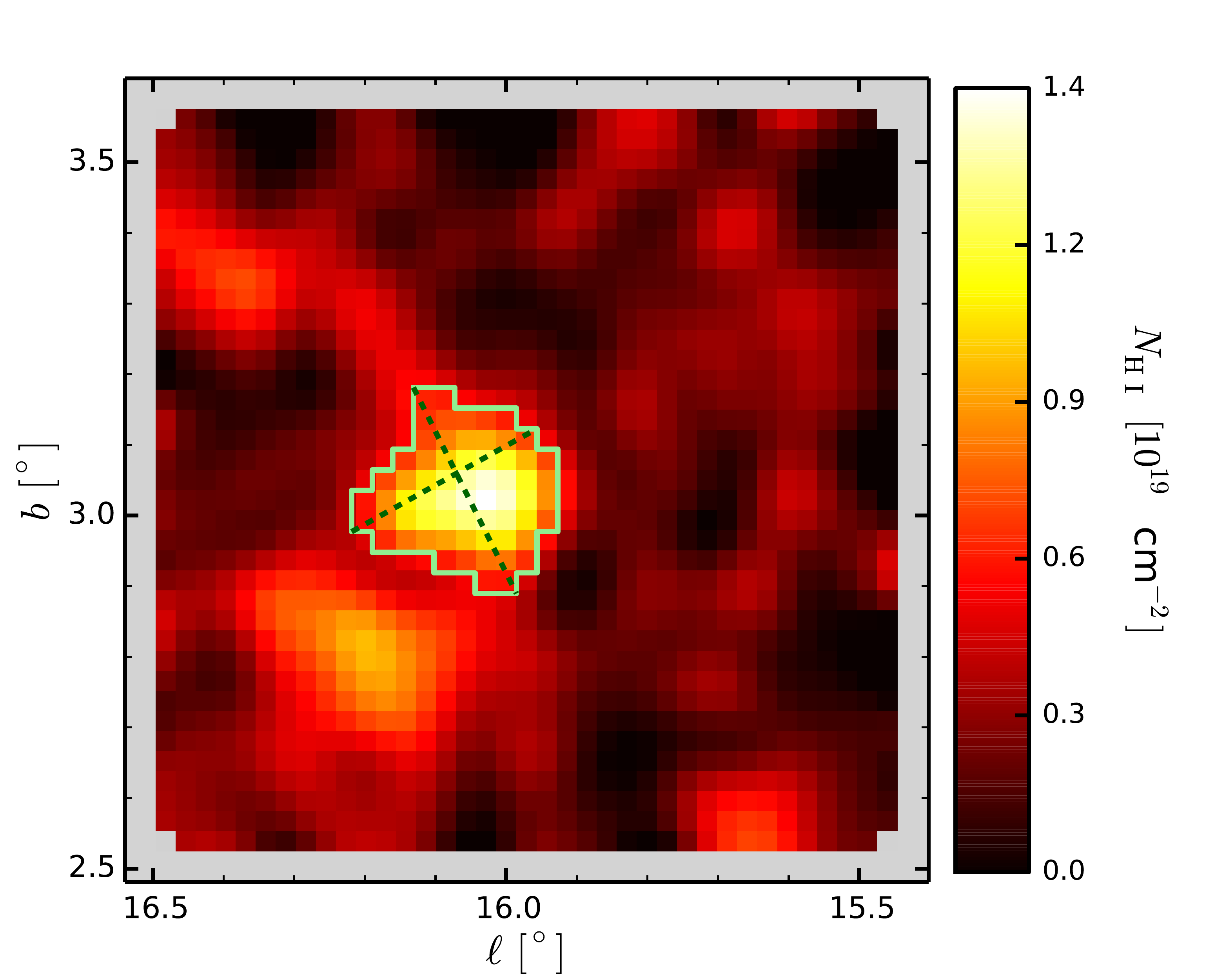}
        }

    \vspace{-1.3cm}

    \subfloat[][]{
        \centering
        \hspace{0.0cm}
        \includegraphics[width=0.5\textwidth]{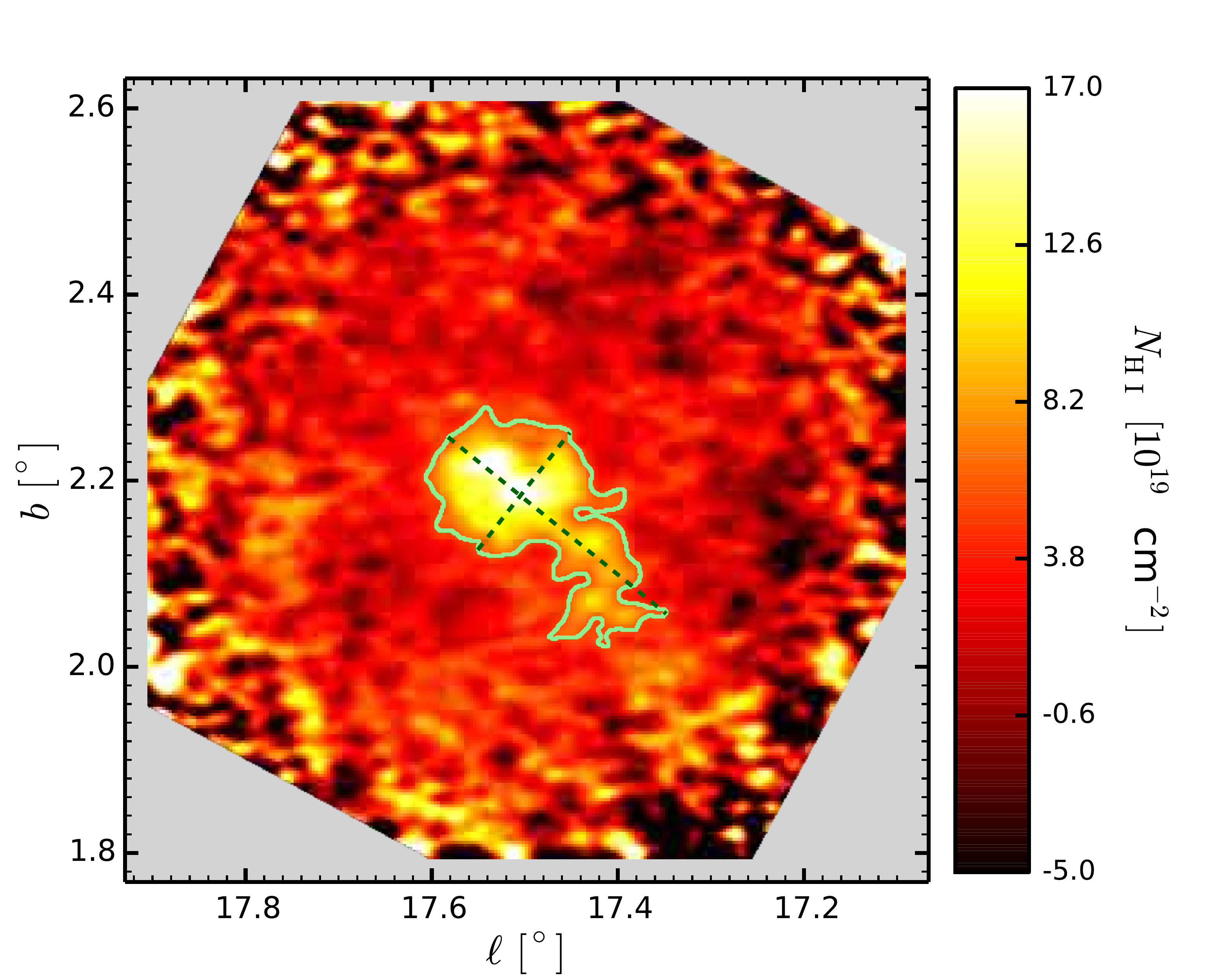}
        \hspace{-0.1cm}}
    \subfloat[][]{
        \centering
        \includegraphics[width=0.5\textwidth]{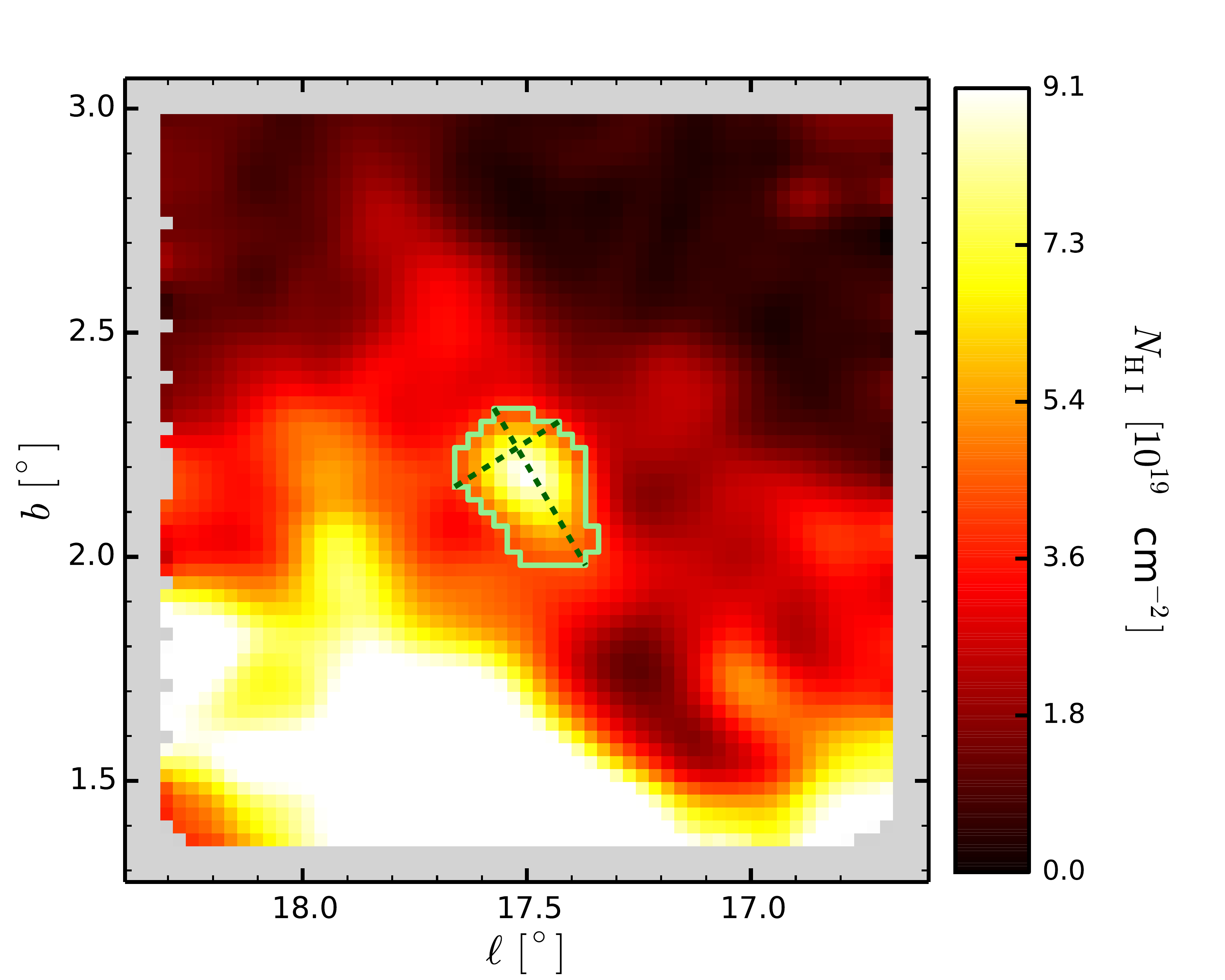}
        }

     \vspace{0cm}

\vspace{-0.5cm}
\caption{Contours for determination of size and mass of clouds based on VLA+GBT (left) and GBT only (right). Top: G$16.0+3.0$, contour levels \NHI~=~$2.0 \times 10^{19}$~\cmm (VLA+GBT) and $5.5 \times 10^{18}$~\cmm (GBT); bottom: G$17.5+2.2$, contour levels \NHI~=~$5.6 \times 10^{19}$~\cmm (VLA+GBT) and $4.4 \times 10^{19}$~\cmm (GBT). The dashed lines designate the axes: major (the longest distance between two contour points) and minor (the longest distance between contour points in the direction, perpendicular to the minor axis).}
\label{fig:MassContours1}
\end{figure}

\begin{figure}
\centering
    \vspace{-2cm}

    \captionsetup[subfigure]{labelformat=empty}
    \subfloat[][]{
        \centering
        \hspace{0.0cm}
        \includegraphics[width=0.5\textwidth]{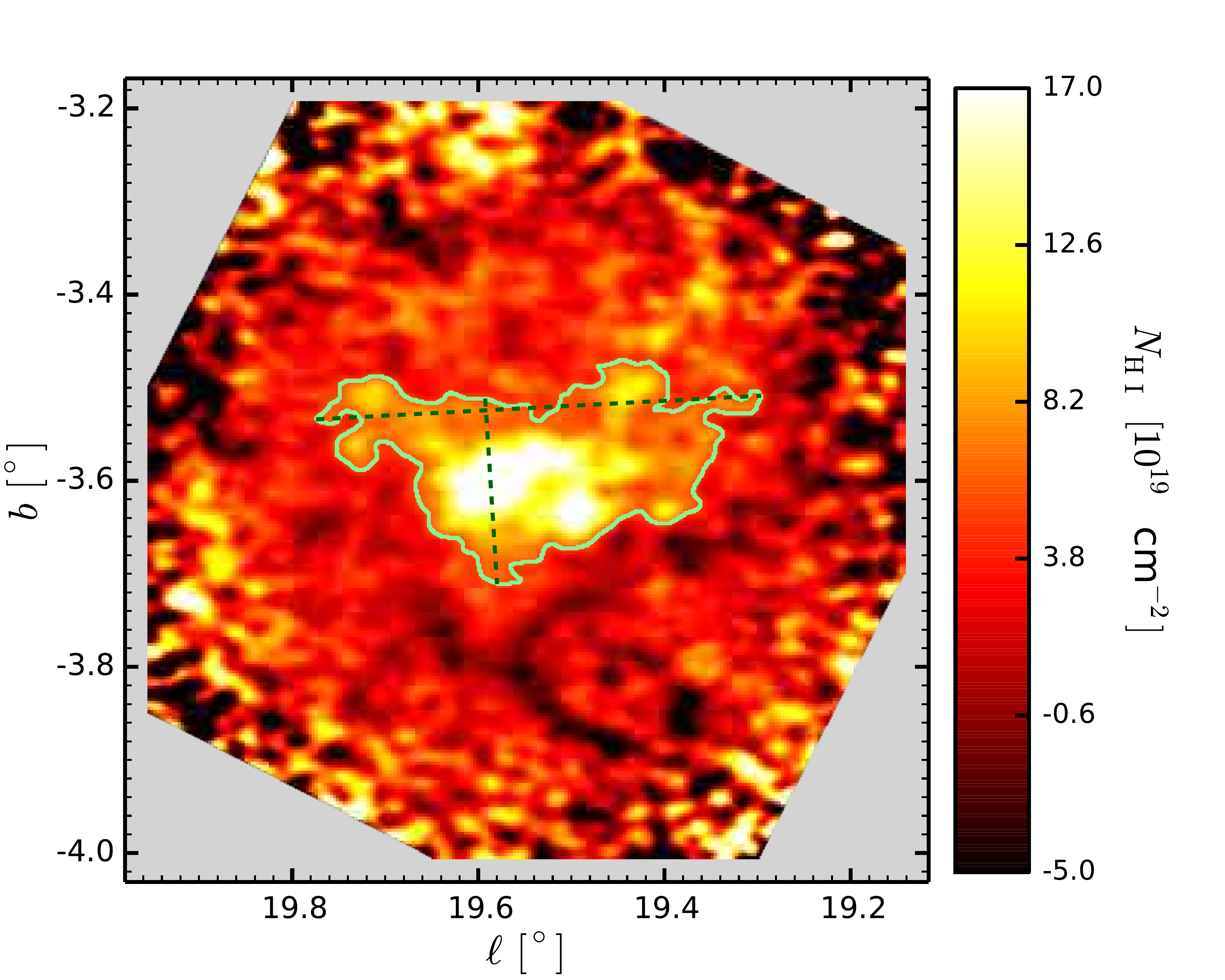}
        \hspace{-0.1cm}}
    \subfloat[][]{
        \centering
        \includegraphics[width=0.5\textwidth]{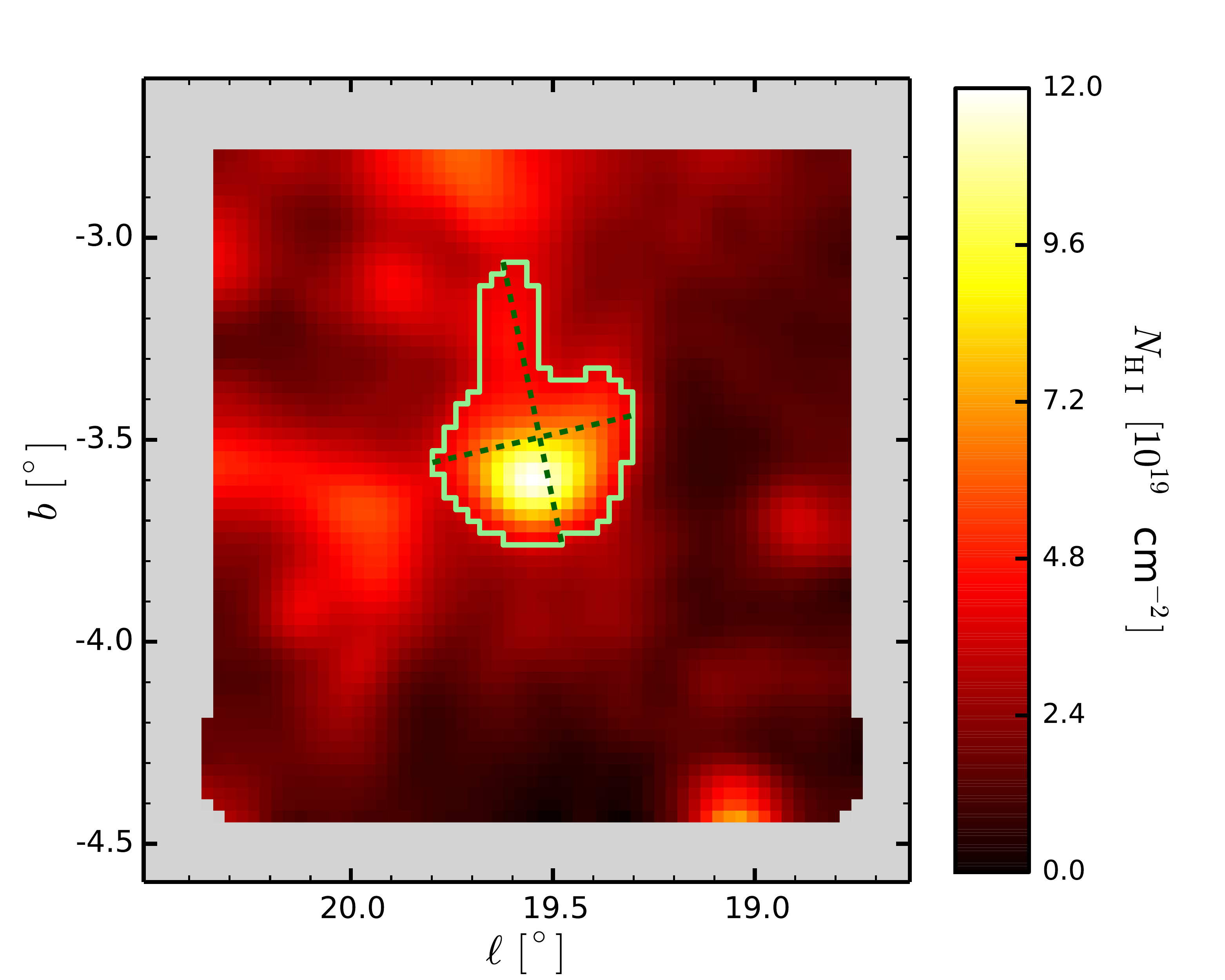}
        }

    \vspace{-1.3cm}

    \subfloat[][]{
        \centering
        \hspace{0.0cm}
        \includegraphics[width=0.5\textwidth]{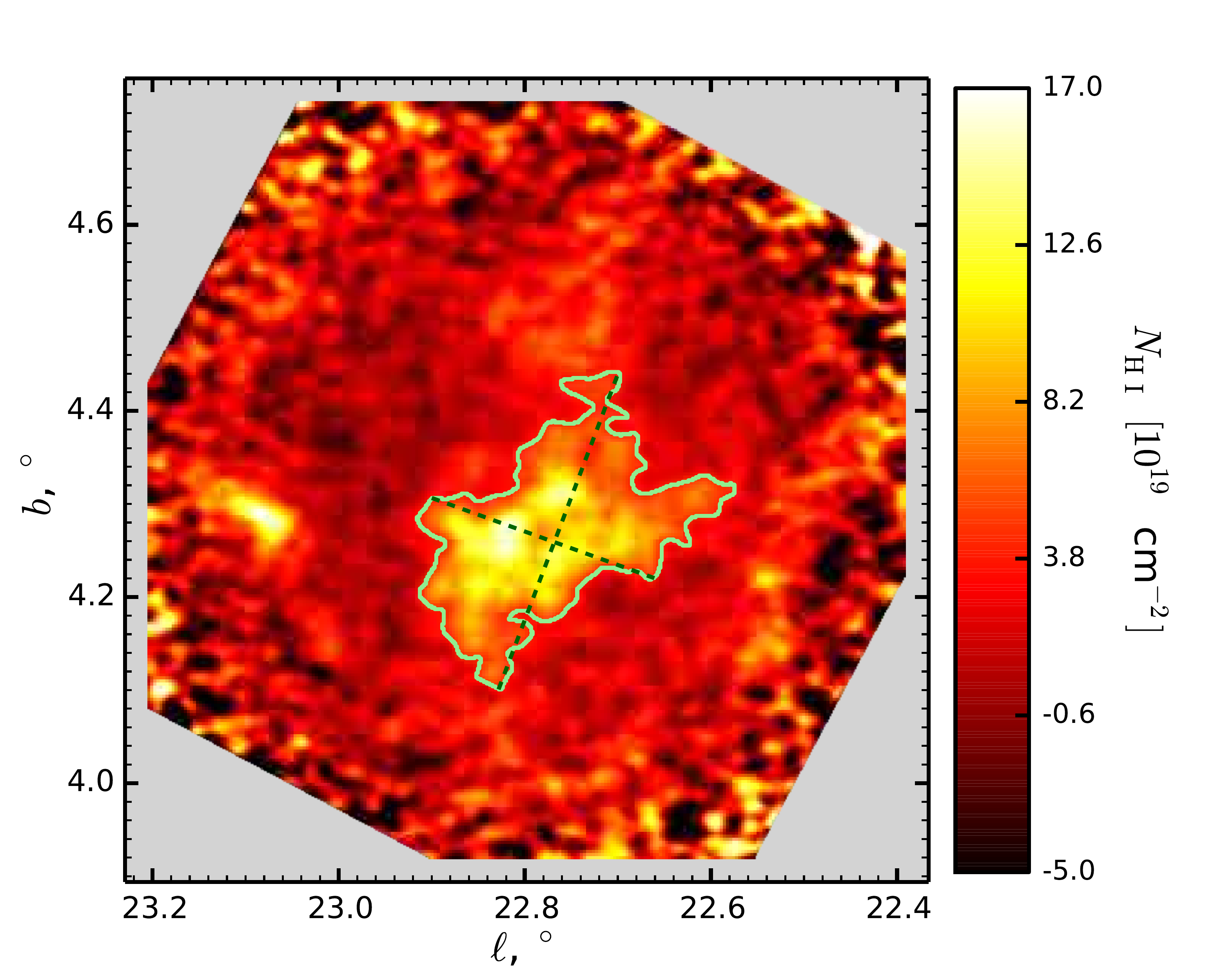}
        \hspace{-0.1cm}}
    \subfloat[][]{
        \centering
        \includegraphics[width=0.5\textwidth]{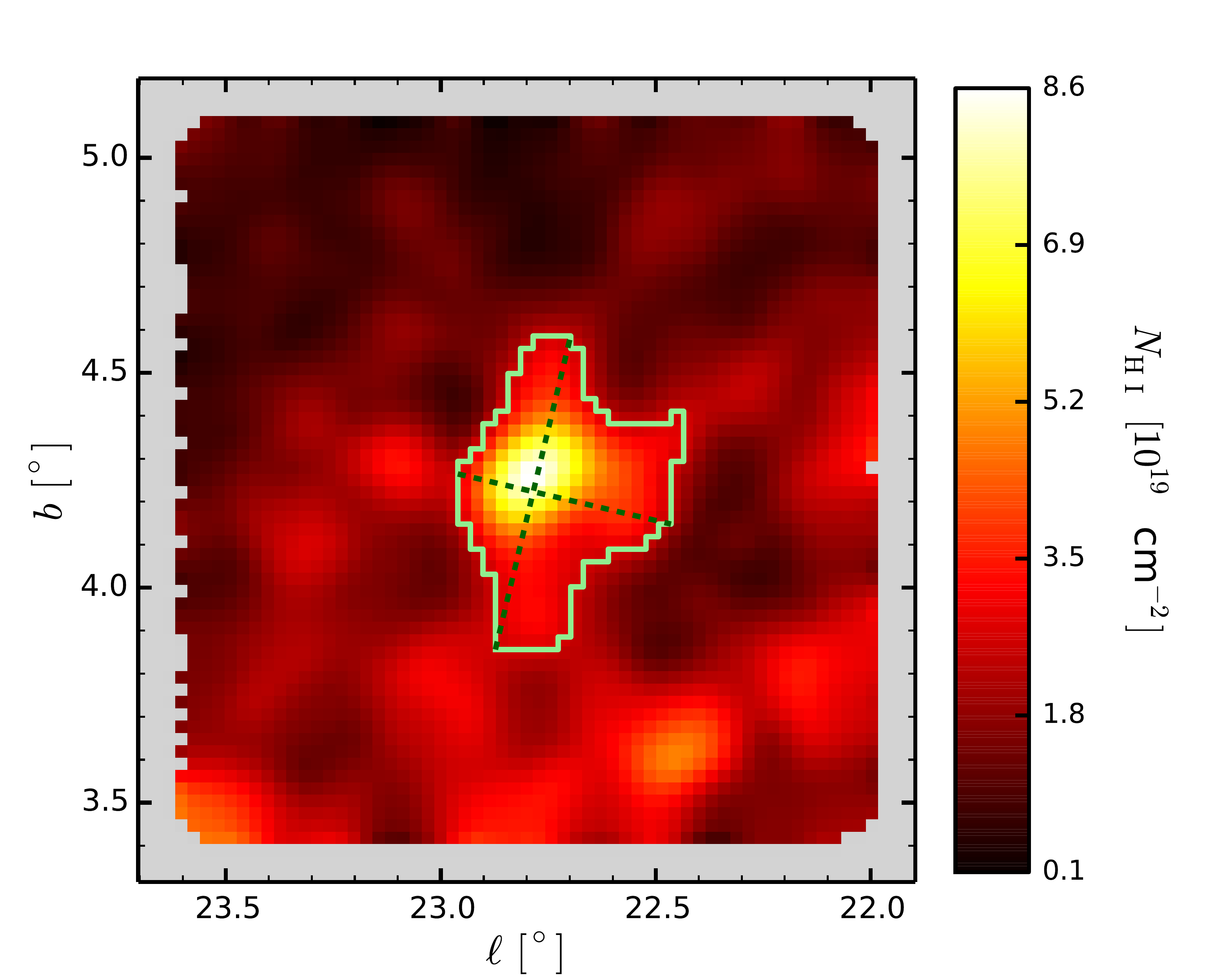}
        }

     \vspace{-1.3cm}

     \subfloat[][]{
        \centering
        \hspace{0.0cm}
        \includegraphics[width=0.5\textwidth]{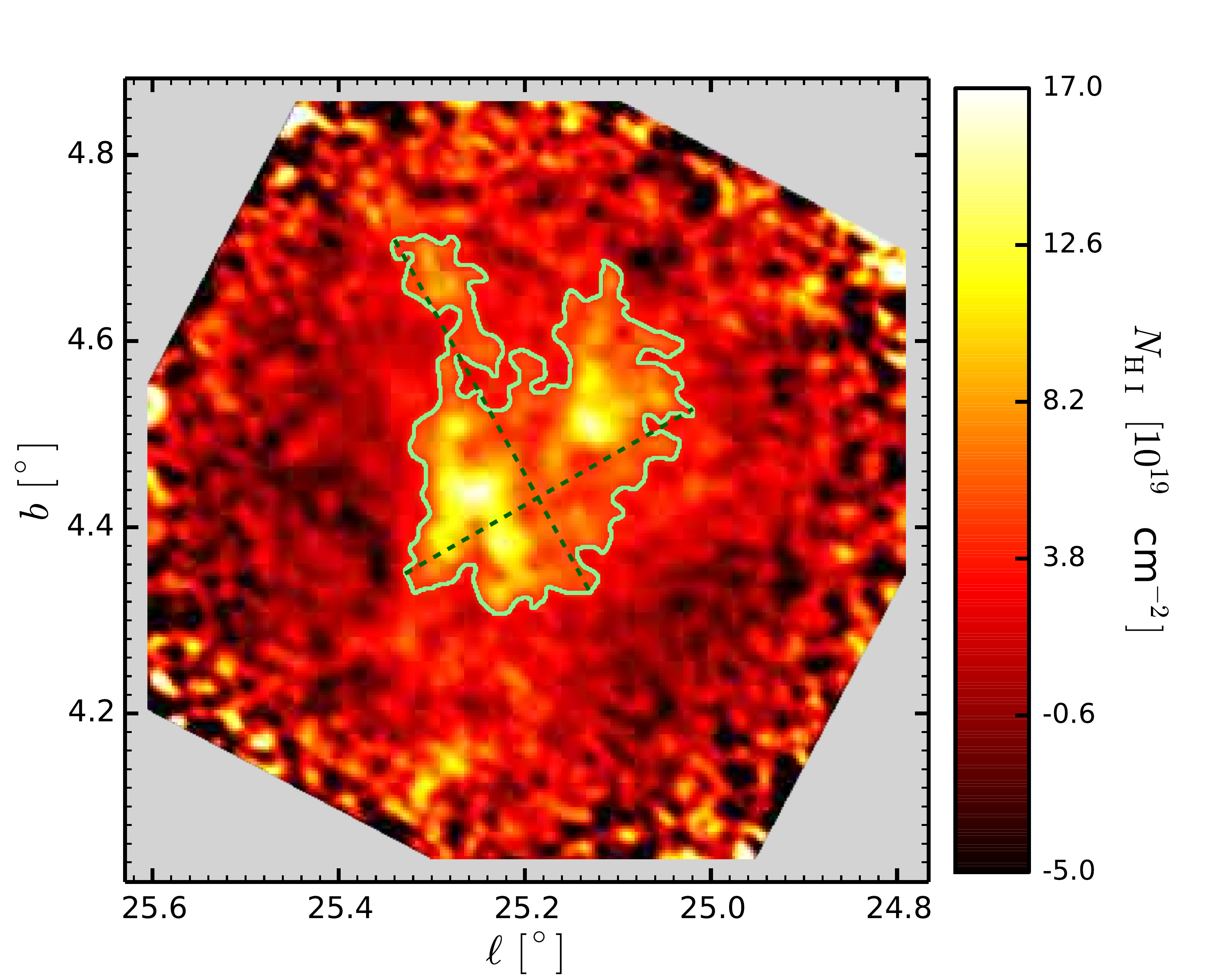}
        \hspace{-0.1cm}}
    \subfloat[][]{
        \centering
        \includegraphics[width=0.5\textwidth]{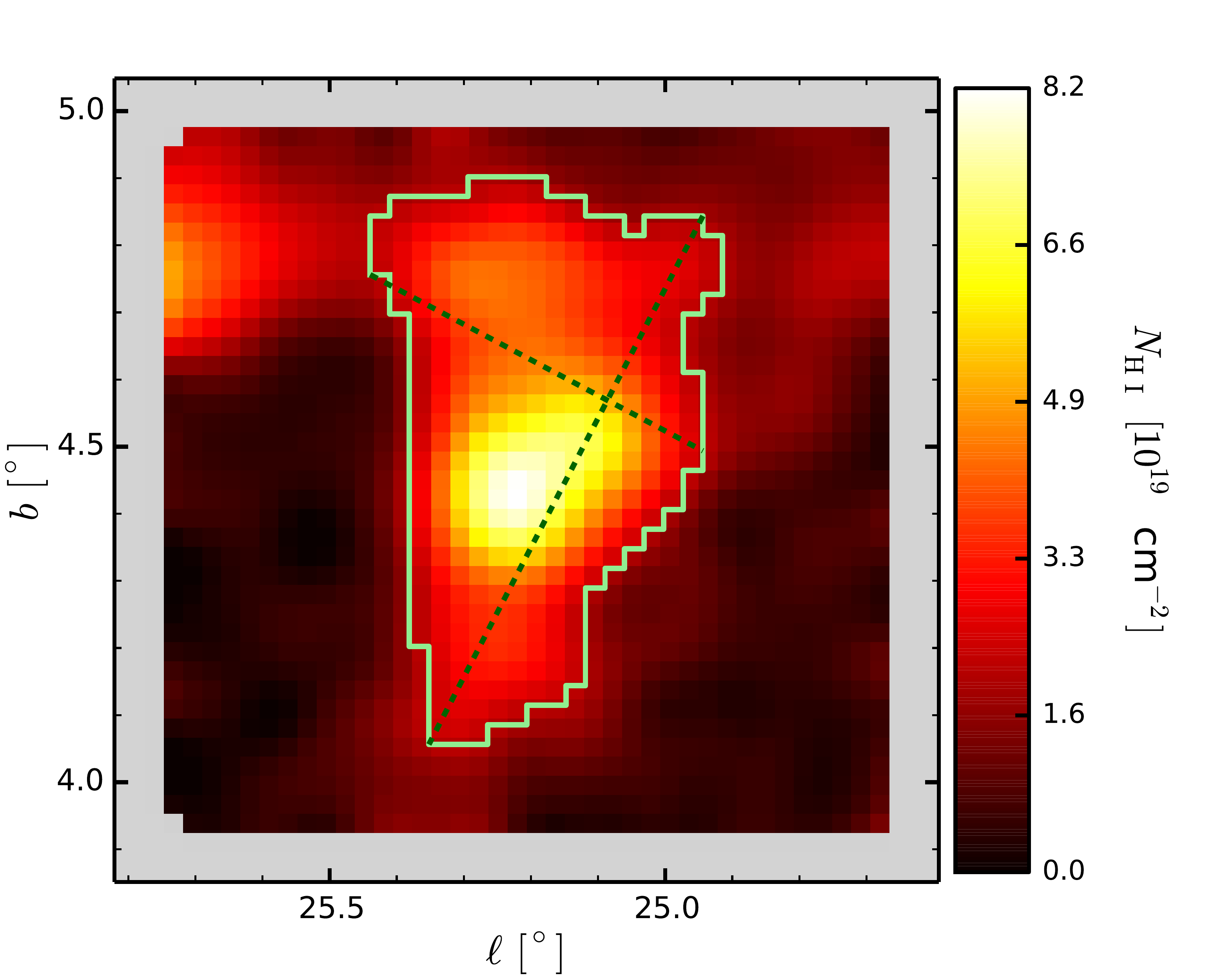}
        }
\vspace{-0.5cm}
\caption{Contours for determination of size and mass of clouds based on VLA+GBT (left) and GBT only (right).
Top: G$19.5-3.6$, contour levels \NHI~=~$5.5 \times 10^{19}$~\cmm (VLA+GBT) and $3.9 \times 10^{19}$~\cmm (GBT);
middle: G$22.8+4.3$, contour levels \NHI~=~$4.1 \times 10^{19}$~\cmm (VLA+GBT) and $2.6 \times 10^{19}$~\cmm (GBT);
bottom: G$25.2+4.5$, contour levels \NHI~=~$4.3 \times 10^{19}$~\cmm (VLA+GBT) and $2.2 \times 10^{19}$~\cmm (GBT).}
\label{fig:MassContours2}
\end{figure}

\begin{figure}
\centering
    \vspace{-3cm}

    \captionsetup[subfigure]{labelformat=empty}
    \subfloat[][]{
        \centering
        \hspace{-0.7cm}
        \includegraphics[width=0.5\textwidth]{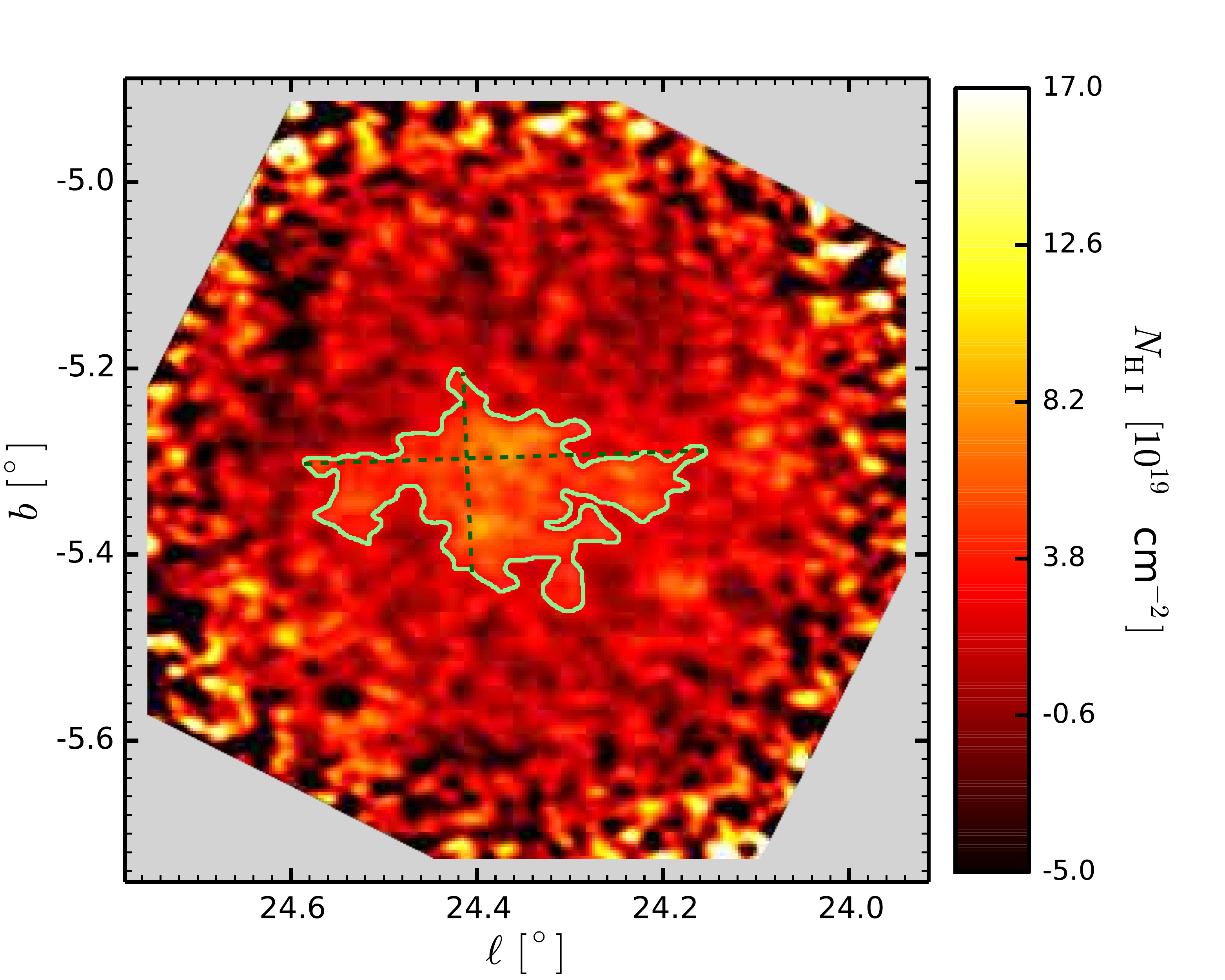}
        \vspace{2.0cm}
        \hspace{-0.1cm}}
    \subfloat[][]{
        \centering
        \includegraphics[viewport = +20 350 1150 650, width=0.8\textwidth]{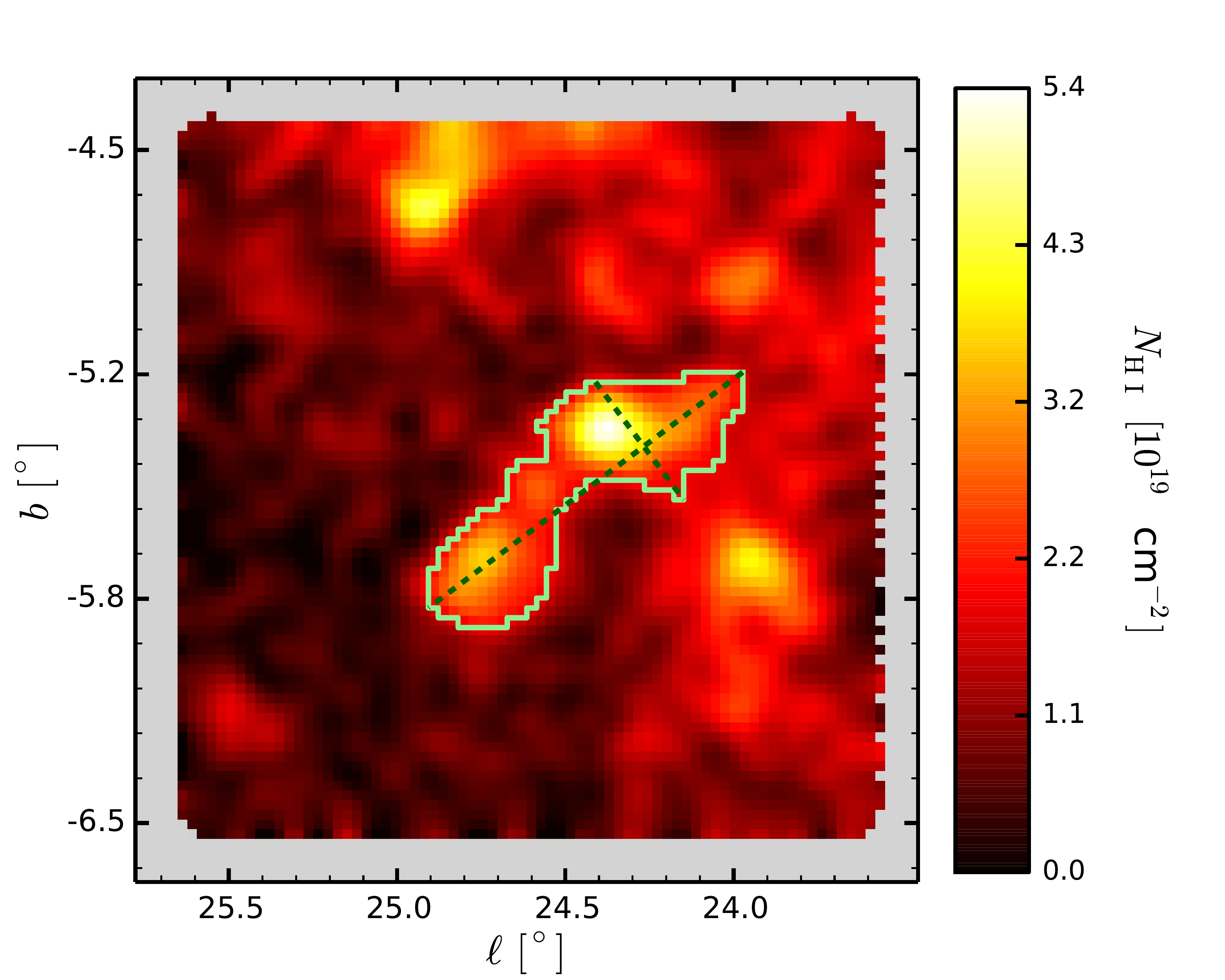}
        }

    \vspace{-1.3cm}

    \subfloat[][]{
        \centering
        \hspace{-9.0cm}
        \includegraphics[width=0.5\textwidth]{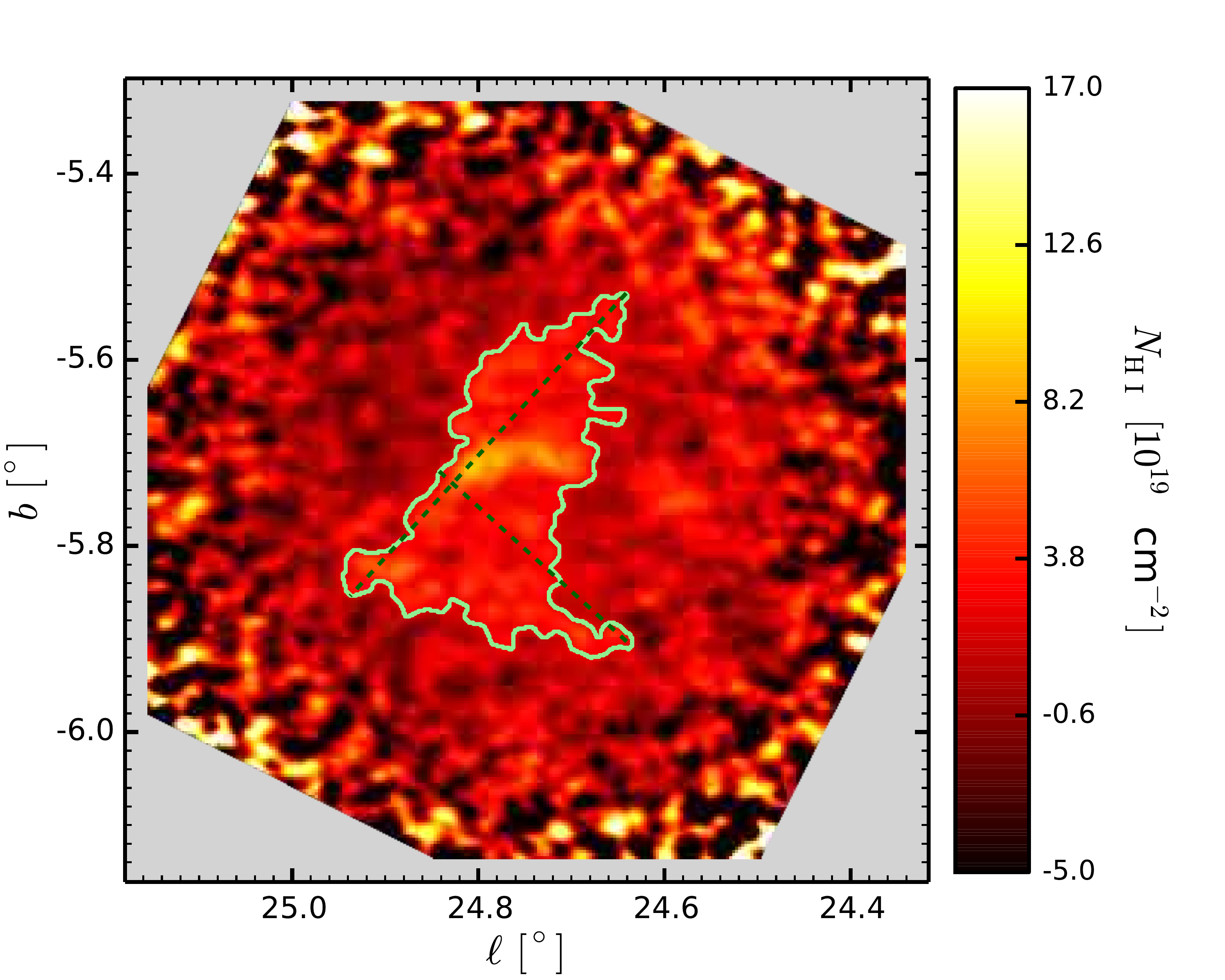}
        }
\vspace{-0.5cm}
\caption{Contours for determination of size and mass of clouds based on VLA+GBT (left) and GBT only (right).
Top left: G$24.3-5.3$ (VLA+GBT), contour at \NHI~=~$3.1 \times 10^{19}$~\cmm;
bottom left: G$24.7-5.7$ (VLA+GBT), contour at \NHI~=~$2.3 \times 10^{19}$~\cmm;
right: Group of G$24.3-5.3$ and G$24.7-5.7$ (GBT), contour at \NHI~=~$1.9 \times 10^{19}$~\cmm.}
\label{fig:MassContours3}
\end{figure}

\begin{figure}
\centering
    \vspace{-2cm}

    \captionsetup[subfigure]{labelformat=empty}
    \subfloat[][]{
        \centering
        \hspace{0.0cm}
        \includegraphics[width=0.5\textwidth]{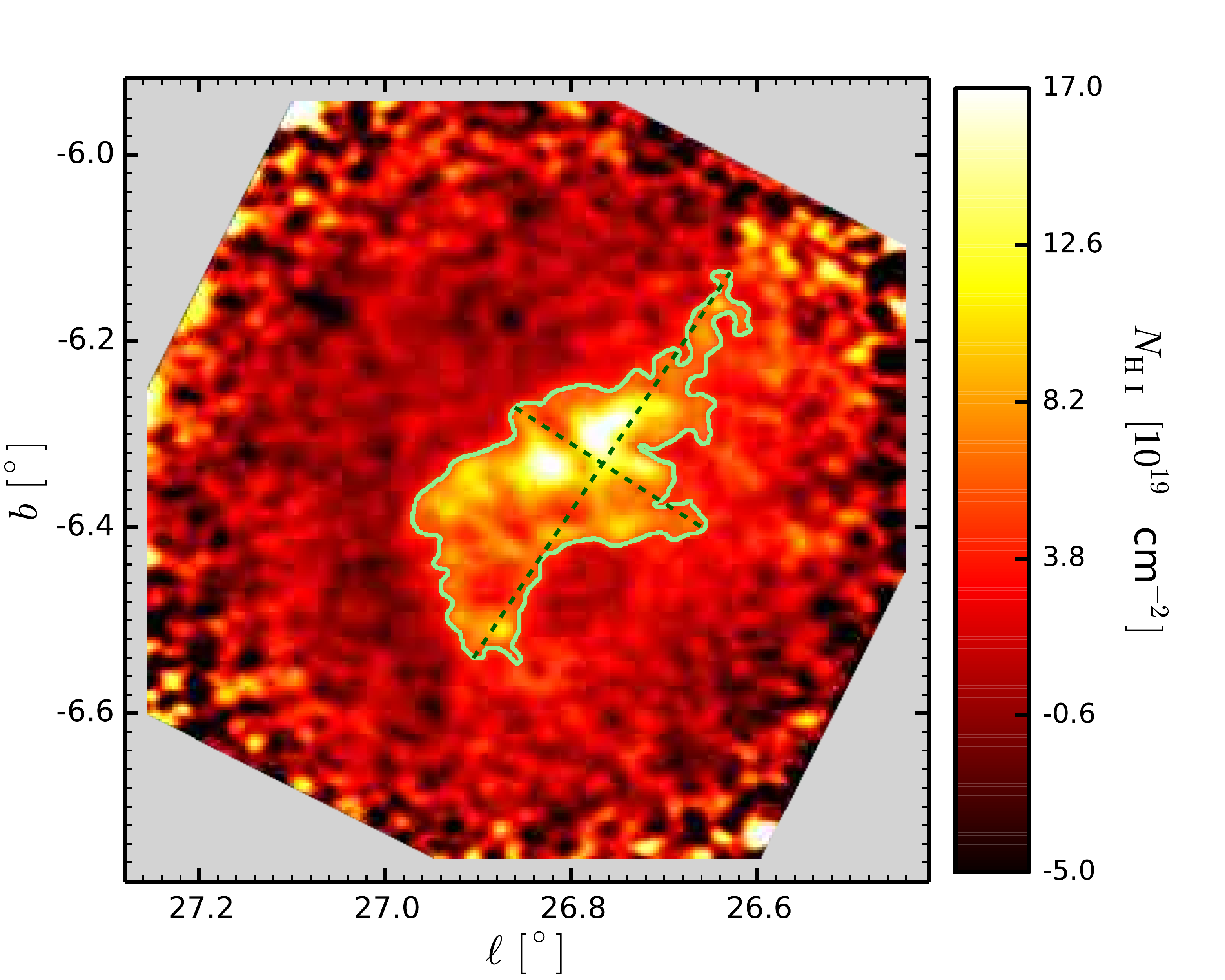}
        \hspace{-0.1cm}}
    \subfloat[][]{
        \centering
        \includegraphics[width=0.5\textwidth]{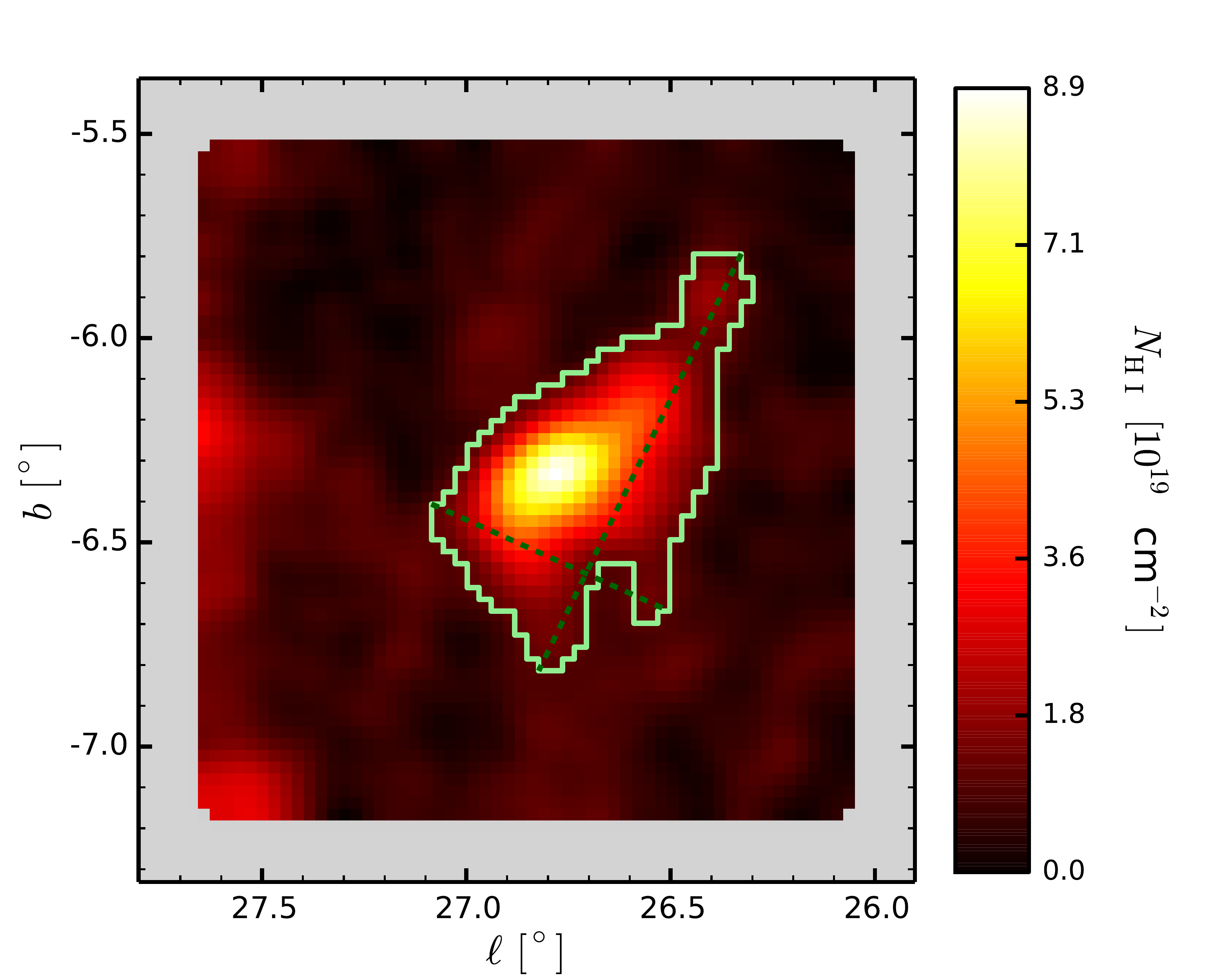}
        }

    \vspace{-1.3cm}

    \subfloat[][]{
        \centering
        \hspace{0.0cm}
        \includegraphics[width=0.5\textwidth]{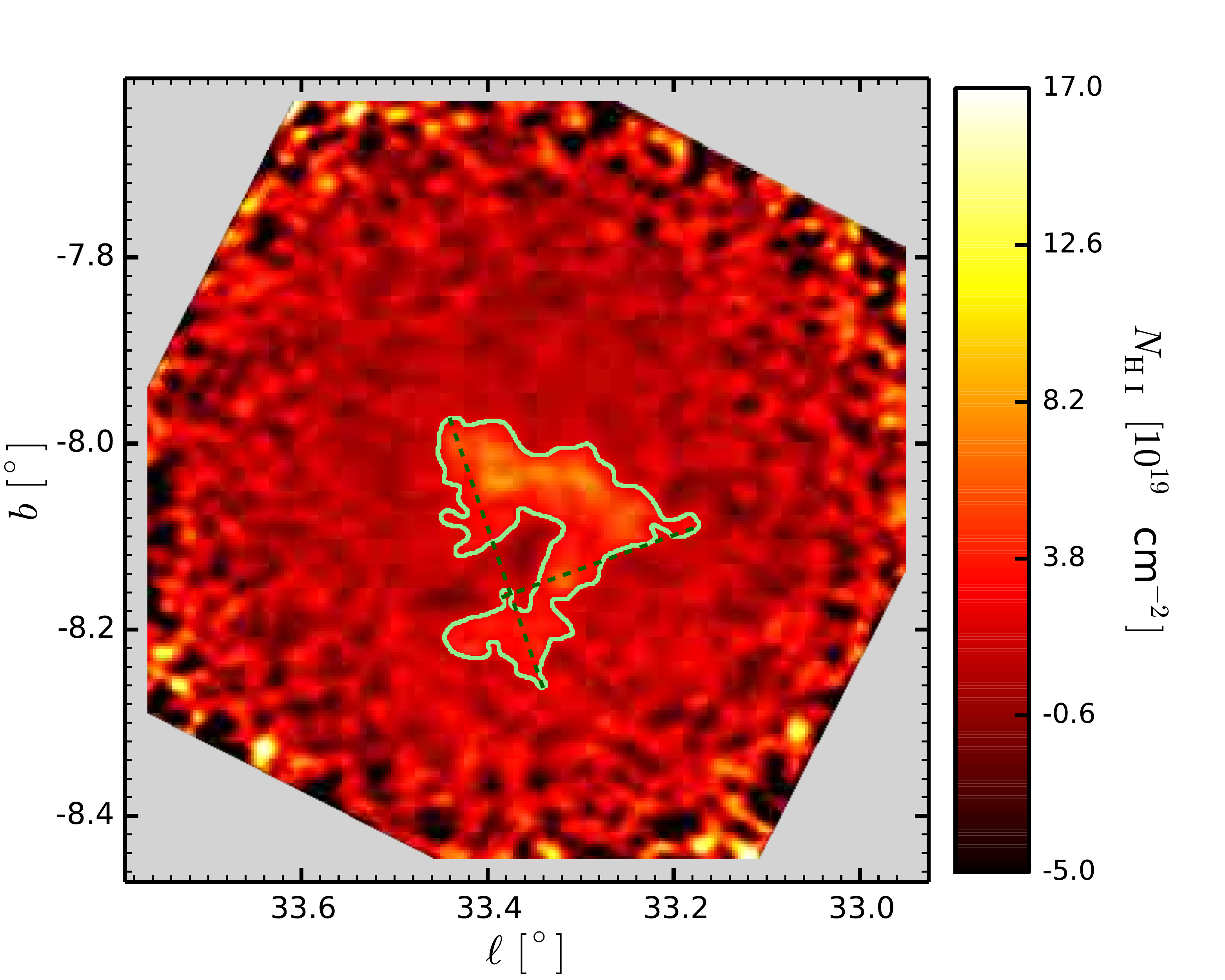}
        \hspace{-0.1cm}}
    \subfloat[][]{
        \centering
        \includegraphics[width=0.5\textwidth]{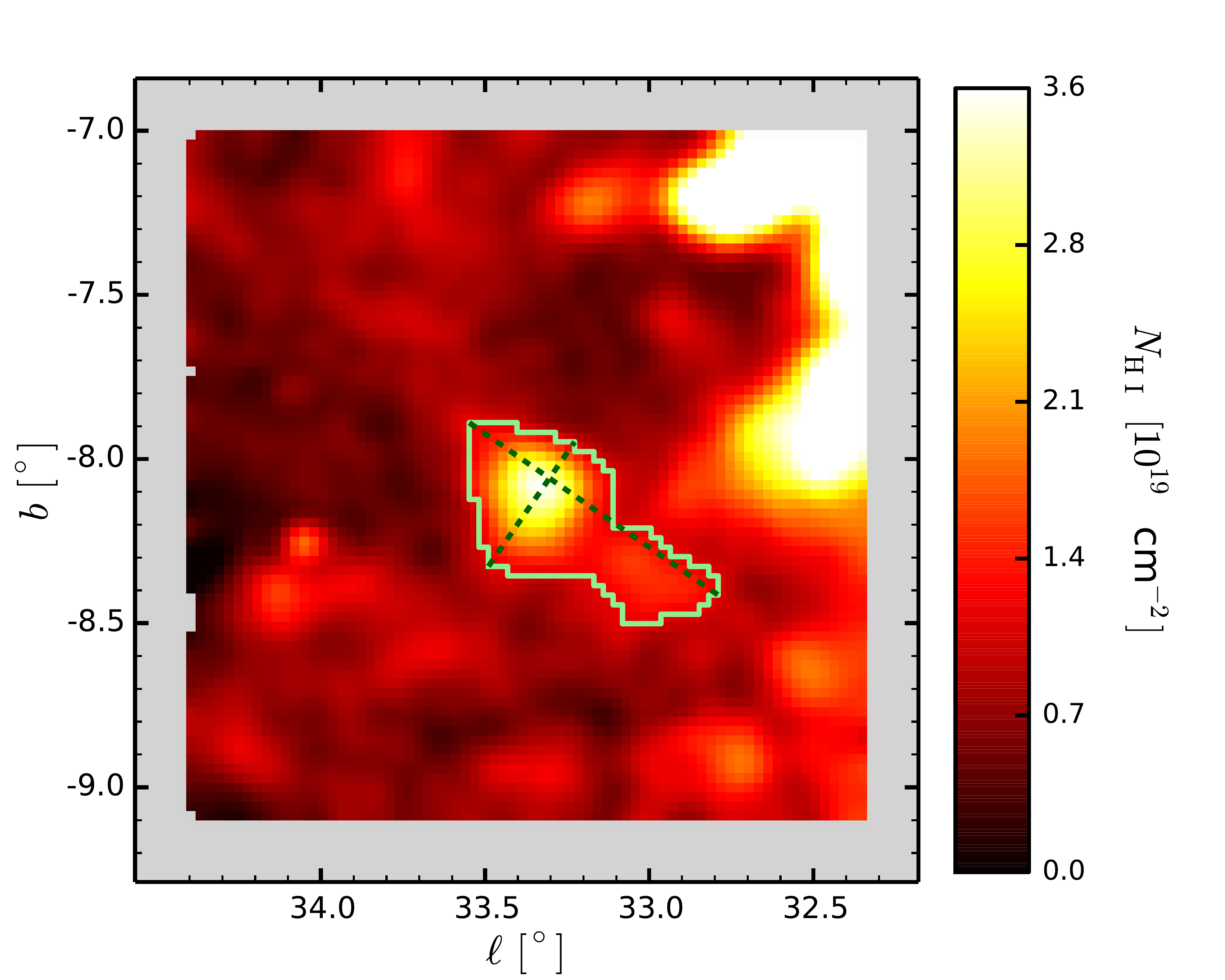}
        }

     \vspace{-1.3cm}

     \subfloat[][]{
        \centering
        \hspace{0.0cm}
        \includegraphics[width=0.5\textwidth]{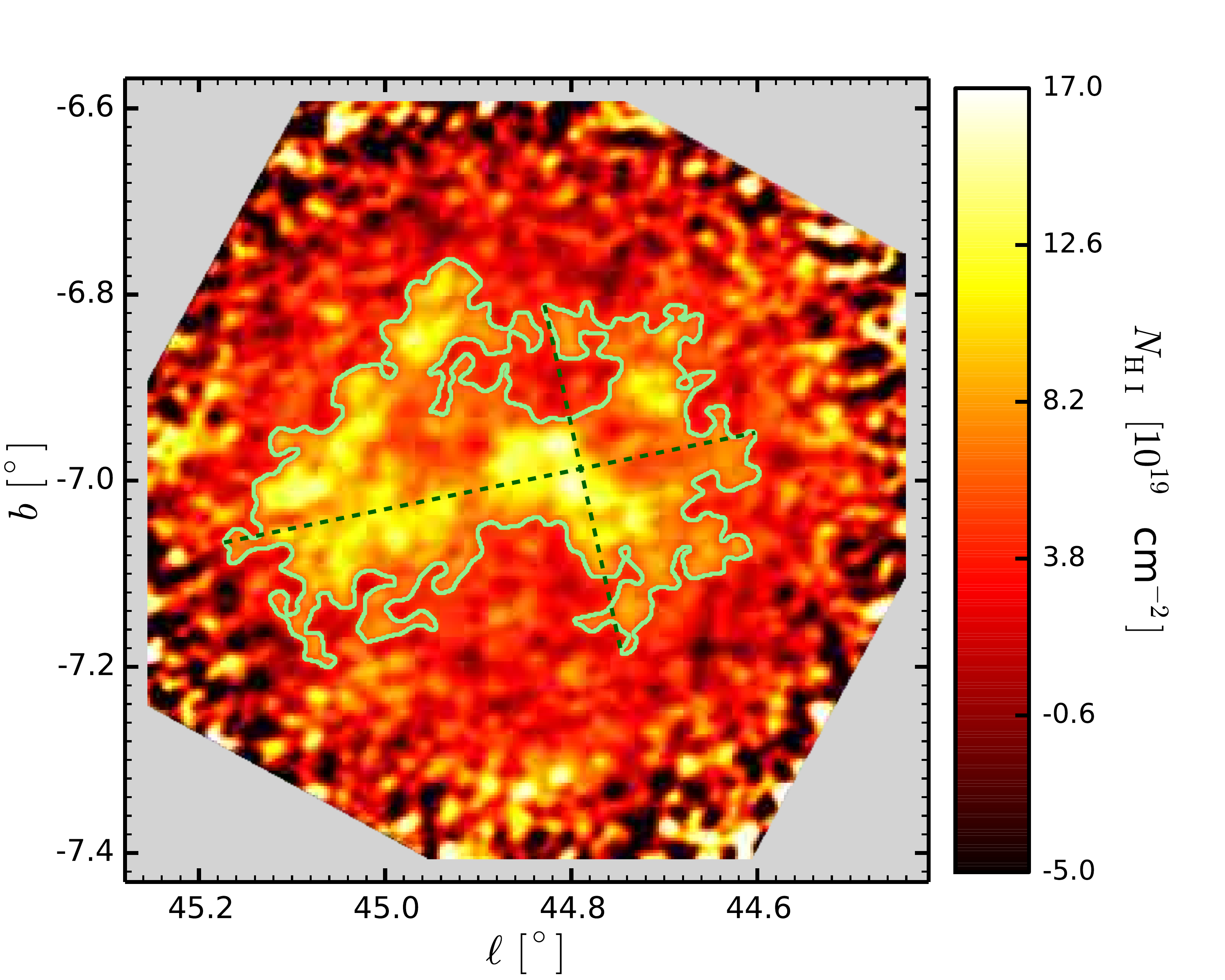}
        \hspace{-0.1cm}}
    \subfloat[][]{
        \centering
        \includegraphics[width=0.5\textwidth]{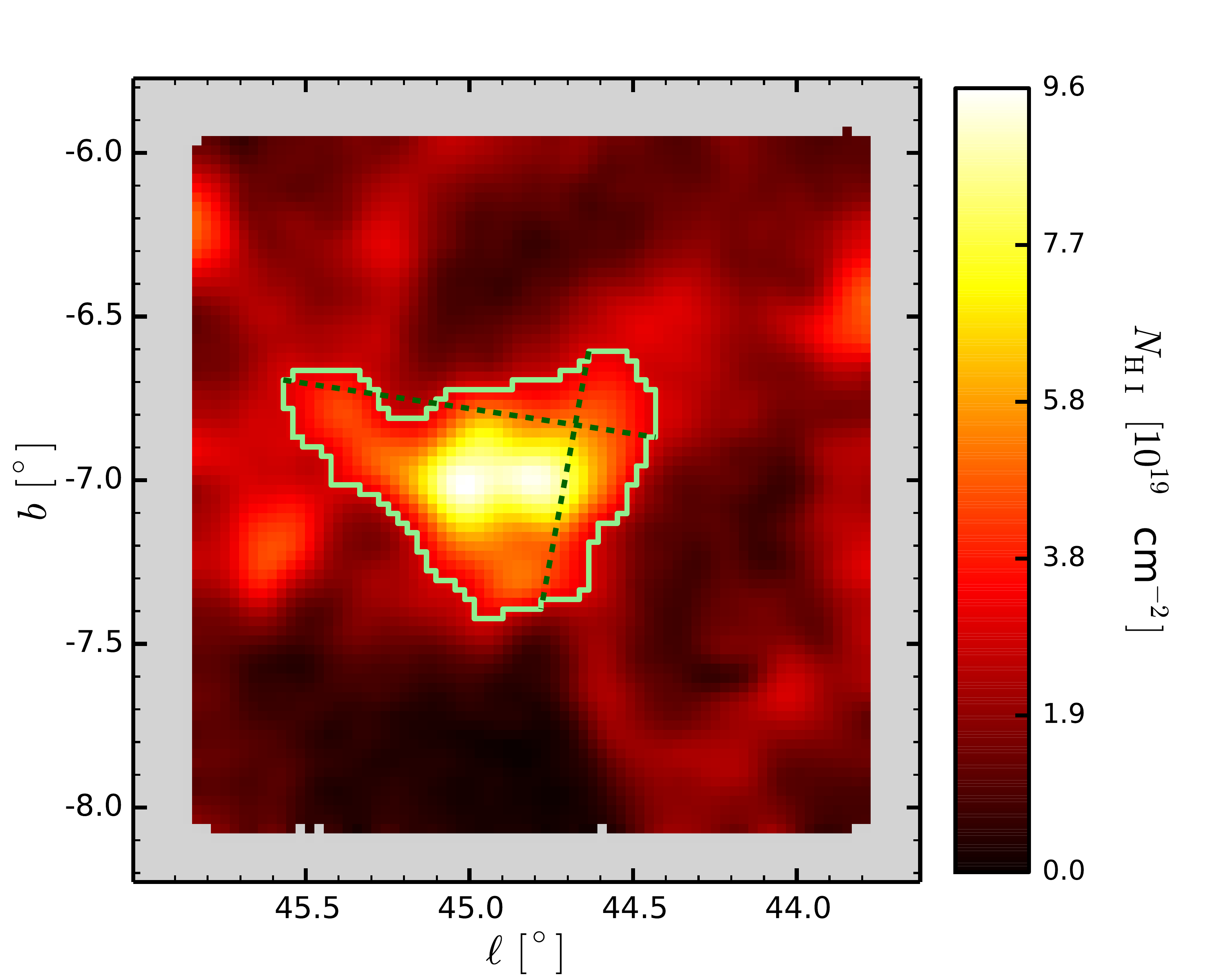}
        }
\vspace{-0.5cm}

\caption{Contours for determination of size and mass of a cloud based on VLA+GBT (left) and GBT only (right):
 Top: G$26.9-6.3$, contours at \NHI~=~$5.3 \times 10^{19}$~\cmm (VLA+GBT) and $1.1 \times 10^{19}$~\cmm (GBT);
 middle: G$33.4-8.0$, contours at \NHI~=~$2.5 \times 10^{19}$~\cmm (VLA+GBT) and $1.2 \times 10^{19}$~\cmm (GBT);
 bottom: G$44.8-7.0$, contours at \NHI~=~$5.5 \times 10^{19}$~\cmm (VLA+GBT) and $3.1 \times 10^{19}$~\cmm (GBT).}
\label{fig:MassContours4}
\end{figure}

In most cases the clouds do not display clear boundaries. In order to determine meaningful sizes and masses we have employed contours of constant \NHI. In each case a contour was chosen to encompass most of the visible cloud structures in the VLA+GBT maps (bottom left panels of Figures~\ref{fig:160_HImaps}, \ref{fig:175_HImaps} etc.). The resulting contours are shown in the left panels of Figures~\ref{fig:MassContours1} -- \ref{fig:MassContours4}. The \HI masses inside these contours for each cloud are listed in column 4. For comparison we have also drawn contours in the GBT maps at much lower \NHI values. The resulting contours are shown in the right panels of Figures~\ref{fig:MassContours1} -- \ref{fig:MassContours4} and the derived properties are listed in Table~\ref{tab:gbtderivedprops}.

For each contour the major axis (the longest distance between two contour points) and the minor axis (the longest distance between contour points in the direction, perpendicular to the minor axis) are determined. The length of these axes are listed as $D_{maj} \times D_{min}$ in column 5 of Table~\ref{tab:derived-props} and column 3 of Table~\ref{tab:gbtderivedprops}.

The volume density in column 6 of Table~\ref{tab:derived-props} and column 4 of Table~\ref{tab:gbtderivedprops} is the total \HI mass divided by the volume given by $\case{4}{3} \pi r_e^3$ where the effective radius  $r_e \equiv \onehalf (D_{maj} D_{min})^{\onehalf}$.

\subsection{Cloud Cores}
\label{subsec:cores}

Table~\ref{tab:core-props} gives estimates of the properties of the cloud cores, the denser regions of each cloud, determined by analyzing the \HI emission at a higher value of \NHI~ around the column density peak of each cloud (column 2).  The mass, size and number density of the enclosed area is given in cols.~3--5. The FWHM in Table~\ref{tab:observed-props} can be used to limit the kinetic temperature --- this value, T$_{\mathrm{limit}}$ is given in col.~6.  Finally, multiplying values in columns 5 and 6 we can get a rough estimate of the core pressure in each cloud (column 7). For the clouds with two velocity components the total number density was split proportionally to the corresponding column densities.

\subsection{Discussion}
\label{subsec:discussion}

We have produced high angular-resolution 21cm \HI maps of ten clouds that lie in
the boundary between the disk and the halo in the inner Galaxy.   This paper presents the data
and the reduction methods necessary to insure accurate results.
We defer a detailed discussion of the cloud properties to a separate publication, but some general comments
can be made, for these clouds are unique samples of the neutral interstellar medium.

The disk-halo clouds, with masses of many hundreds of \Msun\
and locations many hundreds of pc from the Galactic plane,
are orders of magnitude denser than their surroundings.

The following rough estimates are made with the assumption of a spherical cloud of pure monoatomic hydrogen at constant number density $n$, its mass density $\rho = m_H n$. By the cloud ``Size'' we understand its diameter and take that $T_{\mathrm{limit}}$ determined based on the emission line FWHM, represents its true kinetic temperature $T$.

Under these conditions the sound-crossing time can be expressed as:
\be
t_{\mathrm{\,sound}} = \frac{\mathrm{Size}}{c_s} \approx 2 \; \mathrm{Myr} \cdot \frac{\mathrm{Size}}{\mathrm{FWHM}}.
\ee
For clouds' dense cores of sizes $\sim$~10~pc this is just a few Myr. But for whole clouds, with sizes $\sim$~100~pc, $t_{\mathrm{\,sound}}$ may reach 60 Myr.

On the other hand, the free-fall time of gravitational collapse:
\be
t_{\mathrm{\,ff}} = \sqrt{\frac{3\pi}{32 G \rho}} \approx \; 50 \; \mathrm{Myr} \cdot n^{-\frac{1}{2}}.
\ee
For the highest cloud core density $\lbr n \rbr \approx 10$~\cmmm, $t_{\mathrm{\,ff}} \approx 16$~Myr, but for other cores it is 20--50~Myr and for whole clouds this time scale often exceeds 100~Myr. So in all cases $t_{\mathrm{\,sound}} \ll t_{\mathrm{\,ff}}$ and thus the clouds are not gravitationally bound.

It is instructive also to compare $t_{\mathrm{\,sound}}$ with the time $t_z$ of vertical fall to the Galactic plane at cloud's location. Using a simple analytical expression of \citet{Wolfire1995b} for the $z$-component of the Galactic gravitational acceleration $g_z(R,\,z)$, one can see that for $|z| \lesssim 200$~pc the acceleration is close to linear: $g_z(R,\,z) \approx g^{\,\prime}_z(R) \cdot z$. The values of $g^{\,\prime}_z$ range from $(10.8\;\mathrm{Myr})^{-2}$ at $R=8.5$~kpc to  $(5.4\;\mathrm{Myr})^{-2}$ at $R=2.5$~kpc. Since motion with such acceleration is harmonically periodic, the free-fall time is just $\frac{\pi}{2}\,g^{\,\prime}_z\,^{-1/2}$. For higher $|z|$ where $g_z(R,\,z)$ flattens, these values can be used as lower limits:
\be
t_z \gtrsim \frac{\pi}{2}\,\left( g^{\,\prime}_z(R) \right)^{-\frac{1}{2}},
\ee
equal to 8 -- 17~Myr for 2.5~kpc~$<\,R\,<$~8.5~kpc. In~\citet{PhD} a more precise ballistic calculation was done for 2~kpc~$<\,R\,<$~5.3~kpc and a much greater distance from the plane $z \approx 3.4$~kpc using Walter Dehnen's {\em GalPot} package \citep{DehnenGalPot}, obtaining $t_z \approx 30$~Myr, which can be used as an upper limit for all the disk-halo clouds.

For the cloud cores the condition $t_{\mathrm{\,sound}} \ll t_z$ is true, so the clouds have time to respond to local physical conditions. The clouds must have come to internal pressure equilibrium, although the pressure may have contributions from turbulence on a range of scales, and possibly magnetic field and cosmic ray pressures as well. But since  $t_{\mathrm{\,sound}} \gtrsim t_z$ for the larger cloud structure, the overall density distribution of the cloud does not have time to dissipate in the low pressure of the halo over the time it takes the cloud to rise or to fall back to the disk.

Similarly one can estimate the clouds' Jeans masses as
\be
M_J = \left ( \frac{5 k_B T}{G m_H} \right )^{\frac{3}{2}} \; \left ( \frac{3}{4 \pi \rho} \right )^{\frac{1}{2}}\approx \; 9400 \; M_{\odot} \cdot \mathrm{FWHM}^{\,3} \, n^{-\frac{1}{2}},
\ee
which is two to three orders of magnitude larger than their observed gas masses.

All this implies that the clouds are dynamic entities whose properties
must reflect their history as well as conditions at their current locations
\citep[e.g.,][]{KoyamaOstriker2009,Saury2014}.
While the appearance of many of the disk-halo clouds suggests that they are interacting with their local environment producing the steep gradients in \NHI\ or asymmetric shapes, we find little
correlation between location in the Galaxy and fundamental cloud properties,
with one clear exception, G$16.0+3.0$ discussed below.

In theories of the interstellar medium the local pressure is often the controlling factor in the structure of
neutral clouds,  and over a wide range of
conditions in the Galaxy it is expected that neutral clouds could consist of two phases, one warm and one cold
\citep{FGH1969,MO1977,Wolfire1995a,Jenkins2012}.
Four of the clouds studied here have line profiles indicating the presence of two components with different
temperatures at the location of their peak \NHI.
It is unlikely that this results from confusion of unrelated material as  all two-component
clouds are located at $|z|\geq590$ pc from the Galactic plane. The components typically have similar but not
 identical values of $\VLSR$.   The narrower line component contains between $30\%$
and $50\%$ of the total \NHI, about the mass fraction expected from some simulations \citep{Saury2014}.
However,   clouds may have two phases not easily separable in their emission profiles as the
cold gas may have large velocity fluctuations that blend it with the warmer emission
\citep{Vazquez-Semadeni2012,Saury2014}.

Cloud G$16.0+3.0$ is particularly interesting, as it
may lie within  the area around the Galactic nucleus excavated by a hot wind: the ``Fermi Bubble''
\citep[e.g.,][]{Bland-Hawthorn2003,Su2010}.  The boundaries of the region effected by the
wind are not well-defined, especially
at low latitudes, but the G$16.0+3.0$ cloud has such distinctive properties as to suggest that its
environment is different from that of the other clouds.  Within the hot wind the pressure is
many times larger than the typical ISM \citep{Bland-Hawthorn2003,Carretti2013};
this is expected to force clouds into a purely cold phase \citep{FGH1969,Gatto2014}.
The properties of G$16.0+3.0$, with its high density, narrow line width, and compact structure,
are consistent with this interpretation. Moreover, G$16.0+3.0$ is notably smaller then the other
clouds, but its size is similar to that of the \HI clouds found to be entrained in the
nuclear wind \citep{McClure-Griffiths2013}.
In contrast, the nearby cloud G$17.5+2.2$ is nearly indistinguishable from the other
clouds studied here.  It might be as close as 135~pc to G$16.0+3.0$, but is more likely at least 1~kpc
away given the uncertainties in our assignment of distances.  Neither cloud shows
kinematic anomalies suggesting  that they have been accelerated by the nuclear wind \citep{McClure-Griffiths2013}.

We can also compare the properties of our disk-halo clouds with the disk clouds of \citet{Stil2006} observed at similar resolution, but all at $z \approx 0$. Only one of their clouds, $59.67-0.39+60$ has an analogous morphology and size, with a few times larger mass and average density compared to the clouds of our survey. The properties of all their other clouds are close to our cloud cores, only in some cases displaying a few times larger average density.

These disk-halo \HI structures allow us to study unconfused interstellar clouds in a variety of locations;
this may lead to a better understanding of physical conditions that have
hitherto been manifest only in ensemble averages. When the GASKAP survey \citep{GASKAP} is done, and even more when the full SKA is ready, studies like this one will reveal many more such clouds, and the techniques developed here will be useful for understanding their properties.


\acknowledgements
This research was supported in part by the Australian Research Council through grant DP110104101 to the University of Tasmania. YP has started working on this research project while he held a predoctoral fellowship at the NRAO and then did a part of it while being employed at the Joint Institute for VLBI in Europe (Dwingeloo, the Netherlands). We thank the anonymous referee for comments and suggestions that helped to improve this paper.

Facilities: \facility{GBT}, \facility{VLA}


\begin{deluxetable}{cccccc}
\tablecaption{Cloud Properties from GBT Observations
\label{tab:table0}}
\tablehead{
\colhead{Name}  &  \colhead{T$_b$\tablenotemark{1}}  & \colhead{V$_{LSR}$\tablenotemark{1}}
 & \colhead{FWHM}  &\colhead{Distance}  &
 \colhead{z} \\
              &\colhead{ (K)}     & \colhead{(km s$^{-1}$)} & \colhead{(km s$^{-1}$)}   & \colhead{(kpc)} &
 \colhead{(pc)}   \\
\colhead{(1)} & \colhead{(2)} & \colhead{(3)} & \colhead{(4)} & \colhead{(5)} & \colhead{(6)}  \\
}
\startdata
G$16.0+3.0$     &  1.1  & +143 &  5.4   & $8.2\pm1.0$ &  $+430\pm50$ \\
G$17.5+2.2$     &  1.6  & +139 & 16.2   & $8.1\pm1.1$ &  $+310\pm40$ \\
G$19.5-3.6$     &  3.5  & +121 & 13.1   & $8.0\pm1.6$ &  $-500\pm100$  \\
G$22.8+4.3$     &  3.1  & +137 &  7.8   & $7.9\pm1.4$ &  $+590\pm100$ \\
G$24.3-5.3$     &  1.3  & +124 & 13.5   & $7.8\pm1.5$ &  $-720\pm140$ \\
G$24.7-5.7$     &  1.8  & +127 &  8.9   & $7.8\pm1.5$ &  $-770\pm150$ \\
G$25.2+4.5$     &  2.6  & +147 & 10.4   & $7.7\pm1.6$ &  $+610\pm130$ \\
G$26.9-6.3$     &  3.4  & +123 &  8.8   & $7.6\pm1.7$ &  $-840\pm190$ \\
G$33.4-8.0$     &  1.1  & +102 &  5.5   & $7.2\pm2.0$ & $-1000\pm280$ \\
G$44.8-7.0$     &  3.3  &  +94 &  9.0   & $6.1\pm2.5$ &  $-740\pm300$ \\
 \enddata
\end{deluxetable}

\begin{deluxetable}{l c c c c}
\tablecaption{Parameters of the VLA Observations
\label{tab:VLAobserve}}
\tablehead{
  \colhead{Cloud} & \colhead{ Pointing} &  \colhead{Integration}
  & \colhead{Central} & \colhead{Calibrators}\\
  \colhead{ } & \colhead{ Coordinates } &  \colhead{Time,}
  & \colhead{Velocity,} & \colhead{ }\\
  \colhead{  }      & \colhead{ (J2000) }       & \colhead{minutes}  &  \colhead{\kms} & \\}
\startdata
   G$16.0+3.0$ & $18^h08^m42.850^s$ $-13\arcdeg 35\arcmin 17\arcsec$   & 131  & +140 & 1833-210, 1331+305           \\
   G$17.5+2.2$ & $18^h14^m46.672^s$ $-12\arcdeg 39\arcmin 31\arcsec$   & 105  & +135 & 1833-210, 1331+305           \\
   G$19.5-3.6$ & $18^h39^m46.681^s$ $-13\arcdeg 32\arcmin 41\arcsec$   &  90  & +120 & 1833-210, 1331+305           \\
   G$22.8+4.3$ & $18^h17^m20.803^s$ $-06\arcdeg 59\arcmin 31\arcsec$   & 143  & +120 & 1812-068, 1331+305           \\
   G$24.3-5.3$ & $18^h54^m52.557^s$ $-10\arcdeg 03\arcmin 07\arcsec$   & 222  & +125 & 1833-210, 1331+305           \\
   G$24.7-5.7$ & $18^h57^m05.347^s$ $-09\arcdeg 52\arcmin 41\arcsec$   &  64  & +125 & 1833-210, 1331+305           \\
   G$25.2+4.5$ & $18^h21^m24.210^s$ $-04\arcdeg 49\arcmin 06\arcsec$   & 172  & +145 & 1812-068, 1833-210,          \\
               &                                                  &      &      & 1331+305                     \\
   G$26.9-6.3$ & $19^h03^m07.240^s$ $-08\arcdeg 17\arcmin 17\arcsec$   & 215  & +122 & 1833-210, 0137+331,          \\
               &                                                  &      &      & 1331+305                     \\
   G$33.4-8.0$ & $19^h20^m52.118^s$ $-03\arcdeg 16\arcmin 43\arcsec$   & 102  & +102 & 1939-100, 1331+305           \\
   G$44.8-7.0$ & $19^h38^m25.318^s$ $+07\arcdeg 17\arcmin 16\arcsec$   & 105  &  +94 & 1941-154, 1331+305           \\
\enddata
\end{deluxetable}

\begin{deluxetable}{l c c c}
\tablecaption{Synthesized beam sizes \label{tab:beams}}
\tablehead{
\colhead{Cloud} & \colhead{FWHM} & \colhead{Gain} & \colhead{$\lbr \mathrm{FWHM} \rbr$} \\
\colhead{} & \colhead{(\arcsec)} & \colhead{(K/Jy)} & (pc) \\}
\startdata
    G$16.0+3.0$        &  $89.5 \times 52.6$ &  129 & 2.7 \\
    G$17.5+2.2$        &  $85.7 \times 56.5$ &  126 & 2.7 \\
    G$19.5-3.6$        &  $84.6 \times 58.0$ &  124 & 2.7 \\
    G$22.8+4.3$        &  $72.9 \times 56.1$ &  149 & 2.4 \\
    G$24.3-5.3$        &  $75.6 \times 56.2$ &  143 & 2.5 \\
    G$24.7-5.7$        &  $77.0 \times 59.8$ &  132 & 2.6 \\
    G$25.2+4.5$        &  $71.2 \times 55.6$ &  154 & 2.4 \\
    G$26.9-6.3$        &  $73.5 \times 57.3$ &  145 & 2.4 \\
    G$33.4-8.0$        &  $73.0 \times 58.3$ &  143 & 2.3 \\
    G$44.8-7.0$        &  $68.5 \times 57.7$ &  154 & 1.9 \\
\enddata
\end{deluxetable}


\begin{deluxetable}{l c c}
\tablecaption{Noise levels at the field center
\label{tab:sigma_0}}
\tablehead{
  \colhead{Cloud} & \colhead{ $\sigma_0$}    &  \colhead{$\sigma(N_{HI})$\tablenotemark{1}} \\
  \colhead{  }      & \colhead{(K)}       & \colhead{(10$^{18}$\ cm$^{-2}$)}   \\}
\startdata
   G$16.0+3.0$        & 0.45    & $3.2$\\
   G$17.5+2.2$        & 0.58    & $4.1$\\
   G$19.5-3.6$        & 0.51    & $3.7$\\
   G$22.8+4.3$        & 0.55    & $3.9$\\
   G$24.3-5.3$        & 0.46    & $3.3$\\
   G$24.7-5.7$        & 0.48    & $3.5$\\
   G$25.2+4.5$        & 0.42    & $3.0$\\
   G$26.9-6.3$        & 0.42    & $3.0$\\
   G$33.4-8.0$        & 0.43    & $3.1$\\
   G$44.8-7.0$        & 0.41    & $3.0$\\
\enddata
\tablenotetext{1}{In assumption of 40 channels of the total width of 25 \kms.}
\end{deluxetable}


\begin{deluxetable}{l c c c c}
\tablecaption{Propagation of Noise Through the Imaging
\label{tab:noise_prop}}
\tablehead{
\colhead{Cloud} & \colhead{$\lbr\sigma_{\mathrm{dirty}}\rbr$} & \colhead{$\lbr\sigma_{\mathrm{dirty}}\rbr$} &
\colhead{ $\lbr\sigma_{\mathrm{clean}}\rbr$ } & \colhead{ $\lbr\sigma_{\mathrm{clean}}\rbr$} \\
 & \colhead{(mJy/beam)} & \colhead{(K)} & \colhead{(mJy/beam)} & \colhead{(K)} \\}
\startdata
G$16.0+3.0$  & 3.3 &  0.42 &  3.6   & 0.46  \\
G$26.9-6.3$  & 2.7 &  0.39 &  2.9   & 0.42  \\
G$33.4-8.0$  & 2.8 &  0.40 &  3.0   & 0.42  \\
G$44.8-7.0$  & 2.6 &  0.41 &  2.7   & 0.42  \\
\enddata
\end{deluxetable}


\begin{deluxetable}{ccccccc}
\tabletypesize{\small}
\tablecaption{Measured Cloud Properties \label{tab:observed-props}}
\tablehead{\colhead{Cloud} & \colhead{R} & \colhead{z} & \colhead{T$_b$} & \colhead{$\VLSR$} & \colhead{FWHM}
&  \colhead{Peak N$_{HI}$}  \\
&\colhead{ (kpc)} & \colhead{(pc)} & \colhead{(K)} & \colhead{(\kms)}& \colhead{(\kms)} &
 \colhead{($10^{20}$ cm$^{-2}$)}  \\
\colhead{(1)} & \colhead{(2)} & \colhead{(3)} & \colhead{(4)} & \colhead{(5)} & \colhead{(6)} & \colhead{(7)} \\
}
\startdata
G$16.0+3.0$ & 2.34 &  $+430$  & $12.3\pm0.3$ & $143.6\pm0.1$ & $4.7\pm0.1$  & 1.2  \\
G$17.5+2.2$ & 2.56 &  $+310$  & $6.2\pm0.3$  & $138.3\pm0.4$ & $18.2\pm1.0$ & 2.1  \\
G$19.5-3.6$ & 2.84 &  $-500$  & $14.6\pm0.4$ & $118.8\pm0.1$ & $9.9\pm0.3$  & 2.9  \\
G$22.8+4.3$ & 3.29 &  $+590$  & $4.9\pm0.4$  & $135.2\pm0.2$ & $5.8\pm0.6$  & 1.7  \\
            &      &          & $3.2\pm0.4$  & $131.7\pm0.8$ & $24.3\pm2.2$ &      \\
G$24.3-5.3$ & 3.50 &  $-720$  & $2.7\pm0.1$  & $125.6\pm0.5$ & $17.4\pm1.1$ & 0.9  \\
G$24.7-5.7$ & 3.55 &  $-770$  & $7.8\pm0.3$  & $128.0\pm0.1$ & $5.5\pm0.2$  & 1.0  \\
G$25.2+4.5$ & 3.62 &  $+610$  & $6.3\pm0.2$  & $144.5\pm0.2$ & $13.4\pm0.5$ & 1.6  \\
G$26.9-6.3$ & 3.85 &  $-840$  & $13.0\pm0.4$ & $120.3\pm0.1$ & $3.0\pm0.1$  & 2.2  \\
            &      &          & $5.2\pm0.3$  & $124.1\pm0.3$ & $14.5\pm0.6$ &      \\
G$33.4-8.0$ & 4.68 & $-1000$  & $3.5\pm0.3$  & $101.9\pm0.2$ & $ 4.3\pm0.5$ & 0.8  \\
            &      &          & $1.7\pm0.3$  & $101.0\pm0.8$ & $ 20.2\pm3.1$&      \\
G$44.8-7.0$ & 5.99 &  $-740$  & $6.8\pm0.4$  & $93.4\pm0.1$  & $6.1\pm0.4$  & 1.7  \\
            &      &          & $2.3\pm0.3$  & $99.4\pm1.1$  & $18.7\pm1.5$ &      \\
\enddata
\end{deluxetable}

\begin{deluxetable}{lccccc}
\tablecaption{Derived Cloud Properties
\label{tab:derived-props}}
\tablehead{
\colhead{Cloud}  &  \colhead{R}  & \colhead{z}   & \colhead{$M_{HI}$\tablenotemark{1}}  & \colhead{Size\tablenotemark{1}\tablenotemark{2}}  & \colhead{$\lbr n \rbr$}  \\
              &\colhead{ (kpc)}     & \colhead{(pc)}  & \colhead{(\Msun)} & \colhead{(pc)}  & \colhead{(cm$^{-3}$)} \\
\colhead{(1)} & \colhead{(2)} & \colhead{(3)} & \colhead{(4)} & \colhead{(5)} & \colhead{(6)} \\
}
\startdata
G$16.0+3.0$ &  $2.34\pm0.20$  & $+430\pm50$   & $45^{+12}_{-10}$     & $16\times10$ (0.12)   & $1.73\pm0.20$  \\
G$17.5+2.2$ &  $2.56\pm0.23$  & $+310\pm40$   & $440^{+130}_{-110}$  & $43 \times 23$ (0.14) & $1.10\pm0.15$ \\
G$19.5-3.6$ &  $2.84\pm0.42$  & $-500\pm100$  & $830^{+370}_{-300}$  & $67 \times 28$ (0.20) & $0.80\pm0.16$ \\
G$22.8+4.3$ &  $3.29\pm0.29$  & $+590\pm100$  & $550^{+210}_{-180}$  & $50 \times 35$ (0.18) & $0.59\pm0.11$ \\
G$24.3-5.3$ &  $3.50\pm0.31$  & $-720\pm140$  & $300^{+130}_{-100}$  & $59 \times 29$ (0.19) & $0.33\pm0.06$ \\
G$24.7-5.7$ &  $3.55\pm0.30$  & $-770\pm150$  & $290^{+120}_{-100}$  & $60 \times 38$ (0.19) & $0.21\pm0.04$ \\
G$25.2+4.5$ &  $3.62\pm0.34$  & $+610\pm130$  & $690^{+320}_{-260}$  & $58 \times 48$ (0.21) & $0.37\pm0.08$ \\
G$26.9-6.3$ &  $3.85\pm0.36$  & $-840\pm190$  & $600^{+300}_{-240}$  & $66 \times 32$ (0.22) & $0.48\pm0.11$ \\
G$33.4-8.0$ &  $4.68\pm0.41$  & $-1000\pm280$ & $160^{+100}_{-80}$   & $38 \times 28$ (0.28) & $0.36\pm0.10$ \\
G$44.8-7.0$ &  $5.99\pm0.50$  & $-740\pm300$  & $840^{+830}_{-550}$  & $62 \times 41$ (0.41) & $0.51\pm0.21$ \\
 \enddata
\tablenotetext{1}{Clouds' sizes and masses determined based on contours shown in the left panels of Figs.~\ref{fig:MassContours1}--\ref{fig:MassContours4}.}
\tablenotetext{2}{Fractional uncertainties in both dimensions are identical and are given by the value in parentheses. }
\end{deluxetable}

\clearpage

\begin{deluxetable}{lccccc}
\tablecaption{Wide-Field Cloud Properties Based on GBT Data
\label{tab:gbtderivedprops}}
\tablehead{
\colhead{Cloud}&\colhead{$M_{HI}$\tablenotemark{1}}&\colhead{Size\tablenotemark{1}\tablenotemark{2}}&\colhead{$\lbr n \rbr$}\\
 &\colhead{(\Msun)}&\colhead{(pc)}&\colhead{(cm$^{-3}$)}\\
\colhead{(1)} & \colhead{(2)} & \colhead{(3)} & \colhead{(4)} \\
}
\startdata
G$16.0+3.0$ & $83^{+21}_{-19}$      & $47\times43$ (0.12)   &  $0.07\pm0.01$ \\
G$17.5+2.2$ & $750^{+220}_{-190}$   & $57 \times 39$ (0.14) &  $0.56\pm0.08$\\
G$19.5-3.6$ & $1920^{+840}_{-690}$  & $100 \times 71$ (0.20)&  $0.25\pm0.05$\\
G$22.8+4.3$ & $1370^{+530}_{-440}$  & $103 \times 70$ (0.18)&  $0.17\pm0.03$\\
G$24-5$\tablenotemark{3} & $1150^{+480}_{-400}$ & $158 \times 59$ (0.19)& $0.10\pm0.02$\\
G$25.2+4.5$ & $1690^{+780}_{-630}$  & $119 \times 75$ (0.21)& $0.16\pm0.03$\\
G$26.9-6.3$ & $1570^{+780}_{-620}$  & $151 \times 85$ (0.22)& $0.08\pm0.02$\\
G$33.4-8.0$ & $540^{+340}_{-260}$   & $115 \times 58$ (0.28)& $0.08\pm0.02$\\
G$44.8-7.0$ & $2560^{+2530}_{-1670}$&$122 \times 85$ (0.41) & $0.19\pm0.08$\\
\enddata
\tablenotetext{1}{Clouds' sizes and masses determined based on contours shown in the right panels of Figs.~\ref{fig:MassContours1}--\ref{fig:MassContours4}.}
\tablenotetext{2}{Fractional uncertainties in both dimensions are identical and are given by the value in parentheses. }
\tablenotetext{3}{Total for the group of three clouds, containing G$24.3-5.3$ and G$24.7-5.7$.}
\end{deluxetable}

\clearpage

\begin{deluxetable}{lcccccc}
\tablecaption{Cloud Core Properties
\label{tab:core-props}}
\tablehead{
\colhead{Cloud}  & \colhead{contour \NHI} & \colhead{$M_{HI}$}  & \colhead{Size\tablenotemark{1}}
                 & \colhead{$\lbr n_c \rbr$} & \colhead{T$_{\mathrm{limit}}$} & \colhead{T$_{\mathrm{limit}} \cdot \lbr n_c \rbr$} \\
                 & \colhead{($10^{20}$ cm$^{-2}$)} & \colhead{(\Msun)}   & \colhead{(pc)}
                 & \colhead{(cm$^{-3}$)} & \colhead{(K)} & \colhead{($10^3$ K cm$^{-3}$)}\\
\colhead{(1)}    & \colhead{(2)}       & \colhead{(3)}   & \colhead{(4)} & \colhead{(5)} & \colhead{(6)} & \colhead{(7)}\\
}
\startdata
G$16.0+3.0$  &0.2&$45^{+12}_{-10}$& $16\times10$ (0.12)& $1.7\pm0.2$  & $480\pm20$    & $0.8\pm0.14$\\
G$17.5+2.2$  &1.5&$66^{+19}_{-17}$& $14\times6 $ (0.14)& $6.7\pm0.9$  & $7200\pm790$  & $48\pm12$\\
G$19.5-3.6$  &2.2&$60^{+26}_{-22}$& $10\times6 $ (0.20)& $10.0\pm2.0$ & $2100\pm130$  & $21\pm6$\\
G$22.8+4.3$  &1.2&$57^{+22}_{-18}$& $11\times9 $ (0.18)& $4.5\pm0.8$  &               & \\
             & &                     &                 & $1.2\pm0.2$  & $730\pm150$   & $0.9\pm0.3$\\
             & &                     &                 & $3.3\pm0.6$  & $13000\pm2300$& $42\pm15$\\
G$23.1+4.3$\tablenotemark{2}
             &0.9&$38^{+15}_{-12}$   & $10 \times 6$ (0.18) & $6.4\pm1.2$ &           & \\
             & &                     &                 & $2.2\pm0.4$  & $590\pm200$   & $1.3\pm0.7$\\
             & &                     &                 & $4.2\pm0.8$  & $7300\pm2200$ & $30\pm14$\\
G$24.3-5.3$  &0.6&$45^{+19}_{-16}$   & $15 \times 14$ (0.19) & $1.2\pm0.2$   & $6600\pm830$ & $7.6\pm2.4$\\
G$24.7-5.7$  &0.5&$53^{+22}_{-18}$   & $21 \times 7$  (0.19) & $2.3\pm0.4$   & $660\pm50$   & $1.5\pm0.4$\\
G$25.2+4.5$  &1.2&$57^{+26}_{-21}$   & $15 \times 7$ (0.21)  & $4.1\pm0.9$   & $3900\pm290$ & $16\pm5$\\
G$26.9-6.3$  &1.3&$54^{+27}_{-21}$   & $10 \times 7$ (0.22)  & $7.2\pm1.6$   &              & \\
             & &                     &                       & $2.5\pm0.5$   & $200\pm13$   & $0.5\pm0.14$\\
             & &                     &                       & $4.7\pm1.1$   & $4600\pm380$ & $22\pm7$\\
G$33.4-8.0$  &0.5&$47^{+30}_{-22}$   & $24 \times 8$ (0.28)  & $1.4\pm0.4$   &              & \\
             & &                     &                       & $0.4\pm0.1$   & $400\pm90$   & $0.2\pm0.1$\\
             & &                     &                       & $1.0\pm0.3$   & $8900\pm2700$& $9\pm5$\\
G$44.8-7.0$  &1.2&$54^{+53}_{-35}$   & $15 \times 7$ (0.41)  & $3.9\pm1.6$   &              & \\
             & &                     &                       & $1.9\pm0.8$   & $810\pm110$  & $1.6\pm0.8$\\
             & &                     &                       & $2.0\pm0.8$   & $7600\pm1200$& $15\pm9$\\
\enddata
\tablenotetext{1}{Fractional uncertainties in both dimensions are identical and are given by the value in parentheses.}
\tablenotetext{2}{A small separate cloud at the high-longitude side of the G$22.8+4.3$ field. First velocity component: T$_b$ = $6.4\pm0.9$~K, $\VLSR$ = $129.3\pm0.3$~\kms, FWHM = $5.2\pm0.9$; second velocity component: T$_b$ = $3.4\pm0.6$~K, $\VLSR$ = $135.1\pm1.7$~\kms, FWHM = $18.3\pm2.7$.}
\end{deluxetable}

\clearpage

\clearpage
\begin{figure}
    \centering
    \captionsetup[subfigure]{labelformat=empty}
    \subfloat[][]{
        \centering
        \hspace{-1cm}
        \includegraphics[width=0.6\textwidth]{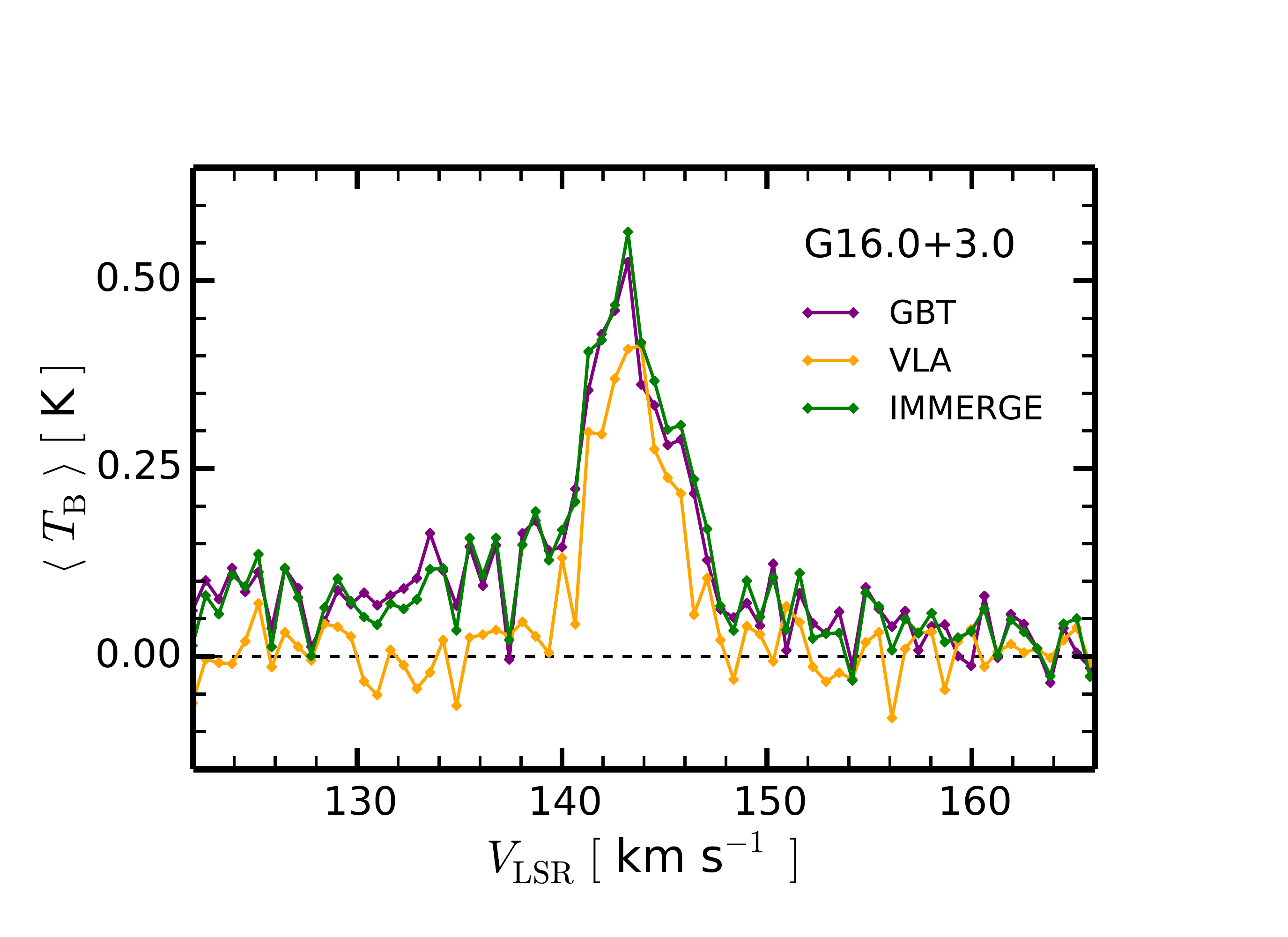}
        \hspace{-1.5cm}}
    \subfloat[][]{
        \centering
        \includegraphics[width=0.6\textwidth]{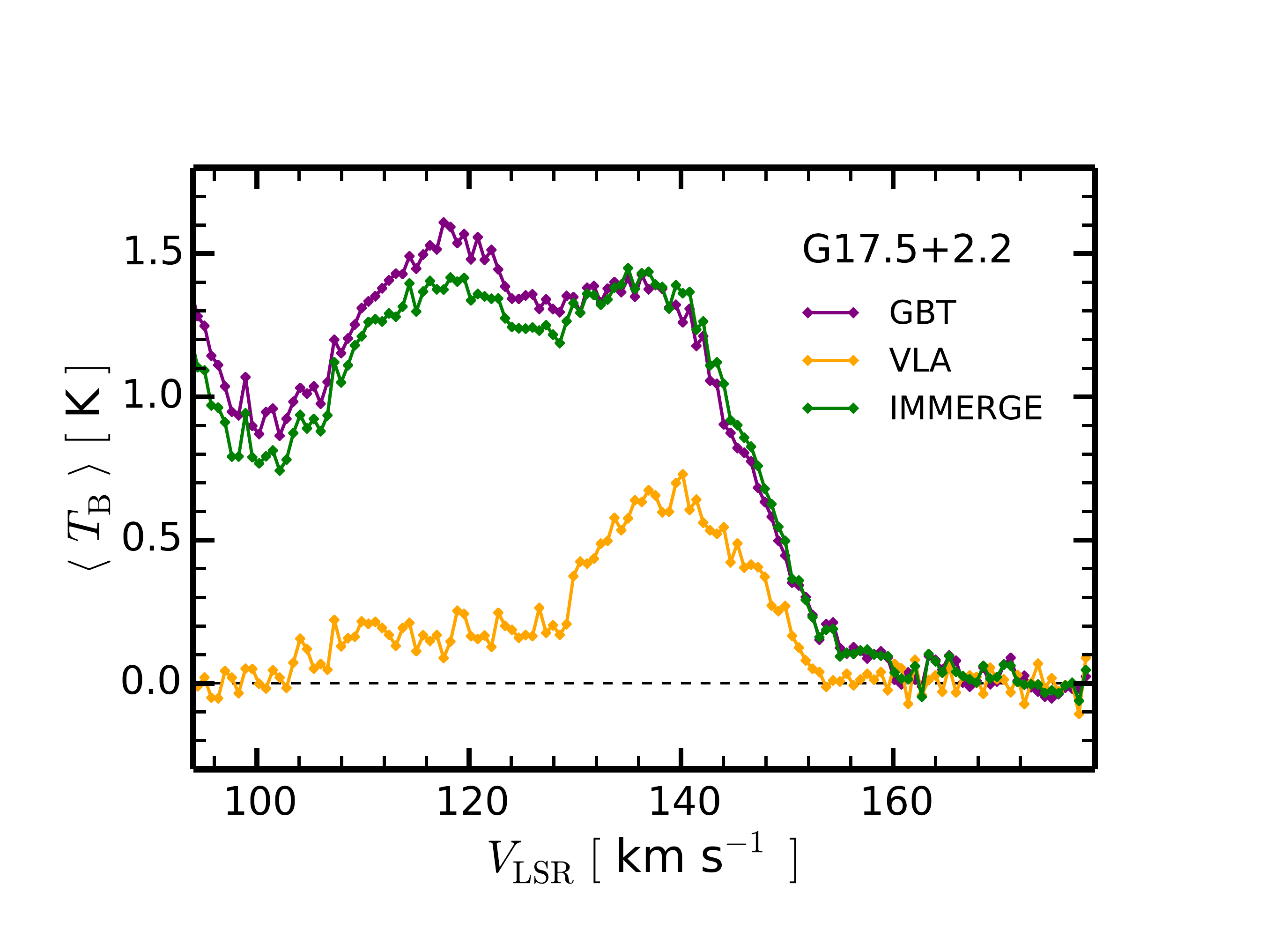}
        }

    \vspace{-2cm}

    \subfloat[][]{
        \centering
        \hspace{-1cm}
        \includegraphics[width=0.6\textwidth]{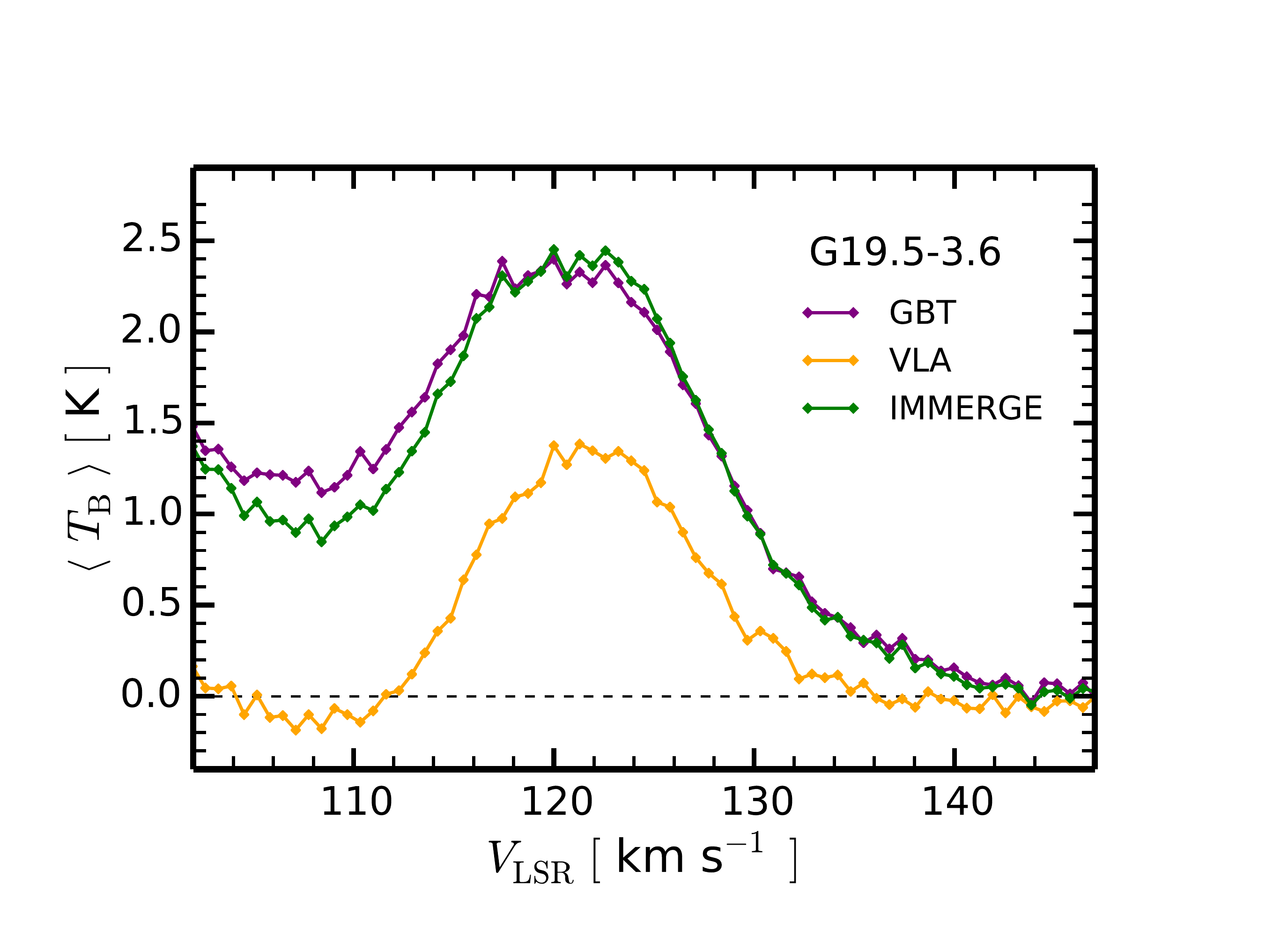}
        \hspace{-1.5cm}}
    \subfloat[][]{
        \centering
        \includegraphics[width=0.6\textwidth]{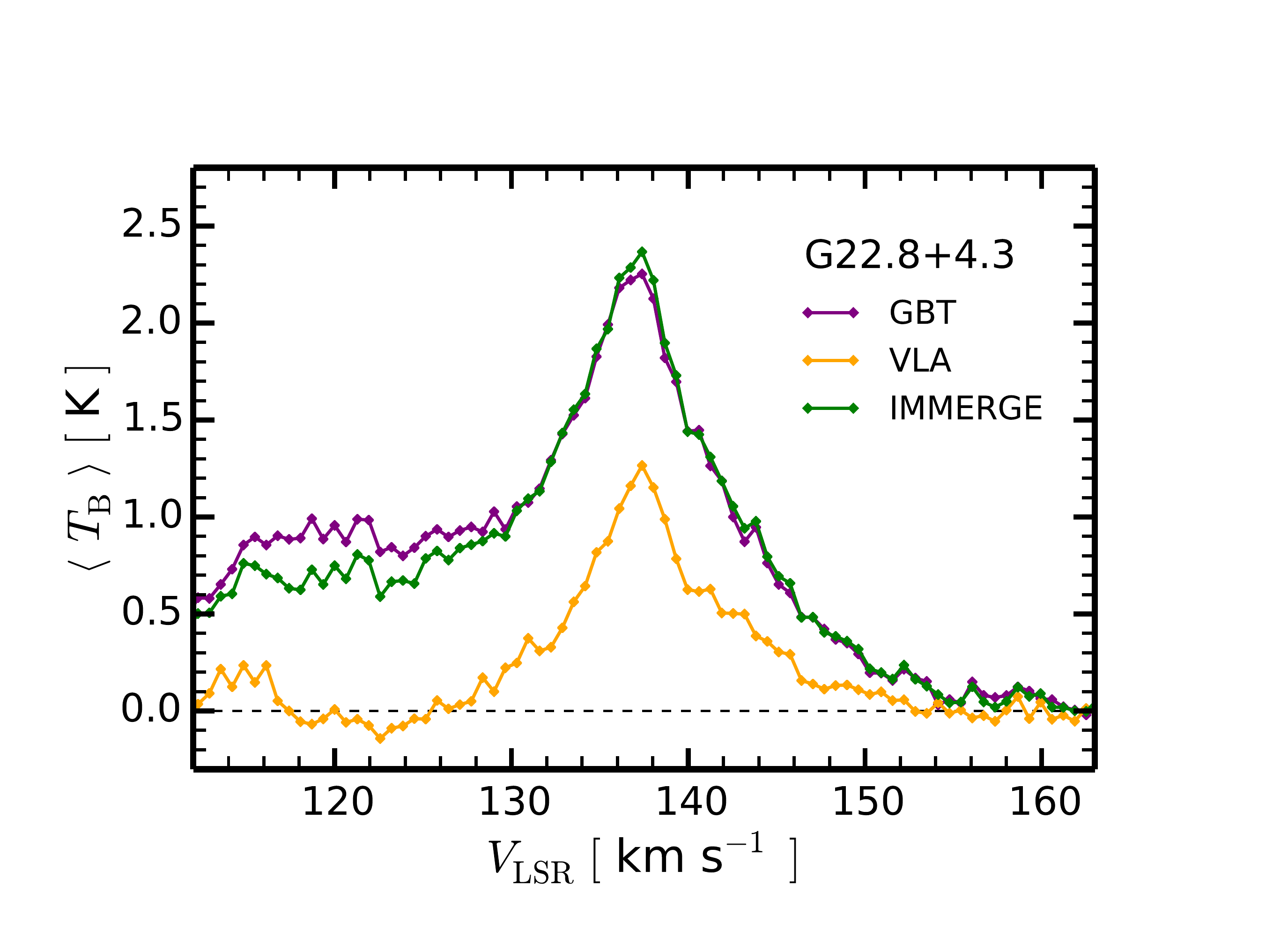}
        }
\caption{Page 1 of 3 --- Mean line profiles in brightness temperature T$_b$, at the center of each field
for  clean VLA data (orange), GBT data (purple) and results of running IMMERGE on
these two images with \fcal=0.87 and no tapering (green). This sum  is
calculated over  all pixels in an 80 by 80 pixel box.
The comparison shows how well the total single-dish flux is recovered
by IMMERGE.
}\label{fig:IMM}
\end{figure}

\clearpage
\begin{figure}
\ContinuedFloat
\centering
    \captionsetup[subfigure]{labelformat=empty}
    \subfloat[][]{
        \centering
        \hspace{-1cm}
        \includegraphics[width=0.6\textwidth]{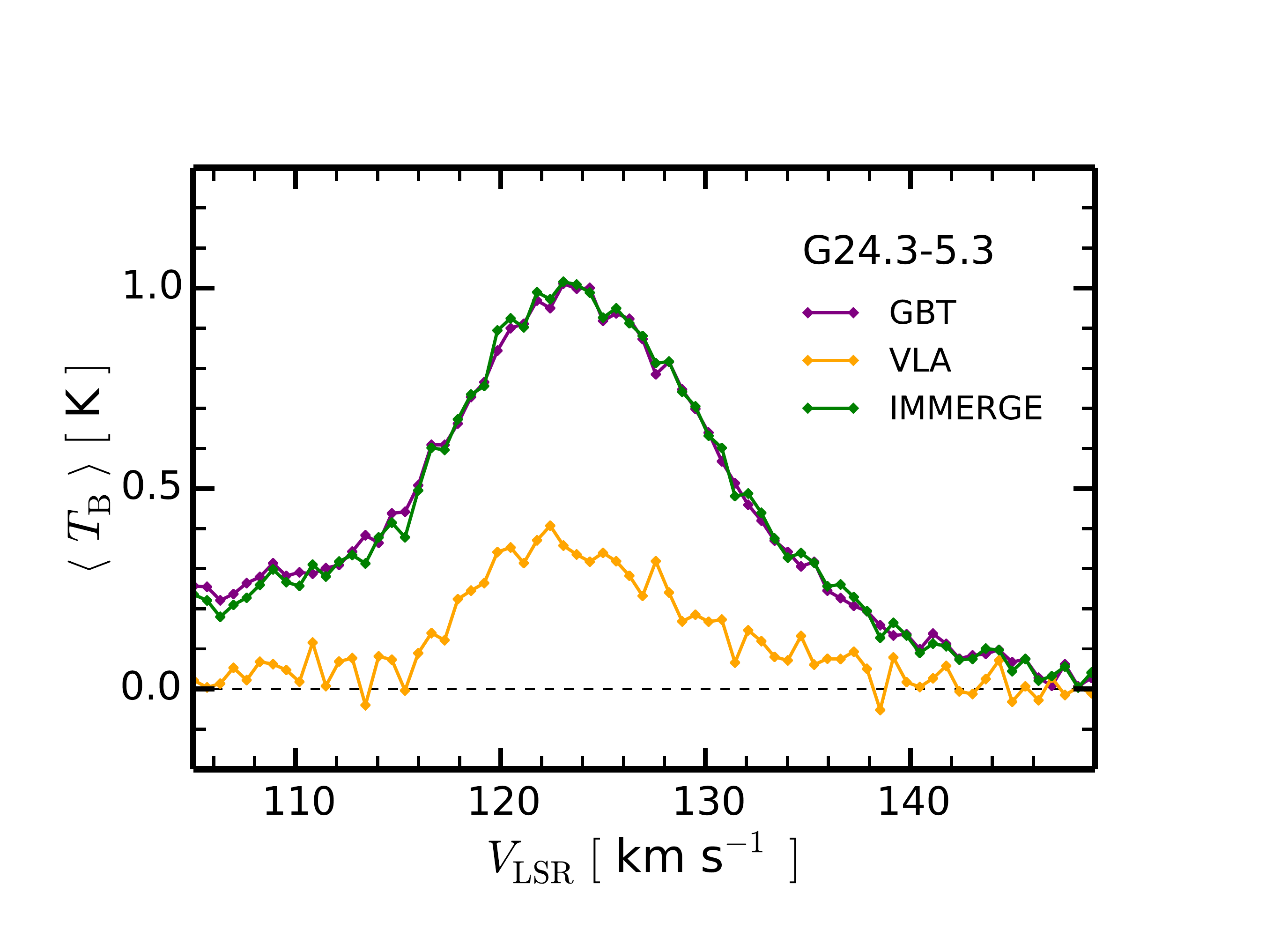}
        \hspace{-1.5cm}}
    \subfloat[][]{
        \centering
        \includegraphics[width=0.6\textwidth]{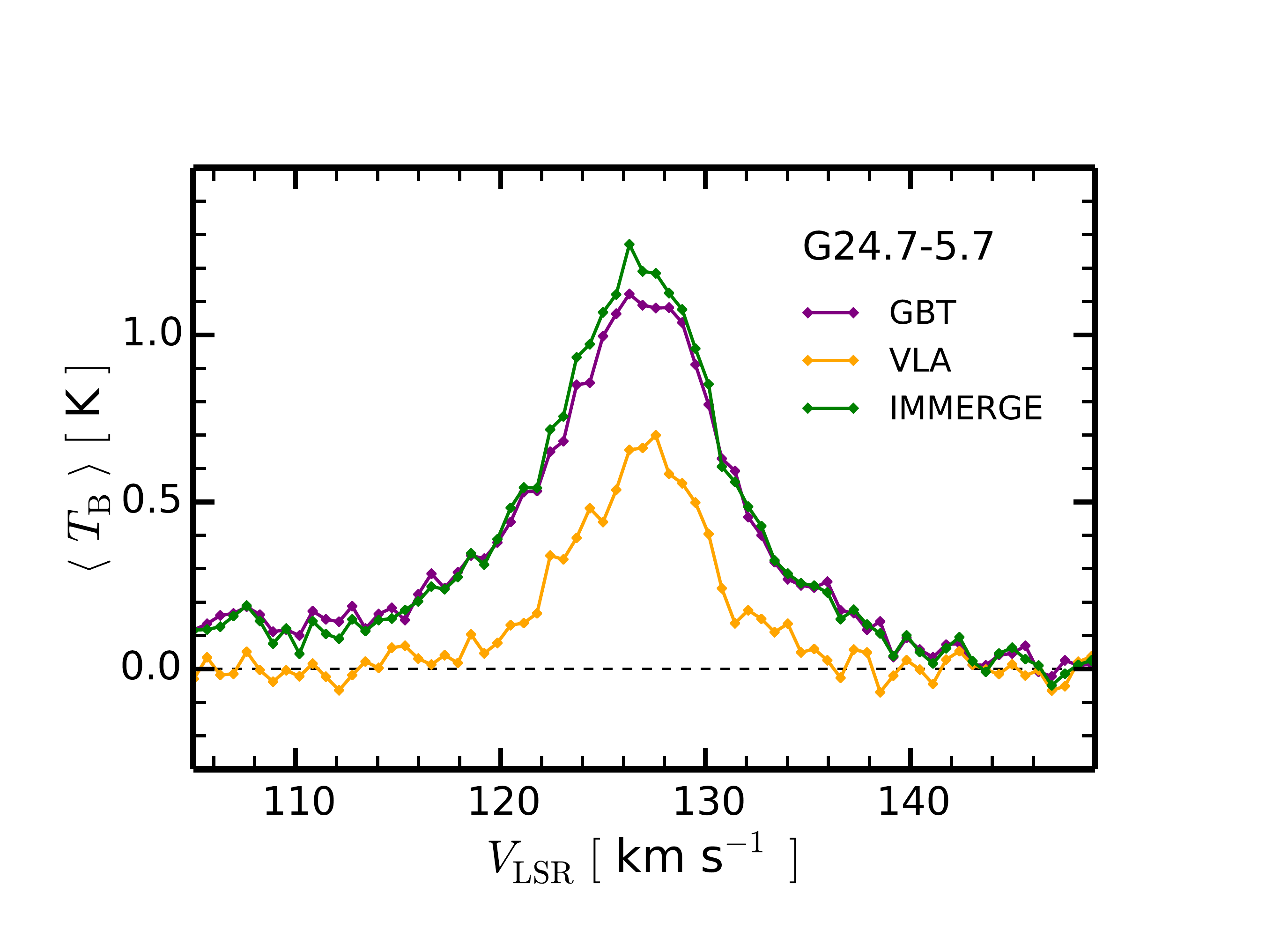}
        }

    \vspace{-2cm}

    \subfloat[][]{
        \centering
        \hspace{-1cm}
        \includegraphics[width=0.6\textwidth]{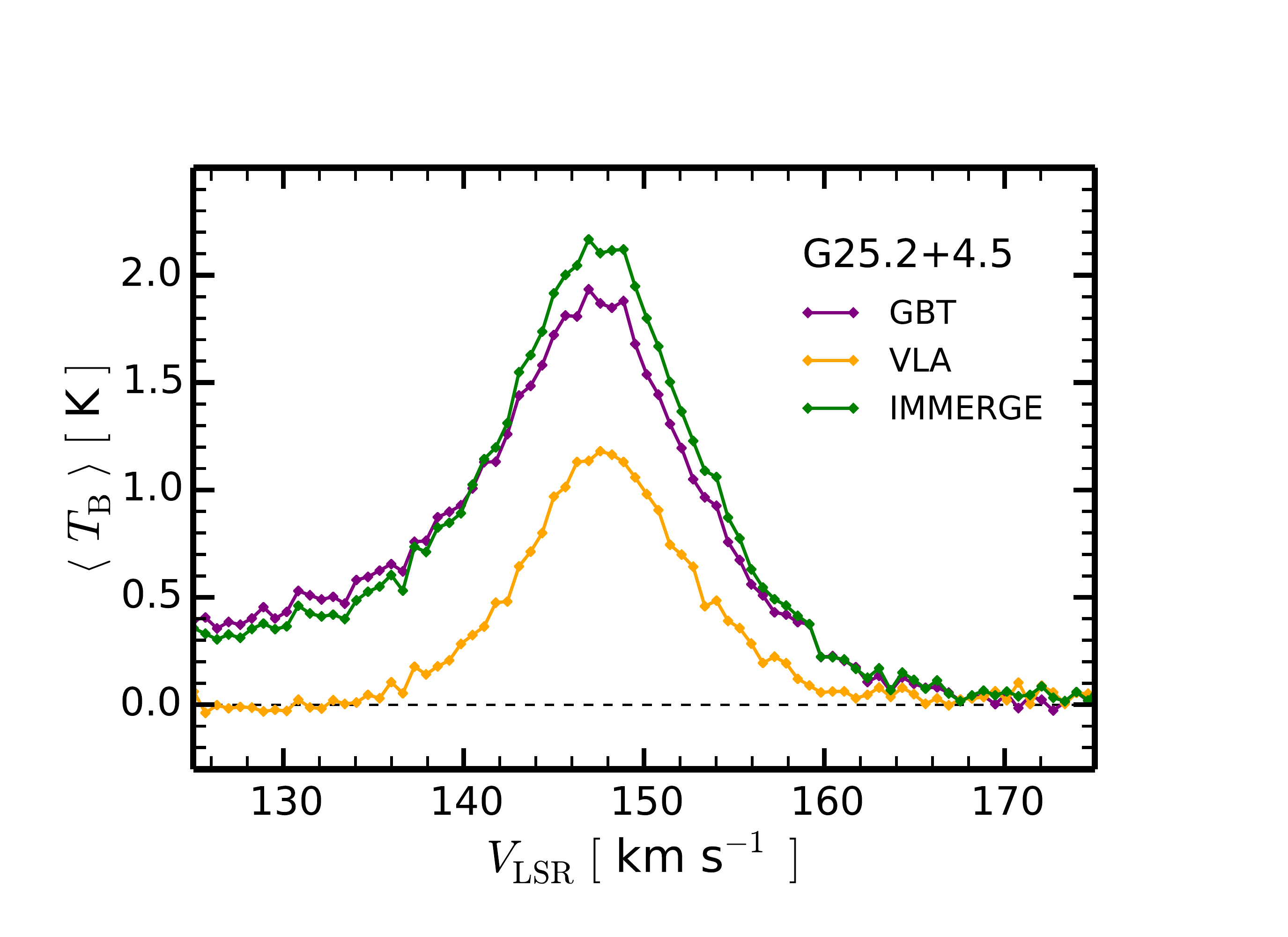}
        \hspace{-1.5
        cm}}
    \subfloat[][]{
        \centering
        \includegraphics[width=0.6\textwidth]{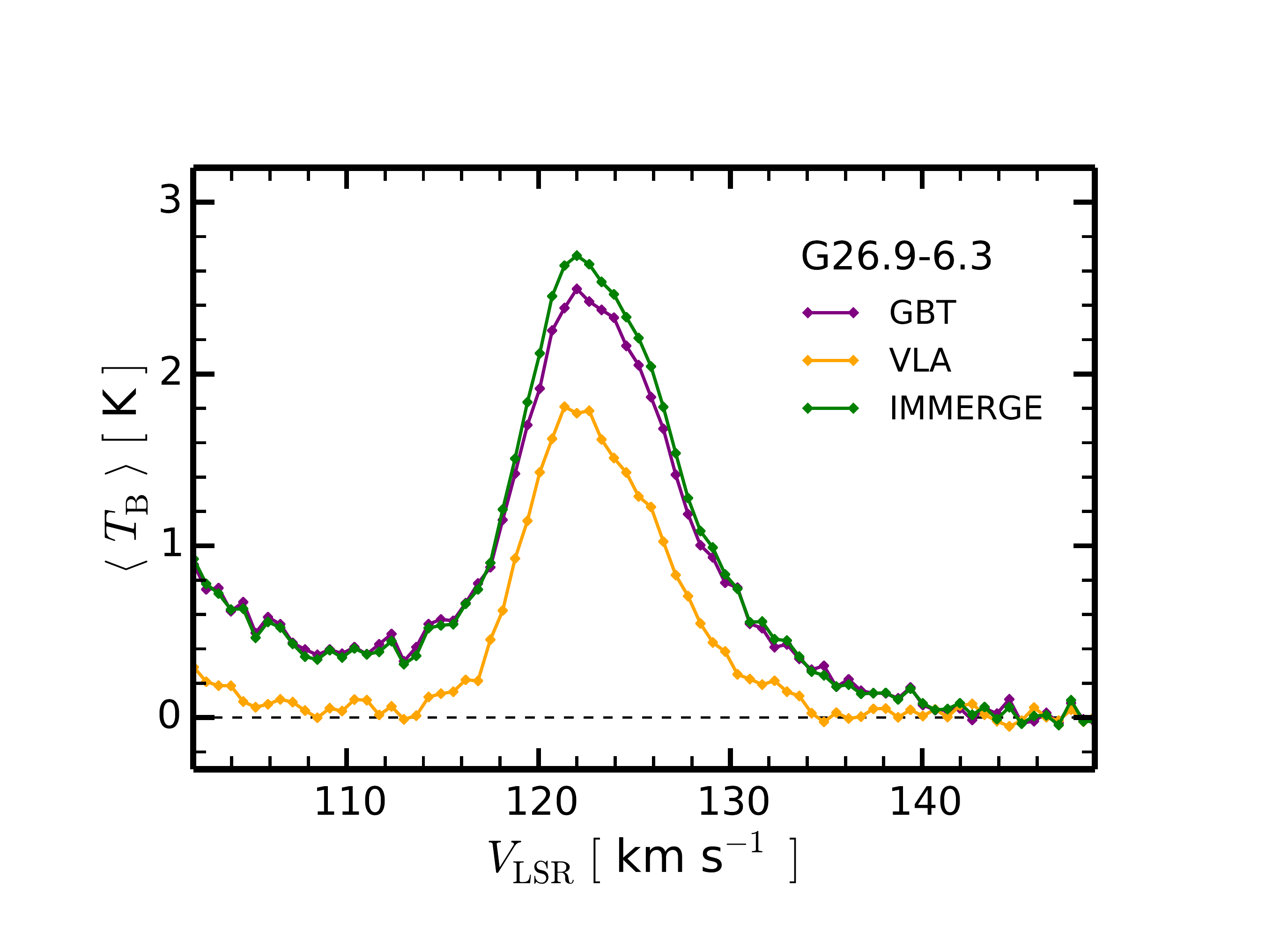}
        }
 \caption{Page 2 of 3 --- Mean line profiles in brightness temperature T$_b$, at the center of each field
for  clean VLA data (orange), GBT data (purple) and results of running IMMERGE on
these two images with \fcal=0.87 and no tapering (green). This sum  is
calculated over  all pixels in an 80 by 80 pixel box.
The comparison shows how well the total single-dish flux is recovered
by IMMERGE.
}\label{fig:IMM}
\end{figure}

\clearpage
\begin{figure}
\ContinuedFloat
\centering
    \captionsetup[subfigure]{labelformat=empty}
    \subfloat[][]{
        \centering
        \includegraphics[width=0.6\textwidth]{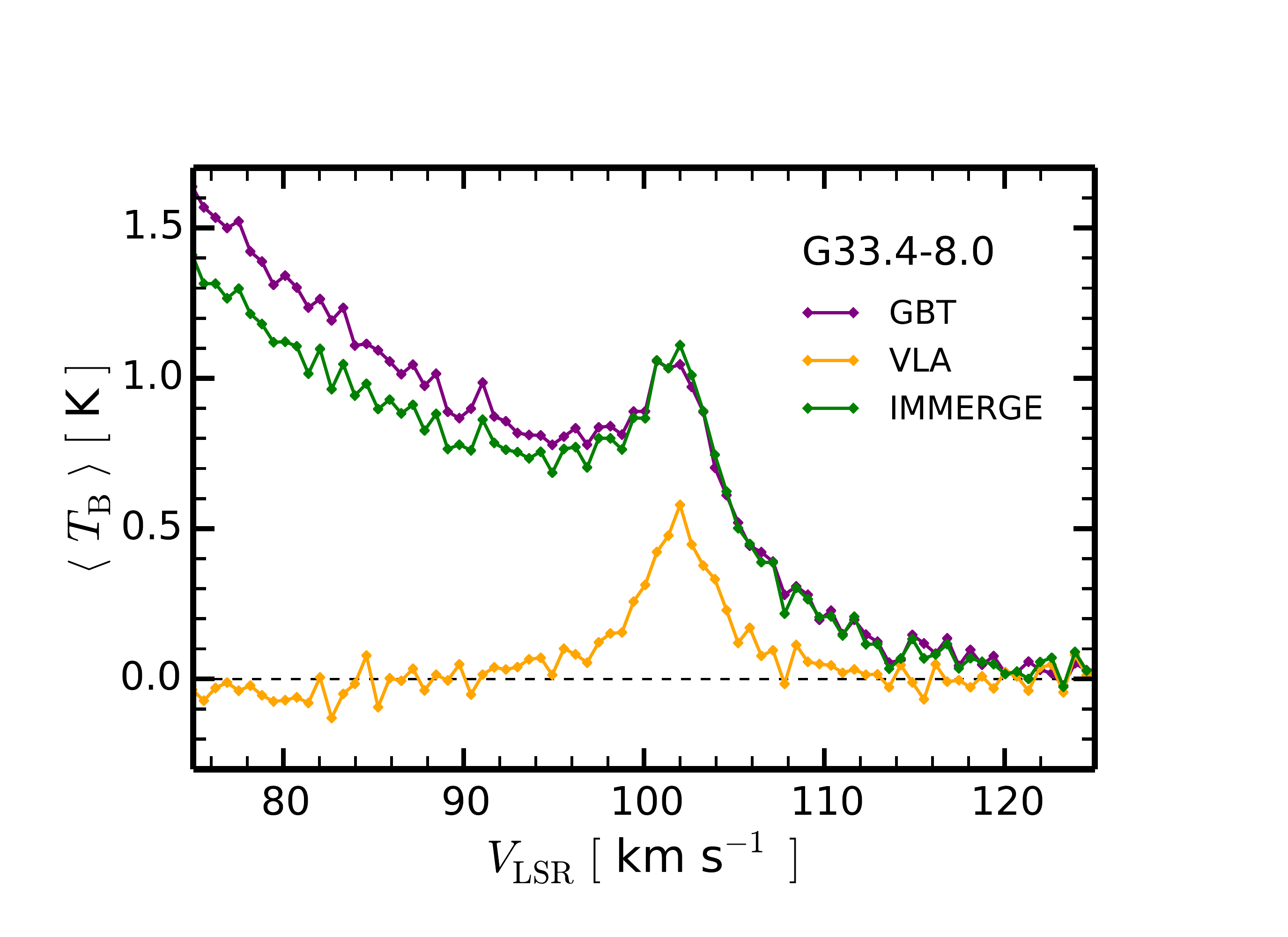}
        }

    \vspace{-2cm}

    \subfloat[][]{
        \centering
        \includegraphics[width=0.6\textwidth]{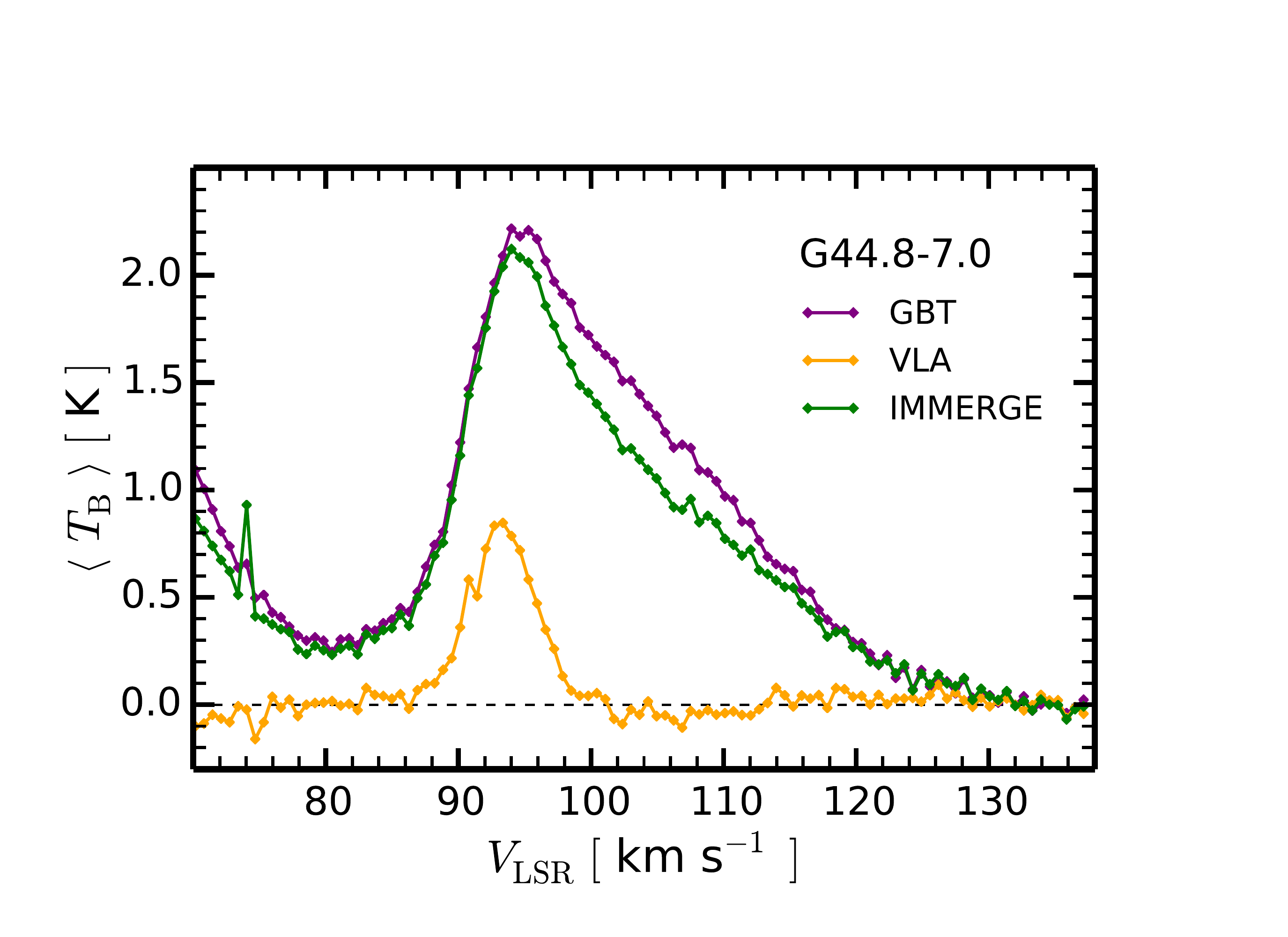}
        }

 \caption{Page 3 of 3 --- Mean line profiles in brightness temperature T$_b$, at the center of each field
for  clean VLA data (orange), GBT data (purple) and results of running IMMERGE on
these two images with \fcal=0.87 and no tapering (green). This sum  is
calculated over  all pixels in an 80 by 80 pixel box.
The comparison shows how well the total single-dish flux is recovered
by IMMERGE.
}\label{fig:IMM}
\end{figure}

\clearpage

\clearpage


\clearpage


\clearpage


\clearpage


\clearpage


\clearpage


\clearpage


\clearpage


\clearpage


\end{document}